\documentstyle[11pt,epsf]{article}

\renewcommand\({\left(}
\renewcommand\){\right)}
\renewcommand\[{\left[}
\renewcommand\]{\right]}

\newcommand\eq[1]{Eq.~(\ref{#1})}
\newcommand\eqs[2]{Eqs.~(\ref{#1}) and (\ref{#2})}
\newcommand\eqss[3]{Eqs.~(\ref{#1}), (\ref{#2}) and (\ref{#3})}
\newcommand\eqsss[4]{Eqs.~(\ref{#1}), (\ref{#2}), (\ref{#3})
and (\ref{#4})}

\newcommand\pa{\partial}

\newcommand\ee{\end{equation}}
\newcommand\be{\begin{equation}}
\newcommand\eea{\end{eqnarray}}
\newcommand\bea{\begin{eqnarray}}

\newcommand\sunit{\,\mbox{s}}

\newcommand\km{\,\mbox{km}}

\newcommand\TeV{\,\mbox{TeV}}
\newcommand\GeV{\,\mbox{GeV}}
\newcommand\MeV{\,\mbox{MeV}}
\newcommand\keV{\,\mbox{keV}}
\newcommand\eV{\,\mbox{eV}}
\newcommand\Mpc{\,\mbox{Mpc}}

\newcommand\mpl{M_{\rm P}}
\newcommand\MPl{M_{\rm P}}
\newcommand\Mpl{M_{\rm P}}
\newcommand\mpltil{\Lambda\sub{UV}}

\newcommand\lsim{\mathrel{\rlap{\lower4pt\hbox{\hskip1pt$\sim$}}
    \raise1pt\hbox{$<$}}}
\newcommand\gsim{\mathrel{\rlap{\lower4pt\hbox{\hskip1pt$\sim$}}
    \raise1pt\hbox{$>$}}}

\def\dslash{\not{\hbox{\kern-2pt $\partial$}}}
\def\Dslash{\not{\hbox{\kern-4pt $D$}}}
\def\Oslash{\not{\hbox{\kern-4pt $O$}}}
\def\Qslash{\not{\hbox{\kern-4pt $Q$}}}
\def\pslash{\not{\hbox{\kern-2.3pt $p$}}}
\def\kslash{\not{\hbox{\kern-2.3pt $k$}}}
\def\qslash{\not{\hbox{\kern-2.3pt $q$}}}
 \newtoks\slashfraction
 \slashfraction={.13}
 \def\slash#1{\setbox0\hbox{$ #1 $}
 \setbox0\hbox to \the\slashfraction\wd0{\hss \box0}/\box0 }
 
\def\eeq{\end{equation}}
\def\beq{\begin{equation}}
\def\smallfrac#1#2{\hbox{${\scriptstyle#1} \over {\scriptstyle#2}$}}

\def\half{{\scriptstyle{1\over 2}}}

\def\calp{{\cal P}}
\def\calr{{\cal R}}

\newcommand\bfk{{\bf k}}

\newcommand\bfr{{\bf r}}

\newcommand\bfx{{\bf x}}

\newcommand\sub[1]{_{\rm #1}}

\newcommand\Tr{{\rm Tr}\,}
\newcommand\minf{M\sub{inf}}

\begin{document}

\begin{titlepage}
\begin{flushright}
LANCS-TH/9720, FERMILAB-PUB-97/292-A, CERN-TH/97-383,
OUTP-98-39-P
\\hep-ph/9807278\\
(After final proof-reading, March 1999)
\end{flushright}
\begin{center}
{\Large \bf 
Particle Physics Models of Inflation and\\ the 
Cosmological Density Perturbation\\}
\vspace{.3in}
{\large\bf  David H.~Lyth$^{\dagger}$ and Antonio 
Riotto $^{*,}$\footnote{On leave of absence {}from Theoretical 
Physics Department, University of Oxford,U.K.}
\\}
\vspace{.4 cm}
{\em $^{\dagger}$Department of Physics,\\
Lancaster University,\\
Lancaster LA1 4YB.~~~U.~K.\\
{\tt E-mail: d.lyth@lancaster.ac.uk}}\\
\vspace{.4 cm}
{\em $^*$  CERN, Theory Division,\\
CH-1211, Geneva 23, Switzerland.\\
{\tt E-mail: riotto@nxth04.cern.ch}}
\end{center}

\vspace{.6cm}
\begin{abstract}
\noindent
This is a review of particle-theory models of inflation, 
and of their predictions for
the primordial density perturbation that is thought to be the origin 
of structure in the Universe. It contains mini-reviews of 
the relevant observational cosmology, of elementary field theory
and of supersymmetry, that may be of interest in their own right.
The spectral index $n(k)$, specifying the 
scale-dependence of the spectrum of the curvature perturbation,
will be a powerful discriminator between models, when it is 
measured by Planck with accuracy $\Delta n\sim 0.01$. 
The usual formula for $n$ is derived, as well as its less familiar 
extension to the case of a multi-component inflaton; in both cases
the key ingredient is the separate evolution of 
causally disconnected regions of the Universe.
Primordial gravitational waves
will be an even more powerful discriminator if they are observed, since 
most models of inflation predict that they are completely negligible.
We treat in detail the new wave of models, which are 
firmly rooted in modern particle theory and have supersymmetry as a 
crucial ingredient. The review is addressed to both astrophysicists
and particle physicists, and each
section is fairly homogeneous regarding the assumed 
background knowledge.

\end{abstract}

\vspace{.6cm}

\centerline{ To appear in {\it Physics Reports}}

%%%%%%%%%%%%%%%%%%%%%%%%%%%%%%%%%%%%%%%%%%%%%%%%%%%%%%%%%%%%%%%%%%%%%%
\end{titlepage}
%%%%%%%%%%%%%%%%%%%%%%%%%%%%%%%%%%%%%%%%%%%%%%%%%%%%%%%%%%%%%%%%%%%%%%
%%%%%%%%%%%%%%%%%%%%%%%%%%%%%%%%%%%%%%%%%%%%%%%%%%%%%%%%%%%%%%%%%%%%%%
                                                                         
\tableofcontents
\newpage
\setcounter{page}{2}
\section{Introduction}

\label{s1}

We do not know the history of the observable Universe before the epoch
of nucleosynthesis, but it is widely believed that there was an
early era of cosmological inflation
\cite{abook,kt,LL2,LL3}. During this era, the Universe was filled
with a homogeneous scalar field $\phi$, called the inflaton field, 
and essentially nothing else. The potential $V(\phi)$ dominated the 
energy density of the Universe, decreasing slowly with time
as $\phi$ rolled slowly down the slope of $V$.

The attraction of this paradigm is that it can 
set the initial conditions for the subsequent hot big bang, which 
otherwise have to be imposed by hand. One of these is 
that there be no unwanted relics (particles or topological defects
which survive to the present and contradict observation).
Another is that the initial density parameter should have the value 
$\Omega=1$ to very high accuracy, to ensure that its present value
has at least roughly this value. There is also the requirement
that the Universe be homogeneous and isotropic to high accuracy.

All of these virtues of inflation were noted when it was first proposed
by Guth in 1981 \cite{guth},\footnote
{Guth's paper gave inflation its name, and for the first time spelled
out its virtues in setting initial conditions. 
Earlier authors had contemplated the possibility of inflation,
as reviewed comprehensively in Reference \cite{olive}.}
and very soon a more dramatic one was also noticed
\cite{hawking,starob82,guthpi}.
Starting with
a Universe which is absolutely homogeneous and isotropic at the classical
level, the inflationary expansion of the Universe will
`freeze in' the vacuum fluctuation of the inflaton field so that 
it becomes an essentially classical quantity. 
On each comoving scale, this happens soon after horizon exit.\footnote
{A comoving scale $a/k$ is said to leave the horizon when $k=aH$,
where $a(t)$ is the scale factor of the Universe and $H=\dot a/a$ is the 
Hubble parameter.}
Associated with this vacuum fluctuation is
a primordial energy density perturbation, which survives after inflation and 
may be the origin of all structure in the Universe. In particular,
it may be responsible for the observed
cosmic microwave background (cmb) anisotropy and for the large-scale
distribution of galaxies and dark matter. 
Inflation also generates
primordial gravitational waves as a vacuum fluctuation, which may contribute
to the low multipoles of the cmb anisotropy.

When it was first proposed in 1982, this 
remarkable paradigm received comparatively little attention.
For one thing observational tests were weak, and for another
the inflationary density perturbation 
was not the only candidate for the origin of structure.
In particular, it seemed as if 
cosmic strings or other topological defects might do the job instead.
This situation changed dramatically in 1992, when COBE measured
the cmb anisotropy on large angular scales \cite{cobe1}, and 
another dramatic change is now in progress with
the advent of smaller scale measurement. Subject to confirmation of the
latter, it seems that the paradigm of slow-roll inflation 
is the only one not in conflict with 
observation. 

The inflaton field perturbation, except in contrived
models, has practically zero mass and negligible interaction. As a 
result, the 
primordial density perturbation is 
gaussian; in other words, its fourier components $\delta_\bfk$
are uncorrelated and have random phases.
Its spectrum $\calp_\calr(k)$, defined roughly 
as the expectation value of
$|\delta_\bfk|^2$ at the epoch of horizon exit, 
defines all of its stochastic properties.\footnote
{To be precise, $\calp_\calr$ is the spectrum of a quantity 
$\calr$ to be defined later, which is a measure of
the spatial curvature seen by comoving observers.}
The shape of the spectrum is conveniently defined by 
the spectral index $n(k)$, defined as
\be
n(k) -1\equiv d\ln\calp_\calr/d\ln k \,,
\label{nofpr}
\ee
Slow-roll inflation predicts a slowly-varying spectrum, corresponding
to $|n-1|$ significantly below 1.
In some models of inflation, $n(k)$ is practically constant on cosmological 
scales, leading to the alternative definition
\bea
\calp_\calr (k) &\propto& k^{n-1}, \\
\calp\sub{grav}(k)  &\propto& k^{n\sub{grav}}.
\eea
The gravitational wave amplitude is also predicted to 
be gaussian, again with a primordial spectrum $\calp\sub{grav}(k)$
which is slowly varying.

The spectra $\calp_\calr(k)$ and $\calp\sub{grav}(k)$
provide the contact between theory and observation. 
The latter is negligible except 
in a very special class of inflationary models, and we shall learn a lot
if it turns out to be detectable. For the moment, observation gives only the
magnitude of $\calp_\calr(k)$ at the  scale $k^{-1}\sim 10^3\Mpc$
(the COBE normalization $\frac25\calp_\calr^{1/2}=
1.91\times 10^{-5}$) plus a 
bound on its scale dependence corresponding to 
$n=1.0\pm 0.2$. 

The observational constraint $\calp_\calr^{1/2}\sim 10^{-5}$
was already known 
when inflation was proposed, and was soon seen to rule out
an otherwise viable model.
Since then, practically all models have been constructed 
with the constraint in mind, so that its power has not always been 
recognized; the huge class of models which it rules out have simply 
never been exhibited. 

The situation regarding the spectral index is quite different.
The present result $n(k)=1.0\pm 0.2$
is
only mildly constraining for inflationary 
models, its most notable consequence being to rule out
`extended' inflation in all except very contrived versions.
But this situation is going to improve 
in the forseeable 
future, and after Planck flies 
in about ten years we shall probably know $n(k)$
to an accuracy $\Delta n\sim 0.01$. As this article demonstrates,
such an accurate number will consign 
to the rubbish bin of history
most of the proposed models of inflation.

What do we mean by a model of inflation? 
Before addressing the question we should be very clear about one thing.
Observation, notably the COBE measurement of the cmb anisotropy,
tells us that {\em when our Universe leaves the horizon}\/\footnote
{A comoving scale $a/k$ is said to be outside the horizon when
it is bigger than $H^{-1}$. Each scale of interest leaves the horizon at 
some epoch during inflation and enters it afterwards. 
The density perturbation on a given scale is essentially generated 
when it leaves the horizon. The comoving scale corresponding to the
whole observable Universe (`our Universe')
is entering the horizon at roughly the present epoch.}
the potential $V(\phi)$  is far below the Planck scale.
To be precise,
$V^{1/4}$ is no more than a few times $10^{16}\GeV$, and it may be many 
orders of magnitude smaller. 
Subsequently, 
there are at most
60 $e$-folds of inflation, and only these
have a directly observable effect. On the other hand, the history of our
Universe begins with $V$ presumably at the Planck scale, and to avoid fine
tuning
inflation should also begin then. This `primary inflation', which may or 
may not join smoothly to the 
the last 60 $e$-folds, cannot be investigated by observation and is of 
comparatively little interest. It will not be treated in this review.
So for us, a `model of inflation' is a model of inflation that applies
after the observable Universe leaves the horizon. It is a model of
`observable', as opposed to `primary', inflation

So what {\em is} meant by a `model of inflation'? The phrase is 
actually used
by the community in two rather different ways. 
At the simplest level a `model of inflation' is taken to mean
a form for the potential, as a function of the fields giving a 
significant contribution to it.
In single-field models there is just the inflaton field 
$\phi$ (defined as the 
one which is varying with time)
whereas in hybrid inflation models most of the potential comes
{}from a second field $\psi$ which is fixed until the end of inflation.
In both cases, one ends up by knowing $V(\phi)$, and the field value
$\phi\sub{end}$ at the end of inflation. This
allows
one to calculate the spectrum $\calp_\calr(k)$, and in particular the
 spectral index $n(k)$. In some cases the prediction for $n(k)$ depends only
on the shape of $V$, as is illustrated in the  table 
on page \pageref{t:1}.
One can also calculate the spectrum $\calp\sub{grav}$
of gravitational waves,
but in most models they are too small to ever be detectable.

At a deeper level, one thinks of a `model of inflation' as something 
analogous to the Standard Model of particle interactions.
One imagines that Nature has chosen some extension of the 
Standard Model, and that the relevant scalar fields are part of 
that model. In this sense a `model of inflation' is
more than merely a specification of the
the potential of the relevant fields. It will provide answers
to at least some of the following questions.
Have the relevant fields and interactions already 
been invoked in some other context, and if so are the parameters
required for inflation compatible with what is already known?
Do the relevant fields have
gauge interactions? If so, are we dealing with
the Standard Model interactions, GUT interactions, or
interactions in a hidden sector?
Is the potential the
classical one, or are quantum effects important?
In the latter case, are 
we dealing with perturbative or non-perturbative effects?

Of course, it would have been wonderful if inflation already dropped
out of the Standard Model, but sadly that is not the case. 
Perhaps more significantly, it is not the case either 
for minimal supersymmetric extensions
of the Standard Model.\footnote
{By `minimal' we mean in this context extensions invoking only
the supersymmetric partners of fermions and gauge bosons,
with a reasonably simple supersymmetric 
extension of the Higgs sector. The simplest possible extension is called
the Minimal Supersymmetric Standard Model (MSSSM).}

Taken in either sense, inflation model building
has seen a recent renaissance. In this 
article, we review the present status of the subject,
taking seriously present thinking about what is likely to lie beyond the
Standard Model. In particular, we take seriously the idea that 
supersymmetry (susy) is relevant. At the fundamental level, susy is 
supposed to be local, corresponding to 
supergravity. When considering particle interactions in the vacuum,
in particular predictions for what is seen at colliders and 
underground detectors, global susy usually provides a good 
approximation to supergravity. But, as we shall discuss in detail,
global susy is {\em not} in general a valid approximation during 
inflation. This remains true no matter how low the energy scale,
and no matter how small the field values, a fact ignored over the years
by many authors.

Being only a symmetry, supersymmetry does not completely define the form of the 
field theory. In fact, in a supergravity theory the number of couplings
that need to be specified is in principle infinite (a non-renormalizable
theory).
For guidance about the form of field theory, 
one may look  to string theory. Taking it to denote the whole class of 
theories that give field theory as an approximation, string theory
comes in many versions, but the two most widely studied are
weakly coupled heterotic string theory 
\cite{bookstring,iban,font2,dkl,flt,dkl2,ant,iban2,kobayashi}
and
Horava-Witten M-theory \cite{mtheory,horwit,noy,low}.
In our present state of knowledge, this gives reasonably detailed 
information in the regime where all field values are $\ll\mpl$,
but almost nothing about the 
regime where some field value is $\gg\mpl$.\label{allfield}\footnote
{$\mpl$ is the Planck mass, defined in units $\hbar=c=1$
by $\mpl=(8\pi G)^{-1/2}=2.4\times 10^{18}\GeV$.}
Accordingly, `models' of inflation in the sense of forms for the potential
can at present be 
be promoted to particle-physics models only in this regime.

\begin{figure}
\centering
\leavevmode\epsfysize=5.3cm \epsfbox{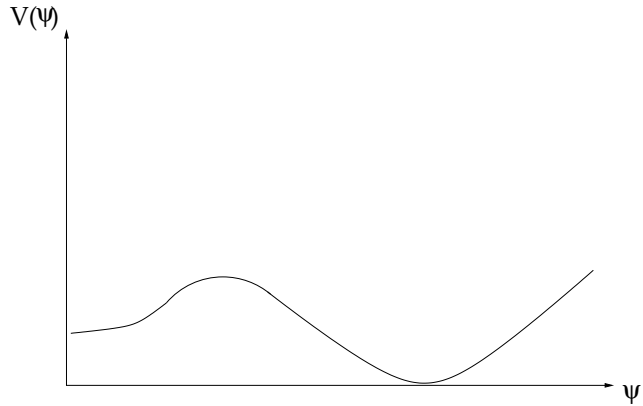}\\
\caption[vvac]{The Old Inflation potential}
\label{f:vvac} 
\end{figure}

Let us briefly review the history.
In Guth's model of 1981 \cite{guth}, some field that we shall call $\psi
$ is trapped at the origin, in a local minimum of the 
potential as illustrated in Figure \ref{f:vvac}. Inflation ends when it 
tunnels through the barrier, and descends quickly to the minimum of the 
potential which represents the vacuum. 
It was soon noted that 
this `old inflation' is not viable because bubbles of the new phase 
never coalesce. 

In 1982, Linde \cite{new1} and Albrecht and Steinhardt
\cite{new2} proposed the first viable model of inflation, which 
has been the archetype for all subsequent models. 
Some field $\phi$, which we shall call the inflaton, is
slowly rolling down a rather flat potential $V(\phi)$. In the 
`new inflation'
model proposed in the above references, the potential has a maximum
at the origin as in the full line of Figure \ref{f:double}, and
inflation takes place near the maximum. It ends when $\phi$
starts to oscillate around the minimum, which again represents the 
vacuum.

The `new inflation' model was a model
 in both senses of 
the word, specifying both the form of the potential and its possible
origin in a Grand Unified
Theory (GUT) theory of particle physics. 
The vacuum expectation value (vev) of the
inflaton, was originally taken to be 
at the GUT scale.  Later it was raised to the Planck scale
(`primordial inflation) which weakened the connection with the GUT. 
Most of the models proposed in this first
phase of particle-theory model building
were very complicated, and are not usually mentioned nowadays.
They were complicated because they worked under two restrictions, which 
have since been abondoned. First, 
the inflaton was required start out
in thermal equilibrium (though Linde pointed out at an early stage that 
this is not mandatory \cite{nontherm}). 
Secondly, they worked almost exclusively with the paradigm of 
single-field inflation, as opposed to the hybrid inflation paradigm that
we shall encounter in a moment.

\begin{figure}
\centering
\leavevmode\epsfysize=5.3cm \epsfbox{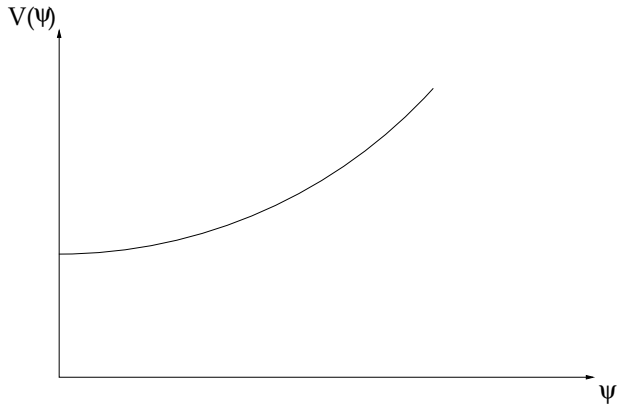}\\
\caption[vinf]{\label{f:vinf} During hybrid inflation, the potential $V(\psi)$
is minimized at $\psi=0$.}
\end{figure}

While this phase of complicated model-building was getting 
under way, Linde proposed \cite{chaotic}
in 1983 that instead the field
might be rolling towards the origin, with a field value
much bigger than $\mpl$. He proposed
a monomial
potential, say $V\propto \phi^2$ or $\phi^4$, which was 
supposed to hold right back to the Planck epoch when $V\sim\mpl^4$. At 
that epoch, $\phi$ was supposed to be a chaotically varying 
function of position.\footnote
{For this reason, inflation 
with such a  potential is usually called `chaotic inflation'.
We shall use the phrase `monomial inflation', because
the hypothesis of chaotic initial conditions has no necessary
connection with the form of the potential during observable inflation.
A wide variety of 
other monotonically increasing functions will also inflate at $\phi\gg\mpl$,
but they are seldom considered because there is too much freedom.}
Working out the field dynamics, one finds that the
observable Universe leaves the horizon when $\phi\sim 10\mpl$, and inflation 
ends when $\phi\sim \mpl$. Such big field values make it practically 
impossible to make a connection with particle physics. After some years,
the monomial paradigm became the favoured one,
and the search for a connection 
with particle physics was largely abandoned.

The seeds for the present renaissance of model-building
were laid around 1990. First, in 1989,
La and Steinhardt proposed what they called `extended inflation'
\cite{extended}.
Its objective was to implement `old' inflation by providing a mechanism 
for making the bubbles coalesce at the end of inflation.
This mechanism was simply to add a slowly-rolling inflaton field
$\phi$ to the original new inflation model, which
makes the Hubble parameter decrease significantly with time.
It invoked the extension of gravity known as Brans-Dicke theory,
and for this reason it was called extended inflation.
The original version conflicted with present-day tests of General 
Relativity, but more complicated versions were soon constructed
that avoided this problem. This paradigm was 
practically killed  in 1992, by the COBE  detection \cite{cobe1} 
the cmb anisotropy. There was no sign of the bubbles
formed at the end of inflation, yet all except very contrived 
versions of the paradigm required that there should be.

Going back to the historical development, 
it was known that like many 
extended gravity theories, extended inflation 
can be re-formulated as an
Einstein gravity theory.
Working from the beginning with Einstein gravity, 
Linde \cite{l90} and Adams and Freese 
\cite{AdamsFreese} proposed in 1991 a crucial change in the idea behind 
extended inflation;
until the end of inflation, tunneling is completely 
impossible (not just relatively unlikely) because 
the trapped field $\psi$ has a coupling to the slowly-rolling inflaton
 field $\phi$. During inflation, this
changes the potential of the trapped field so that it becomes like
the 
one shown in Figure \ref{f:vinf}.
Only at the end of inflation is the final form of Figure \ref{f:vvac}
achieved, and only then does tunnelling take place. In this model,
the bubbles can coalesce very quickly, and be completely invisible in 
the microwave background as required by observation.

The logical end is perhaps not hard to guess; in
1991, Linde \cite{LIN2SC} 
dispensed with the bubbles altogether, by eliminating the dip of the 
potential at the origin. 
At the end of inflation, the field $\psi$ now reverts
to its vacuum value without any bubble formation, so that there is 
a second-order phase transition, instead of the first-order one of the 
original model. This final paradigm is 
known as hybrid inflation. 
It has lead to the renaissance of inflation model building, firmly rooted
in the concepts of modern particle theory,
which is the focus of the present review.

The actual beginning of the renaissance can be traced to 
a paper in 1994 \cite{cllsw}. It contained 
the crucial observation that during hybrid inflation, 
the inflaton field $\phi$ is typically much less than $\mpl$.
As a result, contact with particle theory again becomes a realistic 
possibility. At the same time though, the above paper
emphasized that a generic supergravity theory will fail to inflate
no matter how small are the field values, because the inflaton
mass is too big. Much effort has since been devoted to
finding ways around this problem.

In addition to the small field value, hybrid inflation
has another good feature.
In single-field models the curve $V(\phi)$ must first support 
inflation, and then cease to support it so that inflation ends.
There are only a few simple functions that achieve this, if one excludes
field values much bigger than $\mpl$.
In the hybrid case,
the job of ending inflation is done by the other field $\psi$,
which greatly increases the range of simple possibilities.

We end this introduction with an overview of the present article,
and a list of its omissions.
The article is addressed to a wide audience, including both cosmologists and 
particle physicists. 
To cope with this problem, we have tried to make each section 
reasonably homogeneous regarding the background knowledge that is taken 
for granted, while at the same time allowing considerable variation {}from 
one section to another.
Section  \ref{s2} focusses on the cosmological quantities,
that form a link between a model of inflation and observation.
Section \ref{s3} gives the basics of the slow-roll paradigm of 
inflation, showing how the cosmological quantities are calculated.
Section \ref{s10} is a specialized one, explaining how to derive
the usual prediction of slow-roll inflation, and how to generalize
it to the case of a multi-component inflaton.
Section \ref{s4} summarizes some of the basic ideas of modern particle
theory, which have been used in inflation model-building.
Those with a background in particle theory will skip through it fairly 
 quickly. Using these ideas, Section \ref{s8} reviews `models' of
inflation, taken to mean forms for the potential
that have the general form suggested by particle theory.

Section \ref{s5} summarizes those aspects of
supersymmetry which are most relevant for inflation model-
building.
It is addressed mainly to those who already have some 
understanding of
that subject. As we explain there, the tree-level potential in
a supersymmetric theory is the sum of an `$F$-term' and a 
`$D$-term'.
The terms have very different properties and in all models of 
inflation
so far proposed one or other dominates.
Section \ref{s18} deals with models of inflation where the 
$F$-term
dominates, and Section \ref{s19} with those where the $D$-term
dominates. 
We conclude in Section \ref{s9}.

The above list of topics is formidable, but still not exhaustive.
Let us mention the main omissions.

While the 
paradigm of slow-roll inflation is broadly necessary, in order to
account for the near
scale-independence of the primordial spectrum $\calp_\calr(k)$,
brief interruptions of slow-roll are sometimes contemplated.
So are sharp changes in the direction of slow-roll, of the
kind described in Section \ref{s10}. In both cases, the
effect is to generate a sharp
feature, in the otherwise smooth primordial spectrum.
At the time of writing there is no firm observational
evidence for such a feature, and we mention only briefly the
models that would predict one.

We shall not discuss the pre-big-bang idea, that a bounce at the
Planck scale can do the job of inflation. 
In contrast with inflation, this paradigm provides no natural 
explanation of the near scale-independence of the spectrum
of the primordial curvature perturbation, encoded by the result
$n\simeq 1$. In slow-roll inflation this result is an automatic
consequence of the near time-independence of the Hubble parameter,
but no analogous quantity appears in the pre-big-bang paradigm.

Globally supersymmetric models
using complicated particle physics, in particular
a Grand Unified Theory (GUT) are not mentioned much.
Like some simpler models that we do mention, these models 
usually lack any specific mechanism for controlling the
supergravity corrections. 
Except for a brief mention of
monomial potentials, models invoking field values
much bigger than $\mpl$ are not mentioned. (All know models involving 
non-Einstein gravity are of this type.)

We do not consider the rather special inflationary potentials that
can give an open Universe (negative spatial
curvature) through bubble formation \cite{open,o1,o2,o3}, since little 
attention has so far been paid to these in the context of particle theory.
We are basically focussing on the usual case,
that $\Omega$ has been driven
to 1 long before our Universe leaves the horizon during inflation,
making its present value (including the contribution of any cosmological
constant) also 1. However, most of what we do continues to apply if 
that is not the case, which arguably might happen for any form of the
inflationary potential.\footnote
{With any potential, one can assume that $\Omega$ is fine-tuned to be
small at the Planck scale \cite{ewanopen}, or else 
that the Universe is created at a finely-tuned point in 
field space \cite{ht98,l98,v98,bl}. As usual, 
one can consider eliminating such 
fine-tuning by the anthropic principle.}

We assume that the primordial density perturbation
generated by the vacuum fluctuation of the inflaton is solely 
responsible for large scale structure, except possibly for 
a gravitational wave signal in the cmb anisotropy.
This means that we ignore anything coming {}from topological defects,
as well as the isocurvature density perturbation that could in principle 
be generated by the vacuum fluctuation of a non-inflaton field like
the axion. Subject to confirmation {}from further observations, it
looks as though 
such things cannot be entirely responsible for large scale structure,
so indeed the simplest thing is to assume that they are entirely 
absent.

Finally, we are considering only models of {\em inflation},
not of the subsequent cosmology. In particular, we are not considering 
the reheating process by which the scalar field
is converted into
hot radiation. We are not considering the preheating process
that might exchange energy between 
scalar fields before reheating \cite{kls,r1,r2,r3,r4,r5,r6,r7,r8,r9}.
And we are definitely not considering baryogenesis, dark matter, or 
unwanted relics such as moduli.
All of these phenomena are likely to involve fields, and 
interactions, that play no role during inflation.
We generally set $\hbar=c=1$, and we define the Planck mass by
$\mpl=(8\pi G)^{-1/2}=2.4\times 10^{18}\GeV$.

\section{Observing the density perturbation (and gravitational waves?)}
\label{s2}

The vacuum fluctuation of the inflaton field generates
a primordial energy
 density perturbation, and the vacuum fluctuation in the 
transverse traceless part of the metric
generates gravitational waves. 
In this section we explain briefly how the primordial density 
perturbation, and the gravitational waves,
are related to what is actually observed. 

\subsection{The primordial quantities}

In the unperturbed Universe, the 
separation of comoving observers\footnote
{Both in the perturbed and unperturbed Universe, a comoving observer is 
defined as one moving with the flow of energy. Such observers
measure zero momentum density at their own positions.}
is proportional to the 
scale factor of the Universe $a(t)$, and we 
normalize it to 1 at the present epoch.
The Hubble parameter is $H=\dot a/a$, and its 
present value $H_0=100h\km\sunit^{-1}\Mpc^{-1}$
with $h$ probably in the range $0.5$ to $0.7$. 
The corresponding 
Hubble distance is $cH_0^{-1}=3000h^{-1}\Mpc$,
which is roughly the size of the observable Universe.\footnote
{In the following we shorten `observable Universe' to `Universe'.
The unknown regions outside it are referred to as the `universe' with
capitalization.}

Instead of the physical Cartesian coordinates $\bfr$ it is more convenient
to use coordinates $\bfx$ such that $\bfr=a(t)\bfx$. 
Then the coordinate
position of a comoving observer is time-independent, in the unperturbed 
Universe. The Fourier expansion of a perturbation $g(\bfx,t)$
is made inside a large comoving box, whose coordinate size $L$ 
should be a few orders
of magnitude bigger than that of the observable Universe. (On bigger scales
it would not be justified to assume a homogeneous, isotropic
universe.) The Fourier expansion is
\be
g(\bfx,t) =\sum_\bfk g_\bfk e^{i\bfk\cdot\bfx}.
\ee
For mathematical purposes it is convenient to consider the limit
of an infinite box,
\be
g(\bfx,t) = \int d^3\bfk g(\bfk,t) e^{i\bfk\cdot\bfx}.
\ee
where $(L/2\pi)^3 g_\bfk \to (2\pi)^{-3/2} g(\bfk)$.
A useful wave of specifying the 
physical wavenumber $k/a$ is to give its present value $k$.

During inflation, $aH$ increases with time, and a comoving scale
$a/k$ is said to leave the horizon when $aH/k=1$. After inflation
$aH$ decreases, and the comoving scale is said to
enter the horizon when $aH/k=1$.
For cosmologically interesting scales, horizon entry occurs long after
nucleosynthesis. We shall occasionally refer to the long era between
horizon exit and horizon entry as the {\em primordial}\/ era.
As we shall see, the evolution of the perturbations during the 
primordial era is simple, because causal processes cannot operate.

Leaving aside gravitational waves, there is only one independent 
primordial perturbation, because everything is generated {}from the vacuum 
fluctuation of the inflaton field.
(We are considering the usual case of 
a single-component inflaton field.)
Instead of the inflaton field perturbation, it is actually
more conveniently to 
consider
a quantity $\calr(\bfk)$,
defined by\footnote
{The quantity we are calling $\calr$ 
was defined first by Bardeen Ref.~\cite{bardeen80}, who
called it $\phi_m$. It was called ${\cal R}_m$ by Kodama and Sasaki
Ref.~\cite{kodamasasaki}, and
we drop the subscript following
\cite{ewanopen,LL1,stly}. Later it was called
$\zeta$ by Mukhanov et al \cite{mfb}, which is the other commonly used
notation at present. It is a factor $\frac32k^{-2}$
times the quantity $\delta K$ of Ref.~\cite{lyth85,lyth85b}.
On the scales far outside the horizon where it is constant
(the only regime where it is of interest)
it coincides with the quantity $\xi/3$ of Ref.~\cite{bst}
and the quantity $\xi$ of Ref.~\cite{c3}.
On these scales, it is also equal to $-C\Phi$, where $\Phi$ is
the commonly 
used `gauge invariant potential' \cite{bardeen80}, and $C$ is a factor
of order unity which is constant both during radiation domination
and during matter domination. During the latter epoch, $C=5/3$.}
\be
{\cal R}(\bfk)= \frac14 (a/k)^2 R^{(3)}(\bfk),
\ee
where $R^{(3)}$ is the spatial
curvature scalar seen by comoving observers.
Unlike the inflaton field perturbation, it is 
time-independent during the primeval era, and it continues
to be well-defined after the inflaton field  disappears.

A gravitational wave corresponds to a spatial metric perturbation
$h_{ij}$ which is traceless, $\delta^{ij} h_{ij}=0$, and transverse,
$\pa_i h_{ij}=0$.  This means that each Fourier component is of the
form
\be
h_{ij}=h_+ e^+_{ij} + h_\times e^\times_{ij}.
\label{hij}
\ee
In a
coordinate system where $\bfk$ points along the $z$-axis, the
nonzero components of the polarization tensors are defined by
$e^+_{xx}=-e^+_{yy}=1$ and $e^\times_{xy}=e^\times_{yx}=1$.
The two independent quantities $h_{+,\times}$ are time-independent
well outside the horizon.

Inflation generates gaussian fluctuations.\footnote
{To be more precise, the gravitational wave amplitude is
certainly gaussian, and so is the curvature 
perturbation if the inflaton field fluctuation 
$\delta\phi$ is Gaussian. The latter is true if $\delta\phi$ is a practically
free field, which is the case in practically all models of 
inflation. 
The gaussianity is inherited by all of the 
perturbations as long as they remain small.}
This means that for each perturbation $g(\bfx)$,
at fixed $t$, the Fourier components are uncorrelated except for
the expectation values
\be
\langle g^*(\bfk) g(\bfk') \rangle = \delta^3(\bfk-\bfk') \frac
{2\pi^2}{k^3} \calp_g(k).
\label{specdef}
\ee
The quantity $\calp_g(k)$ is called the spectrum of $g(\bfx)$, 
and it determines all of 
its stochastic properties.

The primordial perturbations consist of the three independent quantities
$\calr$, $h_+$ and $h_\times$, and {}from
rotational invariance the last two have the same spectrum,
\be
\calp_{h_+}=\calp_{h_\times}\equiv \calp\sub{grav}/2.
\ee
We therefore have two independent spectra $\calp_\calr$ 
and $\calp\sub{grav}$, determined in  a slow-roll model of inflation by
the formulas described in the next section.
We shall see that they have 
at most mild scale dependence, and this is consistent with 
observation. 

The spectral index $n(k)$ of the curvature perturbation (\eq{nofpr})
is a crucial point of comparison between theory and observation, and
the same will be true of $n\sub{grav}(k)$ if the gravitational waves are
detectable.

\subsection{The observable quantities}

{}From these primordial quantities, one can calculate the 
observable quantities, provided that one knows enough about
the nature and
evolution of the unperturbed Universe after relevant scales
enter the horizon. Since cosmological 
scales enter the horizon well after nucleosynthesis, one 
indeed has the necessary information, up to uncertainties in the 
Hubble parameter, the nature and amount of dark matter, the epoch
of reionization and the magnitude of the cosmological constant.\footnote
{In principle the reionization epoch can be calculated in terms of the 
other parameters, through the 
abundance of 
early rare objects, but present estimates are fairly crude.}
The observed quantities can be taken to be
the matter density contrast $\delta\equiv\delta\rho/\rho$
(observed through the distribution and motion of the galaxies),
and the cmb anisotropy. The latter consists of the temperature
anisotropy $\Delta T/T$, which is already being observed, and two Stokes
parameters describing the polarization which will be observed 
by the MAP \cite{map}
and Planck \cite{planck} satellites. It is convenient to make multipole
expansions so that one is dealing with the temperature anisotropy
$a_{\ell m}$, and the polarization anisotropies
$E_{\ell m}$ and $B_{\ell m}$.\footnote
{The polarization multipoles are defined with respect to spin-weighted
spherical harmonics, to ensure the correct transformation of the Stokes
parameters under rotation about the line of sight.}

Except for the density perturbation on scales where gravitational
collapse has taken place, the observable quantities are related to
the primordial ones through linear, rotationally invariant,
transfer functions. 
For the density perturbation,
\be
\delta(\bfk,t) = {\cal T}(k,t) \calr(\bfk).
\ee
It can be observed both at the present, and 
(by looking out to large
distances) at earlier times. 
The corresponding spectrum is
\be
\calp_\delta(k,t) = {\cal T}^2(k,t) \calp_\calr(k).
\ee
For the cmb anisotropy, ignoring the 
gravitational waves, 
one has
\bea
a_{\ell m} &=& \frac{4\pi}{(2\pi)^{3/2}}
\int  {\cal T}_\Theta(k,\ell) \calr_{\ell m}(k)
k dk, \\
E_{\ell m} &=& \frac{4\pi}{(2\pi)^{3/2}}
\int  {\cal T}_E(k,\ell) \calr_{\ell m}(k)
k dk, \\
B_{\ell m} &=& 0.
\eea
Here, 
the multipoles of $\calr$ are related to its Fourier components by
\be
\calr_{\ell m}(k) = ki^\ell \int \calr(k,\hat\bfk)
Y_{\ell m}(\hat\bfk) d\Omega_\bfk.
\ee
which is equivalent to the usual spherical expansion. 
They are uncorrelated except for the expectation values
\be
\langle g_{\ell m}^*(k) g_{\ell' m'} (k')
\rangle =  \frac{2\pi^2}{k^3} \calp_g(k)
\delta(k-k') \delta_{\ell \ell'} \delta_{mm'} \,.
\label{glm}
\ee
As a result, the multipoles of the cmb anisotropy are uncorrelated, 
except for the expectation values 
\bea
\langle a^*_{\ell m} a_{\ell' m'}\rangle
&=& C(\ell) \delta_{\ell\ell'}\delta_{m m'}, \\
\langle a^* _{\ell m} E_{\ell' m'} \rangle
&=& C\sub{cross}(\ell )\delta_{\ell\ell'}\delta_{m m'}, \\
\langle E^*_{\ell m} E_{\ell' m'}\rangle
&=& C_E(\ell)
\delta_{\ell\ell'}\delta_{m m'}.
\eea
where
\bea
C(\ell) &=& 4\pi
\int^\infty_0 {\cal T}^2_\Theta(k,\ell)\calp_\calr(k) \frac{dk}{k},
\label{coft}
\\
C\sub{cross}(\ell) &=& 4\pi
\int^\infty_0 {\cal T}_\Theta(k,\ell)
{\cal T}_E(k,\ell) \calp_\calr(k) \frac{dk}{k},
\label{ccroft}\\
C_E(\ell) &=& 4\pi
\int^\infty_0 {\cal T}^2_E(k,\ell) \calp_\calr(k) \frac{dk}{k}.
\label{ceoft}
\eea

The gravitational waves give contributions to the $C$'s which have
a similar form, now with a nonzero $C_B$ defined analogously to 
$C_E$.\footnote
{There is no cross term 
involving $B_{\ell m}$ because it would be odd under the parity
transformation. (The vacuum state is parity invariant, and so is
the Thompson scattering process responsible for the polarization.)}
We shall not give their precise form, but note for future reference
that they fall off rapidly above $\ell\sim 100$.
The reason is that larger $\ell$ correspond to 
scales smaller than the horizon at
photon decoupling; on such scales
the amplitude of the gravitational waves has been reduced {}from its 
primordial value by the redshift.

We should comment on the meaning of the `expectation value',
denoted by $\langle\cdots\rangle$. At the fundamental level, it denotes the
quantum expectation value, in the state that corresponds to
the vacuum during inflation. This state does not correspond to
a  definite perturbation $g(\bfx)$ (because it does not correspond to
definite $g(\bfk)$), so it is a superposition of possible universes.
As usual, this Schr\"odinger's cat paradox
does not prevent us {}from comparing with observation. 
We simply make the hypothesis that our Universe is a typical one,
of the superposition defined by the quantum state.
Except for the low multipoles of the cmb anisotropy, this makes 
observational quantities sharply defined, since they involve
a sum over the practically continuous variables $\bfk$ and $\ell$.
For the low multipoles the expected difference between the observed
$|a_{\ell m}|^2$ and $\langle |a_{\ell m}|^2\rangle$ (called cosmic variance)
needs to be taken into account, but the hypothesis that we live in a 
typical universe is still a very powerful one.

For the density perturbation, the comparison of the above prediction 
with observation has been a major industry for many years. Since
1992 the same has been true of the cmb temperature anisotropy.
Perhaps surprisingly, the result of all this effort is easy to summarize. 

Observation is consistent with the inflationary prediction 
that the curvature perturbation is gaussian, with a smooth spectrum.
The spectrum is accurately measured by COBE at the scale
$k\simeq 7.5 H_0$ (more or less the center of the range explored
by COBE).
Assuming that gravitational waves are negligible, it is
\cite{bunn96}
\be
\delta_H\equiv (2/5)\calp_\calr^{1/2}
=1.91\times 10^{-5}.
\label{cobe}
\ee
with an estimated 9\% uncertainty at the 1-$\sigma$ level.
In writing this expression, we introduced the quantity $\delta_H$
which is normally used by observers.

Assuming that the spectral index is roughly constant 
over cosmological scales, observation constrains it to something like 
the range \cite{LL3}
\be
n= 1.0\pm 0.2 \,.
\label{nobs}
\ee

Gravitational waves have not so far been seen in the cmb anisotropy
(or anywhere else).
Observation
is consistent with the hypothesis that they
account for a significant 
fraction (less than 50\% or so) of the mean-square cmb multipoles
at $\ell \lsim 100$. In quantifying their effect, it is useful to 
consider the quantity $r$ defined in the next section. Up to
a numerical factor it is $\calp\sub{grav}/\calp_\calr$,
and the factor is chosen so that in 
an analytic approximation due to Starobinsky \cite{starob},
\be
r = C\sub{grav}(\ell)/C_\calr(\ell)
\ee
for $\ell $ in the central COBE range.\footnote
{A common alternative is to define $r$ by setting $\ell =2$ 
in Starobinsky's calculation. This increases $r$ by a 
factor 1.118 compared with the above definition.}
(Here $C_\calr$ is the contribution of the curvature perturbation
given by \eq{coft}, and $C\sub{grav}$ the contribution of gravitational
waves.)
We are saying that present observations require $r\lsim 1$ or so.
According to an accurate calculation \cite{bunn96}, 
the relative contribution of 
gravitational waves to the COBE anisotropy is actually $0.75r$, reducing 
the deduced value of $\delta_H$ 
by a factor $\simeq(1+ 0.75r)^{-1/2}$ compared with \eq{cobe}.

What about the future? The magnificent COBE normalization 
will perhaps
never to be improved, but this hardly matters since at present
an understanding of even its order of magnitude is a major theoretical
challenge. Much more interesting is the situation with the spectral
index. The Planck satellite will probably measure $n(k)$ with an accuracy
of order $\Delta n\sim 0.01$, which as already mentioned will be a powerful
discriminator between models of inflation.
The same satellite will also either
tighten the limit on gravitational waves to
$r\lsim 0.1$, or detect them. This last figure is unlikely to be 
improved by more than an order of magnitude in the forseeable future. 

The Planck satellite probes a range $\Delta \ln k\simeq 6$,
and will measure the scale-dependence $dn/d\ln k$ if it is bigger than
a few times $10^{-3}$.

We have emphasized the cmb anisotropy because of the promised high
accuracy, but it will never be the whole story. 
It can directly probe only the scales $10\Mpc \lsim k^{-1} 
\lsim 10^4\Mpc$, where the upper limit is the size of the observable 
Universe, and the
lower limit is the thickness of the last-scattering `surface'.
At present it probes only the upper half of this range,
$100\Mpc \lsim k^{-1} \lsim 10^4\Mpc$.\footnote
{A very limited constraint is provided on much bigger scales
through the 
Grishchuk-Zeldovich \cite{gz,gbgz} effect, which we shall not discuss.}
Galaxy surveys probe the range 
$1\Mpc\lsim k^{-1}\lsim 100\Mpc$, providing a useful overlap
in the future.
The range 
$1\Mpc \lsim k^{-1} \lsim 10^4\Mpc$ is usually taken to be the 
range of `cosmological' scales.
If a signal of early reionization is seen 
in the cmb anisotropy, it will provide an estimate of the spectrum 
on a significantly smaller 
scale, $k^{-1}\sim 10^{-2}\Mpc$. Alternatively, the absence of a signal
will provide a rough upper limit on this scale.

On smaller scales still, information on the spectrum of the 
primordial density perturbation is sparse, and consists entirely of upper
limits. The most useful 
limit, {}from the viewpoint of constraining models of inflation,
is the one on the smallest relevant scale which is the one leaving the
horizon just before the end of inflation. It has been considered
in Refs.~\cite{jim,lisa,toniandrew}, and for a scale-independent 
spectral index corresponds to $n\lsim 1.3$.

\section{The slow-roll paradigm}
\label{s3}

Inflation is defined as an era of
repulsive gravity,
$\ddot a>0$, which is equivalent to $3P<-\rho$ where $\rho$ is the 
energy density and $P$ is the pressure.
As noted earlier, we are concerned only with the era of 
`observable inflation',
which begins when the observable Universe leaves the horizon, since
memory of any earlier epochs has been wiped out.

During inflation the density parameter $\Omega$ is driven towards
1. Subsequently it moves away {}from 1, and its present value is equal to its
value at the the beginning of observable inflation. We are taking
that value to be close to 1, which means that $\Omega$ is close to 1 
during observable inflation. This gives the energy density $\rho$ 
in terms of the Hubble parameter,
\be
3\mpl^2 H^2 = \rho \,.
\label{crit98}
\label{critical}
\ee

During observable inflation, the energy density and pressure are 
supposed to be dominated by scalar fields.
Of the fields that 
contribute significantly to the potential, the inflaton field
$\phi$ is by definition
the only one with significant time-dependence, leading to
\bea
\rho &=& \frac12\dot\phi^2 + V 
\label{rho98}\\
P &=& \frac12\dot\phi^2 - V \,.
\label{P98}
\eea
(We make the usual assumption that $\phi$ has only one component, 
deferring the general case to Section \ref{s10}.)

The evolution of $\phi$ is given by
\be
\ddot\phi+3H\dot\phi=-V' \,,
\label{phiddot}
\ee
where an overdot denotes $d/dt$ and a prime denotes $d/d\phi$.
This is equivalent to the continuity equation
$\dot\rho=-3H(\rho+P)$, which with \eq{critical} is equivalent to
\be
\dot H = -\frac12 \dot\phi^2/\mpl^2 \,.
\label{hdot98}
\ee

\subsection{The slowly rolling inflaton field}

While cosmological scales are leaving the horizon, the 
slow-roll paradigm of inflation  \cite{kt,abook,LL2,LL3}
is practically mandatory
in order to account for the near scale-invariance of spectrum of the
primordial curvature perturbation. 

The inflaton field $\phi$ is supposed to be
on a region of the potential which
satisfies the flatness conditions
\bea
\epsilon &\ll& 1 \label{flat1} \\
|\eta| &\ll& 1 \label{flat2} \,,
\eea
where
\bea
\epsilon&\equiv &\frac12\mpl^2(V'/V)^2,\\
\eta&\equiv &\mpl^2 V''/V.
\eea
Also, it is supposed that the exact evolution \eq{phiddot} can be
replaced by the slow-roll approximation
\be
\dot\phi=-\frac{V'}{3H}.
\label{slowroll}
\ee

The flatness conditions and the slow-roll approximation
are the basic equations, needed to derive the standard prediction for 
the density perturbation and the spectral index. 
For potentials satisfying the flatness conditions, the slow-roll 
approximation is typically valid for a wide range of initial conditions
(values of $\phi$ and $\dot\phi$ at an early time). 

The first flatness condition $\epsilon\ll 1$
 ensures that $\rho$ is close to $V$ and is slowly varying.\footnote
{In what follows,
we say that a function of time satisfying $|d\ln f/d\ln a|\ll 1$
is
`slowly varying'. For a function of wavenumber $k$, `slowly varying'
will mean the same thing with $a$ replaced by $k$.}
As a result $H$ is slowly varying,
which implies that one can write
$a\propto e^{Ht}$ at least over a Hubble time or so.

The second flatness condition $|\eta|\ll 1$ is actually a consequence of the 
first flatness condition plus the slow-roll approximation $3H\dot\phi=
-V'$. Indeed, differentiating the latter one finds
\be
\frac{\ddot\phi}{H\dot\phi} = \epsilon - \eta \,,
\label{ddot}
\ee
and from \eq{phiddot} the slow-roll approximation is equivalent to
$|\ddot\phi|\ll H|\dot\phi|$.

A crucial role is played by the number of Hubble times
$N(\phi)$ of inflation, still remaining when $\phi$ has a given value.
{}From some time $t$ to a fixed later time $t_2$, the
number of Hubble times is
\be
N(t) \equiv \int^{t_2}_t H(t) dt\,.
\ee
The small change satisfies
\be
dN  \equiv -H\,dt(=-d\ln a) \,.
\label{dndef}
\ee
During slow-roll inflation, 
\be
\frac{dN}{d\phi}=-\frac{H}{\dot\phi}=\frac{V}{\mpl^2V'}
\left( = \pm \left(\sqrt {2\epsilon}\mpl\right)^{-1} \right)
\,.
\label{Nrelation}
\label{dndphi}
\ee
The number of $e$-folds of slow-roll inflation, remaining
at a given epoch, is
\be
N(\phi) = \int^\phi_{\phi_{\rm end}} \mpl^{-2}\frac V{V'} d\phi \,,
\label{nint}
\ee
where $\phi\sub{end}$ marks the end of slow-roll inflation.

\subsection{The slow-roll predictions}

\label{srpred}

In this subsection and the next, as well as in Section \ref{s10},
we discuss predictions 
for $\calp_\calr$, $n$ and $dn/d\ln k$.
More material can be found in references \cite{LL1,LL2,stly,LL3}.

Two basic assumptions are made.
One is that the inflaton field perturbation
$\delta\phi$ has negligible interaction with other fields.
This is equivalent to 
the validity during inflation of linear cosmological 
perturbation theory, in other words to the procedure of keeping only
terms that are linear in the perturbations \cite{LL3}.

The other essential assumption is that well
before horizon exit, when the particle concept makes sense,
the relevant Fourier modes of $\delta\phi$ have zero occupation number.
This 
vacuum assumption is more or less mandatory, since too many 
particles would give significant pressure and spoil inflation
\cite{LL1}.

As a result of these assumptions, 
the primordial curvature perturbation is gaussian,
with stochastic properties that are completely defined by its
spectrum $\calp_\calr(k)$.

In this subsection, we make the usual assumption that the slow-roll
paradigm is valid.

\subsubsection{The spectrum}

The perturbation $\delta\phi$ is best defined on spatially flat 
hypersurfaces. Then, in the slow-roll limit $\dot H\to 0$,
one can ignore the effect of the metric perturbation 
\cite{LL2,LL3}, and $\delta\phi$ satisfies
\be
(\delta\phi)\,\ddot{} + 3H(\delta\phi)\,\dot{} + \[V'' + \(\frac k a \)^2
\] \delta\phi = 0 \,.
\ee
The flatness condition (\ref{flat2})
ensures that the mass-squared $2V''$ 
is negligible until at least a few Hubble times
after horizon exit. 
This means that $\delta\phi$ can be treated as a massless
free field. 
A few Hubble times after horizon exit, its vacuum fluctuation
can be regarded as a classical quantity, and its spectrum is then
\be
\calp_\phi = (H/2\pi)^2.
\label{calphi}
\ee
The corresponding
curvature perturbation is given by $\calr = (-H/\dot\phi) \delta\phi
$ (valid in linear perturbation theory independently of 
slow-roll).
Using \eqs{slowroll}{flat1}, this is
equivalent to
\be
\frac4{25}\calp_\calr(k)\equiv
\delta_H^2(k) = \frac1{75\pi^2\mpl^6}\frac{V^3}{V'^2}
=
\frac1{150\pi^2\mpl^4}\frac{V}{\epsilon}.
\label{delh}
\ee
In this expression, the
potential and its derivative are evaluated at the epoch of 
horizon exit for the scale $k$, which is defined by $k=aH$.\footnote
{\eq{delh} becomes valid only a few Hubble times
after horizon exit, but its right hand side is
slowly varying
and we might as well evaluate it actually {\em at} horizon exit.
The difference this makes is of the 
same order as the error in \eq{delh}.}

This prediction, for the spectrum of 
$\calr(k)$ a few Hubble times after horizon exit, is of no use as it 
stands. But one can show that
$\calr(k)$ 
is time-independent between that epoch and the approach of 
horizon entry long after inflation ends. As we saw in Section
\ref{s2}, this allows one to calculate observable quantities.

Comparing \eq{delh} with the value \eq{cobe}
deduced {}from the COBE observation of the cmb 
anisotropy gives\footnote
{This relation ignores any gravitational wave contribution, but there is 
no point in including their effect in the present context. The reason is 
that 
the prediction for $\delta_H$ that is being used has an error of at least
the same order. If necessary one could include \cite{stly,recon}
the effect of the gravitational waves using the more accurate 
formula \eq{pbetter}.}
\be
\mpl^{-3} V^{3/2}/V' = 5.3\times 10^{-4}.
\label{cobenorm}
\ee
This relation provides a useful constraint
on the parameters of the potential. It can be written in the equivalent 
form
\be
V^{1/4}/\epsilon^{1/4}=.027\mpl=6.7\times 10^{16}\GeV.
\label{vbound}
\ee
Since $\epsilon$ is much less than 1, the 
inflationary energy scale $V^{1/4}$
is at least a couple of orders of magnitude
below the Planck scale \cite{lyth85}.

The scale leaving the horizon at a given epoch is directly related to 
the number $N(\phi)$ of $e$-folds of 
slow-roll inflation, that occur after the epoch 
of horizon exit. Indeed, since $H$ is slowly varying we have
$d\ln k =d(\ln (aH) )\simeq d \ln a = H dt$. From the definition
\eq{dndef} this gives
\be
d\ln k = -dN(\phi)\,,
\label{dlnkdN}
\ee
and therefore
\be
\ln(k\sub{end}/k)= N(\phi) \,,
\ee 
where $k\sub{end}$ is the scale leaving the horizon at the end of 
slow-roll inflation.
As we shall see, this 
relation is very useful when working out the prediction for a given form 
of the potential.

This is a good place to insert a historical footnote, about the origin
of the slow-roll prediction for $\calp_\calr$. As we noted 
already, it comes in two parts.
One is the formula 
\eq{delh}
for $\calp_\calr$ a few Hubble times after horizon exit, 
and the other is the statement
that $\calr$ (hence $\calp_\calr$) is time-independent while $k$ is well
outside the horizon. 

Both parts were,
in essence, given at about the same time in 
Refs.~\cite{hawking,starob82,guthpi,bst}.
(Related work \cite{mc81} had been done earlier.)
To be precise, these authors gave results which become more or less
equivalent
after the 
spectrum has been defined, though that last step was not 
explicitly made and except for the last work only a 
particular potential is discussed.\footnote
{The last three works give results equivalent 
to the one we quote, and 
the first gives a result which is approximately the same.}
Soon afterwards the results were given again, this time with an explicitly
defined spectrum \cite{lyth85}.

Strictly 
speaking none of these five derivations is completely satisfactory.
The first three make simplifying assumptions. Regarding the constancy of
$\calr$, all except the third assume
something equivalent to it without adequate proof. 
We discuss the constancy of $\calr$ in 
Section \ref{s10}. Regarding \eq{delh},
none of these early derivations
properly considers the effect of the inflaton field
perturbation on the metric, but as we noted already that turns out to 
be negligible.

\subsubsection{The spectral index}

We have an expression for $\calp_\calr(k)$ in terms of $V$
and $V'$, and we want to calculate 
the spectral index defined by
$n(k)-1\equiv d\calp_\calr/d\ln k$. 
{}From \eqs{dndphi}{dlnkdN},
\be
\frac{d\phi}{d\ln k}
=-\mpl^2 \frac{V'}{V} \,,
\label{nexpression}
\ee 
where, as always, $k=aH$. 
We shall need the following expressions
\bea
\frac{d\epsilon}{d\ln k} &=& 2\epsilon\eta-4\epsilon^2 
\label{deps} \\
\frac{d\eta}{d\ln k} &=& -2\epsilon\eta + \xi^2
\label{deta} \\
\frac{d\xi^2}{d\ln k} &=& -2\epsilon\xi^2+\eta\xi^2 + \sigma^3
\label{dxi} \,,
\eea
where
\bea
\xi^2 &\equiv& \mpl^4\frac{V'(d^3V/d\phi^3)}{V^2} \label{xidef} \\
\sigma^3&\equiv& \mpl^6\frac{V'^2(d^4V/d\phi^4)}{V^3} \label{sigdef}
 \,.
\eea
Following for instance \cite{blp}, we have introduced
respectively the square and the cube of a 
quantity, even though the quantity itself never appears in an equation.
As we shall see, this is a convenient device. Also, in the case
$V'\propto \phi^p$, with $p\neq1$ or $2$,
one has $|\eta|\sim|\xi|\sim|\sigma|$.
The hierarchy can be continued
\cite{blp}, each new equation introducing a new quantity
$\mpl^{2n}V'^{n-1}  (d^{n+1}V/d\phi^{n+1})$.

Using \eqs{deps}{delh}
one finds
\cite{LL1,davis,salopek}
\be
n-1=-6\epsilon+2\eta \,,
\label{n2}
\ee
and using \eqs{deps}{deta} \cite{running},
\be
\frac{dn}{d\ln k}= -16\epsilon\eta
+24\epsilon^2 + 2\xi^2 \,.
\label{running}
\ee

Practically all 
models proposed so far (Section \ref{s8})
have  $V'\propto \phi^p$
or $V'\propto \phi^p \ln\phi $, and in most cases
one also has $\phi\ll\mpl$. Then 
$\epsilon\sim (\phi/\mpl)|\eta|$ is negligible, and one can write
\bea
n-1&=&2\eta \,, \label{n3} \\
\frac{dn}{d\ln k} &=& 2\xi^2 \,.
\label{running3}
\eea

More generally, one can argue
that $\epsilon$ is small irrespectively of the form of 
the potential, provided that $\phi\ll \mpl$.
To see this, take the cosmological range of scales to span four decades, 
corresponding to $\Delta \ln k\simeq 9$.
This corresponds to $9$ $e$-folds of 
inflation.
 In slow-roll inflation $\epsilon$ has negligible variation over one 
$e$-fold and in typical models it has only small 
variation over the
9 $e$-folds. Taking that to be the case, 
and assuming that
$\phi\ll\mpl$,
one learns {}from 
(\ref{Nrelation}) that
$\epsilon \ll \frac12\times (1/9)^2=6\times 10^{-3}$.

\subsubsection{Error estimates for the slow-roll predictions}

In deriving the prediction for $\calp_\calr$ we used the flatness 
conditions $\epsilon\ll 1$ and $|\eta|\ll 1$, as well as the slow-roll 
approximation whose fractional error is $\epsilon-\eta$
(\eq{ddot}). As a result one expects $\calp_\calr$ to pick up fractional 
errors of order $\epsilon $ and $\eta$,
\be
\frac {\Delta \calp_\calr}{\calp_\calr} = O(\epsilon,\eta) \,.
\ee
Using \eqss{deps}{deta}{dxi} one therefore expects
\bea
n-1 &=& 2\eta-6\epsilon + O(\xi^2) 
\label{nerror} \\
\frac{dn}{d\ln k} &=& -16\epsilon\eta
+24\epsilon^2 + 2\xi^2 + O(\sigma^3) \label{dnerror}
\,.
\eea
In the first expression we ignored 
errors that are quadratic in $\epsilon$ and $\eta$, because barring 
cancellations the corresponding fractional errors are small by virtue of 
the flatness conditions $\epsilon\ll 1$ and $|\eta|\ll 1$. In 
the second expression we ignored errors that are cubic in $
\epsilon$, $\eta$ and $\xi$. Barring cancellations, the 
accuracy of the prediction for $n-1$ requires
\be
|\xi^2|\ll\max(\epsilon,|\eta|) \label{flat3} \,,
\ee
and the accuracy of the prediction for its derivative requires
in addition
\be
|\sigma^3| \ll\max(\epsilon^2,\epsilon|\eta|,|\xi^2)|\,.
\label{flat4}
\ee

\subsection{Beyond the slow-roll prediction}

\label{moreacc}

The slow-roll predictions given in the last subsection are very 
convenient, because they involve only $V$ and its low derivatives
evaluated at the epoch of horizon exit. The use of slow-roll is not
however 
mandatory; on the contrary, one can obtain 
\cite{mukhanov,sasaki} predictions using 
essentially no assumptions beyond linear perturbation theory. 

In linear perturbation theory, the
quantity $u=a\delta\phi$ satisfies the following 
exact  equation 
\be
\frac{\pa^2 u}{\pa\tau^2} + \( k^2
-\frac1 z \frac{d^2z}{d\tau^2} \) u = 0 \,.
\label{exacteq98}
\ee
Here, $\tau$ is conformal time defined by $d\tau = dt/a$,
and
\bea
z &\equiv & a\dot\phi/H \\
\frac{d^2z}{d\tau^2} &=& 2a^2H^2 \( 1 +
\epsilon_H + \frac32 \delta+ \frac12\delta^2 + \frac12\epsilon_H
\delta+\frac1{2H}\frac{d\epsilon_H}{dt} + \frac1{2H} \frac{d\delta}
{dt} \) \,,
\label{dzdtau}
\eea
where
\bea
\epsilon_H &\equiv& \frac12\frac{\dot\phi^2}{H^2} 
=-\frac{\dot H}{H^2} \\
\delta&\equiv& \frac{\ddot\phi}{H\dot\phi}  \,,
\eea
and an overdot denotes $d/dt$.

It is convenient to set $\tau=0$ at the end of slow-roll inflation.
In the extreme slow-roll limit
$\dot H=0$, this corresponds to
\be
\tau= -1/(aH) \,.
\ee
One  assumes that
inflation
is near enough slow-roll that $k|\tau|\gg 1$ a few Hubble 
times before horizon exit, and $k|\tau|\ll 1$ a few Hubble times
after. 
Then, there is a solution $u=w$ of \eq{exacteq98}
which satisfies
\be
w=(2k)^{-1/2} e^{-ik\tau} \,,
\label{initial98}
\ee
a few Hubble times
before horizon exit.
A few Hubble times after horizon exit
this solution has the behaviour
\be
w/z \to {\rm \, constant} \,.
\label{final98}
\ee
One can show that the
spectrum of $\calr$ is then given by
\be
\calp_\calr (k) = \frac{k^3}{2\pi^2 z^2} \vert w (k) \vert ^2 \,,
\ee

Given an inflationary trajectory defined by 
$a(\tau)$ and $\dot\phi(\tau)$, this
method gives a practically  unique, and accurate, result in all 
reasonable cases. 
The trajectory in turn follows from the potential practically
independently of the initial conditions, if slow-roll becomes very 
accurate at some early epoch. 

We noted earlier that in 
the regime where the slow-roll predictions for $\calp_\calr$,
$n-1$ {\em and} $dn/d\ln k$ are approximately valid, 
the four flatness conditions \eqsss{flat1}{flat2}{flat3}{flat4}
are also valid. In that case, 
the `exact' solution yields an improved version of the slow-roll
predictions for $\calp_\calr$ and $n-1$
\cite{stly}. Let us see how this goes.

\eqss{ddot}{deps}{deta} and the flatness conditions give the approximation
\be
\frac{d^2z}{d\tau^2} = 2a^2H^2 \( 1 + \epsilon_H + \frac32\delta
\) \,,
\ee
with $\epsilon_H$ and $\delta$ slowly varying on the Hubble timescale.
This leads to the approximation \cite{stly}
\be
\calp^{1/2}_\calr(k) =
\[1-\(2C+1\)\epsilon_H - C\delta\] \frac{H^2}{2\pi|\dot\phi|} \,,
\label{calpstly}
\ee
where $C=-2+\ln 2 +b \simeq -0.73$, with $b$ the Euler-Mascheroni
constant. As always, the right hand side is evaluated at $k=aH$.

We want an expression involving $V$ and its derivatives.
Substituting \eq{slowroll} into \eq{rho98} gives 
\be
\frac{3\mpl^2 H^2}{V} = 1+\frac13\epsilon \,,
\label{himp}
\ee
and substituting \eq{ddot} into \eq{phiddot} gives
\be
-\frac{3H\dot\phi}{V'} =  1-\frac13\epsilon + \frac13\eta \,.
\label{hphidotimp}
\ee
These are improvements in the slow-roll formulas, 
valid to linear order in $\epsilon$ and $\eta$.
Squaring the last equation gives
\be
\frac{\epsilon_H}{\epsilon } = 1-\frac23\epsilon + \frac13\eta \,,
\ee
and \eq{ddot} is
\be
\delta = \epsilon-\eta \,.
\ee
Inserting these four expressions into 
\eq{calpstly} gives
\be
\delta_H=\frac25\calp_\calr^{1/2} (k)
=\frac1{5\sqrt3\pi\mpl^3} \( \frac{V^{3/2}}{|V'|} \) 
\[ 1- \(2C+\frac16\)\epsilon +
\(C-\frac13 \) \eta + O\(\xi^2 \) \] \,.
\label{pbetter}
\ee

The fractional 
error in this improved expression for $\calp_\calr$ 
is expected to be of order $O(\xi^2)$, plus terms quadratic in
$\epsilon$ and $\eta$ that we did not display. The $\xi^2$ term will 
be present, because it contributes to the variation per Hubble time
of $\eta$ (\eq{deta}) which is being ignored.

Using $k=aH$ with \eqss{hdot98}{himp}{hphidotimp} gives the improved formula
\be
\frac{d\phi}{d\ln k} = \sqrt{2\epsilon} \( 1 +\frac13\epsilon
+\frac13\eta \)
\,.
\ee
This leads to
\be
\frac12(n-1) = -3\epsilon + \eta - \(\frac53 + 12 C\)\epsilon^2
+ \(8C-1\) \epsilon\eta +\frac13 \eta^2
-\(C-\frac13\) \xi^2 + O(\sigma^3) \,.
\label{nbetter}
\ee

The fractional error of order $\sigma^3$ comes from differentiating the 
error of order $\xi^2$ in \eq{pbetter} ($d\xi^2/d\ln k$ is given by
\eq{dxi}). Contrary to what is stated in \cite{kv}, the 
actual coefficient of $\sigma^3$ cannot be evaluated without going
back to the exact equation. There will also be error terms 
cubic in $\epsilon$, $\eta$ and $\xi$, that we do not display.

The improved solution becomes exact in the case of power-law
inflation ($a\propto \phi^p$) when $\epsilon_H=-\delta$
is constant, and in the case of $V=V_0\pm\frac12m^2\phi^2$
in the limit $\phi\to 0$ when $\epsilon\to 0$ and $\delta$ becomes
constant.

In some models, the improvement is big enough to measure 
{\em with fixed values of the parameters in the potential}. But in the 
cases that have been examined to date, this 
change can be practically cancelled by varying the parameters.
As a result, the improvement is probably going to be useful only if
gravitational waves are detected (Section \ref{s:gravwave}).

\subsection{The number of $e$-folds of slow-roll inflation}

\label{nefolds}

A model of inflation will give us an inflationary potential $V(\phi)$,
and a prescription for the value $\phi_{\rm end}$ of the field at
the {\em end} of slow-roll
inflation. This is not enough to work out the prediction
for $\calp_\calr(k)$, 
because we need to know the value of $\phi$ when a given cosmological 
scale $k$ leaves the horizon. Using \eq{nint}, we can do this if we
know the number $N(\phi)$ of $e$-folds of slow-roll
inflation taking place after 
that epoch. The model will give $d\ln k/d\phi$, 
(through \eq{nexpression}) so we need this information for just one
cosmological scale.

For definiteness, let us consider the
scale
$k^{-1}=H_0 ^{-1}= 3000h^{-1}\Mpc$, which is the 
biggest cosmological scale of interest.\footnote
{The absolute limit of direct observation is $2H_0^{-1}$,
the distance to the particle horizon in a flat, matter-dominated 
Universe. Since
 the prediction is made for a randomly placed observer
 in a much bigger  patch,
 bigger scales in principle contribute to it,
but sensitivity rapidly decreases outside our horizon.
Only if the spectrum increases sharply on very large scales
\cite{gz,gbgz} might there be a significant effect.
This Grishchuk-Zeldovich effect is not present in any model
of inflation that has been proposed so far.}
As this is more or less the scale probed by COBE, we denote it by a 
subscript COBE.
The number of $e$-folds of inflation after this scale leaves the horizon 
is
\be
N\sub{COBE}= \ln(a_{\rm end}/a\sub{COBE}) \,.
\ee
Since this scale is the one  entering the horizon now,
 $a\sub{COBE}H\sub{COBE}=a_0H_0$ where the subscript 0 indicates the present
epoch. This leads to 
\be
N\sub{COBE}=
\ln\left(\frac{a_{\rm end}H_{\rm end}}{a_0H_0}\right)
-\ln\left(\frac{H_{\rm end}}{H\sub{COBE}}\right).
\ee
The second  term  will be given by the model of slow-roll
inflation and is usually
 $\lsim 1$; for simplicity let us 
ignore it. The first term  depends on 
the evolution of the scale factor between the end of 
slow-roll inflation and the present. 

Assume first that slow-roll inflation gives way promptly to
matter domination ($a\propto t^{2/3}$), 
which is followed by a radiation dominated era ($a\propto t^{1/2}$)
lasting until the present matter dominated era begins. Then 
one has \cite{LL2,LL3}
\be 
N\sub{COBE}=62-\ln(10^{16}\GeV/V_{\rm end}^{1/4}) 
-\frac13\ln(V_{\rm end}^{1/4}/\rho_{\rm reh}^{1/4}),
\label{N-k}
\ee
($\rho_{\rm reh}$ is the `reheat' temperature, when radiation domination 
begins.)
With $V^{1/4}\sim 10^{16}\GeV$ and instant reheating this gives
$N\sub{COBE}\simeq 62$, the biggest possible value. 
In fact, $\rho_{\rm reh}$ should probably be no bigger than
$10^{10}\GeV$ to avoid too many gravitinos \cite{subir}, and using that value
gives $N\sub{COBE}=58$, perhaps the biggest reasonable value.
With $V^{1/4}=10^{10}\GeV$, the
lowest scale usually considered, one finds $N\sub{COBE}=48$ 
with instant reheating,
and $N\sub{COBE}=39$ if reheating is delayed to just before nucleosynthesis. 

The smallest cosmological  scale that will be directly probed in the forseeable
future is perhaps 
six orders of of magnitude lower than $H_0^{-1}$, which corresponds to
 replacing $N\sub{COBE}$ by $N\sub{COBE}-6\ln10=N\sub{COBE}-14$. 

The estimates for $N\sub{COBE}$ are valid only
if there is no additional inflation,
after slow-roll inflation ends. In fact, there are least two 
possibilities for additional
inflation. One is that 
slow-roll gives way
smoothly to a significant amount of fast-roll inflation.
This does not happen in most models, but it does happen in the rather
attractive model described in Sections \ref{lisamodel} and
\ref{ss:loopcorr}. Its effect is to reduce $N\sub{COBE}$ by some amount
$N\sub{fast}$, which is highly model-dependent.\footnote
{The quantity $N\sub{fast}$ defined in this way is not identical with 
the number of $e$-folds of fast-roll inflation, since $H$ is not 
constant during such inflation. But the latter provides
a rough approximation to $N\sub{fast}$ if slow-roll 
is only marginally violated, as in Section \ref{lisamodel}.}
The other possibility is that there is a separate,
late era of thermal inflation, as described in 
footnote \ref{ordtherm} of Section \ref{hybthermal}.
The minimal assumption of one bout of thermal inflation will
reduce $N\sub{COBE}$ by $N\sub{thermal} \sim 10$. 

We want slow-roll inflation 
to generate structure on all cosmological scales. Taking the 
smallest one to correspond to $N\sub{COBE}-15$, 
and remembering that without thermal inflation $N\sub{COBE}$
is in the range $40$ to $60$,
we learn that the amount of additional inflation must
certainly satisfy
\be
N\sub{fast} + N\sub{thermal} < 25{\rm \ to\ }45\,.
\ee
In many models of inflation, $n(k)$ is strongly dependent on $N(\phi)$
at the epoch of horizon exit (see for instance the table 
on page \pageref{t:1}). Then
a more stringent limit upper limit may come from the requirement
that $|1-n|<0.2$.

>From now on, we shall usually denoted $N\sub{COBE}$ simply by $N$.
The more generally quantity $N(\phi)$, referring to an arbitrary
field value, will always have its argument $\phi$ displayed.

\subsection{Gravitational waves}

\label{s:gravwave}

Inflation also generates
gravitational waves, with two independent components
$h_{+,\times}$. Perturbing the Einstein action, one
finds that each of quantities
$(\mpl/\sqrt 2)h_{+,\times}$ has the same action as a massless scalar
field. It follows that $h_{+,\times}$ are independent gaussian
perturbations, whose spectrum on scales far outside the horizon
has the time-independent value
\cite{starob82gw,rubakov}
\be
\calp\sub{grav} (k) = \frac 2{\mpl^2} \left( \frac H{2\pi} \right)^2.
\ee
As usual, the right hand side is evaluated at the epoch of horizon
exit $k=aH$.
According to the analytic approximation mentioned earlier
\cite{starob}, 
the relative contribution 
$C\sub{grav}(\ell)/C_\calr(\ell)$,
of gravitational waves to the low multipoles,
is equal to 
\be
r\equiv 12.4\epsilon.
\label{rofeps}
\ee
We are using $r$ {\em defined by this equation}
as a convenient measure of the relative importance of 
the gravitational waves.

Using the slow-roll conditions, the 
spectral index is
\be
n\sub{grav} = -2\epsilon \,.
\label{consist2}
\ee
This is the fourth quantity we calculated {}from the 
three quantities $V$, $\epsilon$ and $\eta$, so it will provide
a consistency check if gravitational waves are ever detected.

We noted earlier that the primordial gravitational waves will 
not be detectable by Planck unless
$r\gsim 0.1$, and are unlikely to be detected in the forseeable future 
unless $r\gsim 0.01$. Most models of inflation give a much 
smaller value \cite{mygwave}. To see why, note first that the waves are 
significant only up to $\ell\sim 100$, corresponding to 
the first 4 or so $e$-folds of inflation after our Universe leaves the 
horizon. {}From \eq{Nrelation}, this means that 
the field variation is at least of 
order the Planck scale, 
\be
\Delta\phi \simeq 4 \sqrt{2\epsilon} \mpl= 0.5 \mpl (r/0.1)^{1/2}.
\ee

Afterwards, we have say $\sim 50$ $e$-folds more
inflation, which will increase the total
$\Delta\phi$. In models where $\epsilon$ increases with time
this gives
\be
\Delta\phi\gsim \frac{25}{4}\mpl (r/0.1)^{1/2} \,.
\ee
Then detectable gravitational waves require $\Delta\phi\gsim
2$ to $6\mpl$, placing the inflation model out of
theoretical control. In models where $\epsilon $ decreases with time,
the extra change in $\phi$ need not be significant, making it possible
to generate detectable gravitational waves in models
with $\Delta\phi\gsim 
0.2$ to $0.5\mpl$.
Of the models proposed so far in the framework of particle theory, only 
tree-level hybrid inflation is of the latter type ($V=V_0+\frac12m^2\phi^2$
with the first term dominating, or the same thing with a higher power
of $\phi$.) But in most versions of hybrid inflation the field is
small, the only exception so far being reference \cite{lr}.

Another viewpoint is to look at the COBE normalization \eq{vbound}.
It can be written
\be
V^{1/4} = (2.0\times 10^{16}\GeV) (r/0.1)^{1/4} \,,
\ee
so detectable waves require $V^{1/4}\gsim 1\times 10^{16}\GeV$.
Such a big value is the exception rather than the rule
for existing models.

We conclude that a detectable gravitational wave signal is 
unlikely. If such a signal is present, 
\eqsss{delh}{n2}{rofeps}{consist2}
and more accurate
versions of them will allow one to deduce $V(\phi)$ and its low
derivatives.
This is the `reconstruction' programme \cite{recon}. Note that it
will estimate $V(\phi)$ only on the limited portion of the trajectory
corresponding to the ten or so $e$-folds occurring while 
cosmological scales leave the horizon.

\subsection{Before observable inflation}

\label{before}

The only 
era of inflation 
that is directly relevant for observation is the one beginning when 
the observable Universe leaves the horizon.
This era of `observable' inflation will undoubtedly be preceded by
more inflation, but all memory of earlier inflation is lost apart {}from
the starting values of the fields at the beginning of observable
inflation. Nevertheless, one ought to try to understand the earlier era
if only to check that the assumed starting values are not ridiculous.

A complete history of the Universe will 
presumably start when  the energy density is at the Planck 
scale.\footnote
{We discount, for the moment, the fascinating possibility that 
additional space dimensions open up well below the Planck scale.
We also do not consider the idea that a complete (open or closed)
inflating universe
is created by a quantum process, with energy density already far below 
the Planck scale \cite{ht98,l98,v98,bl}. A more modest 
proposal \cite{open} is that our Universe is located within a bubble,
which nucleated at a low energy scale \cite{open}; but the universe 
within which that bubble originated is still supposed to have begun at 
the Planck scale.}
(Recall that $V^{1/4}$ is at least two orders of magnitude 
lower during observable inflation.) The usual 
hypothesis is that the scalar fields
at that epoch take on chaotically varying values as one moves around
the universe, inflation occurring in patches where conditions are 
suitable \cite{chaotic,abook}. The observable Universe is located 
in one of these patches, and from now on we consider only it.

One would indeed  like to  start the descent {}from the Planck scale
  with an era of inflation,
for at least two reasons. One, which applies only to the case of
positive spatial curvature, is
to avoid having the Universe collapse in a few Planck times (or
fine-tune the initial density parameter $\Omega$).
The other, which applies in any case, 
is to have an event horizon so that the
homogeneous patch within which we are supposed to live is
 not eaten up by its inhomogeneous
surroundings. However, there is no reason to suppose
that this initial era of inflation is of the slow-roll variety.
The motivation for slow-roll comes {}from the observed 
fact that $\delta_H$ is almost scale-independent, which applies only 
during the relatively brief era when cosmological scales are 
leaving the 
horizon. In the context of supergravity,
where achieving slow-roll inflation requires rather delicate 
conditions, it might be quite attractive to suppose that
non-slow-roll inflation takes the Universe down {}from the Planck scale with 
slow-roll setting in only much later. A well known potential 
that can give non-slow-roll inflation is
$V\propto \exp(\sqrt{2/p}\phi/\mpl)$, which gives
 $a\propto t^p$ and corresponds to non-slow-roll inflation in
the regime where $p$ is bigger than 1 but not much bigger.

Well before observable inflation,
it is possible to have an era of `eternal inflation'
during which
the motion of the inflaton field is dominated by the
quantum fluctuation.\footnote
{Eternal inflation taking place at 
large field values is discussed in detail in Ref.~\cite{eternal,eternal1}. 
The corresponding phenomenon for inflation near a 
maximum was noted earlier 
by a number of authors.}
The condition for this to occur is
that the predicted spectrum $\calp_\calr$ be 
formally bigger than 1 \cite{stochastic}.

With all this in mind, let us ask what might precede observable 
inflation, with a view to seeing what initial conditions for the latter
might be reasonable.
Going back in time, one might find a smooth
inflationary trajectory going all the way back to an era when
$V$ is at the Planck scale 
(or at any rate much bigger than its value during observable
inflation). In that case
the inflaton field will probably be decreasing during inflation.
Another natural possibility is for the inflaton to 
find itself near a maximum of the potential 
before observable inflation 
starts.
Then there may be eternal inflation followed by 
slow-roll 
inflation.
If the maximum is a fixed point of the symmetries it is quite natural for 
the field to have been driven there by its interaction with other 
fields. Otherwise it
could arrive there by accident, though this 
is perhaps only reasonable if
the distance {}from the maximum to the minimum
is $\gsim \mpl$
(see for instance Ref.~\cite{natural3} for an example).
In this latter case, the fact that eternal inflation occurs near the 
maximum may help to enhance the probability of inflation starting 
there \cite{nontherm}.
If the maximum is a fixed point,
the inflaton field might 
be placed there through a coupling with another field,
with that field initially inflating \cite{izawa2}.\footnote
{The potential will be something like \eq{fullpot1},
with $\psi$ the field corresponding to observable inflation.
One initially has hybrid inflation, but in contrast with the usual case
the destabilized field takes so long to roll down that it becomes
the single inflaton field of observable inflation.}
Alternatively, it may be 
that the inflaton field is placed at the origin
through thermal corrections to the potential
\cite{new1,new2}, but this mechanism is
difficult to implement.

In summary, two kinds of initial condition seem reasonable. One is to 
have the inflaton moving towards the origin, the idea being that 
the field value is initially at least of order $\mpl$. The other is 
to have the inflaton moving away {}from a maximum of the potential, 
preferably located at the origin. We emphasize that these are 
just speculations; to make a definite statement, one needs a definite 
model going back to the Planck scale.

\section{Calculating the curvature perturbation generated by inflation}
\label{s10}

This section is somewhat specialized, and may be omitted by the general 
reader. It concerns the calculation of the spectrum
$\calp_\calr$ of the primordial curvature perturbation $\calr$.
We first consider the standard case of a single-component inflaton;
essentially all of the models considered in the text are of this kind.
Then we explain the concept of a multi-component inflaton, and
see how to extend the calculation to that case.

In both cases we use an approach that has only recently been 
developed \cite{salopek95,ewanmisao,LL3}, though its starting point
can already be seen in the first calculations
\cite{hawking,starob82,guthpi}. This starting point consists of the 
following assumption. During any era of the early Universe,
the  evolution of the relevant quantities 
along each comoving worldline is practically the same as in an 
unperturbed Universe, after smoothing on a comoving scale that is well 
outside the horizon (Hubble distance $H^{-1}(t)$).\footnote
{`Smoothing' on a scale $R$ means that one replaces (say) 
the energy density $\rho({\bf x})$ 
by $\int d^3x' W(|{\bf x}'-{\bf x}|) \rho({\bf x}')$ with 
$W(y)\simeq 1$ for $y\lsim R$ and
$W\simeq 0$ for $y\gsim R$. A simple choice is to take
$W=1$ for $y<R$ and $W=0$ for $y>R$
(top-hat smoothing).}

The assumed condition seems very reasonable. There
needs to be {\em some} smoothing scale that makes the perturbations
negligible or it would not make sense to talk about an unperturbed
Universe.  The horizon scale will be big enough, unless there is
dramatic new physics on a much bigger scale, and the absence of an
observed Grishchuk--Zel'dovich effect \cite{gz}
or tilted
Universe effect \cite{tilted} more or less assures us that there
is no such scale. 

We shall see how a comparison of the
evolution of different comoving regions 
provides a simple and powerful technique for
calculating the density perturbation. 
As we discuss later, this approach is quite different {}from the
usual one of writing down, and then solving, a closed set of equations
for the perturbations in the relevant degrees of freedom
(for instance the components of the inflaton field during inflation).
Roughly speaking the present approach replaces the sequence `perturb then solve'
by the far simpler sequence `solve then perturb', though it is actually 
more general than the other approach. 
For the case of a single-component inflaton it gives 
a very simple, and completely general, proof of 
the constancy of $\cal R$ on scales well outside the horizon.
For the multi-component case it allows one to follow the evolution of
$\cal R$, knowing only the evolution of the {\em unperturbed}
universe corresponding to a given value of the initial inflaton field.
So far
it has been applied to three
multi-component models \cite{salopek95,davidjuan,glw}.

\subsection{The case of a single-component inflaton}

\label{a1}

We begin with a derivation of the usual result for the single-component 
case. The assumption about the evolution along each comoving worldline
is invoked only at the very end,
when it is used to establish the constancy of $\cal R$
which up till now has only been demonstrated for special cases.
Otherwise the proof is the standard one \cite{LL2,LL3}, but it provides a 
useful starting point for the multi-component case.

A few Hubble times 
after horizon exit during inflation, when $\calr(k,t)$ 
can first be regarded as a classical quantity, its spectrum 
can be calculated using the relation
\cite{kodamasasaki,LL2,LL3}\footnote
{In \cite{LL2} there is an incorrect minus sign on the right hand side.}
\be
{\cal R}({\bf x})=H\Delta\tau({\bf x}),
\label{r1}
\ee 
where $\Delta\tau$ is the separation of the comoving hypersurface
(with curvature $\cal R$)
{}from a spatially flat one coinciding with it on average.
The relation is generally true, but we apply it at an epoch
a few Hubble times after horizon exit during inflation.

On a comoving hypersurface the inflaton field $\phi$
is uniform, because the momentum density $\dot\phi {\bf \nabla}\phi$
vanishes. It follows that
\be
\Delta\tau({\bf x})=-\delta\phi({\bf x})/\dot\phi,
\ee
where $\delta\phi$ is defined on the  flat hypersurface.
Note that the comoving hypersurfaces become singular (infinitely
distorted) in the slow-roll limit $\dot\phi\to 0$, so that 
to first order in slow-roll any non-singular choice of hypersurface 
could actually be used to define $\delta\phi$.

The spectrum of $\delta\phi$ 
is calculated by assuming that well before horizon exit
(when the particle concept makes sense) $\delta\phi$ is a 
practically massless free field in the vacuum state. Using the flatness and
slow-roll conditions one finds, a few Hubble times after horizon 
exit, the famous result \cite{LL2,LL3}
${\cal P}_\phi=(H/2\pi)^2$, which leads to the 
usual formula (\ref{delh}) for the spectrum.

However, this result refers to $\cal R$ a few Hubble times after horizon
exit, and we need to check that $\cal R$ remains constant until
the radiation dominated era where we need it. To  calculate the rate of 
change of $\cal R$ we proceed as follows \cite{lythmuk,LL2,LL3}.

In addition to the energy density $\rho(\bfx,t)$ and 
the pressure $P(\bfx,t)$, we consider a locally defined
Hubble parameter 
$H(\bfx,t)=\frac13D_\mu u^\mu$ where $u^\mu$ is the four-velocity of 
comoving
worldlines and $D_\mu$ is the covariant derivative.
(The quantity $3H$ is often denoted by $\theta$ in the literature.)
The Universe is sliced into comoving hypersurfaces, and each 
quantity is split into an average (`background') plus a 
perturbation, 
\be
\rho({\bf x},t)=\rho(t) +\delta\rho({\bf x},t)
\ee
and so on. (We use the same symbol for the local and the background 
quantity since there is no confusion in practice.)
As usual, $\bf x$ is the Cartesian position-vector of a
comoving worldline and $t$ is the time. To first order,
perturbations `live' in unperturbed spacetime, since the inclusion of
the 
perturbation in the space time metric when describing the 
{\em evolution} of
a perturbation would be a second order effect.\footnote
{This includes the case that the perturbation being evolved is itself
a perturbation in the metric, such as the gravitational wave amplitude
or the spatial curvature perturbation $\cal R$.}

We ignore the anisotropic stress of the early Universe, since it is 
unlikely to affect the constancy of $\calr$ \cite{LL3}.
The locally defined quantities satisfy 
\cite{hawkingellis,lyth85,lythmuk,LL2,LL3}
\be
H^2(\bfx,t)=\mpl^{-2}\rho(\bfx,t)/3+\frac23\nabla^2 {\cal R}.
\label{localfr}
\ee
The laplacian acts on comoving hypersurfaces.
This is the Friedmann equation except that 
$K(\bfx,t)\equiv-(2/3)a^2\nabla^2\cal R$ need not be constant.
The evolution along each worldline is
\bea
\frac{d\rho(\bfx,t)}{d\tau}&=&-3H(\bfx,t)(\rho(\bfx,t)+P(\bfx,t)),
\label{cont}\\
\frac{d H(\bfx,t)}{d\tau} &=& -
H(\bfx,t)^2 - \frac12 \mpl^{-2} (\rho(\bfx,t)+3P(\bfx,t))
-\frac13\frac{\nabla^2\delta P}{\rho+P}.
\label{raych}
\eea
Except for the last term these are the same as in an unperturbed 
universe. If that term vanishes $\cal R$ is constant, but otherwise one finds
\be
\dot{\cal R}=-H\delta P/(\rho +P).
\label{rdot}
\ee
In this equation we have in mind that $\rho$ and $P$ are 
the unperturbed quantities, depending only on $t$,
though as we are working to first order in the perturbations it would 
make no difference if they were the locally defined quantities.

According to \eq{rdot}, $\cal R$ will be constant if $\delta P$ is negligible.
We now show that this is so, by first demonstrating that 
$\delta\rho$ is negligible, and then using the new viewpoint
to see that $P$ will be a practically 
unique function of $\rho$ making $\delta P$ also negligible.

{}From now on, we work with Fourier modes, represented by the same symbol,
and replace $\nabla^2$ by
$-(k/a)^2$.
Extracting the perturbations {}from
Eq.~(\ref{localfr}) gives 
\be
2\frac{\delta H}{H}=\frac{\delta\rho}{\rho } -\frac23\left(
\frac{k}{aH}\right)^2 {\cal R}.
\label{perturbfre}
\ee
This allows one to calculate the evolution of $\delta \rho$ {}from
Eq.~(\ref{cont}), but we have 
to remember that the proper-time separation of the hypersurfaces
is position-dependent. Writing
$\tau({\bf x},t)=t+\delta\tau({\bf x},t)$ we have
\cite{kodamasasaki,lythmuk,lyst} 
\be
\delta(\dot\tau)= -\delta P/(\rho+P).
\label{deltataudot}
\ee
Writing $\delta\rho/\rho\equiv (k/aH)^2 Z$ 
one finds \cite{lythmuk}
\be
(fZ)'=f (1+w) {\cal R}.
\label{zexpr}
\ee
Here a prime denotes $d/d(\ln a)$ and 
$f'/f\equiv(5+3w)/2$ where $w\equiv P/\rho$.
With $w$ and $\cal R$ constant, and dropping a decaying mode, this gives
\be
Z=\frac{2+2w}{5+3w} \cal R.
\ee
More generally, integrating Eq.~(\ref{zexpr})
will give $|Z|\sim |\cal R|$ for any reasonable variation of
$w$ and $\cal R$. Even for a bizarre variation there is no scale 
dependence in either $w$ (obviously) or in $\cal R$ (because 
Eq.~(\ref{penult}) gives it in terms of $\delta P$, and we will see that
if $\delta P$ is significant it is scale-independent).
In all cases
$\delta\rho/\rho$ becomes negligible on 
scales sufficiently far outside the horizon.\footnote
{As it stands, this analysis fails if a single oscillating field
dominates (as might happen just after inflation) because $1+w$ then passes
through zero. In that case one can consider $(1+w)\calr$. Combining
\eqs{rdot}{zexpr}, one sees that it satisfies a non-singular differential
equation, which means that it will change by a negligible amount during
each of the brief episodes when the right hand side of \eq{rdot} becomes 
non-negligible and formally goes through infinity. During such an 
episode, the comoving hypersurfaces become infinitely distorted and $\calr$
briefly loses its meaning, but what matters is that $\calr$ is 
practically constant except for these episodes.}

The discussion so far applies to each Fourier mode separately, on 
the assumption that the corresponding perturbation is small.
To make the final step, of showing that $\delta P$ is also negligible,
we need to consider the full quantities $\rho({\bf x},t)$ 
and so on. But we still want to consider only scales that are well
outside the horizon, so we suppose that all quantities are smoothed
on a comoving scale somewhat shorter than the one of interest.
The smoothing removes Fourier modes on scales shorter than the 
smoothing scale, but has practically no effect on the scale
of interest. 

Having done this, we invoke the assumption that the evolution of the 
Universe along each worldline is practically the same as in an unperturbed 
universe. In the context of slow-roll inflation, this means that 
the evolution is determined by the  inflaton field
at the `initial' epoch a few Hubble times after horizon exit.
To high accuracy, $\rho$ and $P$  are well
defined functions of the initial inflaton field and
{\em if it has only one component}
this means that they are well defined 
functions of each other. Therefore $\delta P$ will be very small
on comoving hypersurfaces because $\delta \rho$ is.\footnote
{If $k/a$ is the smoothing scale, the 
assumption that the evolution is the same as in an unperturbed
universe with the same initial inflaton field
has in general errors of order $(k/aH)^2$. 
In the single-component case, where $\delta P$ is also of this order,
we cannot use the assumption to actually 
calculate it, but neither is it of any interest.}

Finally, we note for future reference that $\delta H$ is also
negligible because of Eq.~(\ref{perturbfre}).

\subsection{The multi-component case}

So far we have assumed that the slow-rolling inflaton field is 
essentially unique. What does `essentially' mean in this context?
A strictly unique inflaton trajectory would be one lying
in a steep-sided valley in field space. This is not very likely in a 
realistic model. Rather there will be a whole 
family of possible inflaton trajectories, lying in the space of two or 
more real fields $\phi_1$, $\phi_2, \cdots$.  
Usually, though, the different trajectories are completely equivalent,
so that we still have an `essentially' unique inflaton field.
For instance, in many cases
the inflaton field is the modulus of a complex field, with $V$ 
independent of the phase. Each choice of the phase gives a different
but equivalent inflaton trajectory in the space of the complex field.
Also, there may be a field(s) $a$,
unrelated to the inflaton field, which has practically zero mass.
Different choices of $a$ lead to different inflaton 
trajectories in field space, but in the usual case that $a$ has no
cosmological effect these trajectories will again be equivalent.

In both of these cases, one can modify things so that the trajectories
are {\em inequivalent}. In the case of the complex field, it might be
that $V$ is a function of both the real and imaginary parts, call them
$\phi_1$ and $\phi_2$, with $V$ satisfying the flatness conditions
\eqs{flat1}{flat2} as a function of each field separately.
Then there will in general be a family of curved
inflaton trajectories, corresponding to 
the lines of steepest descent, which are inequivalent. 
In this case, it is useful to think of the inflaton as a two-component
object $(\phi_1,\phi_2)$. More generally, there might be a family of 
curved inflaton trajectories in the space of several fields, so that 
there is a {\em multi-component} inflaton.

In the case of an unrelated massless field $a$, that field might
survive and be stable, to become dark matter after it starts to oscillate
about its minimum. The inflaton trajectories are now inequivalent,
but the inequivalence shows up only when the oscillation starts. 
The vacuum fluctuation
of $a$ during inflation then turns into an
{\em isocurvature density perturbation}. Extensions of the standard 
model typically contain a field which can have just these properties,
namely the axion. Postponing until later the discussion of this case,
we continue discussion of the multicomponent case.

Multi-component inflaton models generally have just two components,
and are called double inflation models because the trajectory can
lie first in the direction of one field, then in the direction of the 
other.  They were first proposed in
the context of non-Einstein gravity 
\cite{starob85,d1,d2,d4,d5,d51,d11,noncanon,non1,non2,non3,davidjuan}.
By redefining the fields
and the spacetime metric one can recover Einstein gravity,
with fields that are not small on the Planck scale and
in general non-canonical kinetic terms and a non-polynomial 
potential. Then models 
with canonical kinetic terms were proposed 
\cite{c3,d7,d9,d10,d13,c1,c2,d3,d6,d8,isocurv,salopek95},
with potentials such as
$V=\lambda_1\phi_1^p+\lambda_2\phi_2^q$.
These potentials too inflate in the large-field regime
where theory provides no guidance about the form of the potential.
However there
seems to be no bar to having a multi-component model
with $\phi\ll\mpl$,
and one may yet emerge in 
a well-motivated particle theory setting.
In that case a hybrid model might emerge, though 
the models proposed so far 
are all of the 
non-hybrid type (ie., the multi-component inflaton is entirely 
responsible for the potential).

In this brief survey we have focussed on the era when cosmological 
scales leave the horizon. In the hybrid inflation
model of Ref.~\cite{lisa,glw}, 
the `other' field is responsible for the last several $e$-folds
of inflation, so one is really dealing with a
two-component inflaton (in a non-hybrid model). 
The scales corresponding to the last 
few $e$-folds are many orders of magnitude shorter than
the cosmological scales, but it turns out that the perturbation
on them is big so that black holes can be produced.
This phenomenon was investigated in Refs.~\cite{lisa,glw}.
The second reference also investigated the possible production of
topological defects, when the first field is destabilized.

\subsection{The curvature perturbation}

It is assumed that while cosmological scales are leaving the horizon
all components of the inflaton have the slow-roll behaviour
\be
3H\dot \phi_a = - V_{,a}.
\ee
(The subscript $,a$ denotes the derivative with respect to $\phi_a$.)
Differentiating this
and comparing it with the exact expression $\ddot\phi_a
+3H\dot \phi_a +V_{,a}=0$ gives consistency provided that
\bea
\mpl^2 (V_{,a}/V)^2 &\ll& 1,\\
\mpl^2 |V_{,ab}/V| &\ll& 1.
\label{flat22}
\eea
(The second condition could actually be replaced by a weaker one but let 
us retain it for simplicity.)
One expects slow-roll to hold if these flatness conditions are 
satisfied. Slow-roll plus the first flatness condition
imply
that $H$ (and therefore $\rho$) is slowly varying, giving 
quasi-exponential inflation. The second flatness condition 
ensures that $\dot\phi_a$ is slowly varying.

It is not necessary to assume that
all of the fields continue to slow-roll after cosmological scales 
leave the horizon. For 
instance, one or more of the fields might start to oscillate, while the others
continue to support quasi-exponential inflation, which ends only
when slow-roll fails for all of them.
Alternatively, the oscillation of 
some field might briefly interrupt inflation, which resumes when its 
amplitude becomes small enough. (Of course these things might happen 
while cosmological scales leave the horizon too, but that case 
will not be considered.)

The expression (\ref{r1}) for 
$\cal R$ still holds in the multi-component case. Also, 
one still has 
$\Delta\tau=-\delta\phi/\dot\phi$ if $\delta\phi$ 
denotes the component of the vector $\delta\phi_a$
parallel to the trajectory. A few Hubble times after horizon exit
the spectrum of 
every component of the vector 
$\delta\phi_a$, in particular the parallel one,
is still $(H/2\pi)^2$. If $\cal R$ had no subsequent variation this 
would lead to the usual prediction, but we are considering the case
where the variation is significant.
It is given in terms of $\delta P$ by 
Eq.~(\ref{rdot}), and when $\delta P$ is significant it can be calculated
{}from the assumption that the evolution along each
worldline is the same as for an unperturbed universe with the same
initial inflaton field. This will give 
\be
\delta P=P_{,a}\delta\phi_a,
\ee
where $\delta\phi_a$ is evaluated at the initial epoch and the 
function $P(\phi_1,\phi_2,\cdots,t)$ represents the evolution of 
$P$ in an unperturbed universe. Choosing the basis so that one of 
the components is the parallel one, and remembering that all components
have spectrum $(H/2\pi)^2$, one can calculate 
the final spectrum of $\cal R$. The only input is the evolution of 
$P$ in the unperturbed universe corresponding to a generic initial 
inflaton field (close to the classical initial field).

In this discussion we started with Eq.~(\ref{r1}) for the initial
$\cal R$, and then invoked Eq.~(\ref{rdot}) to evolve it.
The equations 
can actually be combined to give 
\be
{\cal R}=\delta N,
\ee
where $N=\int Hd\tau$ is the number of Hubble times 
between the initial flat hypersurface and the final comoving one on 
which $\cal R$ is evaluated. This remarkable expression was given
in Ref.~\cite{starob85} and proved in Refs.~\cite{salopek95,ewanmisao}.
The approach we are using is close to the one in the
last reference.

The proof that
Eqs.~(\ref{r1}) and (\ref{rdot}) lead to ${\cal R}=\delta N$ 
is very simple. First
combine them to give
\be
{\cal R}({\bf x},t)=H_1\Delta\tau_1({\bf x})-
\int_{t_1}^{t} H(t) \frac{\delta P}{\rho+P},
\label{penult}
\ee
where $t_1$ is a few Hubble times after horizon exit.
Then use Eq.~(\ref{deltataudot}) to give
\be
{\cal R}({\bf x},t)=H_1\Delta\tau_1({\bf x})+
\int_{t_1}^{t} H(t) \delta \dot\tau({\bf x},t)dt.
\label{penult1}
\ee
As we remarked at the end of Section \ref{a1},
$\delta H$ is negligible. As a  result,
this can be written
\be
{\cal R}({\bf x},t)=H_1\Delta\tau_1({\bf x})+
\delta \int_{t_1}^{t} H({\bf x},t) \dot\tau({\bf x},t)dt.
\label{ult}
\ee
Finally redefine $\tau({\bf x},t)$ so that it
vanishes on the
initial {\em flat} hypersurface, which gives the desired 
relation ${\cal R} = \delta N$.

In Ref.~\cite{ewanmisao} this relation is derived using an 
arbitrary smooth interpolation of hypersurfaces
between the initial and final one, rather than by making the sudden jump 
to a comoving one. Then $H$ is replaced by the corresponding quantity 
$\tilde H$ 
for worldlines orthogonal to 
the interpolation (incidentally making $\delta \tilde H$ non-negligible).
One then finds ${\cal R}=\delta\tilde N$. One also finds that the right 
hand side is independent of the choice of the interpolation, as it must 
be for consistency. If the interpolating hypersurfaces are chosen to be
comoving except very near the initial one, $\tilde N\simeq N$ which 
gives the desired formula ${\cal R}=\delta N$.\footnote
{The last step is not spelled out in Ref.~\cite{ewanmisao}.
The statement that $\tilde N$ is independent of the interpolation 
is true only on scales well outside the 
horizon, and its physical interpretation is unclear though it drops out 
very simply in the explicit calculation.}

\subsection{Calculating the spectrum and the spectral index}

Now we derive explicit formulas for the spectrum and the spectral index, 
following \cite{ewanmisao}. 
Since the evolution of 
$H$ along a 
comoving worldline will be the same as for a homogeneous universe
with the same initial inflaton field, $N$ is a function only of this
field and we have
\be
{\cal R}=N_{,a} \delta\phi_a.  
\ee
(Repeated indices are summed over
and the subscript $,a$ denotes differentiation with respect to $\phi_a$.)
The perturbations $\delta\phi_a$ are Gaussian random fields generated by the
vacuum fluctuation, and have a common 
spectrum $(H/2\pi)^2$. The
spectrum $\delta_H^2\equiv (4/25){\cal P}_{\cal R}$
is therefore 
\be
\delta_H^2= \frac{V}{75\pi^2\mpl^2}N_{,a}N_{,a}.
\ee
In the single-component case, $N'=\mpl^{-2} V/V'$
and we recover the usual expression. In the multi-component case
we can always choose the basis 
fields so that while cosmological scales are leaving the horizon
one of them points along the inflaton trajectory, and then its
contribution gives the standard result with the orthogonal
directions giving an additional contribution.
Since the spectrum of gravitational waves is independent of the number 
components (being equal to a numerical constant times $V$)
the relative contribution $r$ of gravitational waves to the cmb
is always {\em smaller} in the multi-component case.

The contribution {}from the orthogonal directions
depends on the whole inflationary potential after the relevant scale
leaves the horizon, and maybe even on the evolution of the energy 
density after inflation as well. This is in contrast to the contribution 
{}from the parallel direction which depends 
only on $V$ and $V'$ evaluated
when the relevant scale leaves the horizon.
The contribution {}from the orthogonal directions will be at most of order
the one {}from the parallel direction provided that all $N_{,a}$ are at 
most of order $\mpl^{-2} V/V'$.
We shall see later that this is a 
reasonable expectation at least if $\cal R$ stops varying after the end 
of slow-roll inflation.

To calculate the spectral index we need the analogue of
Eqs.~(\ref{nexpression}) and (\ref{Nrelation}). Using
the chain rule and $dN=-Hdt$ one finds
\bea
\frac{d}{d\ln k}
&=&-\frac{\mpl^2}{V}V_{,a} \frac{\partial}{\partial\phi_a},\\
N_{,a} V_{,a} &= & \mpl^{-2} V.
\label{nv}
\eea
Differentiating the second expression gives
\be
V_{,a} N_{,ab}+ N_{,a} V_{,ab} = \mpl^{-2}V_{,b}.
\ee
Using these results one finds
\be
n-1 = -\frac{\mpl^2V_{,a}V_{,a}}{V^2}
-\frac2{\mpl^2N_{,a}N_{,a}}
+2\frac{\mpl^2N_{,a} N_{,b} V_{,ab} }
{VN_{,d}N_{,d}}.
\label{multin}
\ee
Again, we recover the single field case using $N'=\mpl^{-2} V/V'$.

Differentiating this expression and setting $\mpl=1$ for clarity
gives
\bea
\frac{dn}{d\ln k}&=&
-\frac {2}{V^3}V_{,a} V_{,b} V_{,ab}
+\frac2{V^4} (V_{,a}V_{,a})^2
+\frac 4 V \frac{(V-N_{,a}N_{,b} V_{,ab} )^2}{(N_{,d}N_{,d})^2}\nonumber\\
&+&\frac2 V \frac{N_{,a} N_{,b} V_{,c} V_{,abc} }{N_{,d}N_{,d}}
+\frac4 V \frac {( V_{,c} -N_{,a} V_{,ac}) N_{,b} V_{,bc} }
{N_{,d}N_{,d}}.
\eea
A correction
to the formula for $n-1$ has also been worked out \cite{nak}. 
Analogously with the single-component case, both this correction 
and the variation of $n-1$ involve the first, second and third 
derivatives of $V$.
Provided that the derivatives of $N$ in the orthogonal directions are 
not particularly big, and barring cancellations, 
a third flatness 
condition 
$V_{,abc}V_{,c}/V^2\ll \max\{\sum_a(V_{,a})^2,\sum_{ab}|V_{,ab}|\}$
ensures that both the correction
and the variation of $n-1$ in  a Hubble time are small.
(One could find a weaker condition that would do the same job.)

These formulas give the spectrum and spectral index of the 
density perturbation, if one knows the evolution
of the homogeneous universe corresponding both to the
classical inflaton trajectory and to nearby trajectories.
An important difference in principle {}from the single-component
case, is that the classical trajectory is not uniquely specified by 
the potential, but rather has to be given as a separate piece of 
information. However, if there are only two components the classical 
trajectory can be determined {}from the COBE normalization of the 
spectrum, and then there is still a prediction for the spectral index.

This treatment 
can be generalized straightforwardly \cite{ewanmisao}
to the case of
non-canonical kinetic terms  described by \eq{hmetric}.
However, in the regime where all fields are $\ll\mpl$
one expects the 
`curvature', associated
with the `metric' $H_{ab}$ in \eq{hmetric}
to be negligible, and then one can recover 
the canonical normalization 
$H_{ab}=\delta_{ab}$ by redefining the 
fields.

\subsection{When will $\cal R$ become constant?}

We need to evaluate $N$ up to the epoch where ${\cal R}=\delta N$ 
has no further time dependence. When will that be?

As long as all fields are slow-rolling, $\cal R$ is constant if and only 
if the inflaton trajectory is straight.
If it turns through a small angle $\theta$,
and the trajectories have not converged appreciably since horizon exit,
the fractional change in $\cal R$ is in fact $2\theta$.\footnote
{Thinking in two 
dimensions and taking the trajectory to be an arc of a circle,
a displacement  $\delta\phi$
towards the center decreases 
the length of the trajectory by an amount $\theta\delta\phi$,
to be compared with the decrease $\delta\phi$ for the same
displacement along the trajectory. (The rms displacements will indeed
be the same if the trajectories have not converged.) The 
speed along the new trajectory is faster in inverse proportion to the 
length since it is proportional to $V'$
and $V$ is fixed at the 
initial and final points on the trajectory. Thus the perpendicular 
displacement increases $N$ by $2\theta$ times the effect of a 
parallel displacement, for $\theta\ll 1$.}
Since slow-roll
requires that the change in the vector $\dot\phi_a$ during 
one Hubble time is negligible, the total angle turned is $\ll N$.
Hence the relative contribution of the orthogonal directions cannot
be orders of magnitude bigger than the one {}from the parallel 
direction, if it is generated during slow-roll inflation.
(In two dimensions the angle turned cannot exceed $2\pi$ of course, but 
there could be say a corkscrew motion in more dimensions.) 
Later slow-roll may fail for one or more of the fields,
with or without interrupting inflation, and things become more 
complicated, but in general there is no reason why $\cal R$ should stop varying
before the end of inflation.

Now let us ask what happens after the end of inflation 
(or to be more precise,
after significant particle production has spoiled the above analysis,
which may happen a little before the end).
The simplest case is if 
the relevant trajectories have practically converged to a 
single trajectory $\phi_a(\tau)$, as in Ref.~\cite{glw}.
Then $\cal R$ will not vary any more (even after inflation is over)
as soon as the trajectory has
been reached. Indeed, setting $\tau=0$ at the end of 
inflation, this unique trajectory corresponds to a post-inflationary universe 
depending only on $\tau$. The fluctuation in the initial field values 
causes a fluctuation $\Delta\tau$ 
in the arrival time at the end of inflation, leading to a 
time-independent ${\cal R}=\delta N
=H_{\rm end}\Delta\tau$. 

What if the trajectory is not unique at the end of inflation?
Immediately following inflation there might be 
a quite complicated situation, with `preheating' 
\cite{kls,r1,r2,r3,r4,r5,r6,r7,r8,r9}
or else the quantum fluctuation of the `other' field in hybrid models
\cite{cllsw} converting most of the inflationary potential energy
into marginally relativistic particles in much less than a Hubble time.
But after at most a few Hubble times
one expects to 
arrive at a matter-dominated era so that $\cal R$ is constant.
Subsequent events will not cause $\cal R$ to vary provided that they 
occur at definite values of the energy density, since again $P$ will 
have a definite relation with $\rho$.
This is indeed the case for the usually-considered events, such as 
the decay of
matter into radiation and thermal
phase transitions
(including thermal inflation).
The conclusion is that it is 
reasonable to suppose that $\cal R$ achieves a constant
value at most a few Hubble times after inflation, which is maintained
until horizon entry except possibly for
the large-scale isocurvature effect mentioned in Section \ref{a8}.
On the other hand one cannot exclude the possibility that one of the 
orthogonal components of the inflaton provides a significant additional 
degree of freedom, allowing $\cal R$ to have additional variation before 
we finally arrive at the radiation-dominated era preceding the present 
matter-dominated era. 

\subsection{Working out the perturbation generated by slow-roll
inflation}

If $\cal R$ stops varying by the end of inflation, the
final hypersurface can be located
just before the end (not necessarily at the very end
because that might not correspond to a hypersurface of constant energy
density). Then, knowing the potential
and the hypersurface in field space that corresponds to the end of
inflation, one can work out $N(\phi_1,\phi_2,\cdots)$
using the equations of motion for the fields, and the expression
\be
3\mpl^2 H = \rho =V+\frac12\frac{d\phi_a}{d\tau}\frac{d\phi_a}{d\tau}.
\ee
To perform such a calculation it is not necessary that
all of the fields continue to slow-roll after cosmological scales leave
the horizon. In particular, the oscillation of
some field might briefly interrupt inflation, which resumes when its
amplitude becomes small enough.
If that happens it may be necessary to take
into account `preheating' during the interruption.

In general all this is quite complicated, but there is one 
case that may be extremely simple, at least 
in a limited regime of parameter space.
This is the case
\be
V=V_1(\phi_1) + V_2(\phi_2) + \cdots
\ee
with each $V_a$ proportional to a power of $\phi_a$.
For a single-component inflaton this gives inflation ending at 
$\phi_{\rm end} \simeq \mpl$, with cosmological scales leaving
the horizon at $\phi\gg\phi_{\rm end}$.
If the potentials $V_a$ are identical we recover that case.
If they are different, slow-roll
may fail in sequence for the different components, 
but in some regime of parameter space
the result for
$N$ (at least) might be
the same as if it failed simultaneously for all components.
If that is the case one can derive simple formulas \cite{d7,salopek95},
provided that cosmological scales leave the horizon
at $\phi_a\gg \phi_a^{\rm end}$ for all components.

One has
\be
Hdt=-\mpl^{-2}\frac{V}{V_1'}d\phi_1=
-\mpl^{-2}\sum_a \frac{V_a}{V_a'} d\phi_a.
\ee
It follows that 
\be
N=\mpl^{-2}\sum_a \int_{\phi_a^{\rm end}}^{\phi_a}\frac{V_a}{V_a'} d\phi_a.
\ee
Since each integral is dominated by the endpoint $\phi_a$, we have 
$N_{,a} = \mpl^{-2}V_a/ V_a'$ and
\be
\delta_H^2 =\frac{V}{75\pi^2\mpl^6} \sum_a \left(\frac{V_a}{V_a'}
\right)^2.
\ee
The spectral index is given by
Eq.~(\ref{multin}),
which simplifies slightly
because $V_{,ab}=\delta_{ab}V_a''$.

The simplest case is 
$V=\frac12m_1^2\phi_1^2+\frac12m_2^2\phi_2^2$.
Then $n$ is given by the following formula
\be
1-n= \frac1 N\left[
\frac{(1+r)(1+\mu^2 r)}{(1+\mu r)^2} + 1\right],
\ee
where $r=\phi_2^2/\phi_1^2$ and $\mu=m_2^2/m_1^2$.
If $\mu=1$ this reduces to the single-component formula
$1-n=2/N$.
Otherwise it
can be 
much bigger, but note that our assumptions will be valid
if at all in a restricted region of the $r$-$\mu$ plane.

\subsection{An isocurvature density perturbation?}

\label{a8}

Following the astrophysics usage, we classify a density perturbation as 
adiabatic or isocurvature with reference to its properties at some epoch 
{\em 
during the radiation-dominated era preceding the present 
matter-dominated era}, while it is
still far outside the horizon.
For an adiabatic density perturbation, the density of each particle 
species is a unique function of the total energy density. 
For an 
isocurvature density perturbation the total
density perturbation vanishes, but those of the individual particle 
species do not. 
The most general density perturbation is the sum of an adiabatic and
an isocurvature perturbation, with $\cal R$ specifying the adiabatic density 
perturbation only.

For an isocurvature perturbation to exist the universe 
has to possess more than the single degree of freedom provided by the 
total energy density.
If the inflaton trajectory is unique, or has become so by the end of 
inflation, there is only the single degree of freedom corresponding to 
the fluctuation back and forth along the trajectory and 
there can be no isocurvature perturbation. 
Otherwise one of the orthogonal fields
can provide the necessary degree of freedom.
The simplest way for this to happen is for the orthogonal field 
to survive, and acquire a potential so that it
starts to oscillate and becomes matter.\footnote
{If the potential  of the `orthogonal' field
already exists during inflation the inflaton
trajectory will have a tiny component in its direction, so that it is 
not strictly orthogonal to the inflaton trajectory. This makes no 
practical difference. In the axion case the potential is usually 
supposed to be generated by QCD effects long after inflation.}
The start of the oscillation 
will be determined by the total energy density, but its amplitude
will depend on the initial field value so there will be an isocurvature 
perturbation. It will be compensated,
for given energy density, by the perturbations in the other species of 
matter and radiation which will continue to satisfy the adiabatic
condition $\delta\rho_m/\rho_m=\frac34\delta\rho_r/\rho_r$.

The classic example of this is the axion field
\cite{abook,kt,myaxion}, which is simple because
the fluctuation in the direction of the axion 
field causes no adiabatic density 
perturbation, at least in the models
proposed so far. The more general case, where one of the components of 
the inflaton may cause both an adiabatic and an 
isocurvature perturbation 
has been looked at in for instance Ref.~\cite{isocurv}, though not in the 
context of specific particle physics.
If an isocurvature perturbation in the non-baryonic dark matter density
exists,
 it must not conflict with observation and this imposes strong 
constraints on, for instance, models of the axion
\cite{myaxion,LIN2SC}. 

An isocurvature perturbation in the density of a species of matter
may be defined by 
the `entropy perturbation' \cite{kodamasasaki,lyst,LL2,LL3}
\be
S= \frac{\delta\rho_m}{\rho_m}-\frac34\frac{\delta\rho_r}{\rho_r},
\ee
where $\rho_m$ is the non-baryonic dark matter density.
Equivalently, $S=\delta y/y$, where $y=\rho_m/\rho_r^{3/4}$.
Since we are dealing with scales far outside the horizon,
$\rho_m$ and $\rho_r$ evolve as they would in an unperturbed 
universe which means that $y$ is constant and so is $S$.
Provided that the field fluctuation is small $S$ will be proportional to 
it, and so will be a Gaussian random field with a nearly flat spectrum
\cite{myaxion,LL2,LL3}.

For an isocurvature perturbation, $\cal R$ vanishes 
during the radiation dominated era
preceding the present matter
dominated era. But on the
very large scales entering the horizon well after matter 
domination, $S$ generates a nonzero $\cal R$ during 
matter domination, namely
${\cal R}=\frac13 S$. A simple way of seeing this, which 
has not been noted before, is through the relation
(\ref{rdot}).
Since $\delta\rho=0$, one has $S=-(\rho_m^{-1}+\frac34\rho_r^{-1})
\delta\rho_r$. Then, using $\delta P=\delta\rho_r/3$, 
$\rho_r/\rho_m\propto a$ and $Hdt=da/a$ one finds
the quoted result by integrating Eq.~(\ref{rdot}).

As discussed for instance in Ref.~\cite{LL2,LL3}, the large-scale
cmb anisotropy coming {}from an isocurvature
perturbation is $\Delta T/T=-(\frac13+\frac 1{15})S$, where 
$S$ is evaluated on the last-scattering surface. The 
second 
term is the Sachs-Wolfe effect coming {}from the curvature perturbation we 
just calculated, and the first term is the anisotropy $\frac14
\delta\rho_r/\rho_r$ just after last scattering (on a comoving 
hypersurface). By contrast the anisotropy {}from an adiabatic perturbation
comes only {}from the Sachs-Wolfe effect, so for a given 
large-scale density 
perturbation the 
isocurvature perturbation gives an anisotropy six times bigger.
As a result
an isocurvature perturbation with a flat spectrum
cannot be the dominant contribution to
the cmb, though one could contemplate a small contribution
\cite{iso}.

\section{Field theory and the potential}
\label{s4}

All models of inflation assume the 
validity of field theory, and in particular the existence of a 
potential $V$ which is a function of the scalar fields.
In this section we discuss, in an elementary way, the form of the scalar 
field potential that one might expect on the basis of particle theory.

\subsection{Renormalizable versus non-renormalizable theories}

\label{rvnon}

A given field theory, like the Standard
 Model or a supersymmetric extension of it, 
is nowadays regarded as an effective theory. Such a theory is
valid when the (biggest) relevant 
energy scale is less than some `ultraviolet cutoff', which we shall denote by
$\mpltil$. 
In the context of collider physics, the relevant energy scale
is usually the collision energy.
In the context of inflation, it is usually 
the value of the 
inflaton field. It will be helpful to keep these two cases
in mind.

In the most optimistic case, $\mpltil$ 
will be the Planck
scale $\mpl$. For field
theory in three space dimensions, 
$\mpltil$ presumably cannot be higher than $\mpl$,
since at that scale the theory will be
invalidated by effects
like the quantum fluctuation of the spacetime metric.
But it might be lower. In weakly coupled heterotic string theory
it is suggested (Section \ref{dilgs})
that $\mpltil$ is the string scale $M\sub{str}
\simeq g\sub{str}\mpl$, 
where $g\sub{str}^2\sim$ 1 to $0.1$ is the gauge coupling at the string scale. 

At high scales, $n$ compactified space dimensions may
become relevant. In that case, 
the biggest possible value of $\mpltil$ is presumably
the Planck scale $M_{4+n}$ for gravity with these extra dimensions.
It is typically lower than $\mpl$, as the following argument shows.
If $R$ is the size of the compactified dimensions (assumed to be all 
equal) the
Newtonian gravitational force $1/(\mpl r)^2$ 
is valid only for $r\gg R$, and for $r\ll R$ it turns into
$1/(M_{4+n} r)^{2+n}$ where $M_{4+n}$ 
is the Planck scale for gravity with the $n$
extra dimensions. Matching these expressions 
at the scale $r\sim R$ one learns that
$(M_{4+n}/\mpl)^{2+n} \sim (\mpl R)^{-n}$. The right hand side is less than 
1, or it would not make sense to talk about the extra dimensions.

At least if one is dealing with a field theory
in which the fields are confined to the three space dimensions,
$M_{4+n}$ may be a useful estimate of the appropriate renormalization 
scale.
This is what happens in Horava-Witten 
M-theory \cite{horwit,mtheory}
(there are two sets of fields, each confined to a different
three dimensional space). There is one extra dimension
(plus much smaller ones that we do not consider), with 
$(M_5/\mpl) \sim 0.1 $ and therefore $(\mpl R)^{-1}\sim 10^{-3}$.
Another proposal (Reference \cite{add,aadd} 
and earlier ones cited there)
invokes $n=2$, $M_6\sim 1\TeV$
and therefore $R\sim 1$\,mm.

If the fields of the inflation model are confined to three space
dimensions, extra dimensions {\em per se}
should make no difference provided that their size is much less than the
Hubble distance during inflation. As one easily verifies, this is 
automatic for
$n\geq 2$, given the condition 
$V^{1/4} < M_{4+n}$ that certainly needs to be imposed.
It will also be the case in
Horava-Witten M theory, since the COBE bound \eq{vbound} requires
$H\lsim 10^{-3}\mpl$.

There is also the proposal that the cutoff is inversely
related to the size $L$ of the region that is to be described.
For instance, \cite{ckn} suggest that one needs 
$\mpltil \lsim \sqrt{\mpl/L}$; with a 
box size a few times bigger than the Hubble distance (used
when calculating the 
vacuum fluctuation of the inflaton) this would give
$\mpltil\sim V^{1/4}$ \cite{ckn}.

All of this refers to the cutoff for a field theory including all
of the fields in Nature. In the context of terrestrial and
 astrophysics, one often considers an an effective theory, 
obtained by integrating out fields with mass
$\gsim \mpltil$.\footnote
{\label{intout} For the present purpose, integrating fields  out 
(of the action) can be taken to mean that the scalar
field potential is minimized with respect to them, at fixed values of 
the fields which are not integrated out.
This gives a well-defined field theory, 
if the motion of the integrated-out fields about this minimum is 
negligible. That will always be the case if their masses
are much bigger than those of the fields that are not integrated out.
It is also the case when the coupling between the two sets of fields 
is of only gravitational strength, even if the integrated-out fields
are not particularly heavy; an example is provided by the
dilaton and bulk
moduli of string theory, which are usually integrated out when considering 
the other (`matter') fields.}
In the simpler context of inflation model-building
though, it is usually supposed that $\mpltil$ is associated with the 
breakdown of field theory itself.
In general, we shall assume that $\mpltil\sim\mpl$,
while recognizing occasionally the important possibility of a 
lower value.

A field theory is specified by the lagrangian (density) $\cal L$,
such that the action is $S=\int d^4x \cal L$. 
It has dimension [energy]${}^4$, and is a function
of the fields and their derivatives with respect to space and
time. 
In a given theory the lagrangian will contain parameters, that 
define the masses of the particles and their interactions.
In a renormalizable theory, 
the number of parameters is finite, even after quantum effects
are included; the Standard Model is such a theory.
Nowadays, a renormalizable theory is regarded as an approximation
to a non-renormalizable one.
The non-renormalizable theory is supposed to be a complete description 
of nature, on energy scales $\lsim\mpl$. 

The non-renormalizable theory contains an infinite number of 
parameters, which may be thought of as summarizing the unknown
Planck-scale physics, and it can be replaced by the renormalizable
theory in any situation 
where $\mpl$ can be regarded as infinite.

We are focussing on supersymmetric theories, which can be either 
renormalizable or non-renormalizable. 
Supergravity, which is presumed to be the version of supersymmetry
chosen by nature, is non-renormalizable.
A simpler version, called global supersymmetry, 
can be renormalizable.
Following the usual practice, we shall take 
the term `global supersymmetry' to denote 
a version that is renormalizable,
with the possible exception of terms appearing in the
the superpotential; see Section \ref{s6}.

 In the usual situations, including most models of inflation,
global supersymmetry is supposed to be valid.
Global supersymmetry may be broken either explicitly or softly
(see below) and both possibilities are considered for inflation models.
An important consideration for inflation model-building
is the fact that soft susy breaking
coming from the underlying supergravity theory (gravity-mediated susy
breaking) has to be weaker than would be 
expected for a generic theory. Several proposals have been made
for achieving this, and at present there
is no consensus about which one is correct. 

\subsection{The lagrangian}

\label{thelag}

The fields can be classified according to the spin of the corresponding 
particles; in the Standard Model one has spin 0 (Higgs),
spin $1/2$ (quarks and leptons) and spin $1$ (gauge bosons).
Fields with these spins are ubiquitous in extensions of the Standard 
Model. There is also the graviton with spin $2$, and 
according to 
supergravity the gravitino with spin $3/2$.
At the particle physics level, a model of inflation consists of the
relevant part of the lagrangian.

The spin-$0$ fields are called { scalar fields}, and 
they are what we need for inflation.
Happily, there are lots of scalar fields in supersymmetric 
extensions of the Standard Model. This is because every spin $1/2$ 
field is accompanied by either a spin 0 or a spin 1 field,
with the first case ubiquitous.

During inflation, only scalar fields exist in the Universe.
At the classical level, their evolution  is determined by that
part of the Lagrangian $\cal L$ containing only the scalar fields.
If only a single, real, scalar field is relevant, the Lagrangian
in flat spacetime is of the form
\be
{\cal L} = \frac12 \partial_\mu \phi \,
        \partial^\mu \phi -V(\phi).
\label{scallag}
\ee
In this expression, $V(\phi)$ is the potential.
The other term is called the kinetic term, and in it
$\partial_\mu$ denotes the spacetime derivative
$\partial/\partial x^\mu$.
Up to a field
redefinition, this is the only Lorentz-invariant expression containing
first derivatives but no higher.
The resulting equation of motion is 
\be
\ddot \phi - \nabla^2\phi + V' (\phi) = 0,
\label{fieldeq}
\ee
where the prime denotes $d/d\phi$.

For a spatially homogeneous field this becomes
\be
\ddot\phi + V'(\phi) = 0 \,.
\ee
This is the same as for a particle moving in one dimension, with 
position $\phi(t)$ and potential $V(\phi)$.

The assumption of flat spacetime corresponds to Special Relativity and 
negligible gravity. In the expanding Universe we need General 
Relativity, describing curved spacetime. Its effect on the field 
equation  is to introduce an extra term $-3H\dot\phi$ on the left hand 
side, so that we get \eq{phiddot}. This is analogous to a friction
term for particle motion. The extra term is significant only in 
the context of cosmology.

With a suitable choice of the origin, a
non-interacting (free) field has the
potential $V=m^2\phi^2/2$ where $m$ is the mass
of the corresponding particle. 
The field equations has a time-independent,
spatially homogeneous, solution $\phi=0$, which represents the vacuum.
Plane waves, corresponding to
oscillations around the vacuum state,
correspond after quantization to non-interacting
particles of the species $\phi$, which have mass $m$.
Self interactions
correspond to 
higher-order terms in $V(\phi)$. In a renormalizable theory,
only cubic and quartic terms are allowed. The cubic term is usually 
forbidden by a symmetry, and dropping it the potential is
\be
V=\frac12 m^2 \phi^2 +\frac14 \lambda\phi^4.
\label{vren}
\ee
It is assumed that $\lambda\lsim 1$, because otherwise the interaction
would become so strong that $\phi$ would not correspond to a physical
particle (the non-perturbative regime).
On the other hand, values of $\lambda$ very
many orders of magnitude less than 1 are not usually envisaged since 
they would represent fine-tuning.

The full potential will have an infinite number of terms,
and including the cubic one for generality one can write
\be
V(\phi)
=V_0 + \frac12 m^2\phi^2 + \lambda_3 \mpl\phi^3 +\frac14\lambda\phi^4
+ \sum_{d=5}^{\infty} \lambda_d\mpl^{4-d}\phi^d
+\cdots \,.
\label{power}
\ee
The non-renormalizable ($d>4$) 
couplings $\lambda_d$ are generically of order 1, though they may be 
suppressed in a supersymmetric theory as we shall discuss.

All this extends to the case of several scalar fields $\phi$, $\psi$, etc.
With two fields, the simplest lagrangian density is
\be
{\cal L}= \frac12 \partial_\mu \phi \,
        \partial^\mu \phi +
\frac12 \partial_\mu \psi \,
        \partial^\mu \psi 
	- V(\phi,\psi).
\label{2scallag}
\ee
The field equations are
\bea
\ddot \phi - \nabla^2\phi + \frac{\partial V(\phi,\psi)}{\partial \phi} &=& 
0,
\\
\ddot \psi - \nabla^2\psi + \frac{\partial V(\phi,\psi)}{\partial \psi} &=& 
0.
\eea
The extension to further fields is similar. 

The potential as a function of all the fields will be a power series.
With the origin in field space chosen to be the vacuum, as we are 
assuming at the moment, the power series for each field will have the form
\eq{power} (no linear term) provided that the other fields are fixed
at the origin.

It is often appropriate to combine two real fields $\phi_1$ and $\phi_2$ 
into a single complex field, defined by convention as
\be
\phi = \frac1{\sqrt2 } \left( \phi_1+ i\phi_2 \right).
\label{compdef}
\ee
The kinetic term corresponding to \eq{2scallag} is
\be
{\cal L}\sub{kin} = 
\partial_\mu \phi^* \,\partial^\mu \phi.
\label{compkin}
\ee
The use of a complex field is 
particularly appropriate if the potential depends only on $|\phi|$.
Then \eq{vren} is replaced by
\be
V(\phi) = V_0 + m^2|\phi|^2 + \frac14 \lambda |\phi|^4.
\label{vren2}
\ee
Complex fields are, in any case, part of the language of supersymmetry.

With two or more real fields, it is no longer true that the most general
Lorentz-invariant lagrangian density $\cal L$ can be reduced to the above 
form, \eq{2scallag}, 
by 
a field redefinition. For several real fields 
$\phi_n$, the most general kinetic term
involving derivatives is 
\be
{\cal L}\sub{kin} = \sum_{m,n} H_{mn} \partial_\mu \phi_m
\,\partial^\mu \phi_n,
\label{hmetric}
\ee
where $H_{mn}$ is an arbitrary function of the fields.
In a supersymmetric theory, all fields are complex and the most
general kinetic term has the more restricted form
\be
{\cal L}\sub{kin} = \sum_{m,n} K_{mn^*} \partial_\mu \phi_m
\,\partial^\mu \phi_n^* \,.
\label{kmetric}
\ee
where $K_{mn^*}\equiv\pa^2 K/\pa\phi_m\pa \phi_n^*$ 
and $K$ is called the K\"ahler potential.

We recover the canonical expression
\eq{2scallag} only with the canonical choice
if $K_{mn^*}=\delta_{mn}$.  With
more than one field it is not in general possible to recover this form
by a field redefinition. If it is impossible, the space of the fields
is said to be curved.  One 
expects that the 
curvature scale will be of
order $\mpl$, allowing one to choose $K_{mn^*}=
\delta_{mn}$ to high accuracy in the regime 
$|\phi_n|\ll \mpl$.\footnote
{This is the case if the origin $\phi_n=0$ is chosen to 
so that the vacuum values of the fields are $\ll\mpl$. 
As we shall see soon, a different choice is more natural for certain 
fields predicted by string theory.}

A non-canonical kinetic term modifies the field equations, so that the
slope of the potential no longer has its usual significance. 
Canonical normalization is assumed when describing slow-roll 
inflation.

\subsection{Internal symmetry}

\subsubsection{Continuos and discrete symmetries}

\label{intern}

In addition to Lorentz invariance, the action will usually be invariant
under a group of transformations acting exclusively on the fields, with
no effect on the spacetime indices. This is called an { internal 
symmetry}. 

Consider first the case of a single real field $\phi$, with
$V$ a function of $\phi^2$ as for example in \eq{vren}.
Then there is invariance under the $Z_2$ group $\phi\to-\phi$.
Invariance under a group 
like this, which has only discrete elements, is called a discrete 
symmetry. 

Now consider the case of a single complex field, with
$V$ depending on $|\phi|$ as for example in \eq{vren2}.
Then there is invariance under the $U(1)$ group
\be
\phi\to e^{i\chi}\phi \mbox{\ \ \ \ $\chi$\ arbitrary} \,,
\label{mattersym}
\ee
with $\chi$ an arbitrary real number. 
This is the case for \eq{vren2}.
Alternatively, there might be invariance under the $Z_N$ group
\be
\phi\to e^{i\chi}\phi \mbox{\ \ \ \ $\chi=2\pi n/N$} \,,
\label{zn}
\ee
with $n$ an arbitrary integer. In the limit
where the integer $N$ goes to infinity, the $U(1)$ group
is recovered. Invariance under a continuous group like $U(1)$ is 
said to be a continuous symmetry.

A given symmetry group acts on some of the fields, but not on others.
The action of a given $Z_N$ or $U(1)$ on the full set of fields may be given 
by
\be
\phi_n\to e^{iq_n\chi} \phi_n \,, \label{qndef}
\ee
which defines the charge $q_n$ of each field under the given symmetry.

In these expressions, the origin in field space has been taken to be the 
fixed point of the symmetry group.
The gradient of the potential vanishes at the fixed point, which 
therefore represents a maximum, minimum or 
saddle point of the potential.

In a supersymmetric theory, it is usual to take all scalar fields to be 
complex. (Each of them is the partner of a spin-half field that has
two components, corresponding to the two possible spin values.)
If such a theory emerges from  string theory, there are 
two kinds of field. The most numerous, usually called matter fields,
transform under groups built out
of $U(1)$'s 
(continuous symmetries) and $Z_N$'s (discrete symmetries).
As in \eqs{mattersym}{zn} there is a unique fixed point in field space, which 
is generally chosen as the origin. For a given $U(1)$ (say)
the transformation can be brought into the form \eq{mattersym} 
with a suitable choice of the directions in field space that 
define the $\phi_n$, but in general this cannot be done for all of them 
simultaneously. We then have a non-abelian group (one whose elements do not 
commute) such as $SU(N)$.

In addition to the matter fields, there are special fields
namely the dilaton $s$, \label{dilmention}
and certain fields called bulk moduli.
In the example we shall discuss in Sections \ref{sugfstr}
and \ref{ss:sstring}
there 
are three of the latter, $t_I$ with $I=1$ to 3. 
The dilaton and bulk moduli are charged under discrete symmetry
groups that are not built out of \eq{mattersym},
and the most convenient choice of origin for these fields 
is not the fixed point of these symmetry groups.

A $U(1)$ symmetry is said to be global, if $\chi$ in \eq{mattersym}
is independent of spacetime position. This is mandatory if no gauge
field transforms under the $U(1)$, 
because then the spacetime derivatives in the kinetic 
term inevitably spoil the symmetry. 
If gauge fields have a suitable 
transformation under the  $U(1)$, 
we can allow $\chi$ to depend on
position, because 
the change in the kinetic term is cancelled by a change 
in the part of the action involving the gauge fields.
The symmetry is then said to be a
{ local symmetry}, or a { gauge symmetry}.
An example is the 
electromagnetic gauge field (electromagnetic potential)
$A_\mu$. This generalizes to non-abelian groups. 
There is an electromagnetic-like interaction associated with each 
gauge symmetry.
The Standard Model is invariant under the gauge symmetry group
$SU(3)\sub C \otimes SU(2)\sub{L} \otimes SU(1)_Y$,
the factors corresponding respectively to the colour,
left-handed electroweak and hypercharge interactions.

Generalizing from \eq{qndef},
the fields not affected by a given symmetry group are said to be 
uncharged under the group, or to be singlets.\footnote
{The latter terminology originated with the case of non-abelian groups,
where each field charged under the group is necessarily part of a 
multiplet of charged fields.}
It is usually supposed that every field is charged 
under some symmetry, though the opposite possibility of a 
`universal singlet' is sometimes considered \cite{nus}.
\label{unsin}

\subsubsection{Spontaneously broken symmetry and vevs}

\label{s:vevs}

Any minimum of the potential represents a possible vacuum state,
with the scalar fields having the time-independent value
corresponding to the minimum. (Such values are indeed solutions of the
field equation \eq{fieldeq}).
In the examples encountered so far there is a unique minimum, but
matters can be more complicated.
           
\begin{figure}
\centering
\leavevmode\epsfysize=5.3cm \epsfbox{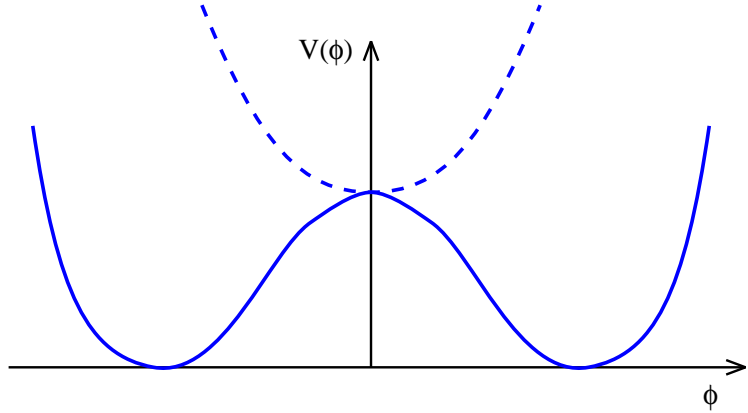}\\
\caption[double]{\label{f:double} 
The full line illustrates 
schematically
the potential \eq{mexhat}. The dashed line shows the same potential
with the sign of $m^2$ reversed (symmetry restoration).
This Figure is taken from reference \cite{LL3}.}
\end{figure}

As a simple example, consider \eq{vren} with
the sign of the mass
term reversed,
\be
V(\phi) = V_0 -\frac12m^2\phi^2 + \frac14\lambda \phi^4.
\label{mexhat}
\ee
It has the same $Z_2$ symmetry as the original potential, 
corresponding to invariance under $\phi\to-\phi$. But
as shown in Figure \ref{f:double}, the minimum at the origin is replaced 
by minima at $\phi=\pm(m/\sqrt\lambda)$. Taking, say,
the  positive sign, one can define a new field $\tilde\phi = \phi- 
(m/\sqrt\lambda)$. Then, if the constant 
$V_0$ is chosen appropriately, 
one has near the minimum 
\be
V= \frac 12 \tilde m^2 \tilde\phi^2 + A\tilde\phi^3 + B\tilde
\phi^4,
\label{exp98}
\ee
where $\tilde m=\sqrt 2 m$ and we are not interested in the precise
values of $A$ and $B$. The minima
represent possible vacuum expectation values (vevs) 
of the field. Each of them represents a possible vacuum of the theory,
around which are small oscillations corresponding (after quantization)
to particles.
The oscillations correspond to an almost-free field if the cubic and
quadratic terms in \eq{exp98}
are small. (It turns out that the criterion
for this is $\lambda\lsim 1$, which as in the previous case one 
assumes to be valid.) On the other hand, the original $Z_2$
symmetry will not be evident in this almost-free field theory,
and one says that it has been { spontaneously broken}.

The vev of a field is denoted by angle brackets, so that in the above
case one has $\langle \phi \rangle = \pm m/\sqrt\lambda$.

Now consider \eq{vren2} with the sign of the mass term reversed,
\be
V(\phi) = V_0 - m^2|\phi|^2 + \frac14 \lambda |\phi|^4.
\label{vren3}
\ee
The vacuum now consists of the circle
$|\phi|= \langle|\phi|\rangle \equiv 2m/\sqrt\lambda$.
About any point in the vacuum, there
is a `radial' mode of oscillation corresponding to the one
we already considered, plus an `angular' mode 
with zero frequency. 

For a global symmetry,
 the
particle corresponding to the angular mode
is called the { Goldstone boson} of the symmetry, while the particle
corresponding to the radial mode has no particular name.
As we discuss in a moment, continuous global symmetries are usually 
broken, so that their Goldstone bosons acquire mass and become
pseudo-Goldstone bosons. Examples are the pion (corresponding to the 
chiral symmetry of QCD) and the axion (corresponding to the hypothetical
Peccei-Quinn symmetry that is proposed to ensure the $CP$ invariance of 
QCD).

For a gauge symmetry, the particle corresponding to the
radial mode is called a Higgs particle,
while the would-be Goldstone boson loses its identity to become one of the 
degrees of freedom of the gauge boson. This case,
generalized to the $SU(2)$ group, occurs in the electroweak sector of 
the Standard Model, and supersymmetric generalizations of it. 
More Higgs fields occur in GUT models.

The field which spontaneously breaks the symmetry, that we have denoted
by $\phi$, need not be one of the elementary fields 
appearing in the lagrangian. Instead it can be a product of these
fields, called a condensate. The fields can be spin half, so if all
symmetry-breaking scalars were condensates one would have no need
of elementary scalars. 
The pion field is a condensate, and in some models so is the 
axion. In this case there need be no particle 
corresponding to the radial mode.

Higgs fields are usually taken to be elementary,
because this is the simplest possibility.
The desire to have elementary scalar fields
is one of the  most important motivations for supersymmetry.

The above discussion applies to matter fields, but a similar one applies
to any internal symmetry and in particular to the dilaton and bulk 
moduli. The general criterion for spontaneous symmetry breaking is that 
the vacuum (the minimum of the potential)
does not correspond to a fixed point of the symmetry; as a result of 
this there is more than one copy of the vacuum, different copies being 
related by the symmetry.

\subsubsection{Explicitely broken global symmetries}

Global (but not gauge) symmetries can be explicitly broken. This means 
that the action is not precisely invariant under the symmetry group.

Consider first the $Z_2$ symmetry acting on a real field,
$\phi\to -\phi$. It is broken if one adds to the potential
\eq{vren} or \eq{mexhat} an odd term.

Now consider a global $U(1)$ acting on a complex field, according to
\eq{mattersym}. It is broken if one adds to the
the potential \eq{vren2} or \eq{vren3} a
term that depends on the phase of $\phi$.
For instance, there might be a contribution of the form
\be
\Delta V = \lambda_d \mpl^{4-d} \(\frac{\phi^d+\phi^{*d}}2\)
\,. \label{expbr}
\ee
Instead of being generated from the tree-level potential in this way,
$\Delta V(\theta)$ can come from a non-perturbative effect
(to be precise, an instanton). 

With explicit breaking, a
Goldstone boson acquires mass, to become
a pseudo Goldstone boson. This case occurs in QCD, where the
pion is a pseudo-Goldstone boson. The axion (if it exists)
is also a pseudo-Goldstone boson.
If we write $\phi=\langle|\phi|\rangle e^{i\theta}$, the
canonically normalized pseudo-Goldstone boson field is
$\psi\equiv \sqrt2|\phi|\theta$. Its potential $V(\theta )$ has period 
$2\pi/N$ where $N$ is some integer. For $N\geq2$,
the original $U(1)$ symmetry \eq{mattersym} has been broken down to
the residual symmetry $Z_N$ \eq{zn}. For $N=1$, there is no residual symmetry.

In the above example
\be
\Delta V(\theta ) = \lambda_d \mpl^4\(
 \frac{\langle|\phi|\rangle}{\mpl}\) ^d \cos(d\theta) \label{pgbpot} \,.
\ee
Defining the zero of $\psi$ to be at a minimum of $V$, 
one finds 
\be
m_\psi^2 = d^2\lambda_d \mpl^2 \( \frac{\langle |\phi| \rangle}{\mpl}
\)^{d-2} \,.
\label{mpbg}
\ee

Provided that $m_\psi$ is much less than
$m_\phi$, the radial part of $\phi$ which has the latter mass
can remain practically at the vev while $\psi$ oscillates.
For much bigger values this becomes impossible, and we have completely 
lost the original symmetry.

It is usually supposed that all continuous global symmetries are 
approximate. One reason is that this seems to be the case for field 
theories derived 
\label{glvdis}
from string theory \cite{choikim,banks88,dine}.
In contrast, they typically 
contain many discrete symmetries \cite{witten85}.

\subsubsection{The restoration of a spontaneously broken
internal symmetry}

\label{s:symres}

In the early Universe, the scalar fields will be displaced {}from their
vacuum expectation values (vevs). In particular, a field with a nonzero vev,
corresponding to a spontaneously broken symmetry, may
have zero value in the early Universe. Then the symmetry is restored
at early times. This may happen during inflation, and also during
the subsequent hot big bang.

A simple example which can illustrate both cases is the following
potential, involving real fields $\phi$ and $\psi$.
\bea
V&=&V_0-\frac12 m_\psi^2 \psi^2 + \frac14\lambda\psi^4 
 + \frac12 m^2 \phi^2 +\frac12\lambda'\psi^2\phi^2 \\
&=&\frac14\lambda(M^2-\psi^2)^2 +\frac12m^2\phi^2
+\frac12\lambda'\psi^2\phi^2. 
\label{fullpot1}
\eea
Comparing the two ways of writing the potential, 
one sees that the parameters are related by
\bea
m_\psi^2&=&\lambda M^2, \\
V_0 &=& \frac14\lambda M^4 = \frac14 M^2 m_\psi^2. 
\eea
The minimum of $V$, corresponding to the vacuum, is at
$\phi=0$ and $\psi=M$. The latter field has a nonzero vev, spontaneously 
breaking the discrete symmetry $\psi\to-\psi$.
But now suppose that, in the early Universe, 
$\phi^2$ has a nonzero value, bigger than a critical value
$\phi\sub{c}^2= m_\psi^2/\lambda'$. 
Then the minimum with respect to $\psi$
lies at $\psi=0$, and the symmetry is restored. With the relabelling
$\psi\to\phi$, this is
illustrated by the dashed line in Figure \ref{f:double}.

In an appropriate region of parameter space, 
the fields can be in thermal equilibrium 
at temperature $T\gsim \phi\sub{c}$, making $\phi^2$ typically of order
$T^2$. Then the symmetry $\psi\to-\psi$ is restored for 
$T\gsim \phi\sub{c}$, and 
spontaneously breaks as $T$ falls below that  value. 

Alternatively, $\phi$ might be the inflaton. Then the symmetry 
$\psi\to-\psi$ is 
restored until $\phi$ falls below $\phi\sub{c}$, after which it
spontaneously breaks. If $V_0$ dominates, this signals the end of 
inflation and we have hybrid inflation
\cite{LIN2SC}. Even if it does not, the change might correspond to
a feature in the spectrum $\calp_\calr$,
or topological defects
\cite{Kofman,vishniac,kofpog,yokoyama,kbhp,c3,lyth90,hodpri,nagasawa}.
(This is an alternative to the 
familiar Kibble mechanism \cite{tom} of defect formation, 
which applies if the symmetry is 
restored by thermal effects.)

\subsection{The true vacuum and the inflationary vacuum}

The different vacua, that occur when
a symmetry is spontaneously broken, are physically equivalent,
and are simply referred to as {\em the} vacuum.\footnote
{For simplicity we are supposing that the vacuum so-defined is unique.
In general the potential might have another minimum (or set of minima
related by a spontaneously broken symmetry) in which $V$ has a 
different value; or there might be three or more minima with
different values of $V$. In these cases,
it is not clear whether 
the vacuum corresponding to our Universe (the one with $V=0$)
must be the global minimum
(the one with the lowest value of $V$).
If it is not the global 
minimum, the lifetime for
tunneling to the latter
should presumably be much bigger than the age of the Universe.
Examples of multiple vacua are shown in Figures \ref{f:loopb} and
\ref{f:loopc}.}

During inflation, the spatially-averaged inflaton field $\phi$ is not at a 
minimum of the potential, 
and it varies slowly with time. The spatially-averaged 
non-inflaton fields mostly
adjust themselves to be at the instantaneous minimum of the potential
with $\phi$ at the current value, which may or may not be
the same as the vacuum value. Others may have extremely flat 
potentials, giving negligible motion for the spatial average.

These spatially-averaged fields provide a classical background, around which
are the quantum fluctuations described by quantum field theory.\footnote
{The averaging is to be done over the comoving box within which 
quantum field theory is formulated. It should be large compared with the
comoving scale presently equal to the 
size of the observable Universe, but it is neither necessary nor 
desirable to make it exponentially bigger.}
The classical background may be taken to be constant, because
the variation of $V$ is slow
on the Hubble timescale. It defines an effective vacuum 
for a quantum field theory,
which we shall call the inflationary vacuum.
To emphasize the distinction, we shall often call the actual
vacuum, corresponding to the minimum of the potential, the true vacuum.

In some applications, such as when calculating the vacuum fluctuation of
the inflaton field, it is necessary to formulate quantum field theory
in the setting of curved spacetime (the expanding Universe).
The main difference though, between the inflationary vacuum
and the true one, is the value of the 
vacuum energy density $V$. In the true vacuum it is practically zero
($|V|^{1/4}\lsim 10^{-3}\eV$, corresponding to the bound on the 
cosmological constant). During inflation it is big.

 {}From this perspective {\em two separate searches are in progress,
for quantum field theories beyond the Standard Model}. 
There is the search for the field theory 
that applies in the true vacuum, 
and the search for the field theory that applies during inflation.
In some proposals these theories are very different, whereas in others
they are almost the same. Roughly speaking, the former proposals predict
that the inflationary energy scale is
$V^{1/4}\gg 10^{10}\GeV$, and the latter predict that it is in the range
$10^5\GeV \lsim V^{1/4}\lsim 10^{10}\GeV$. 

\subsection{Supersymmetry}

\label{ss:susy1}

Practically all viable extensions of the Standard Model invoke 
supersymmetry. The main reason is that they invoke fundamental
scalar fields, which look natural only in the context of
supersymmetry. Indeed, supersymmetry eliminates the quadratic
divergences in the mass $m^2$ of fundamental light scalar fields,
$\delta m^2\sim \Lambda\sub{UV}^2$, $\Lambda\sub{UV}$ 
being the scale beyond which
the low energy theory no longer applies. 
In about ten years, the 
Large Hadron Collider (LHC) at CERN will either discover supersymmetry, if 
it has not been 
discovered before then, or practically kill it. In the latter eventuality 
the task of understanding whatever {\em is} observed at the LHC will take 
precedence over such relatively trivial matters as inflation model-building, 
so let us suppose optimistically that supersymmetry is valid.

We shall consider supersymmetry in detail in Section \ref{s5},
but let us note a few important points. 
Supersymmetry is an extension of Lorentz invariance, and therefore not 
an internal symmetry. It relates bosons and fermions.
In the `$N=1$' version generally adopted, there are `chiral' 
supermultiplets each containing a complex scalar field (spin 0)
plus a chiral fermion (spin $1/2$) field, and `gauge' supermultiplets
each containing a gauge field (spin 1) and a gaugino (spin $1/2$) field.
Each Standard Model particle has an undiscovered superpartner; 
there are squarks and sleptons with spin 0, Higgsinos with spin $1/2$ 
and gauginos with spin $1/2$. (It turns out that at least two Higgs 
fields are required.) 

Supersymmetry is expected to be local as opposed to global, 
and local supersymmetry is called 
supergravity because it automatically incorporates gravity. 
In $N=1$ supergravity, the
graviton (spin 2) is accompanied by the gravitino
(spin $3/2$). Global supersymmetry provides a good
approximation to supergravity for most purposes.

To decide between different possible forms of field theory, and in
particular supergravity, one may look to a hopefully more fundamental 
theory like weakly coupled string theory or Horava-Witten M-theory.

Unbroken supersymmetry would require that each particle has the same mass 
as its partner. This is not observed, so supergravity 
is spontaneously broken in the true vacuum.
(A local symmetry cannot be explicitly broken.)
The scale of this breaking 
is conveniently 
characterized by a scale $M\sub S$, related to the gravitino mass
$m_{3/2}$ by
\be
{M\sub S}^2 =\sqrt3 \mpl m_{3/2} .
\label{msdef1}
\ee

To have a viable phenomenology, the spontaneous breaking
is supposed to occur in a 
`hidden sector' of the theory, communicating only weakly with the
`visible' sector containing particles with Standard Model 
interactions. 
In the visible sector, one has for most purposes global supersymmetry
with explicit breaking of a special kind, called `soft supersymmetry breaking'.
Soft susy breaking must give the squarks and sleptons masses
\be
\widetilde{m} \sim 100\GeV\:\: {\rm to }\:\: 1\TeV.
\label{msdef}
\ee
(Gauginos may also have such masses, or they may be 
lighter.)
This typical `soft' mass $\widetilde{m}$ is an important parameter for model
building. It cannot be much above $1\TeV$ or susy would not 
do its job of allowing us to understand the existence of the 
Standard Model Higgs field. Nor can it be much less than $100\GeV$, 
or the squarks and sleptons would have been observed.

The relation between $M\sub S$ and $\widetilde{m}$ is model-dependent.
In a class of theories known as gravity-mediated 
one has
\be
M\sub{S} \simeq \sqrt{\widetilde{m} \mpl}
\sim 10^{10}\:\: {\rm to}\:\: 10^{11} \GeV.
\label{I}
\ee
(For definiteness we usually take $10^{10}\GeV$ in what follows.)
Then $m_{3/2}\sim \tilde m$.
In another class, called gauge-mediated, $M\sub S$ can be anywhere
between $10^6\GeV$ and $10^{11}\GeV$, corresponding to
$1\keV\lsim m_{3/2} \lsim 1\TeV$.

All this refers to the true vacuum. During inflation,
susy is also necessarily broken. In most models the 
mechanism of susy breaking during inflation has nothing to do with
the mechanism of susy breaking in the true vacuum (and is much simpler).
In an interesting class of models, the mechanism is supposed to be the 
same. As a rough guide, inflation models with
\label{inbr98}
$V^{1/4}\gg M\sub S$ fall into the first class, while models with 
a lower $V$ fall into the second.

\subsection{Quantum corrections to the potential}

So far we specified  the part of the Lagrangian involving only the 
scalar fields. 
When quantum effects are included, this is not enough to 
describe these fields; 
we need the rest of the lagrangian, that describing 
higher-spin fields that can couple to 
scalar fields.
During inflation, when 
the scalar fields are almost independent of position, these effects can 
be summarized by giving an effective potential $V$ and
(if necessary) an effective kinetic function 
$K_{mn^*}$, which are to be used in the field equation
\eq{fieldeq} or its non-canonical equivalent.
Note that we are using the same symbols for the effective objects
and the ones that appear in the lagrangian.

Quantum effects are determined by the couplings of the fields
(as well as their masses). Gauge couplings (couplings to gauge fields)
are characterized by a dimensionless constant $g$,
or equivalently by $\alpha=g^2/4\pi$. (For electromagnetism,
$g$ is the electron charge and $\alpha$ evaluated at low energy
is the fine structure constant $\alpha\sub{em}=1/137$.)
Couplings not involving gauge fields, called Yukawa couplings,
can again be characterized by dimensionless constants. 
Complex scalar fields with no gauge couplings are called gauge singlets,
and both their radial and angular components (pseudo-Goldstone bosons)
are favourite candidates for the inflaton field.

Quantum effects are
of two kinds; the perturbative effects represented 
by Feynman graphs, and the non-perturbative effects represented by 
things like instanton contributions to the path integral.
This separation is meaningful only if the
relevant couplings are small, in particular if gauge couplings satisfy
$\alpha\lsim 1$. At large couplings the theory is completely non-perturbative.

Gauge couplings are not supposed to be extremely small,
and one should take $g\sim 1$ for crude order of magnitude estimates
(making $\alpha$ one or two orders of magnitude below 1).
For renormalizable Yukawa couplings, values 
a few orders of magnitude below unity are 
generally regarded as reasonable, at least for the renormalizable 
couplings in an effective field theory.

\subsubsection{Gauge coupling unification and the Planck scale}

\label{gaugecoup}

With quantum effects included,
the masses and couplings to be used in the lagrangian
depend 
on the relevant energy scale $Q$. The dependence on $Q$ (called `running')
can be calculated 
through the renormalization group equations (RGE's), and is 
logarithmic.
In the context of collider physics, $Q$ can be taken to be 
the collision energy, if there are no bigger relevant scales
(particle masses).
In the context of inflation, $Q$ can be taken 
to be the value of the 
inflaton field if, again, there are no bigger relevant scales
(particle masses, or values of other relevant fields).

For the Standard Model
there are three gauge couplings, $\alpha_i$ where $i=3,2,1$, corresponding 
respectively to the strong interaction,
the left-handed electroweak interaction and 
electroweak hypercharge.
(The electromagnetic gauge coupling 
is given by $\alpha^{-1}= \alpha_1^{-1}+ \alpha_2^{-1}$.) 
In the one-loop approximation, ignoring the Higgs field,
their running is given by
\be
\frac{d\alpha_i}{d\ln(Q^2)}  = \frac{b_i}{4\pi}\alpha_i^2 \,.
\label{70}
\ee
The coefficients $b_i$ depend on the number of particles with
mass $\ll Q$. Including all particles in the minimal supersymmetric
Standard Model particle gives $b_1= 11$, $b_2=1$ and $b_3=-3$.

Using the values of $\alpha_i$ measured by collider experiments at 
a scale $Q\simeq 100\MeV$, one finds that all three couplings
become equal at a scale \cite{couplingunity,n1,n2,n3,n4}\footnote
{To be precise, $\frac53\alpha_1=\alpha_2=\alpha_3=
\alpha\sub{GUT}$, the factor
$5/3$ arising because the historical definition of 
$\alpha_1$ is not very sensible. In passing we note that the unification 
fails by many standard deviations in the absence of supersymmetry,
which may be construed as evidence for supersymmetry and 
anyhow highlights the remarkable accuracy of the experiments leading to 
this result.}
$Q=M\sub{GUT}$, where
\be
M\sub{GUT} \simeq 2\times 10^{16}\GeV.
\ee
The unified value is 
\be
\alpha\sub{GUT}\simeq 1/25.
\ee
One explanation of this remarkable
experimental result may be that there is a Grand Unified Theory (GUT), 
involving a higher symmetry with a single gauge coupling,
which is unbroken above the scale $M\sub{GUT}$.
Another might be that field theory becomes invalid above the 
unification scale, to be replaced by something like
weakly coupled string theory 
or Horava-Witten M-theory, which 
is the source of unification. At the time of writing there is no
consensus about which explanation is correct \cite{dienesrev}.

\subsubsection{The one-loop correction}

\label{onel}

The perturbative part of the effective potential is given by a
sum of terms, corresponding to the number of loops in the 
Feynman graphs. The no-loop term is called the 
tree-level term (because the Feynman graphs look like trees)
and it has the power-series form \eq{power}.

In any given situation, one can usually choose the 
renormalization scale $Q$ so that the loop corrections are small.
Then, the 1-loop correction typically dominates, and only it
has so far been considered in connection with inflation model-building.

We now discuss the
form of the 1-loop correction, initially making the choice
$Q=\mpl$. In a supersymmetric theory, in the usual case that
$\phi$ is much bigger than the masses of the particles in the loop,
two cases arise.

If the relevant part of the Lagrangian is supersymmetric,
corresponding to spontaneous susy breaking, 
the loop correction is typically of the approximate form
\be
\delta V \simeq V c \ln (\phi/\mpl) \,,
\label{spont}
\ee
where $V$ is the tree-level potential.
In this expression, $c$ is related to the dimensionless coupling
$g$ (between $\phi$ and the field in the loop) by something like
$c\sim g^2/(8\pi^2)$.
It is often called a loop suppression factor,
because in a typical situation each additional loop gives another factor
$c$. 
If the field in the loop is a gauge supermultiplet, 
$c$ is expected to be of order $10^{-1}$ to $10^{-2}$.
If it contains a chiral supermultiplet, 
$g$ is a Yukawa coupling and $c$ might be of the same order, or 
it might be a few orders of magnitude 
smaller. 

The other case typically arises if there is soft
susy breaking in the relevant sector.
The loop correction
in this case is of the approximate form
\be
\delta V \simeq  \frac12 c \mu^2 \phi^2 \ln (\phi/\mpl).
\label{soft}
\ee
Assuming that we are dealing with a flat direction the complete 
potential is now
\be
V=V_0 + \frac12 m^2(\phi) \phi^2 + \cdots \,,
\label{vrunning}
\ee
where
\be
m^2(\phi) = m^2 +  c\mu^2 \ln(\phi/\mpl) \,,
\label{mcrude}
\ee
and the dots represent non-renormalizable terms.

\begin{figure}
\centering
\leavevmode\epsfysize=5.3cm \epsfbox{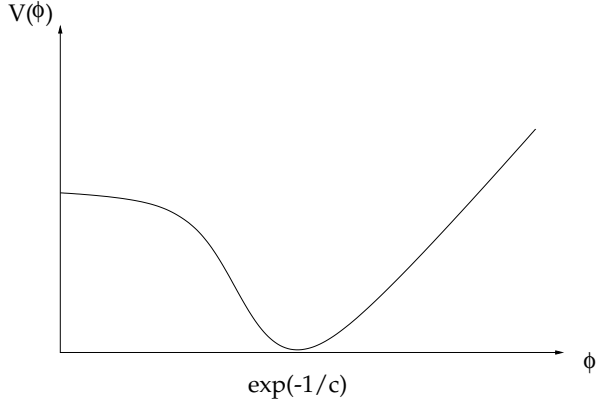}\\
\caption[loopa]{\label{f:loopa} 
A non-perturbative loop correction generates a minimum in the 
potential. The minimum corresponds to $\phi\sim \exp(-1/c)\mpl$,
where $c\ll 1$ is a loop suppression factor.}
\end{figure}

\begin{figure}
\centering
\leavevmode\epsfysize=5.3cm \epsfbox{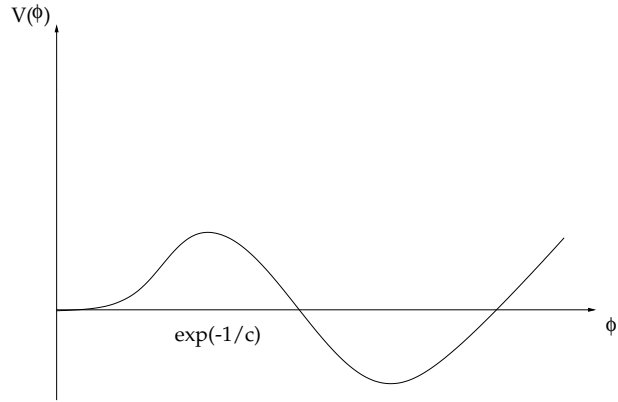}\\
\caption[loopb]{\label{f:loopb} 
A non-perturbative loop correction generates a maximum in the potential,
at a value $\phi\sim \exp(-1/c)\mpl$ hierarchically smaller than the 
Planck scale. Non-renormalizable terms generated a minimum,
at a bigger $\phi$ which may or may not be of order $\mpl$.
There is another minimum at $\phi=0$, which typically corresponds to a 
bigger value of $\phi$. In the true vacuum, $V$ vanishes.
As shown in the graph, the true vacuum may be at $\phi=0$.}
\end{figure}

\begin{figure}
\centering
\leavevmode\epsfysize=5.3cm \epsfbox{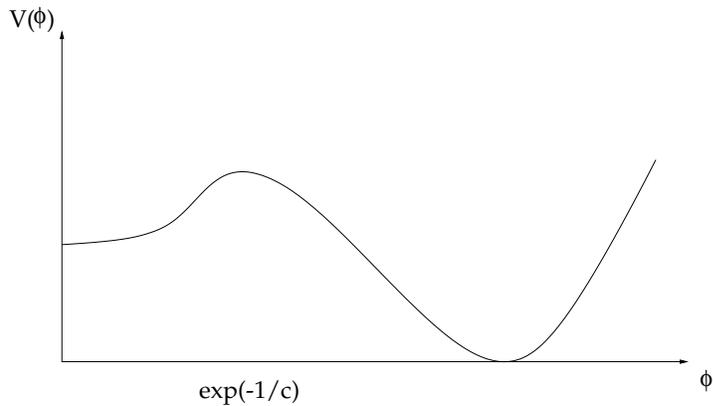}\\
\caption[loopc]{\label{f:loopc} 
Alternatively, the true vacuum may be at the minimum with
nonzero $\phi$.}
\end{figure}

\begin{figure}
\centering
\leavevmode\epsfysize=5.3cm \epsfbox{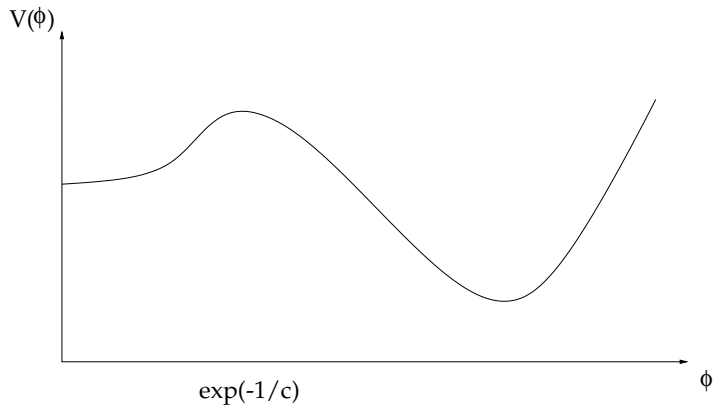}\\
\caption[loopd]{\label{f:loopd} 
A third possibility is that neither of the minima correspond to
the true vacuum. Rather, it lies in another field direction,
`out of the paper'.}
\end{figure}

Let us consider a typical case, where 
the parameter $\mu$ is of order $m$.
Because the loop 
suppression factor $c$ is $\ll 1$, $V$ will have a maximum
or minimum at $\phi\sim e^{-1/c}\mpl$.
The minimum occurs if the mass-squared is positive at the Planck scale.
This case is illustrated in Figure \ref{f:loopa}.
The maximum 
occurs if the mass-squared is negative at the Planck scale.
In that case there is a minimum at $\phi=0$, and another at some
some value 
$\phi\sub{min}\gsim \exp(-1/c)\mpl$ which is
determined by the non-renormalizable 
terms of the tree-level potential. 
Typically, $\phi\sub{min}$ is the global minimum.
If $\phi=0$ is our vacuum, $V$ vanishes there as shown in Figure
\ref{f:loopb}. In that case, the lifetime for tunneling to the 
global minimum had better be much longer than the age of 
the Universe. If $\phi=0$ is not our vacuum,
one can have either of 
the situations shown in Figures \ref{f:loopc} and \ref{f:loopd}.
We shall see in Section \ref{ss:loopcorr} how they permit one to construct 
models of inflation.

A notable feature of these expressions is that they
generate a scale many orders or magnitude less than $\mpl$
without fine-tuning. This occurs because the loop suppression factor,
say a couple of orders of magnitude below 1, is exponentiated.
This phenomenon is known as
dimensional transmutation, and optimistically 
one may suppose that with its help 
all mass scales can be generated more or less 
directly {}from the Planck scale.

For an accurate calculation of the potential $V(\phi)$
we should abandon the choice
$Q=\mpl$ for the renormalization scale. With a general scale $Q$,
the potential becomes
\be
V(Q,\phi) = V_0 + \frac12 m^2(Q) \phi^2 + \frac12
c(Q) \mu^2(Q)\ln (\phi/Q) .
\ee
At a given value of $\phi$, the 1-loop correction vanishes if we set
$Q\simeq \phi$. The 2-loop and higher corrections are then hopefully small, 
and we obtain \eq{vrunning} with $m^2(\phi)\equiv
m^2(Q\simeq\phi)$ now given by the RGE's instead of by \eq{mcrude}.
The RGE for $m^2$ is
\be
\frac{d m^2(Q)}{d\ln Q} = c(Q) \mu^2(Q) \label{mrge} \,.
\ee
Those of $c$ and $\mu$ will also be first order differential equations,
and $m(Q)$ is determined by solving the equations simultaneously
as in Section \ref{ss:loopcorr}. If $c$ and $\mu$ have negligible 
running we recover \eq{mcrude}. 

Being a physical quantity, $V(Q,\phi)$ should actually be independent of
$Q$, so that $\pa V/\pa Q$ vanishes. \label{dvdq}
Including only 1-loop corrections, this 
is guaranteed at the point $Q=\phi$ by the
RGE of $m(Q)$.
Well away from the point $Q=\phi$, $\pa V/\pa Q$ becomes
significantly different from zero if we include only the 1-loop 
correction. 
This is what one would expect, since the 1-loop correction then becomes 
big which indicates that the 2-loop and higher corrections 
need to be included.

\subsection{Non-perturbative effects}

\subsubsection{Condensation and dynamical supersymmetry breaking}

\label{cdsb}

Since $b_3$ is negative, the QCD coupling $\alpha_3(Q)$ 
increases 
as $Q$ is decreases, and it becomes of order 1 at the scale
$Q=\Lambda\sub{QCD}\sim 100\MeV$.\footnote
{For $Q\lsim 100\GeV$,
the value of $b_3$ changes as massive particles cease to be
effective, but it remains negative.}
On smaller scales we are in
the non-perturbative regime. In this regime, 
quarks and gluons bind into hadrons, and should no longer be 
included in a perturbative calculation. Also, products of two
quark fields acquire nonzero vevs,
\be
\langle q\bar q\rangle \sim \Lambda\sub{QCD}^3,
\ee
(spin $1/2$ fields have the energy dimension $3/2$, whereas scalar 
fields have dimension 1).

Note that the large hierarchy between the GUT scale, and the scale
$\Lambda\sub{QCD} \sim 100\MeV$ corresponding to $\alpha_3=1$, 
is generated naturally by the RGE's. Indeed, {}from \eq{70}
\be
\frac{\Lambda\sub{QCD}}{M\sub{GUT} }=
\exp\left(-\frac 1{b_3 c} \right)
\label{73}
\ee
with 
\be
c=\frac {\alpha\sub{GUT}}{2\pi}.
\label{loopsup}
\ee

With a view to generating supersymmetry breaking, it is usually supposed
that the behaviour exhibited by QCD occurs also 
for some other gauge interaction.
The particles with this interaction should not possess the 
Standard Model interactions, and correspond to the hidden sector 
mentioned earlier.
One can again have spin-1/2 condensates $\langle \lambda\bar \lambda \rangle
\sim \Lambda\sub c^3$, where $\lambda$ can be either a 
chiral fermion field as in QCD, or a gaugino field.
The condensation scale $\Lambda\sub c$ of the hidden sector may be
far bigger than $\Lambda\sub{QCD}$.

\subsubsection{A non-perturbative contribution to the potential}

\label{anonp}

Above the condensation scale, the 
effect on the potential is to introduce a term 
like $\Lambda_c^{4+p}/\phi^p$.
In a flat direction, it can be stabilized by 
a non-renormalizable tree-level term $\phi^{4+m}/\mpl^m$,
to generate a large vev  given by
\be
\langle \phi \rangle \sim \left(\frac{\Lambda_c}\mpl\right)
^{\frac {4+p}{4+p+m} } \mpl \sim e^{-1/c} \mpl .
\ee
(To obtain the final equality, we used the generalization of
\eq{73}.)

\subsection{Flatness requirements on the tree-level inflation
potential}

\label{keepflat}

So far our discussion of the potential has been quite general.
Now we want to specialize to the case 
where $\phi$ is the inflaton field.
We shall formulate conditions on 
the tree-level potential that ought to be satisfied in any
model of inflation, and ask how they can be satisfied in a 
supersymmetric theory.

During inflation, the tree-level potential with all other fields fixed
will be of the form \eq{power}. The mass-squared (and for that matter
the coefficients of higher-order terms) can have either sign;
equivalently we can make the convention that all coefficients are 
positive and there is a plus or minus sign in front of them. We adopt the 
latter convention so that $V=V_0\pm \frac12m^2\phi^2 \cdots $.

In \eq{power} the origin has been chosen as a point where
$V'$ vanishes, and before proceeding we want to comment on this choice.
As mentioned earlier (page \pageref{dilmention}) string-derived
field theories contain matter fields on the one hand, and
the dilaton and bulk moduli on the other.
\label{originchoice}

In the space of the matter fields, the
origin is usually chosen to be the
(unique) fixed point of the internal symmetries. The derivatives
of $V$ vanish there. 
In most models of inflation, the inflaton is supposed to be the radial 
part of a matter field, with this choice of origin.
Then $V'$ vanishes at the origin, provided that any other
matter fields coupling to the inflaton vanish during 
inflation. 

If there are nonzero matter fields coupling to the inflaton, or
if the inflaton is 
a pseudo-Goldstone boson (corresponding essentially to the real 
part of a matter field with a displaced origin), 
we simply define the origin as a point where $V'$ vanishes.

Finally we come to the case that
the inflaton is the real or imaginary part of 
a bulk modulus or the dilaton, with some choice of the origin.
For these fields the usual choice of origin is not at all useful,
so we again choose the origin as a point where $V'$ vanishes.
In this case we expect $\phi$ to 
be of order $\mpl$ during inflation, 
whereas if $\phi$ is a matter field it is usually much 
smaller.

Assuming canonical normalization of the fields,
inflation requires that the potential satisfies the flatness conditions
$\epsilon\ll 1$ and $|\eta|\ll 1$, where $\epsilon\equiv
\frac12\mpl^2(V'/V)^2$ and $\eta\equiv \mpl^2 V''/V$
(\eqs{flat1}{flat2}). 
As mentioned in Section \ref{thelag}, canonical normalization 
is not expected to hold if $\phi\gsim \mpl$, but should be 
a good approximation if $\phi$ is significantly below $\mpl$. 
In what follows, we assume at least approximate canonical normalization,
and $\phi\lsim \mpl$.\footnote
{A different case, where inflation happens at $\phi\sim\mpl$
and 
the kinetic function becoming singular at slightly higher values,
is discussed briefly in Section
\ref{kininf}.}

We want to see how the two flatness
conditions constrain the tree-level potential 
\eq{power}. As we have seen, 
quantum corrections have to be added to the tree-level expression.
They may give a significant or even dominant contribution to the 
slope of the potential. But it is reasonably to assume that 
this contribution does not
accurately cancel the tree-level contribution, over the whole relevant
range of $\phi$ values (the values corresponding to horizon exit for 
cosmological scales).\footnote
{In this context, we are regarding the use of a running inflaton mass
(Section \ref{hybrunning}) as still a tree-level effect.
Note also that in mutated hybrid inflation (Section \ref{mut})
there is an additional 
contribution to the inflaton potential, coming from the implicit
dependence in $V(\phi)=V(\phi,\psi(\phi))$. 
In that case our discussion can be taken to apply to
$V(\phi,0)$.}
By the same token, one can assume
that there is no accurate cancellation between different terms of the
tree-level potential.

The assumptions that $\phi\lsim \mpl$ and that there are no
cancellations lead to considerable simplification.
The flatness conditions require
$V\simeq V_0$, since if a single term were to dominate $V$
the flatness conditions would certainly be violated.
Also, the second flatness condition implies the first
for each tree-level term.

We will consider a slightly more precise version of the second
flatness 
condition, $|\eta|\lsim 0.1$. (Barring a cancellation between the two
terms in \eq{n2} this follows from
the observational bound \eq{nobs}.) With the stated assumptions
this gives the following bound on each coefficient
\bea
\frac{\mpl^2 m^2}{V_0} &\lsim & 0.1 \label{f1}\, \\
3\lambda \(\frac{\phi}{\mpl}\)^2 &\lsim & 0.1 \frac{V_0}{\mpl^4}
\label{f2} \,, \\
d(d-1) \lambda_d \( \frac
\phi\mpl \)^{d-2}
&\lsim & 0.1 \frac{V_0}{\mpl^4} \,.
\label{f3}
\eea
These bounds simply say that the contribution of each term to $V$
is at most of order $0.1(\phi/\mpl)^2 V_0/d^2$;
this is essentially the same as our assumption that these contributions 
are $\ll V_0$, 
given our other assumption
$\phi\lsim \mpl$.

The first constraint \eq{f1} gives a constraint on the inflaton mass,
which is independent of the field. 

It looks at first sight
as if the other
inequalities can always be ensured by making
$\phi$ very small, but matters are not so simple because
\eqs{cobenorm}{dndphi} require $\phi$ to vary appreciably.
To be precise, the biggest and smallest scales
probed by COBE differ by a factor of $50$ or so, corresponding
to $\Delta N\simeq 4$. Using \eqs{cobenorm}{dndphi}, we learn that
while these scales are leaving
the horizon,
\be
\frac{V_0}{\mpl^2\phi^2} \lsim
\frac{V_0}{\mpl^2(\Delta \phi)^2} \simeq
\frac{(.027)^2}{2\Delta N}\simeq
2\times 10^{-8}
\label{philb}
\, .
\ee
Putting this into \eq{f2} we find
\be
3\lambda \lsim 2\times 10^{-9} ,.
\label{12lbound}
\ee
Putting it into \eq{f3} gives
\be
d(d-1)\lambda_d \lsim 2\times 10^{-9} \(
\frac{2\times 10^{-8}\mpl^4}{V_0} \)^{\frac{d-4}{2}} \,.
\label{lambound}
\ee

\subsection{Satisfying the flatness requirements in a supersymmetric
theory}

\label{sss:tree}

These constraints are quite strong in the context of received ideas 
about particle theory. 
Consider first the constraint \eq{f1}, on the inflaton mass.
In a globally supersymmetric theory (or a non-supersymmetric
theory)
the constraint poses no particular problem since the mass can be
set to an arbitrarily small value. 
Unfortunately, the corrections to global susy coming from a generic
supergravity theory are not small during inflation; rather, they
give $m^2\gsim V_0/\mpl^2$ 
for every scalar field and in particular for 
the inflaton \cite{cllsw,ewansgrav}.\footnote
{This fact was first recognized in References 
\cite{burt1,coughlan,dinefisch}, but the last two did not 
consider the case of the inflaton. The first, working actually in the 
context of minimal supergravity, took the view that a sufficiently small 
mass will occur through an accidental cancellation.}
Therefore, to construct a model of inflation in the context of 
supergravity, one must either invoke 
an accidental \label{accid}
cancellation, or a non-generic supergravity theory. 
We shall have much 
more to say about the problem of keeping the inflation mass small
in the context of supergravity.

For the constraints on $\lambda$ and $\lambda_d$, we
need to consider separately the case that the inflaton is
the radial part of a matter field (the usual case),
and the case that it is 
a bulk modulus or the dilaton (more precisely, the real or imaginary
part of one of these with respect to some origin).

\subsubsection{The inflaton a matter field}

Consider first the constraint \eq{12lbound} on $\lambda$.
If $\phi$ is a generic field,
one does not expect $\lambda$ to be so small.
But in a globally supersymmetric theory, 
the potential is typically independent of some of the fields,
when the others are held at the origin.
Such fields are called 
`flat directions' (in field space).\footnote
{The direction is flat only when other fields coupling to 
it vanish; for instance if $\phi$ is  a flat direction,
and there is a term 
$\phi^2\psi^2$ in the potential, a nonzero value of $\psi$ will lift the 
flatness. At this point, we should make it clear that the flat direction
can be a linear combination of the fields that one would naturally 
choose; for instance in the above example the natural fields 
(with say definite charges under a $U(1)$ symmetry) might be
$(\phi\pm\psi)$.}
This makes $\lambda=0$ in the globally supersymmetric theory.
When we go the full supergravity theory, we generically find
in a flat direction that 
$\lambda $ is of order $V_0/\mpl^4$.
Then, the flatness condition \eq{f2} is satisfied
provided that $\phi\ll\mpl$.

Now we consider \eq{f3}, omitting the cubic term $d=3$ since it 
is usually forbidden by a symmetry.
Even in a flat direction, the non-renormalizable couplings
$\lambda_d$ are generically of order 1, at least
for $d$ not too large.\footnote
{\label{vld}For large $d$, $\lambda_d$ might be suppressed by a
large $d$-dependent factors. For instance,
if supergravity were obtained by integrating out heavy fields, from some
renormalizable field theory valid on scales bigger than $\mpl$,
then one might expect $|\lambda_d|\sim 1/d!$ \cite{km}.
Such is not
the case, but we are reminded that $d$-dependent factors might be
present when supergravity is matched to say a string theory.
As our estimates of the $\lambda_d$
apply only if 
$d$ is not too large, and are anyhow very rough, 
the factor $d(d-1)$ in \eq{f3} 
cannot be taken seriously, and we set it equal to 1 in what 
follows.}
In that case, \eq{lambound} becomes an upper bound on $V_0$.
For $d=5$ it gives \label{v14}
$V_0^{1/4}<3\times 10^{11}\GeV$,
and for $d=6$ it gives
$V_0^{1/4}\lsim 1\times 10^{14}\GeV$.
For $d\to\infty $ it becomes 
$V^{1/4}\lsim 1.5\times 10^{16}\GeV$ which is anyhow more or less 
demanded by the COBE normalization. (Not a coincidence, as one sees by
examining the argument that led to \eq{lambound}.)

For low $d$ these bounds are violated in many models of inflation.
In these cases, at least some of the 
$\lambda_d$ must be below
the generic value $\lambda_d\sim 1$. But provided that
$\phi$ is well below $\mpl$, this will be needed only for the first few
coefficients, and it is enough to make these of order
$V_0/\mpl^4$. As we
shall see in Section \ref{ss:flat}, that can be achieved in a supersymmetric 
theory by imposing a discrete symmetry.

In some models, notably $D$-term inflation, $\phi$ is of order $\mpl$
rather than much less. In that case \cite{km} {\em all} of
the $\lambda_d$ (as well as $\lambda$)
need to be significantly {\em less} than $V_0/\mpl^4$.
This can be achieved by
imposing an exact $U(1)$ (or higher)
symmetry, but in the usual case that the 
inflaton is a gauge singlet the symmetry would have to be global
and as we noted on page
\pageref{glvdis}
global continuous symmetries
do not seem to be present 
in field theories derived \label{bigphi}
from string theory \cite{choikim,banks88,dine}.
Accordingly, models of inflation with $\phi\sim\mpl$ are at present
quite speculative. One possibility \cite{km} is that the 
the coefficients $\lambda_d$ actually fall off rapidly at large $d$,
so that only the first few need be suppressed.

According to these estimates, 
the power series expansion \eq{power} ceases to be reliable for $\phi\gg
\mpl$. In this regime 
one has
in general no idea what form the potential will take.

\subsubsection{The inflaton a bulk modulus or the dilaton}

If the inflaton $\phi$ is the real or imaginary part of
a bulk modulus (with some choice of origin) 
its potential during inflation will 
be of the form
\be
V=A+B f(x) \,.
\label{vdilmod}
\ee
Here, $x=\phi/\mpl$ and
$f(x)$ and its derivatives are generically of order 1 in magnitude.
Also, $\phi$ is typically of order $\mpl$ during inflation.

The constant term $A$ can be negligible,
or can dominate $V$. If it is negligible, it is clear that the flatness 
conditions $|\eta|\ll1$ and $\epsilon\ll 1$ are 
marginally violated.
(In terms of the coefficients we have $m^2 \sim V_0/\mpl^2$ 
and $\lambda \sim \lambda_d \sim V_0/\mpl^4$.)
If the constant term dominates,
the flatness conditions
are satisfied.

The potentials of the real and imaginary parts of the
dilaton are very model dependent, but they are often 
supposed also to be of the above form, with $A$ negligible.

\section{Forms for the potential; COBE normalizations and predictions 
for $n$}
\label{s8}

At the lowest level, a `model of inflation' is simply a 
specification of the form of the potential relevant during inflation;
this will be $V(\phi)$ for a single-field model, or 
$V(\phi,\psi_1,\psi_2\cdots)$
for a hybrid inflation model.\footnote
{The other, irrelevant, fields
are supposed to give a negligible contribution to $V$. Most of them will
have masses during inflation that are big enough to anchor them at the
vacuum values. (The criterion for this is
$\mpl^2 V''/V\gg 1$ where the prime is the derivative
with respect to the relevant field, or equivalently mass
$\gg H$.)
At the other extreme, some
of the irrelevant fields may correspond to field directions
which are even flatter than that of the inflaton. Such fields
may be displaced far {}from their vacuum values during inflation, with
possibly observable effects. The classic examples are the dilaton
and the bulk moduli
of string theory,
and the QCD axion.}
In this section we give a survey 
of `models' in this sense, that have been proposed in the literature.
The particle theory background will be mentioned only briefly, pending 
the full discussion  of Sections \ref{s18} and \ref{s19}.

The potential of a given model will contain one, 
two or more parameters. Discounting particle theory, these are 
constrained only by observation. The most fundamental constraint is the COBE 
normalization \eq{cobenorm}. The corresponding upper bound was 
known (to order of magnitude) long before the cmb anisotropy was 
actually observed, and was therefore available when inflation was first
proposed. It ruled out the first viable models of inflation \cite{new1,new2}
(or to be precise, required that the dimensionless coupling is 
tiny, \eq{newlb})
and has been imposed as a constraint on all models of inflation 
since then.\footnote
{Before the COBE observation one did not have a precise 
normalization, but the approximate one was known from galaxy surveys.
In the early days one entertained the possibility that the cmb 
anisotropy and large-scale structure had a non-inflationary origin,
in which case the normalization became an upper bound.}
The COBE normalization typically determines the magnitude of the 
potential, as opposed to its shape.

The other important constraint is provided by the 
spectral index $n$, given by 
\eq{n2} or more usually \eq{n3}. The spectral index
can often be calculated just from
the shape of the potential, and is a powerful discriminator between
models.

One can also calculate the scale-dependence of $n$ and the relative
contribution $r$ of gravitational waves. The latter is too small ever to 
observe in most models, but the former may well provide additional 
discrimination in the future.

Without going into detail, we shall try to give some indication of the 
extent to which each form for the potential is attractive in the context
of current ideas about particle theory. In particular, we shall indicate
whether there is a mechanism for keeping the inflaton mass small in the 
context of supergravity (page \pageref{accid}), or whether an accidental
cancellation is invoked.

\subsection{Single-field and hybrid inflation models}

As we already pointed out,
there are two broad classes of `model'. In { \bf single-field}
models, 
the slow-rolling inflaton field $\phi$ gives the dominant contribution 
to the potential, and inflation ends when $\phi$ starts to oscillate about 
its vacuum value. 
In {\bf hybrid} models, the dominant
contribution to the potential
$V$ comes {}from some field $\psi$ which is {\em not}
slow-rolling, but is fixed by its interaction with $\phi$.\footnote
{In mutated hybrid inflation, $\psi$ is a function of
the inflaton field. Then the potential during inflation is $V(\phi,\psi
(\phi))$.
At this point, we should note that
the definition
of a `field' is in principle not unique. However, we
are supposing that the fields can be taken to be canonically
normalized, so that the field-space `metric' $K_{mn^*}$ is Euclidean.
Then, apart {}from the choice of origin,  the choice of fields
corresponds to a choice of orthogonal directions in
field space.
In the context of particle physics
there is usually a naturally
preferred choice (up to gauge transformations) making the definition of
the fields essentially unique in that context. On the other hand, the
`inflaton field' $\phi$ may be a linear combination of
the particle physics fields.}

There are two, very different, kinds of single-field model.
In what are usually called {\em chaotic inflation models} 
$\phi$ is moving towards the origin, and its magnitude during observable 
inflation is several times
$\mpl$.
In what are usually called
{\em new inflation} models, $\phi$ 
is moving away from the origin, and during observable inflation
its magnitude is at most of order 
$\mpl$.\footnote
{We shall generally avoid 
the terms `chaotic' and `new', since
they are also used to indicate initial conditions
long before observable inflation starts (respectively chaotically 
varying fields, and fields in thermal equilibrium).}
In hybrid models, $\phi$ may be moving in either direction, but
its magnitude is again at most of order $\mpl$. 

Both in single-field and hybrid inflation, one will
have a potential $V(\phi)$ during inflation, which depends on one or 
more parameters. One will also  know the value $\phi\sub{end}$ at the 
end of slow-roll inflation.\footnote
{In single-field models it always
corresponds to the failure of one of the flatness conditions
(\eqs{flat1}{flat2}). In hybrid inflation, this may be the case,
or alternatively it may correspond to $\phi$ arriving at the critical
value $\phi\sub c$ at which the non-inflaton field is destabilized.}
Given this information, the recipe for obtaining
the predictions is simple.
\begin{itemize}
\item Calculate the number of $e$-folds $N(\phi)$ to the end of 
slow-roll inflation
using \eq{nint}. In many cases, this integral is insensitive to
$\phi\sub{end}$ in which case the predictions are independent of
that quantity.
\item The value of $N(\phi)$ when the observable Universe
leaves the horizon, denoted simply by $N$ with no argument,
depends on the history of the Universe after slow-roll inflation 
ends. We saw in Section \ref{nefolds}, that a reasonable estimate
is $N\sim 50$, unless there is significant inflation after slow-roll
inflation ends. Using this, or a lower, estimate of $N$,
calculate the corresponding slow-roll parameters $\eta$ and
$\epsilon$.
\item Use $\epsilon$ to impose the COBE normalization \eq{cobenorm}
on the model.
\item See if there are significant gravitational waves.
As discussed in Section \ref{s:gravwave}, this requires
$\epsilon\gsim 0.01$ which is hardly ever satisfied. We shall mention 
gravitational waves only in the rare models where they are significant.
\item Check that $\epsilon\ll |\eta|$. If it is, the 
full expression $n-1=2\eta-6\epsilon$ may be replaced
by $n-1= 2\eta$. As discussed after \eq{n3}, this is usually the case,
and we shall mention the full expression for $\eta$ only for those rare 
models where it is needed.
Using one expression or the other, calculate $n$.
As shown in the  table on page \pageref{t:1}, it often depends 
only on the shape of the 
potential.
\item Check to see if $n$ has significant variation on cosmological 
scales, corresponding to $N-10\lsim N(\phi) \lsim N
$.\end{itemize}

\subsection{Monomial and exponential potentials}

\label{monom}

Now we begin our survey of models, starting with single-field 
models and going on to hybrid models. We start with the simplest 
potential of all. It is 
\be
V= \frac12 m^2\phi^2 \,.
\ee
Almost as simple are 
$V=\frac14\lambda\phi^4$,
and $V=\lambda M_{\rm Pl}^{4-p}\phi^p$ with $p/2\geq 3$.
These monomial potentials
were proposed as the simplest realizations
of chaotic initial conditions (Section \ref{before}) at the Planck scale
\cite{chaotic}. 

Inflation ends 
at $\phi_{\rm end}\simeq p\mpl$, after which $\phi$
starts to oscillate about its vev $\phi=0$.
When cosmological scales leave the horizon
$\phi= \sqrt{2Np}\mpl$. Since the inflaton field is
then of order 1 to $10\mpl$, there is no 
particle physics motivation for a monomial potential.

The model gives  $n-1=-(2+p)/(2N)$ (using the full expression
$n=1+2\eta-6\epsilon$), and gravitational waves 
are big enough to be 
eventually observable with $r=2.5p/N=5(1-n)-2.5/N$. 
The COBE normalization \eq{cobenorm} 
corresponds to $m=1.8\times 10^{13}\GeV$ for the quadratic case.
For $p=4,6,8$ it gives respectively
$\lambda=2\times 10^{-14}$, $\lambda=8\times 10^{-17}$,
$\lambda=6\times 10^{-20}$ and so on.
The COBE normalization gives $V^{1/4}\sim 10^{16}\GeV$.
The same prediction is obtained for a more complicated potential,
provided that it is proportional to $\phi^p$ during cosmological
inflation,
and in particular $\phi$ could have a nonzero vev $\ll\mpl$
\cite{shafi1,l1,l2,kls}. 

Inflation at $\phi\gg\mpl$ which ends at $\phi\sim\mpl$
is the prediction of a wide variety of monotonically increasing 
potentials \cite{hodges,blumhodges}, but they are seldom considered
because there is too much freedom and no guidance from particle theory.

The limit of a high power is  an exponential potential,
of the form $V=\exp(\sqrt{2/q}\phi)$. This gives
 $\epsilon=\eta/2=1/q$
which lead to
$n-1=-2/q$ and $r=10/q$.
This is the case of 
`extended inflation',
where the basic interaction involves non-Einstein gravity but the
exponential potential occurs after transforming to Einstein gravity
\cite{extended,extended1}.
However, simple versions of this proposal are ruled out by observation,
because  the end of inflation corresponds to a first order phase
transition, and in order for the bubbles not to spoil the cmb
isotropy one requires $n\lsim 0.75$. With the effect of gravitational 
waves included, this strongly contradicts observation
\cite{LL1,green}.

\subsection{The paradigm $V=V_0+\cdots$}

{}The models we have just considered are the only ones that have
$\phi$ well in excess of $\mpl$. {\em  
In all of the other models that we shall
described it is assumed that $\phi\lsim \mpl$ during observable 
inflation.} 
As a result of this condition, the potential is always of the form
$V=V_0+\cdots$, with the constant $V_0$ dominating.
To avoid repetition we shall take all this for granted in what follows.

\subsection{The inverted quadratic potential}

\label{invq}

Another simple potential leading to inflation is
\cite{bingall,nontherm,natural,ovrut,natural2,kumekawa,%
natural3,natural4,izawa,paul,kinney}
\be
V=V_0-\frac12m^2\phi^2 +\cdots \,,
\label{natural}
\ee
with the constant $V_0$ dominating.
We shall call this the `inverted' quadratic potential, to distinguish it 
{}from the same potential with the plus sign which comes {}from the simplest 
version of hybrid inflation.
The dots indicate the effect of higher powers, that are supposed 
to come in after cosmological scales leave the horizon.

This potential gives $1-n=2\eta=2\mpl^2m^2/V_0$. If $m$ and $V_0$ are regarded 
as free parameters, the region of parameter space permitting slow-roll
inflation corresponds to $1-n\ll 1$. Thus $n$ is indistinguishable {}from
1 except on the edge of parameter space. However,
there are two reasons why the edge might be regarded as 
favoured. One is the fact that 
a generic supergravity theory gives $m^2\sim V_0/\mpl^2$.
Since slow-roll inflation 
requires $|\eta|\ll 1$, either $\eta$ is somewhat reduced {}from its 
natural value by accident, or it is suppressed because the theory has a 
non-generic form. One might argue that $\eta$ should be as big as 
possible in models that rely on an accident, corresponding to 
$n$ significantly different from 1.

The other reason for expecting $n$ to be significantly below 1, which is 
specific to this potential, has to do with the position of the minimum,
$\phi\sub m$. If the inverted quadratic form
for the potential holds until $V_0$ ceases to dominate, 
one expects
\be
\phi\sub m\sim \frac{V_0^{1/2}}{m}
=\left(\frac{2}{1-n} \right)^{1/2} \mpl.
\ee
(This is also an estimate of $\phi_{\rm end}$ in that case.)
To have any hope of understanding the potential within the  context of particle
theory, $\phi\sub m$ should not be more than a few times
$\mpl$, which requires $n$ to be well below 1. 

The second reason for expecting $n$ to be significantly below 1
does not hold if  the potential steepens drastically
soon after cosmological scales leave the horizon, as in the model
at the end of Section \ref{cubhigh}, or if inflation ends through
a hybrid mechanism as in Section \ref{invhyb}.
In some of these models the first reason does not hold either, and
$n$ is in fact indistinguishable from 1.

The COBE normalization \eq{cobenorm} for the inverted quadratic
potential is
\be
\frac{2}{1-n} \frac{V_0^{1/2}}{\mpl\phi} = 5.3 \times 10^{-4}
\label{invcob} \,.
\ee
The field $\phi$ is evaluated when COBE scales leave the horizon,
$N$ $e$-folds before the end of slow-roll inflation
at some epoch $\phi\sub{end}$. It is given by
$\phi = \phi_{\rm end} e^{-x}$ where 
\be
x\equiv N|1-n|/2<5 \,.
\label{xofnn}
\ee
(The bound comes from $N<50$ and $|1-n|<0.2$. At the moment we are 
dealing with the case $n<1$ but we shall use the variable $x$ also for 
the case $n>1$.) This gives
\be
\frac{V_0^{1/2}}{\mpl^2}= 5.3\times 10^{-4} \frac{1-n}{2} 
e^{-x}\frac{\phi_{\rm end}}{\mpl} \,. \label{vofn}
\ee
If the inverted quadratic form holds until $V_0$ ceases to dominate,
$\phi_{\rm end}\gsim \mpl$, and $V_0^{1/4}\gsim 1\times 10^{15}\GeV$.
If it fails earlier, as in the two cases mentioned, $V_0$ can be
much lower.

Since the field variation is bigger than $\mpl$, this 
type of model is unattractive in the context of particle theory.
Let us consider the proposals that have been made.

\paragraph{Modular inflation}
If $\phi$ is the (real or imaginary part of the)
dilaton or a bulk modulus of string theory, and other fields are not 
significantly displaced from their vacuum values,
its potential will be given by \eq{vdilmod} with $A$ negligible,
$V=B f(\phi/\mpl)$, with $f(x) $ and its derivatives roughly 
of order 1 
in the regime $|x|\lsim 1$.\footnote
{The case of $A$ dominating would correspond to hybrid inflation, which 
we are not considering at the moment.}
In that regime one expects the flatness parameters
$\eta\equiv\mpl^2V''/V $
and $\epsilon \equiv \mpl^2(V'/V)^2/2$ to be both roughly of order 1, 
and they might both 
be  significantly
below 1 near some value of $\phi$ so that slow-roll inflation can occur
there. 
One favours the case that this value would be a maximum of the potential
so that `eternal' inflation would set the initial condition.
Then the potential will be of the inverted quadratic form.
So far, investigations using specific models 
\cite{bingall,natural2,macorra,p97berk}
have actually concluded that viable inflation does not occur.\footnote
{Ref.~\cite{bento} claims to have been successful, but 
an analytic calculation of that model reviewed in Ref.~\cite{ournew} 
finds that it is not viable.}

\paragraph{Radial part of a matter field}
Alternatively one could take 
$\phi$ to be the radial part of a matter field,
but this is problematic in the context of string theory
for the reasons discussed on page \pageref{bigphi}.

\paragraph{Angular part of a matter field}

Instead of taking $\phi$ to be radial part of a matter field,
one might take it to be a pseudo-Goldstone boson, corresponding to the 
angular part of a matter field whose radial part is fixed.
This was first proposed 
in Ref.~\cite{natural}, and dubbed 
`natural' inflation. It has subsequently been considered by several 
authors \cite{natural3,ovrut,natural2,natural4,glw}.
One might think that this proposal avoids the problem mentioned on
page \pageref{bigphi} but this turns out not to 
be the case. The potential of the pseudo-Goldstone boson, coming say 
from instanton effects, is
typically of the form
\be
V(\phi) = V_0 \cos^2(\phi/M).
\label{gbpot}
\ee
where $\frac12 M/\sqrt2$ is the magnitude of the corresponding complex
field. Near the top of the potential, inflation takes place and
to sufficient accuracy we have an inverted quadratic potential
with $m^2=2V_0/M^2$, and $1-n=4(\mpl/M)^2$, and to have viable inflation 
we need $M$ to be significantly bigger than $\mpl$.
>From Eq.~(144), non-renormalizable terms will then give a `correction'
$\Delta V\gg V(\phi)$ unless they are suppressed to all orders. The
difficulty of understanding such a suppression is precisely the problem
stated in Section 5.9.1.
        
\subsection{Inverted higher-order potentials}

\label{cubhigh}

If the quadratic term is heavily suppressed or absent, one will have
\be
V\simeq V_0 ( 1-\mu \phi^p +\cdots)
\label{higher}
\ee
with $p\geq 3$. For this potential one expects that the integral
(\ref{nint}) for $N$ is dominated by the limit
$\phi$ leading to \cite{kinney} 
\be
\phi^{p-2}=[p(p-2)\mu N \mpl^2]^{-1}
\label{phiend}
\ee
and
\be
n\simeq1-2\left(\frac{p-1}{p-2}\right)\frac1 N
\label{ncubhigh}
\ee

It is easy to see that the integral is indeed dominated by the $\phi$ limit,
if higher terms in the potential (\ref{higher}) become significant only when
$V_0$ ceases to dominate at 
$\phi^p\sim \mu^{-1}$. Then, in 
the regime where $V_0$ dominates,
$\eta=[(p(p-1)\mpl^2/\phi^2]\mu\phi^p$, and if this expression becomes of order
1 in that regime inflation presumably ends soon after. 
Otherwise inflation ends when $V_0$ ceases to 
dominate. At the end of inflation one therefore has 
$\mpl^2\mu\phi_{\rm end}^{p-2}\sim 1$ if $\phi_{\rm end}\ll 
\mpl$, otherwise one has $\mu\phi_{\rm end}^p\sim 1$.
(We are supposing for simplicity that $p$ is not enormous, and dropping 
it in these rough estimates.)
The integral
(\ref{nint}) is dominated by the limit
$\phi$ provided that 
\be 
N\mpl^2\mu\phi_{\rm end}^{p-2}\gg 1.
\label{criterion1}
\ee
This is always
satisfied in the first case, and is satisfied in the second case
provided that $\phi_{\rm end}\ll \sqrt N \mpl$ which we shall assume.
If higher order terms come in more quickly than we have supposed, or if
inflation ends through a hybrid inflation mechanism then
$\phi_{\rm end}$ will be smaller than these estimates, and 
one will have to see whether the criterion (\ref{criterion1})
is satisfied. If it is satisfied, 
the COBE normalization \eq{cobenorm} is \cite{kinney} 
\be
5.3\times 10^{-4}= (p\mu\mpl^p)^{\frac1{p-2}} [N(p-2)]^{\frac{p-1}{p-2}}
V_0^{\frac12} \mpl^{-2} \,.
\label{cobecubhigh}
\ee

For $p=4$, this becomes a bound on the dimensionless coupling 
$\lambda$ defined by $V=V_0-\frac14\lambda\phi^4+\cdots$, which is 
independent of $V_0$;
\be
\lambda = 3\times 10
^{-15}(50/N)^3 \,.
\label{newlb}
\ee
Such a tiny number can hardly be a fundamental parameter, but
it can be generated if $\lambda$ is a function of some
heavy fields which are integrated out as in the example of
Section \ref{ss:gmed}. 

A practically equivalent form for the potential
is
\be
V=V_0 + \frac14\lambda \phi^4 \log(\phi/Q) 
\label{newinf98}
\,.
\ee
The logarithim comes from the loop correction ignoring $\phi$'s 
supersymmetric partner. This was the first viable model of inflation
\cite{new1,new2} (see also \cite{singlet}). The
constraint $\lambda\sim 10^{-15}$ presumably rules out the model
if $\lambda$ is a fundamental parameter though 
there is a dissenting view \cite{langbein}.
In any case, the model does not survive with 
supersymmetry, since the fermionic partner of $\phi$ then
gives an equal and opposite loop contribution
(Section \ref{sss:loop}).

A dynamical mechanism for suppressing the mass-squared term has been 
proposed \cite{grahamnew}. The potential is
\be
V=V_0(1+\beta\phi^2\psi - \gamma\phi^3 +\cdots),
\ee
where $\psi$ is another field. Then, with $\beta$ and $\gamma$ of order 1 in 
Planck units, and initial values 
$\psi\sim \mpl$ and $\phi\simeq 0$ one can check that
the quadratic term is driven to a negligible value
before cosmological inflation 
begins.
For this proposal to work, the mass $m_\psi$ has to have a negligible
effect, which requires $m_\psi^2\ll V_0/\mpl^2$. As with the 
inflaton mass, this is violated in a generic supergravity theory.
In Reference \cite{grahamnew} $\psi$ is supposed to be a 
pseudo-Goldstone boson, but as we noted earlier this is not an 
\label{grahamprop}
attractive mechanism for keeping the mass small in the context of string
theory. 

The above proposal gives $\phi$ a vev of order $\mpl$.
Some particle-physics motivation for a vev $\ll \mpl$
is given in
 Refs.~\cite{natural4,kinney}, though not in the context of
supergravity.

One could contemplate models in which
more than one power of $\phi$ is significant while 
cosmological scales leave the horizon, but this requires a delicate 
balance of coefficients. Models of this kind were also discussed a long 
time ago \cite{primordial,olive}, again with a vev of order $\mpl$,
but their motivation was in the context of setting the initial value of 
$\phi$ through thermal equilibrium and has disappeared with the 
realization that this `new inflation' mechanism is not needed.

A more recent proposal is described in Section \ref{ss:gmed}.
It gives 
\begin{equation}
V(\phi)\simeq V_0-\frac{m^2}{2}\phi^2-\frac{\lambda}{4}\phi^4,
\end{equation}
This gives 
\be
-V' = m^2 \phi +\lambda \phi^3+ \cdots \,
\label{vprgmed}
\ee
and the two terms are equal at
$\phi=\phi_*\equiv m/\sqrt\lambda$.
It is supposed that the first term dominates while cosmological
scales are leaving the horizon, but that the second term dominates 
before the end of inflation. For an estimate of \eq{nint}, one can 
keep only the first term of \eq{vprgmed} when the integration variable
is less than $\phi_*$, and 
only the second term when it is bigger. In the latter case, one can
also take the integral to be dominated by its lower limit $\phi_*$.
This gives
\be
\frac{\phi}{\phi_*}\simeq \exp\left(\frac{1}{2} - x\right),
\ee
where $x$ is defined by \eq{xofnn}.
The COBE normalization \eq{invcob} then gives
\be
\lambda =  3\times 10^{-15} (50/N)^3 (2x)^3 \exp(1-2x) \,.
\ee
Using the constraint 
\be
\frac12 < x <5 \,,
\ee
this becomes
\be
4\times 10^{-16} < \lambda < 3\times 10^{-15} \,.
\ee
In this model, the tiny 
value of $\lambda$ occurs because it is of the form $F/\mpl^2$,
where $F$ is a function of fields that have been 
integrated out.

\subsection{Another form for the potential}

\label{kininf}

Another potential that has been proposed is
\be
V\simeq V_0(1- e^{-q\phi/\mpl})
\label{another}
\ee
with  $q$ of order 1.
This form is supposed to 
apply in the regime where $V_0$ dominates, which is $\phi\gsim \mpl$.
Inflation ends at $\phi_{\rm end}\sim \mpl$, and when 
cosmological scales leave the horizon one has
\bea
\phi&=& \frac1q \ln(q^2 N) \mpl,\\
n-1&=&-2\eta= -2/N.
\label{r2pot}
\eea
This potential is mimicked by $V=V_0(1-\mu \phi^{-p})$ with
$p\to\infty$ (Table 1).
Gravitational waves are negligible.
The COBE normalization \eq{cobenorm} is now $V_0^{1/4} \simeq 7\times
10^{15}\GeV$.

This potential occurs in 
what one might call
non-minimal inflation
\cite{ewansgrav}. Here, the 
original potential is not particularly flat, but the kinetic term
given by \eq{kmetric}
becomes singular at a field value of order $\mpl$
leading to a flat potential after converting to a 
canonically-normalized inflaton field.
Suppose, for example, that $K$ is given by 
\eqs{single1}{single2}, and purely for convenience suppose that
$t+t^*=\mpl$ (it is expected to be of this order). Suppose also that
all other fields vanish except for some field $\phi_1$, and set 
$\mpl=1$. Then 
$K=-3\ln(1-|\phi_1|^2)$,
and assuming that $V$ is independent of the phase of $\phi_1$
it is easy to show \cite{ewansgrav} that the 
potential is given by \eq{another} with $q=\sqrt 2$ and
the canonically normalized field
\be
\phi = {\rm tanh\,}^{-1} \sqrt2|\phi_1| -\frac1{\sqrt 2} \ln\frac
{2 dV/d|\phi_1|}{V}\vert_{|\phi_1|=1}
\ee

Another derivation \cite{rsq,burt} modifies Einstein gravity
by adding a large $R^2$ term to the usual $R$ term, but with a 
huge
coefficient,
and a third \cite{vplanck} uses a variable Planck mass. 
In both cases,
after transforming back to Einstein gravity one obtains the above form
with $q=\sqrt{2/3}$. These proposals too invoke large field values,
making it difficult to see how $V$ can be sufficiently small
(and how the kinetic terms can be almost canonical, as is assumed).

\subsection{Hybrid inflation}

We now turn to hybrid inflation models. In these models, the 
slowly rolling inflaton field $\phi$ is 
not the one responsible for most of the energy density.
That role is 
played by another field $\psi$, which is held in place
by its interaction with the inflaton field
until the latter falls below 
a critical
value $\phi\sub{c}$.
When that happens $\psi$ is destabilized
and inflation ends.

This paradigm has proved very fruitful, since its 
introduction by Linde \cite{LIN2SC} in 1991.
Early treatments of it are references
\cite{LL2} (1993), \cite{LIN2SC2,cllsw,silvia,qaisar,wang}
(1994), \cite{ewansgrav,mutated,lazpan,dave}
(1995), and \cite{lisa,dvaliloop,glw,ournew,bindvali,halyo}
(1996); by now it is the standard paradigm of inflation.

In a related class of models the inflaton field is rolling away
from the origin, and inflation ends when it {\em rises} above
some critical value $\phi\sub c$.  
This paradigm, now known as inverted hybrid inflation, is less useful
as we shall discuss in Section \ref{invhyb}. It was
introduced by  Ovrut and
Steinhardt \cite{burt2} in 1984, but has received little attention.

Note that the essential feature of hybrid inflation is the 
{\em dominance} of the potential, by the field that is held fixed.
Potentials
of the form proposed by Linde had been considered earlier
by several 
authors, starting with Kofman and Linde \cite{Kofman}.
But they presumed 
the parameters to be such that the other field
gives only a small contribution to the potential.
As we noted at the end of Section \ref{s:symres}, 
such models are interesting because they
might produce topological defects, or a 
feature in the spectrum, but they are not hybrid inflation 
models.\footnote
{The earlier model of \cite{enqvist1,enqvist2} seems also to be of this kind.}

\subsection{Hybrid inflation with a quadratic potential}

We begin with the case that 
the potential during inflation has 
the simplest possible tree-level form,
\be
V=V_0+ \frac12 m^2\phi^2.
\label{vord}
\ee
The first term is supposed to dominate, and inflation occurs provided
that the condition
\be
m^2 \ll V_0/\mpl^2
\label{phicond}
\ee
is at least marginally satisfied (this is the condition $\eta\ll 1$).
We shall assume unless otherwise stated that 
$\phi\ll\mpl$, so that $\epsilon\ll\eta$ and 
\be
n-1=2\eta=2\mpl^2 m^2/V_0 \,.
\ee

By itself, the above potential has no mechanism for ending inflation,
since the flatness parameters $\epsilon$ and $\eta$ become smaller
as $\phi$ decreases. Inflation is supposed to end through a hybrid 
inflation mechanism as described in a moment, 
when $\phi$ falls below some critical value
$\phi\sub c$. 
When the observable Universe leaves 
the horizon
\be
\frac{\phi}{\phi\sub{c}}=e^x \,,
\label{phih}
\ee
where $x$ is given by \eq{xofnn}.
At least  with the two prescriptions for $\phi\sub c$ discussed below,
\eq{phih} is consistent with the assumption $\phi\ll\mpl$.

We emphasize at this point that the loop correction, ignored 
when one considers this potential, often dominates in reality.
Several examples will be given later.

Proceeding with the assumption of a tree-level potential,
the COBE normalization \eq{cobenorm} is
\be
\frac2{n-1} \frac{V_0^{1/2}}{\mpl\phi} =5.3 \times 10^{-4} \,,
\label{cobenormhyb} 
\ee
or
\be
\frac{V_0^{1/2}}{\mpl^2}= 5.3\times 10^{-4} \frac{n-1}{2} 
e^{x}\frac{\phi_{\rm c}}{\mpl} \,. 
\label{cobenormhyb2} 
\ee

To work out $\phi\sub c$, we need to include the non-inflaton 
field $\psi$ that is responsible for $V_0$. 
The full potential for the original model 
\cite{LIN2SC} is \eq{fullpot1} that we already considered.
\bea
V&=&V_0-\frac12 m_\psi^2 \psi^2 + \frac14\lambda\psi^4 
 + \frac12 m^2 \phi^2 +\frac12\lambda'\psi^2\phi^2 \\
&=&\frac14\lambda(M^2-\psi^2)^2 +\frac12m^2\phi^2
+\frac12\lambda'\psi^2\phi^2. 
\label{fullpot}
\eea
Comparing the two ways of writing the potential, 
one sees that the parameters are related by
\bea
m_\psi^2&=&\lambda M^2, \label{rel1} \\
V_0 &=& \frac14\lambda M^4 = \frac14 M^2 m_\psi^2. 
\label{rel2}
\eea
This gives 
\be
\phi\sub{c}^2=m_\psi^2/\lambda'=\lambda M^2/\lambda'.
\label{phic}
\ee

It is useful to define
\be
\eta_\psi\equiv \frac{m_\psi^2\mpl^2}{V_0} \,.
\ee
To have inflation end promptly
when $\phi$ falls below $\phi\sub c$, as is assumed in this model,
one needs $\eta_\psi$
significantly bigger than 1.\footnote
{This is obvious if $\psi$ remains homogeneous, but the same result can
actually be established \cite{cllsw} even without that assumption.}
In terms of $\eta_\psi$, the COBE normalization becomes
\be
\lambda' = 2.8\times 10^{-7} e^{2\eta N} \eta^2 \eta_\psi \,.
\label{lamband}
\ee

A different prescription \cite{lisa} is to replace the 
renormalizable 
coupling $\frac14\lambda\psi^2\phi^2$ by a non-renormalizable coupling
\be
\frac12\psi^2\phi^4/\mpltil^2\,.
\label{otherphipsi}
\ee
The COBE normalization is now
\be
\frac{V_0^{1/2}}{\mpl\mpltil} = 2.8 \times 10^{-7}
e^{2\eta N} \eta^2 \sqrt\eta_\psi \,.
\label{mnorm}
\ee
If one takes $\mpltil$ significantly below $\mpl$, the flatness
conditions on the potential discussed in Section
\ref{keepflat} may become more stringent.\footnote
{The scale $\mpltil$ is presumably supposed to come from integrating
out some sector of the full theory. The non-renormalizable terms 
relevant for $\phi$ may or may not have the same effective scale
$\mpltil$. See Section 
footnote \ref{rvnon}.}

With $\phi\sub{c}$ given
by either of these prescriptions, \eq{cobenormhyb2} implies
\cite{cllsw}
a limit
$n\lsim 1.3$ (assuming that 
$V_0$ dominates the potential and that $M\lsim \mpl$).
In that case, the present observational limit $|n-1|<0.2$
is more or less 
predicted.

Different prescriptions
will be considered in Sections
\ref{mut}, \ref{fromd} and \ref{stevemod}, and on page
\pageref{gstrofv}. In the last case, $\phi$ is of order $\mpl$
when the observable Universe leaves the horizon.

\subsection{Masses from soft susy breaking}

\label{lisamodel}

When hybrid inflation is implemented in a
supersymmetric theory, the slope of the potential is often dominated
by a loop correction. But there are cases where
a tree-level slope $\frac12m^2\phi^2$
can dominate and we mention one of them now.

The crucial feature of the model \cite{lisa}
is that 
the parameters
$\eta$ and $\eta_\psi$ are both very roughly of order 1.
(This is what one might expect if the masses $m$ and $m_\psi$
both vanish in the limit of global supersymmetry, and come only from
supergravity corrections.)
The
vev of $\psi$ is therefore roughly $M\sim \mpl$.
It is presumed that this is achieved by replacing the first term of
\eq{fullpot} by a more complicated function of $\psi$, rather than by
making $\lambda$ tiny as would be required by \eq{rel2}.
If $\psi$ is the radial part of a matter field, $\lambda$ is
presumably negligible while the non-renormalizable terms are suppressed.

The simplest thing is to assume that $\psi$ is the
dilaton or a bulk modulus, whose potential is of the form
\eq{vdilmod}. 
Alternatively it might be a matter field with non-renormalizable 
coupling
suppressed to high order as a result of a discrete symmetry.
In any case, $\psi=0$ is presumably a fixed point of the 
relevant
symmetries.

A less crucial feature is the assumption that $V_0^{1/4}$ is very 
roughly of order $10^{10}\GeV$. This is motivated by an 
assumption that there is a gravity-mediated mechanism of 
susy breaking in the true vacuum, which operates also
during inflation with essentially the same strength.

As we have seen, 
the observational constraint $|n-1|<0.2$ actually requires $|\eta|<0.1$.
The reduction of $\eta$ below its natural value of order 1 is supposed 
to come from an accidental cancellation in this model. To minimize the 
cancellation required, one prefers 
$n$ to be significantly 
above 1.\footnote
{Of the six examples 
displayed in the Figures of Ref.~\cite{lisa} only one actually has 
$n$ significantly bigger than 1, and therefore it should be regarded as
a favoured parameter choice.}

With the choice $\eta_\psi\sim 1$ some number $N_\psi$ of
$e$-folds of inflation occur 
after $\phi$ reaches $\phi\sub c$.
As discussed in Section \ref{nefolds}, one
has to require that $N_\psi$ is less than the total number of 
$e$-folds after cosmological scales leave the horizon, since
the fluctuation while $\psi$ is rolling does not generate the
flat spectrum required in this regime.
In fact, it gives a spike in the spectrum \cite{lisa,glw},
and one must 
require that it does not
lead to excessive black hole formation. Typically this reduces the 
already significant upper 
limit on $n$, that follows from the same requirement
in the absence of a spike
\cite{jim,lisa,toniandrew}.

Assuming that $\psi$ remains almost homogeneous after $\phi$
falls below $\phi\sub c$, one can calculate the number of $e$-folds
of inflation that occur while $\psi$ rolls down to its vacuum value
$\psi=M$.\footnote
{In the case $\eta_\psi\ll 1$ one has slow-roll inflation, and the 
homogeneity can be checked by calculating the vacuum fluctuation.
It seems reasonable that is will hold to sufficient accuracy also
if $\eta_\psi\sim 1$.}
The result is 
\cite{ewanloop2}
\be
N_\psi =\frac1{2\eta_\psi} \left(
1+\sqrt{1+\frac{4\eta_\psi}{3} } \right)
\ln\frac{M}{\psi\sub{initial}} \,.
\label{nfast}
\ee
Here 
$\psi\sub{initial}\sim H$ is the initial value of $\psi$,
given by its quantum fluctuation.
Since $V_0^{1/4}$ is supposed to be of order $10^{10}\GeV$,
in this model, $\psi\sub{initial}\sim 10^{-16}\mpl$, leading to
\cite{ewanloop2}
\be
N_\psi \sim \frac{37}{2 \eta_\psi} \left(
1+\sqrt{1+\frac{4\eta_\psi}{3 } } \right).
\label{nfast2}
\ee
Requiring $N_\psi <10$ leads to $\eta_\psi> 8 $,
and requiring $N_\psi < 30 $ leads to $\eta_\psi > 1.7 $.

In this model, the COBE normalization requires 
$\lambda'$ in \eq{lamband}, or
$\mpltil/\mpl$ in \eq{mnorm}, to be a few orders of magnitude below
unity. These small couplings are 
consistent with the assumption that loop corrections
are negligible. On the other hand, the inflaton could still have
large couplings to other fields, which could give a large loop
correction. If that happens, one arrives at the running inflaton 
mass model of Section \ref{hybrunning}.

\subsection{Hybrid thermal inflation}

\label{hybthermal}

Related to the scheme we just described, is a radical proposal
\cite{graham3},
which would have a distinctive observational signature. Its basic 
ingredients are fairly natural, though the particular combination
required may be difficult to arrange.

The idea is to have a hot big bang during the era immediately preceding 
observable inflation, 
with all relevant 
fields in thermal equilibrium as was proposed in the early models of inflation.
(This primordial hot big bang is presumably preceded by more inflation
as described in Section \ref{before}.) 
Let us begin with the 
simplest version of the proposal.
Including the finite temperature $T$, the
potential during inflation is something like
\be
V (\phi,\psi) = V_0 +T^4 
+T^2\psi^2 - \frac12m_\psi^2 \psi^2+ T^2\phi^2 - \Delta V(\phi)
 \,.
\ee
As in the previous case, it is supposed that very roughly
$m_\psi^2\sim V_0/\mpl^2$, corresponding to a true vacuum value
$\psi$
very roughly of order $\mpl$ (but maybe some orders of magnitude less).
The last term, which will determine the motion of the 
the inflaton field $\phi$,
is not specified in detail.

The temperature falls roughly like $1/a$, and an
epoch of what one might call `hybrid thermal inflation' 
begins when the potential is dominated
by $V_0$ 
at $T\sim V_0^{1/4}$, and ends when $\psi$ is destabilized at 
$T\sim m_\psi$.\footnote
{\label{ordtherm} The phenomenon of ordinary thermal inflation was noted
in References \cite{bingall,thermal2}, and discussed in detail in
References \cite{t3,thermal,t2,t1}. Ordinary thermal inflation is 
identical with the phenomenon we are describing now, except that the 
field $\phi$ is not present. Ordinary thermal inflation is
supposed to happen long after ordinary inflation is over,
with the susy breaking scale the same as in the vacuum. This makes
$m_\psi$ a typical soft mass of order $100\GeV$,
and assuming $\psi\ll\mpl$ it makes $V_0^{1/4}\ll 10^{10}\GeV$.}
This lasts for $N\sub{thermal}\sim 10$ 
$e$-folds. After a further $N_\psi$ $e$-folds, given by \eq{nfast2},
$\psi$ arrives at its true vacuum value and inflation ends.
Meanwhile, $\phi$ rolls slowly, and is supposed to be the dominant 
source of the primordial curvature perturbation.
(This last feature would need checking case by case, as the other
field $\psi$ may be significant---see Section \ref{s10}.)

To avoid unacceptable relics of the thermal era, at least a few
$e$-folds of inflation have to occur before the observable Universe
leaves the horizon \cite{LL2,tilted}, which will 
probably use up all of the $e$-folds of thermal inflation.
In that case, we just have a hybrid inflation model with
the unspecified potential $V=V_0 -\Delta V(\phi)$.
(Different from the usual case though, in that the other field $\psi$
is already destabilized when the observable Universe leaves
the horizon.)
However, there could well be several of the other fields $\psi_n$,
taking different numbers of $e$-folds to reach their vacuum values.
As each one does so, a feature in the spectrum could be generated,
because the inflaton mass coming from supergravity may change.
As a more complicated variant of the scheme, one may suppose that
the destabilization of one field affects the stability of another.

\subsection{Inverted hybrid inflation}

\label{invhyb}

One can also construct hybrid inflation models where 
$\phi$ is rolling away from the origin, under the influence
of the inverted quadratic potential \eq{natural}.
A simple potential $V(\phi,\psi)$ which achieves this is
\cite{ournew}
\be
V = V_0 - \frac{1}{2} m_\phi^2 \phi^2 + \frac{1}{2} m_\psi^2 \psi^2
- \frac{1}{2} \lambda \phi^2 \psi^2 + \cdots \,.
\label{first}
\ee
The dots represent terms which give $V$ a minimum where it vanishes,
but which play no role during inflation.
At fixed $\phi$ there is a minimum
at $\psi=0$ provided that
\be
\phi < \phi_{\rm c} = \frac{m_\psi}{\sqrt{\lambda}}.
\ee
A better-motivated potential leading to inverted hybrid inflation
will be described in Section \ref{stevemod}. A more complicated
one appears in Reference
\cite{burt2}, but the inflaton trajectory turns out to be unstable 
\cite{olive}.

Inverted hybrid inflation is characterised by the 
appearance of a negative coupling $-\phi^n\psi^m$, in contrast with the usual
positive coupling $\phi^n\psi^m$. 
Such a negative
coupling, for fields in thermal equilibrium, corresponds to
high temperature symmetry restoration \cite{weinberg}. In the context
of supersymmetry it
is more difficult to arrange than the positive coupling.
In any case, one has to ensure that the 
potential remains bounded from below in its presence.

\subsection{Hybrid inflation with a cubic or higher potential}

Instead of the quadratic potential \eq{vord}, one might consider
a potential
\be
V= V_0 \( 1 + c\phi^p \) \,,
\ee
with $p\geq 3$ (and $c>0$). 

This case is similar to the one that we discuss in some detail 
in Section \ref{dsbinflation}.
One has
\be
\eta = c\mpl^2 p (p-1) \phi^{p-2} \,,
\ee
and inflation is possible \cite{LIN2SC2}
only in the regime $\eta\ll 1$.\footnote
{We are assuming that $V\simeq V_0$ as long as the right hand side of the 
above expression is $\ll 1$. 
As usual, we consider only the case $\phi\lsim \mpl$.}
It is not clear how the inflaton is supposed to get into this
regime.

The number of $e$-folds to the end of inflation is
\be
N(\phi) \simeq \( \frac{p-1}{p-2} \) \( \frac1{\eta(\phi\sub c)}
-\frac1{\eta(\phi)} \) \,.
\ee
For $\phi\gg \phi\sub c$, $N(\phi)$ approaches a constant
\be
N\sub{max} \equiv 
\left({p-1\over p-2}\right) {1 \over
\eta\left(\phi\sub{c}\right)} 
\label{ntot}
\end{equation}
The spectral index is given by
\be
\frac{n-1}{2} = \(\frac{p-1}{p-2} \) \frac1{N\sub{max} - N} 
\,.
\ee

The quartic case has been considered in some detail \cite{dave},
including the regime $\phi\gg\mpl$ that we are ignoring.

One may also consider the case 
where (say) quadratic, cubic and quartic terms are all important 
during observable inflation
\cite{wang}, but that will clearly involve considerable fine-tuning. 

\subsection{Mutated hybrid inflation}

\label{mut}

In both ordinary and inverted hybrid inflation, the other field $\psi$ is 
precisely fixed during inflation. If it varies,
an effective potential $V(\phi)$ can be generated even if the
original potential contains no piece that depends only on $\phi$.
This mechanism was first proposed in Ref.~\cite{mutated}, where it was called 
mutated hybrid inflation. The potential considered was
\be
V = V_0\( 1-\psi/M \) + \frac14\lambda\phi^2\psi^2 + \cdots
\label{mutpot}
\ee
The dots represent one or more additional terms, which
give $V$ a minimum at which it vanishes but play no role during 
inflation.
All of the other terms are significant,
with $V_0$ dominating.
For suitable choices of the parameters inflation takes place
with $\psi$ held at the instantaneous minimum, leading to a potential
\be
V= V_0 \(1-\frac {V_0}{\lambda^2 M^2 \phi^2} \) \,.
\ee
This gives
\be
n-1 = -\frac3{2N} \,,
\ee
and the COBE normalization \eq{cobenorm} is
\be
5.2\times 10^{-4} = (2N)^{3/4} \sqrt\lambda \frac{V_0^{1/4} \sqrt M}{\mpl^{3/2}}
\,. 
\label{cobemut}
\ee

A different version of 
hybrid inflation \cite{lazpan} was called
`smooth' hybrid 
inflation emphasizing that any topological defects associated with 
$\psi$ will never be produced.
In this version, the potential is 
$V=V_0-A\psi^4+B\psi^6\phi^2 +\cdots$.
It leads to 
$V=V_0(1-\mu \phi^{-4})$.

Retaining the original name, the most general mutated hybrid inflation 
model with only two significant terms is \cite{ournew}
\be
V = V_0 - \frac{\sigma}{p} \mpl^{4-p}\psi^p + \frac{\lambda}{q} 
\mpl^{4-q-r}\psi^q \phi^r
+\cdots \,.
\ee
In a suitable regime of parameter space,
$\psi$ adjusts itself to minimize $V$ at fixed
$\phi$, and
$\psi\ll\phi$ so that the slight curvature of the 
inflaton trajectory does not affect the field dynamics. Then,
provided that $V_0$ dominates the energy density, the effective potential 
during inflation is
\be
V=V_0(1-\mu \phi^{-\alpha}),
\label{vmut}
\ee
where
\bea
\mu&=&\mpl^{4+\alpha} \left(\frac{q-p}{pq}\right)
\frac{\sigma^{\frac q{q-p}}\lambda^{-\frac{p}{q-p}}}
{V_0} >0,\\
\alpha&=&\frac{pr}{q-p}.
\eea
For $q>p$, the exponent $\alpha$ is positive as in the examples already 
mentioned, but for $p>q$ it is negative with $\alpha<-1$.
In both cases it can be non-integral, though integer values are the most 
common for low choices of the integers $p$ and $q$.
This potential is supposed to hold until $V_0$ ceases to dominate at
\be
\phi\sub{end}\sim \mu^{1/\alpha} \,,
\label{phiend2}
\ee
after which slow-roll inflation ends.

The situation in 
the regime $-2<\alpha<-1$ is similar to the one that we 
discussed already for the case $\alpha=-2$;
the prediction for $n$ covers a continuous range below 1 because it
depends on the parameters,
but to have a model with $\phi\ll\mpl$ the potential has to be steepened
after cosmological scales leave the horizon. The COBE normalization
in this case is \cite{ournew} 
\be
5.3 \times 10^{-4} 
= \frac{\mpl^{\alpha-2} V_0^{1/2} }{ |\alpha| \mu }
\left[ \mpl^{|\alpha|-2} \phi_{\rm c}^{2-|\alpha|} - 
|\alpha| \left( 2 - |\alpha| \right) \mpl^\alpha \mu N
\right]^{ - \frac{|\alpha|-1}{2-|\alpha|} } \,.
\ee

In the cases $\alpha< -2$ and $\alpha>-1$, the situation is similar
to the the one that we 
encountered in Section \ref{cubhigh} (except for the special cases
$\alpha\simeq -2$ and $\alpha\simeq -1$, which we do not consider). 
In the case $\alpha < -2$, the integral (\ref{nint})
is dominated by the limit $\phi$
provided that $\phi\sub{end}\ll \sqrt N \mpl$, which we assume.
In the case $\alpha>-1$ one has $\phi\sub{end}<\phi$, and
assuming $\phi\ll \mpl$ while cosmological scales leave the horizon
again means that \eq{nint} is dominated by the limit $\phi$.
In all of these cases, the COBE normalization \eq{cobecubhigh}
and the prediction \eq{ncubhigh} are valid, with $p$ replaced
by $-\alpha$.

Of the various possibilities regarding $\alpha$, some 
are preferred over others in the context of supersymmetry.
One would prefer \cite{ournew} $q$ and $r$ to be even if
$\alpha>0$ (corresponding to $q>p$)
and $p$ to be even if $\alpha<0$. Applying this criterion with $p=1
$ or $2$ and $q$ and $r$ as low as possible leads \cite{ournew}
to the original mutated hybrid model, along with the cases
$\alpha=-2$ and $\alpha=-4$ that we discussed earlier
in the context of inverted hybrid and single-field models.

A different example of a mutated hybrid inflation potential is given in
Ref.~\cite{glw}, where $\psi$ is a pseudo-Golstone boson
with the potential (\ref{gbpot}). 

\paragraph{Mutated hybrid inflation with explicit $\phi$ dependence}

So far we have assumed that the original potential has no piece that
depends only on $\phi$. If there is such a piece it has to be added 
to the inflationary potential (\ref{vmut}). If it 
dominates while cosmological scales leave the horizon, 
the only effect that the $\psi$ variation  has 
on the inflationary prediction is to 
determine $\phi\sub{c}=\phi\sub{end}$ through Eq.~(\ref{phiend2}). 

\subsection{Hybrid inflation from dynamical supersymmetry breaking}

\label{dsbinflation}

In Section \ref{anonp}, we noted that non-perturbative effects,
such as those associated with dynamical supersymmetry breaking,
could give a 
potential proportional to $1/\phi^p$ where $p$ is some 
integer,
\begin{equation}
V\left(\phi\right) = V_0 + {\Lambda^{p + 4} \over \phi^p} + 
\cdots \,,
\end{equation}
where the dots represent terms that are negligible during inflation.
This potential 
has been proposed \cite{riottokinney,riottokinney2} as a model of 
inflation. It is convenient to define a dimensionless quantity 
$\alpha\equiv \Lambda^{p+4} \mpl^{-p} V_0^{-1}$, so that
\be
V = V_0 \( 1 + \alpha \( \frac{\mpl}{\phi} \)^p +\cdots \) \,.
\ee
This gives
\be
\eta =\alpha p(p+1) (\mpl/\phi)^{p+2} \,.
\ee
The potential 
 satisfies the flatness conditions in the regime
$\eta\ll 1$.\footnote
{We are assuming that $V\simeq V_0$ as long as the right hand side of the 
above expression is $\ll 1$.} 
Inflation is supposed to end when $\phi$ reaches a critical
value $\phi\sub c$, through some unspecified hybrid inflation mechanism.

The number of e-folds to the end of inflation is
\begin{equation}
N (\phi) \simeq \left({p + 1 \over p + 2}\right) \left({1 \over
\eta\left(\phi\sub{c}\right)} - {1 \over \eta\left(\phi\right)}\right),\quad
\epsilon \ll \eta \,,
\end{equation}

For $\phi
\ll \phi\sub{c}$ $N(\phi)$ approaches a constant
\begin{equation}
N_{{\rm tot}} \equiv  \left({p + 1 \over p + 2}\right) {1 \over
\eta\left(\phi\sub{c}\right)} = {1 \over p \left(p + 2\right)} \alpha^{-1}
\left({\phi\sub{c} \over \mpl}\right)^{p + 2}.
\label{ntot2}
\end{equation}
This is quite an unusual feature.
Most models of inflation have no intrinsic upper limit on the
total amount of expansion that takes place during the inflationary phase,
although only the last $50$ or $60$ e-folds are of direct
observational significance.
Here the total amount of inflation is bounded {}from above, although that
upper bound can in principle be very large.

The COBE normalization \eq{cobenorm} is
\begin{equation}
\delta_H \simeq{\left(p + 2\right) \over  2\pi} 
\sqrt{1 \over 3} \left({V_0^{1/2} \over
\mpl \phi\sub{c}}\right) N_{\rm tot} \left(1 - {N\over N_{\rm
tot}}\right)^{\left(p + 1\right) / \left(p + 2\right)} \,,
\end{equation}
where $N\lsim 50$ corresponds to the epoch when COBE scales leave the 
horizon. 
The spectral index is given by 
\begin{equation}
n - 1 
\simeq \left({p + 1 \over p + 2}\right) {2 \over N_{\rm tot} 
-N }.
\end{equation}
The spectrum turns out to be blue ($n>1$), but  for $N_{\rm tot} \gg 50$
the spectrum approaches scale-invariance
($n=1$). If one takes the
case of $p = 2$ and $\phi\sub{c} \sim V_0^{1/4}$, the COBE constraint 
\eq{cobenorm} 
 is met for $V_0^{1/4} \simeq 10^{11}\ {\rm GeV}$ and $\Lambda
\simeq 10^{6}\ {\rm GeV}$. 

In this class of models, $n$ is indistinguishable from 1
in most of parameter space. A value of $n$ significantly above 1
is however possible for for properly tuned values of the parameters.
Taking $N=50$ and $p=2$,
a spectral index of $n > 1.1$ requires
$N_{\rm tot}$ given by \eq{ntot2} to be less than
$65$. In the context of supergravity, it 
is more comfortable to be in this regime since an accidental 
cancellation is being invoked to avoid the generic contributions
of order 1 to the quantity $2\eta=n-1$.

Such a small amount of inflation could have observationally important
consequences. Also, unlike standard hybrid inflation models, 
dynamical supersymmetric inflation  allows a measurable deviation from a 
power-law spectrum of fluctuations, with a variation in the scalar 
spectral index $|dn / d(\ln k)|$ that may be as large 
as 0.05 \cite{riottokinney2}. 

It is important to note that this upper limit on the
total amount of inflation can potentially lead to difficulties with
initial conditions: 
how does the field end up in the correct region of the potential
with a small enough rate of change to initiate slow-roll?
While this sort of problem with initial conditions is in
fact common to many models of inflation, it is mitigated to a certain
degree by the existence of classical solutions which admit a formally
{\it infinite} amount of inflation.
No such solution exists in this case.
It is reasonable to expect that the field will initially be at small values,
$\phi \ll \left\langle\phi\right\rangle$,
since the term $\phi^{-p}$ in the potential will generically
appear only at scales smaller than $\Lambda$, with
a phase transition connecting the high energy and low energy behaviours.
However, in the absence of a detailed model for this phase transition,
the question of initial conditions remain quite obscure. 

\subsection{Hybrid inflation with a loop correction from spontaneous 
susy breaking}

\label{smallloop}

 The models considered so far work at tree level. This is valid only
if the couplings of the inflaton to other fields are strongly suppressed.
In particular, the inflaton presumably has to be a gauge
singlet (no coupling to gauge fields) since gauge couplings are not 
supposed to be suppressed.

In the absence of supersymmetry, the couplings should indeed be 
suppressed. The reason is that the loop correction is then $\Delta V
\propto \phi^4 \ln(\phi/Q)$ which would spoil inflation
as in \eq{newinf98}. But with supersymmetry, there is no reason to suppose 
that the inflaton couplings are suppressed. 

As we saw in Sections \ref{onel} and \ref{loopc}, the 1-loop 
correction in a supersymmetric theory typically has one of two forms,
$\Delta V\propto\ln(\phi/Q)$ or 
$\Delta V\propto\phi^2\ln(\phi/Q)$. 
We discuss the first form in this subsection, and the second form in the next
one.

This form typically arises if susy is broken spontaneously.
Assuming that tree-level terms are negligible during inflation, the
potential is of the form 
\be
V= V_0\( 1+ \frac{C g^2}{8\pi^2} \ln(\phi/Q) \) \,.
\label{vsmallloop}
\ee
In this expression, $C$ may be taken to be the number of
possible 1-loop diagrams, in other 
words the 
number of fields which have significant coupling to the inflaton.
The other factor $g$ is a typical coupling of these fields
(times a numerical factor of order 1).
It may be a gauge coupling  ($D$-term inflation, Section \ref{s19})
or a Yukawa coupling (Section \ref{linearmod}).
In the former case $C$ might be of order 100,
which as we shall see would be bad news.

In both cases, this potential occurs as 
part of a hybrid inflation model. Depending on the parameters,
inflation ends when either slow-roll fails ($\eta\sim 1$)
or the critical value is reached, whichever is earlier.\footnote
{If slow-roll fails at a value $\phi\sub{end}>\phi\sub c$
inflation will continue until the amplitude of the oscillation
becomes of order $\phi\sub c$. The number of $e$-folds of this type of 
inflation is
$\Delta N\sim \ln (\phi\sub{end}/\phi\sub c)$, which is typically 
negligible.}
However, the precise value of $\phi\sub{end}$
is irrelevant because the integral \eq{nint}
is dominated by the limit $\phi$. It gives
\bea
\phi & \simeq& \sqrt\frac{NC g^2 }{4\pi^2} \mpl \\
&=&11\sqrt{\frac N{50} \frac C{100} \frac{g^2}{1.0} } \mpl \\
&=& 0.2 \sqrt{\frac N{20} C \frac{g^2}{0.1} } \mpl 
\,.
\label{vv}
\eea
This makes $\phi$ comparable with the Planck scale, and maybe bigger.
As we discussed in Section \ref{sss:tree} one needs $\phi\lsim\mpl$
and preferably
$\phi\ll\mpl$, in order to keep the theory under
control and in 
particular to justify 
the assumption of canonical normalization for the fields.
Let us proceed on the assumption that $\phi$ is not too big.

Assuming that the loop dominates the slope, and using \eq{nint},
the flatness parameters are
\bea
\eta &=& -\frac{1}{2N}, \\
\epsilon &=& C\frac{g^2}{8\pi^2} |\eta|.
\eea
The COBE normalization \eq{cobenorm} is 
\be
V^{1/4} = 6.0\(\frac{50}{N} \) ^{\frac 1 4} C^{1/4} g \times 10^{15}\GeV 
\,.
\label{loopcobe}
\ee

The spectral index is given by
\be
1-n=\frac{1}{N}\left(1+ \frac{3C g^2}{16\pi^2} \right) \,.
\ee
Taking the bracket to be close to 1,
and $N$ to be in the range
$25$ to $50$, one obtains the distinctive prediction
$n=.96$ to $.98$.
With $g=1$ and $C=100$, $1-n$ is increased by a factor
$\simeq 2$, but it is clear that anyhow $n$ is close to 1.
This prediction will eventually be tested.

\subsection{Hybrid inflation with a running mass}

\label{hybrunning}

Now we turn to the case, that the loop correction is 
of the form $\phi^2\ln(\phi/Q)$, which typically arises
when susy is softly broken.
Models of inflation invoking such a correction have been proposed by
Stewart \cite{ewanloop1,ewanloop2}.

As we noted in Section \ref{onel},
this type of loop correction 
is equivalent to replacing the inflaton mass by a slowly varying 
(running) mass $m^2(\phi)$. 
At $\phi=\mpl$, the running mass is supposed to have the
magnitude $|m^2|\sim V_0/\mpl^2$, which is the minimum one 
in a generic supergravity theory.
The inflaton is supposed to have couplings (gauge, or maybe
Yukawa) that are not 
too small, and for the most part we assume that $m^2(\phi)$ passes 
through zero before it stops running.\footnote
{The running associated with a given loop
will stop when $\phi$ falls below the mass of the particle 
in the loop.}
Because the couplings are small compared with unity, 
$V'$ then vanishes at some relatively nearby point, which we 
denote by $\phi_*$. 

\subsubsection{General formulas}

It is useful to write 
\be
V(\phi) = V_0 \( 1 - \frac12 \mpl^{-2} \mu^2(\phi) \phi^2 \) \,,
\label{vrunx}
\ee
where
\be
\mu^2(\phi) \equiv -\mpl^2 m^2(\phi)/V_0 \,.
\ee
We are supposing that $V_0$ dominates, since this is necessary 
for inflation in the regime $\phi\lsim \mpl$ where the field
theory is under control.
Then
\bea
\mpl\frac{V'}{V_0} 
&=& - \phi \left[\mu^2 + {1\over 2} {d\mu^2\over dt} 
\right] 
\label{V-prime}\\
\eta\equiv 
\mpl^2\frac{V''}{V_0} &=& - 
\[ \mu^2 + {3\over 2} {d\mu^2\over dt} +
{1\over 2} {d^2\mu^2\over dt^2} \]
\,,
\label{eps-eta}
\eea
where $t\equiv\ln(\phi/\mpl)$.

We 
assume that while observable scales are leaving the 
horizon one can make a linear expansion in $\ln\phi$,\footnote
{This is equivalent to writing $\mu^2=c\ln(\phi/Q)$ as in the table
on page \pageref{t:1}, the free parameter $Q$ then replacing the free
parameter $\phi_*$. In turn, this is equivalent to using a loop 
correction, with
the renormalization scale $Q$ fixed at the point where $m^2$ vanishes.}
\be
\mu^2 \simeq \mu_*^2 + c\ln(\phi/\phi_*) \,,
\label{linear}
\ee
where $|c|\ll 1$ is related to the couplings involved.
This gives
\bea
\mpl \frac{V'}{V_0} &=& c\phi\ln (\phi_*/\phi) 
\label{vp} \\
\eta \equiv \mpl^2 \frac{V''}{V_0} 
&=&
c \[ \ln (\phi_*/\phi) -1 \] \,.
\label{vpp}
\eea
Note that $\mu_*^2=-\frac12c$, and that $\mu^2=0$ 
at $\ln(\phi_*/\phi) = -\frac12$ while $V''=0$ at 
$\ln(\phi_*/\phi) = 1$.

The number $N(\phi)$ of $e$-folds to the end of slow-roll inflation
is given by
\be
N(\phi) = \mpl^{-2}\int_{\phi\sub{end}}^\phi \frac{V}{V'} d\phi
\label{Nfull}
\,.
\ee

Using the linear approximation near $\phi_*$, this gives
\be
N(\phi) = -\frac1c \ln \( \frac c \sigma \ln \frac{\phi_*}{\phi} \) \,,
\label{N}
\ee
or
\be
(\sigma/c) e^{-cN} = \ln(\phi_*/\phi) \,.
\label{ecN}
\ee
Knowing the functional form of $m^2(\phi)$, and the value
of $\phi\sub{end}$, the constant $\sigma$ can be evaluated
by taking the limit $\phi\to\phi_*$ in
the full expression \eq{Nfull}. 
We shall see that in most cases 
one expects
\be
|c|\lsim |\sigma| \lsim 1 \,.
\label{sigmaexpect}
\ee

The spectral index $n=1+2\eta$ is given in terms of $c$ and $\sigma$
by 
\be
\frac{n-1}{2} = \sigma e^{-cN} -c \,.
\label{hybn}
\ee

The COBE normalization is 
\be
\frac{V_0^{1/2}}{\mpl^2} = 5.3\times 10^{-4} \mpl \frac{|V'|}{V_0} \,,
\ee
In our case it is convenient to define a constant $\tau$
by
\be
\ln(\mpl/\phi_*) \equiv \tau/|c|\,.
\ee
Assuming that $|m^2|$ has the typical value $V_0/\mpl^2$
at the Planck scale,
the linear approximation \eq{linear} 
applied at that scale would give $\tau\simeq 1$.
Will the linear approximation apply at that scale?
If {\em all} relevant masses at the Planck scale are of order 
$V_0/\mpl^2$, one expects on dimensional grounds that the linear 
approximation will be valid in the regime $|c\ln(\phi/\phi^*)|\ll
1$. Then the approximation will be just beginning to fail at the Planck 
scale. At least in this case, one expects $\tau$ to be very roughly of 
order 1.

Using the definition of $\tau$, 
\eqs{vp}{ecN} give
\be
\frac{V_0^{1/2}}{\mpl^2} = e^{-\tau/|c|}
\exp \(-\frac\sigma c e^{-cN\sub{COBE}} \) |\sigma | e^{-cN\sub{COBE}} 
\times 5.3\times 10^{-4} \,.
\label{thiscobenorm}
\ee

In these models, the spectral index may be strongly scale-dependent.
In fact, using $d\ln k=-dN$ one finds
\be
\frac{dn}{d\ln k} = 2 c\sigma e^{-cN} 
=2c\(\frac{n-1}{2} +c \) \,.
\label{hybdndk}
\ee
For it to be eventually observable we need $|dn/d\ln k|\gsim
10^{-3}$, and this condition is satisfied in a large part of the 
parameter space. 

Let us discuss the regime of validity of
\eqs{hybn}{hybdndk}, using \eqs{nerror}{dnerror}.
The quantities appearing in these expressions are
\bea
\xi^2 &=&c\sigma e^{-cN} \\
\sigma_V^3 &=& -\xi^4/c = -\xi^2 c\ln(\phi/\phi_*)  \,.
\eea
(We relabelled the quantity $\sigma$ in \eq{sigdef}
as $\sigma_V$.)

\eq{hybn} will be a good approximation if
\be
|\xi|^2\ll |\eta| \,.
\label{xilleta}
\ee
In contrast to the other models
we have discussed 
(where $V'\propto \phi^p$), this condition is not
guaranteed. But in this model, $\xi^2$ is slowly varying.
As a result 
\eq{dxi} (with $\epsilon$ negligible) implies
that the condition will hold except within a few $e$-folds
of a point where $\eta$ changes sign.

The error of order $\xi^2$ just represents a small change in 
the effective value of $\sigma$, which can be cancelled by a small 
change in the underlying parameters (couplings and masses).
The improved slow-roll approximation
\eq{nbetter} shows that the error actually corresponds
to changing $\sigma$ by an amount $1.06 c$.
In the present state of theory the precise 
amount is not of interest. It would become so only if the 
underlying parameters were predicted by something like string theory.

When cosmological leave the horizon, $|\sigma_V^3|\ll|\xi^2|$,
so the slow-roll formula for $dn/d\ln k$ will also be valid.

\subsubsection{The four models}

Four types of inflation model are possible, corresponding to 
whether $\phi_*$ is a maximum 
or a minimum, and whether
$\phi$ during inflation is smaller or bigger than $\phi_*$.

In the case that 
$\phi_*$ is a maximum, one expects the potential to have the form 
shown in Figures \ref{f:loopc} and \ref{f:loopd}. There is a minimum 
at $\phi=0$, and the non-renormalizable terms will ensure that there is
a minimum also at some value $\phi\sub{min}>\phi_*$.
The latter will generally be lower than the one at the origin, and we 
assume that this is the case. 
This lowest minimum represents the true vacuum if $V$ vanishes
there as in Figure \ref{f:loopc}. If instead  $V$ is positive
as in Figure \ref{f:loopd}, the vacuum lies in some other field
direction, `out of the paper'.
In this case, it is supposed that $\phi$ arrives near the maximum 
by tunneling from the minimum that lies on the opposite side.

In the case that $\phi_*$ is a minimum, 
the potential will be like the one in Figure \ref{f:loopa}.
The unique minimum represented by $\phi_*=0$ is the vacuum if $V$
vanishes there (the case shown in Fig.~\ref{f:loopa}). If instead
$V$ is positive at the minimum, 
the vacuum lies in some other field direction.

\paragraph{Model (i); $\phi_*$ a maximum with $\phi<\phi_*$}

This model \cite{ewanloop2,p98laura} corresponds to 
$m^2(\mpl)<0$, 
$c>0$ and $\sigma>0$, with $\phi$ decreasing during inflation.
The spectral index increases as the scale $k^{-1}$ decreases,
and can be either bigger or less than 1.

For inflation to end, the form \eq{vrunx} of $V(\phi)$ must
be modified when $\phi$ falls below some critical value
$\phi\sub c$, presumably through a hybrid inflation mechanism.
On the other hand, if the inflaton mass continues to run 
until $m^2\simeq V_0/\mpl^2$,
{\em slow-roll} inflation will end then.
Let us suppose first that this is the case, and define 
$\phi\sub{fast}$ by
\be
m^2(\phi\sub{fast}) = V_0/\mpl^2 \,.
\ee
This is equivalent to defining $\eta(\phi\sub{fast})=1$,
up to corrections of order $c$ which presumably should not be included 
in a one-loop calculation.
The end of
slow-roll inflation corresponds to $\phi\sub{end}=\phi\sub{fast}$,
and the linear approximation \eq{linear}
gives the rough estimate
$|\ln(\phi\sub{end}/\phi_*)|\sim 1/c$, making $\sigma\sim 1$.

Now consider the case where inflation ends at some value
$\phi\sub c$, with $|m^2(\phi\sub c)|<  V_0/\mpl^2$. If 
the mass is still running
at that point, the linear estimate \eq{N} gives
\linebreak
$\sigma \sim c\ln(\phi_*/\phi\sub c) < 1$. 
Values $\sigma\ll c$ can be achieved
only with $\phi\sub c$ very close to $\phi_*$ which would 
represent fine-tuning.
Therefore we expect in this case
$c\lsim \sigma\lsim 1$.

If the mass stops running before 
$\phi\sub c$ is reached, at some point $\phi\sub{low}$,
then $m^2$ has a constant value $m^2\sub{low}=m^2(\phi\sub{low})$ 
in the regime
$\phi\sub c < \phi <\phi\sub{low}$.
In this regime, some
number $\Delta N$ of $e$-folds of slow-roll inflation occur.
We are assuming that cosmological scales leave the horizon while the 
mass is still running, which requires
\bea
\Delta N &< & N\sub{COBE} - 10 \\
& <& 38 + \ln (V_0^{1/4}/10^{10}\GeV) \,.
\label{delN}
\eea
Retaining the estimate of the previous paragraph for
the $e$-folds of inflation before the mass stops running, the 
constant $\sigma$ to be used in \eq{ecN} will be in the range
\be
c\lsim \sigma \lsim e^{c\Delta N} \,.
\label{bigsigma}
\ee
After imposing observational constraints \cite{p98laura,p98laura2},
one finds that
$ e^{c\Delta N}$ is 
no more than one or two orders of magnitude above
unity. 

\paragraph{Model (ii); $\phi_*$ a maximum with $\phi>\phi_*$}

Like the previous model, this one corresponds to $m^2(\mpl)<0$ and
$c>0$, but now $\sigma<0$ and $\phi$ increases during inflation.
The spectral index is less than 1, and decreases as the scale decreases.

In contrast with the previous case, inflation
can end without any need for  a hybrid inflation mechanism, or a change 
in the form of the potential \eq{vrunx}, if the minimum at $\phi >\phi_*$
is the true vacuum. If the form \eq{vrunx} holds until 
$\phi$ reaches the value $\phi\sub{fast}$ defined by
$\eta(\phi\sub{fast})=-1$, slow roll inflation will end there. 
To leading order in $c$ this corresponds
to\footnote
{This estimate of $\phi\sub{fast}$ assumes that quartic and 
higher terms in the tree-level potential
are negligible at $\phi\sub{fast}$. Assuming that only 
one such term is significant, one easily checks that
the estimate is roughly correct, unless the dimension 
of the term is not extremely large. We do not consider that case,
or the case where more than one term is significant.}
\be
m^2(\phi\sub{fast}) = -V_0/\mpl^2 \,.
\ee
Setting $\phi\sub{end}=\phi\sub{fast}$, and 
using the crude linear 
approximation
one finds $\phi\sub{end}\sim e^{1/|c|}\phi_*\sim \mpl$, and $\sigma\sim 
-1$.

On the other hand, slow-roll inflation might 
end at some 
earlier point $\phi\sub c$. In the true-vacuum case illustrated in
Figure \ref{f:loopc}, this may happen 
through a steepening in the form of $V(\phi)$. Otherwise it may happen 
through an inverted hybrid inflation mechanism. In both cases,
we expect $c\lsim |\sigma| \lsim 1$. 

In contrast with the previous model, this one also makes sense 
if $m^2$ stops running (as $\phi$ decreases)
before it changes sign; in other words,
if it stops running at $\phi\sub{low}$ with $m^2(\phi\sub{low})
<0$, but very small. In this case the maximum of the potential
is at the origin and $\eta$ is small and constant up to $\phi =0$.
The above treatment remains valid if $m^2$ has started to run before 
cosmological scales leave the horizon (remember that in this model,
$\phi$ increases during inflation).
Otherwise, one has a different model that we shall not consider.

\paragraph{Model (iii); $\phi_*$ a minimum with $\phi<\phi_*$}

This corresponds to $m^2(\mpl)>0$, $c<0$ and $\sigma<0$,
and $\phi$ increases during inflation. The spectral index can
be either above or below 1, and it increases as the scale decreases.

Now $|m^2|$ decreases during 
inflation, and slow-roll inflation
ends only when
the potential \eq{vrunx} 
ceases to hold at some value $\phi\sub{end}= \phi\sub c$.
In a single-field model, corresponding to $V$ vanishing at
 the minimum,
this 
can occur through a steepening of the form of the tree-level
potential, as higher powers of $\phi$ become important. Alternatively,
if $V$ is positive at the minimum
it can occur through a hybrid inflation mechanism (inverted hybrid
inflation).

To estimate $\sigma$ in this case, suppose first that 
(as $\phi$ decreases) the mass continues to run until 
$m^2= -V_0/\mpl^2$, and denote the point where this happens
by $\phi\sub{fast}$. Slow roll inflation can then 
only occur in the regime
$\phi\gsim \phi\sub{fast}$. 
It follows that 
\be
\phi\sub {end} \gsim \phi\sub{fast} \,,
\label{phiend3}
\ee
and the 
linear approximation $\phi\sub{fast}\sim e^{-1/|c|}\phi_*$
then gives $|\sigma|\lsim 1$. As before $|\sigma|\gsim |c|$
is required to avoid the fine-tuning 
$\ln(\phi_*/\phi\sub c)\ll 1$.\footnote
{Stewart \cite{ewanloop1} took the view that models (iii) and (iv)
require a fine-tuning of $\phi\sub c$ over the whole range of parameter 
space. As with all views on fine-tuning, this is a matter of
taste.}

If the mass stops running at some point $\phi\sub{low}$,
with $|m^2(\phi\sub{low})|\ll 1$,
 inflation can begin at arbitrarily small field values.
If cosmological scales
start to leave the horizon only after the mass has started to run,
\eq{phiend3} still applies and the estimate for
$\sigma$ is unchanged. We do not consider the opposite case.

\paragraph{Model (iv); $\phi_*$ a minimum with $\phi>\phi_*$}

Like the previous case this one corresponds to $m^2(\mpl)>0$
and $c<0$, but now $\sigma>0$ and $\phi$ decreases during inflation.
The spectral index is bigger than 1, and it decreases as the scale 
decreases.

Everything is the same as in the previous case, except that 
a hybrid inflation mechanism will definitely be needed to end
inflation, since higher-order terms in $\phi$ can hardly become
more important as $\phi$ decreases.
We again expect $|c|\lsim \sigma
\lsim 1$, with the lower limit needed to avoid the
fine-tuning $\ln(\phi\sub c/\phi_*)\ll 1$.
As a result we expect $|c|\lsim \sigma \lsim 1$.

Like Model (iii), this one can still make sense if the mass stops 
running before $\phi_*$ is reached. 
The above treatment applies if cosmological scales leave the horizon
while the mass is still running. We do not consider the opposite case.

\subsubsection{Observational constraints}

In this model, the spectral index can change very significantly
on cosmological scales. The usual constraint $|n-1|<0.2$
may therefore not apply, but as a crude procedure \cite{p98laura2}
one can impose this 
constraint at both $N\sub{COBE}$ and $N\sub{COBE}-10$.
In all four models one finds a viable range of parameter space.

\subsection{The spectral index as a discriminator}

\label{discuss}

The point of contact with observation is the spectral index
$n(k)$. The Planck satellite will measure it with an accuracy
$\Delta n\sim 0.01$ over a range $\Delta\ln k\simeq 6$,
and will measure $dn/d\ln k$ if it exceeds a few times $10^{-3}$.
Let us summarise the predictions of the various models, and see how
well the Planck measurement will discriminate between them.

In most models of inflation, the potential is of the form
$V(\phi)=V_0 + \cdots$, with the constant first term dominating
and $\phi\lsim\mpl$. With certain qualifications stated in the text
(notably a requirement $\phi\ll \mpl$ that needs to be imposed
in certain cases) the spectrum of the
gravitational waves is too small ever to observe. With similar
qualifications, the spectral index for various models
is shown in Tables \ref{t:2} and \ref{t:1}, along with its scale-dependence 
$dn/d\ln k$. 

The simplest cases are $V=V_0\pm \frac12m^2\phi^2$, which give 
a scale-independent spectral index that may or may not be close to 
1.

Next in simplicity come the cases $V=V_0(1- c\phi^{p})$. 
Here $p$ can be an integer $\geq 3$, corresponding to self-coupling of 
the inflaton at tree-level, or it can be in the ranges
$2 <p <\infty$ or $-\infty< p< 1$ (not necessarily an integer) corresponding to
mutated hybrid inflation. Related to these, as far as the prediction
is concerned, are the cases $V=V_0(1-e^{-q\phi})$ 
(Section \ref{kininf})
which corresponds to
$p\to -\infty$ and $V=V_0(1+c\ln(\phi/Q))$ (Section \ref{smallloop})
which corresponds to $p\to 0$.
In all these cases the predictions are
\bea
\frac12(n-1) &=& -\(\frac{p-1}{p-2} \) \frac1 N \\
\frac12\frac{dn}{d\ln k} &=& -\(\frac{p-1}{p-2} \) \frac1 {N^2} \,.
\eea
The second expression can be written
\be
\frac12\frac{dn}{d\ln k} = -\(\frac{p-2}{p-1} \) \(\frac{n-1}{2} \)^2 \,.
\label{october}
\ee
Excluding the cases $p\simeq 1$ and $p\simeq 2$, the factor
$(p-1)/(p-2)$ is of order 1. As a result,
$(n-1)$ is 
far enough below zero to be eventually
observable. The scale-dependence will probably be too small to measure if
$N$ is around 50, but should be observable if $N$ is significantly 
smaller.

Next consider the case
$V=V_0(1+c\phi^{p})$ with $p$ an integer
$\geq 3$ (tree-level self-coupling) or $\leq -1$ (dynamical symmetry breaking).
In these cases there is a maximum possible number of $e$-folds of 
inflation, whose value is unknown. If it is not too big,
$n-1$ may be far enough above zero to eventually detect.
The scale-dependence is given by \eq{october}, and will be observable if
$|n-1|$ is more than a few times $0.01$.
Note that in these models, it is (more than usually)
unclear how the inflaton is supposed to arrive at the 
inflaton part of the potential.

Finally we come to the case of a running inflaton mass
(Section \ref{hybrunning}).
This gives a distinctive prediction for the shape of $n$,
and in contrast with the other models the predicted magnitude of
$dn/d\ln k$ can be of order $(n-1)$. 

\begin{table}
\centering
\begin{tabular}{|cllll|}
\hline
$p$ & \multicolumn{2}{c}{$1-n$} & \multicolumn{2}{c|}{$-10^3dn/d\ln k$} \\
& $N=50$ & $N=20$ & $N=50$ & $N=20$  \\ \hline
$p\to 0$ & $0.02$ & $0.05$ & $(0.4)$ & $2.6$ \\
$p=-2$ & $0.03$ & $0.075$ & ($0.6$) & $3.8$ \\
$p\to \pm \infty$ & $0.04$ & $0.10$ & ($0.8$) & $5.0$ \\
$p=4$ & $0.06$ & $ 0.15$ & ($1.2$) & $5.4$ \\
$p=3$ & $0.08$ & $ 0.20$ & ($1.6$) & $10.0$ \\ \hline
\end{tabular}
\caption[table2]{\label{t:2} Predictions for the spectral index $n$
and its variation $dn/d\ln k$,
are displayed for some potentials of the form
$V_0(1 -c \phi^p)$ that are discussed in the text. 
The variation will be detectable by Planck if $|dn/d\ln k|\gsim 2.0
\times 10^{-3}$.
The case $p\to 0$ corresponds to the potential $V_0(1-c\ln\phi)$, and the case
$p\to -\infty$ corresponds to $V_0(1-e^{-q\phi})$.}
\end{table}

\begin{table}
\centering
\begin{tabular}{|llll|}
\hline
Comments 
& $V(\phi)/V_0$ & $\frac12 (n-1)$ & $\frac12\frac{dn}{d\ln k}$
\\ \hline
Mass term & $1\pm\frac12c \frac{\phi^2}{\mpl^2} $ 
& $\pm c$ & 0 
\\
Softly broken susy &
$1\pm\frac12c \frac{\phi^2}{\mpl^2} \ln \frac\phi Q$ &
$\pm c + \sigma e^{\pm c N} $& $\mp c \sigma e^{\pm c N}$ \\
Spont. broken susy &
$1+c \ln\frac\phi Q$ & $-\frac1{2N} $ &  $-\frac12\frac1{N^2}$ \\
$p>2$ or $-\infty<p<1$ & $1-c\phi^p$ &
$-\(\frac{p-1}{p-2} \) \frac1 N$   &  
$-\(\frac{p-1}{p-2} \) \frac1 {N^2}$\\
(self-coupling or hybrid)  &&&\\
Various models & $1- e^{-q\phi}$ & $-\frac1 N$ & $-\frac1 {N^2}$ \\
$p$ integer $\leq -1$ (dyn.~s.~b.) & $1+c\phi^p$ &
$\frac{p-1}{p-2}\frac1{N\sub{max}-N}$ & 
$-\(\frac{p-2}{p-1}\) \( \frac{n-1}2 \)^2$ \\
or $\geq 3$ (self-coupling) &&& \\
\hline
\end{tabular}
\caption[table]{\label{t:1} Predictions for the spectral index $n(
k)$.
Wavenumber $k$ related to number of $e$-folds $N$
by $d\ln k=-dN$. 
Constants $c$, $q$ and $Q$ are positive 
while $\sigma$ and $p$ can have either sign.
In the first three cases, there is a theoretical constraint
$|c|\ll 1$. In the second case, one expects $|\sigma|\gsim |c|$.}
\end{table}

\section{Supersymmetry}
\label{s5}

\subsection{Introduction}

In the last section we looked at some `models' of inflation,
taken to mean forms for the inflationary potential that look
reasonable from the viewpoint of particle theory.
Now we go deeper,  taking on board present ideas about what might lie 
beyond the Standard Model. The eventual goal is to see whether
deeper considerations favour one form of the potential over another.
We begin by reviewing supersymmetry,
which is the almost universally accepted framework for constructing 
extensions of the Standard Model.

Supersymmetry  can be formulated either as a global or a local symmetry. 
In the latter case it 
includes gravity, and is therefore called supergravity. 
Supergravity is presumably the version chosen by
Nature. 

\subsection{The motivation for supersymmetry}

It is widely accepted that the standard model of
gauge interactions  describing the laws of physics
at the weak scale is extraordinarily successful.
The agreement between  theory and experimental
data is very good. Yet, we believe that  the present
structure is incomplete. Only to mention a few drawbacks, 
the theory has too many parameters, it does not
describe the fermion masses and why the
number of generations is three.
It contains fundamental
scalars, something difficult to reconcile with our current
understanding of non-supersymmetric field theory.
Finally, it does not incorporate gravity.

It is tempting to speculate that a new (but   yet
undiscovered) symmetry,   supersymmetry
\cite{nilles,haberkane,wessbagger,bailinlove}, 
may provide answers to these fundamental questions.
Supersymmetry is the only framework in which we seem to be able
to understand light fundamental scalars. It addresses the question
of parameters:   first, unification of gauge couplings
works much better with than without supersymmetry;
second,
it is easier to attack questions such as
fermion masses in supersymmetric theories, in
part simply due to the presence of fundamental
scalars.  Supersymmetry seems to
be intimately connected with gravity.  
So there are a number of arguments that suggest
that nature might be supersymmetric, and that supersymmetry
might manifest itself at energies of order the weak
interaction scale.  

Is supersymmetry expected to play a fundamental role
at the early stages of the evolution of the
Universe and, more specifically, during inflation? 
The answer is almost certainly yes. For one thing, the mere fact that we 
are invoking scalar fields (the inflaton, and at least one other in the 
case of hybrid inflation) means that supersymmetry is involved.
More concretely, the potential needs to be very flat in the direction of 
the inflaton, and supersymmetry can help here too. We noted earlier
that supersymmetric theories typically possess many flat directions,
in which the dangerous quartic term of the potential vanishes.
It helps in a more general sense too.
 While the necessity of introducing very small
parameters to ensure the extreme flatness of the inflaton potential seems very
unnatural and fine-tuned in  most non-supersymmetric theories, this technical
naturalness may be achieved in supersymmetric models. Indeed,  the
nonrenormalization theorem guarantees that a fundamental
object in supersymmetric theories, the  superpotential,  is not
renormalized to all orders of perturbation theory \cite{grisaru79}.
In other words, the nonrenormalization theorems in unbroken, 
renormalizable 
global supersymmetry guarantee that we can fine-tune
any parameter at the tree level and this
fine-tuning will not be destabilized
by radiative corrections at any order in perturbation theory.
Therefore, inflation in the context of supersymmetric theories
seems, at least technically speaking, more natural than
in the context of non-supersymmetric theories.

\subsection{The susy algebra and supermultiplets}

\label{ss:lag}

We begin with some basics, that apply to both global susy and
supergravity.

In the low-energy regime, phenomenology requires the type
of supersymmetry known as $N=1$ (one generator). This is usually
assumed to be the case also in the higher energy regime relevant
during inflation (though see \cite{juan97}).
In this section, we present some features of
$N=1$ supersymmetric theories, that are likely to be relevant for 
inflation. The reader interested
in more details is referred to the  excellent introductions by  
Nilles \cite{nilles}, Bailin and Love \cite{bailinlove}
and Wess
and Bagger \cite{wessbagger}. Except where stated, we use the 
conventions of Wess and Bagger 
except that some of their symbols
are replaced by more modern ones (for instance, the superpotential is 
denoted by $W$ instead of $P$.)

The basic supersymmetry
algebra is given by 
\beq
\{ Q_{\alpha}, \overline{Q}_{\dot\beta}\}
=2 \sigma^{\mu}_{\alpha\dot\beta} P_{\mu}, 
\eeq
where $Q_{\alpha}$ and $\overline{Q}_{\dot\beta} $are the supersymmetric 
generators (bars stand for conjugate),
 $\alpha$ and $\beta$
run {}from 1 to 2 and denote the two-component Weyl spinors 
(quantities with dotted indices transform under the $(0,\frac{1}{2})$ 
representation of the Lorentz group, while those with undotted indices 
transform under the $(\frac{1}{2},0)$ conjugate representation).
$\sigma^\mu$ is a matrix four vector, $\sigma^\mu=(
-{\bf 1},\vec{\sigma})$ and $P_\mu$ is the generator of
spacetime displacements (four-momentum). 

The chiral and vector 
 superfields are two irreducible
representations of the supersymmetry algebra containing
fields of spin less than or equal to one. 
Chiral fields
contain a Weyl spinor and a complex scalar; vector
fields contain a Weyl spinor and a (massless)
vector.
In superspace
a chiral superfield may be expanded in terms of the 
Grassmann variable $\theta$ \cite{wessbagger}
\beq
\phi(x,\theta)= \phi(x) + \sqrt{2}
\theta \psi(x) + \theta^2F(x). 
\label{chiralfield}
\eeq
Here $x$ denotes a point in spacetime, 
$\phi(x)$ is the complex scalar, $\psi$ the fermion,
and $F$ is an auxiliary field.
As in this expression, we shall generally use the same symbol
to represent a superfield and its scalar component.
Under a supersymmetry transformation with anticommuting
parameter $\zeta$, the component fields transform
as
\bea
\delta \phi &= &\sqrt{2} \zeta \psi, 
\label{atransform} \\
\delta \psi &= &\sqrt{2} \zeta F + \sqrt{2} i
\sigma^{\mu} \bar \zeta \partial_{\mu} \phi, \label{psitransform} \\
\delta F &= &-\sqrt{2}i \partial_{\mu} \psi \sigma^{\mu} \bar
\zeta.
\eea
Here and in the following, for any generic two-component 
Weyl spinor $\lambda$, $\bar\lambda$ indicates the complex conjugate of  
$\lambda$. 
For a gauge theory one has to introduce  
vector superfields and  the physical content is most
transparent in the  Wess-Zumino gauge.  In this gauge and for the 
simplest case of an abelian group $U(1)$, the
vector superfield may be written as
\beq
V=-\theta \sigma^{\mu} \bar \theta
A_{\mu} + i \theta^2 \bar \theta \bar \lambda
-i \bar \theta^2 \theta \lambda + \half
\theta^2 \bar \theta^2 D.
\label{vectorfield}
\eeq
Here $A_{\mu}$ is the gauge field, $\lambda_{\alpha}$
is the gaugino, and $D$ is an auxiliary field.
The analog of the
gauge invariant field strength is a chiral field:
\beq
W_{\alpha} = -i \lambda_{\alpha}
+ \theta_{\alpha}D -\smallfrac i 2
(\sigma^{\mu} \bar \sigma^{\nu} \theta)_{\alpha} F_{\mu \nu}
+ \theta^2 \sigma^{\mu}_{\alpha \dot \beta} \partial_{\mu}
\bar \lambda^{\dot \beta},
\label{wdefinition}
\eeq
where $F_{\mu\nu}=\partial_\mu A_\nu-\partial_\nu A_\mu$,
and where 
$\bar{\sigma}^\mu=(-{\bf 1},-\vec{\sigma})$. 
Regarding the supersymmetry transformations, let us just note
that
\be
\delta\lambda 
= i\zeta D + \zeta \sigma^{\mu} \bar{\sigma}^\nu F_{\mu\nu} .
\label{lamtran}
\ee

Global supersymmetry is defined as invariance under these
transformations with $\xi$ independent of spacetime position,
and local supersymmetry (supergravity) as invariance with $\xi$ depending on 
spacetime position. In the latter case one has to introduce
another supermultiplet containing the graviton and gravitino.

Global supersymmetry need not be 
renormalizable (Section \ref{s6}). 
But the usual convention is that `global supersymmetry'
refers to a theory which is renormalizable, except possibly for the
superpotential $W$ defined below. For the most part we follow that 
convention.\footnote
{One can also consider
the fully non-renormalizable version of global susy, which includes
a non-trivial K\"ahler potential and/or a non-trivial gauge kinetic
function. 
At this point, let 
us make it clear that we are talking about the K\"{a}hler potential,
and the gauge kinetic function, of the fundamental 
lagrangian, giving the tree-level potential.}

As we discuss in Section \ref{s6}, global supersymmetry may be regarded
as a limit of supergravity, in which roughly speaking 
gravity is made negligible by taking $\mpl$ to infinity.
{\em For most purposes} it is
a good approximation if the vevs of all relevant
scalar fields and auxiliary fields are much
less than $\mpl$. (Relevant here means that they have not been 
integrated out (page \pageref{intout}).)
There are however two notable exceptions. 

In the true vacuum,
global susy (whether renormalizable or not)
would predict a large positive value for $V$, instead of the 
practically zero value observed in our Universe.
According to supergravity, a negative contribution of unknown
magnitude should be subtracted from the global susy value.
It is assumed that this value makes $V$ practically zero in the
true vacuum, though one does not understand the origin of this exact 
cancellation. (This is called the cosmological constant problem.)

During inflation, the naive limit $\mpl\to\infty$  makes
no sense \cite{ewanloop2}, because as we saw in Section \ref{s3}
$\mpl$ plays an essential role. 
The approximation of global supersymmetry can be justified
only in special circumstances,
by methods more subtle than simply taking $\mpl$ to infinity.
As we shall see, this is a problem for inflation model-building, because 
a generic supergravity theory does not give a potential that is 
sufficiently flat for inflation. 
By contrast a generic globally 
supersymmetric theory works perfectly well. 

\subsection{The lagrangian of global supersymmetry}

\label{globs}

We focus first on global susy, with the usual restriction that it
be renormalizable except for 
possible non-renormalizable terms
in the superpotential.

To write down the action for a set of chiral superfields, $\phi_{i}$,
transforming in some representation of a gauge 
group $G$, one introduces,  for each gauge
generator, a vector superfield, $V^a$.  
Defining the matrix $V=T^a V_a$, 
where $T^a$ are the hermitian generators of the gauge group 
$G$ in the representation defined by the scalar fields and excluding 
the possible 
Fayet-Iliopoulos term to be discussed later, the
most general
renormalizable lagrangian, written in superspace, is then 
\beq
{\cal L} = \sum_n\int d^4 \theta
\phi_n^{\dagger} e^{ V} \phi_n+ 
{1 \over 4 k} \int d^2 \theta W_{\alpha}^2
+ \int d^2 \theta W(\phi_n) + {\rm h.c.}, 
\label{superspacel}
\eeq
where in the adjoint representation ${\rm Tr} (T^a T^b)=k\delta^{ab}$ and 
$W(\phi_n(x,\theta))$ is a fundamental object known as
superpotential. The corresponding function of the scalar components
$\phi_n(x)$, denoted by the same name and symbol, is
a  holomorphic function of the $\phi_n$.
For simplicity, we shall pretend that there is a 
single gauge $U(1)$ interaction, with coupling constant $g$.
This is adequate since such an interaction is the only one
that we consider in detail. (To be precise, we consider
a $U(1)$ with a Fayet-Iliopoulos term.)
In the case of several $U(1)$'s,  
there are no cross-terms in the potential {}from the $D$-terms, 
i.e. $V_D$ is simply expressed as $\sum_n(V_D)_n$. 

To write this down in terms of component fields, we need the covariant
derivative 
\be
D_\mu = \pa_\mu - \frac{i}{2}g A_\mu. 
\ee
In terms of the component fields, the lagrangian
takes the form: 
\bea
{\cal L} &=& \sum_n\left ( D_\mu \phi_n^* D^\mu \phi_n
+ i D_\mu \bar{\psi}_n\bar{\sigma}^{\mu} 
\psi_n+ \vert F_n\vert^2 \right )\nonumber\\
\label{componentla}
&-&\frac{1}{4} F^2_{\mu \nu} 
-i \lambda \sigma^\mu \partial_\mu 
\bar{\lambda} +\frac{1}{2}D^2 + \frac{g}{2} D\sum_n
q_n\phi_n^*\phi_n\nonumber\\
&-&\left[ i \sum_n\frac{g}{\sqrt{2}} \bar{\psi}_n\bar{\lambda} \phi _n
- \sum_{nm} {1\over 2}{\partial^2 W \over \partial \phi_n
\partial \phi_m} \psi_n\psi_m \right.
\nonumber\\
&+& \left.   
\sum_nF_n\left(\frac{\partial W}{\partial \phi_n}\right)\right]
+{\rm c.c.} \,.
\eea
At the end of the second line, 
$q_n$ are the $U(1)$-charges of the fields $\phi_n$.
The equations of motion for the auxiliary fields $F_n$ and
$D$ are the constraints:
\bea
F_n&=&- \left(\frac{\partial W }{\partial \phi_n}\right)^* \label{fi1} \\
D &=& -\frac{g}{2} \sum_nq_n\vert\phi_n\vert^2 \,.
\label{d1}
\eea

Eq.  (\ref{componentla}) contains the gauge invariant kinetic
terms for the various fields, 
which specify their gauge interactions.
It also contains, after having made use of Eqs. (\ref{fi1}) 
and (\ref{d1}),  the scalar field potential, 
\bea
V &=& V_F + V_D, \label{vfvd} \\
V_F &\equiv& \sum_n|F_n|^2, \label{fi2}\\
V_D &\equiv& \frac{1}{2} D^2.
\label{d2}
\eea
This separation of the potential into an $F$ term and a $D$ term 
is crucial for inflation model-building, especially when it is 
generalized to the case of supergravity.

The potential specifies the masses of the scalar 
fields, and their interactions with each other. The 
first term in the third line
specifies
the interactions of gaugino and scalar fields, 
while 
the second 
specifies the masses of the chiral fermions and their interactions with the 
scalars. All of these non-gauge interactions are called Yukawa 
couplings.

To have a renormalizable theory, $W$ is at most cubic in the fields,
corresponding to a potential which is at 
most quartic. However, one commonly allows $W$ to be of higher 
order, producing the kind of potentials that were mentioned 
in Section \ref{sss:tree}.

>From the above expressions, in particular \eq{fi2}, one sees that 
the overall 
phase of $W$ is not physically significant. An internal symmetry
can either leave $W$ invariant, or alter its phase. The latter
case corresponds to what is called an R-symmetry. \label{holow}
Because $W$ is holomorphic, the internal symmetries restrict its form much 
more than is the case for the actual potential $V$. 
In particular, terms in $W$ of the form $\frac12m\phi_1^2$
or $m\phi_1\phi_2$, which would
generate a mass term $m^2|\phi_1|^2$ in the potential,
are usually forbidden.\footnote
{An exception is the $\mu$ term of the MSSM,
$\mu H_U H_D$, which gives mass to the Higgs fields.}
As a result, scalar particles usually acquire masses
only from
the vevs of scalar fields (ie., from the spontaneous breaking of an 
internal symmetry) and from supersymmetry breaking.
The same applies to the 
spin-half partners of scalar fields, with the former contribution 
the same in both cases. 

In the case of a $U(1)$ gauge symmetry, one can add to the above 
lagrangian what is called a Fayet-Iliopoulos term
\cite{fayil},
\beq
-2\xi \int d^4 \theta ~V.
\label{dterm}
\eeq
This corresponds to adding a contribution $-\xi$ to the $D$ field,
so that \eq{d1} becomes
\be
D= - \frac{g}{2}\sum_n q_n|\phi_n|^2-\xi.
\label{fid1}
\ee
The $D$ term of the potential therefore becomes
\be
V_D = \frac{1}{2}\left(\frac{g}{2}\sum_n q_n |\phi_n|^2+\xi\right)^2.
\label{fivd}
\ee

{}From now on, we shall use a more common notation, where $\xi$ and the charges
are redefined so that
\be
V_D = \frac12 g^2 \( \sum_n q_n |\phi_n|^2 + \xi \)^2 \,.
\ee
This is equivalent to
\be
D =  - g \( \sum_n q_n |\phi_n|^2 + \xi \) \,.
\label{ddef}
\ee
A Fayet-Iliopoulos 
term may be present in the underlying theory {}from the very 
beginning,\footnote
{It is allowed by a gauge symmetry, unless the $U(1)$ is embedded
in some non-Abelian group. $\xi = 0$ can be enforced
by charge conjugation symmetry
which flips all $U(1)$ charges.
Such symmetry is possible in nonchiral theories.}
or appears in the effective theory after some heavy degrees of
freedom have been integrated out. 

It looks  particularly intriguing
that an  anomalous $U(1)$
symmetry  is  usually present in weakly coupled string theories \cite{u(1)A}.
(Anomalous in this context means that $\sum q_n\neq 0$.)
In this case \cite{fi,fi1,fi2} 
\be
\xi = \frac{g\sub{str}^2}{192\pi^2}\:{\rm Tr} {\bf Q}\:\mpl^2 .
\label{xi}
\ee
Here ${\rm Tr} {\bf Q}=\sum q_n$, 
which is typically \cite{font2,kobayashi} of order 100.
One expects the string-scale gauge coupling
$g\sub{str}$ (Section \ref{dilgs}) to be of order $1$ to $10^{-1}$,
making $\xi \simeq 10^{-1}$ to $10^{-2}\mpl$.

In the context of the strongly coupled $E_8 \otimes E_8$ heterotic
string  \cite{horwit}, anomalous $U(1)$ symmetries may appear and have
a nonperturbative origin, related to the presence, after
compactification, of five-branes in the five-dimensional bulk of the
theory. There is, at the moment, no general agreement on the relative
size of the induced Fayet-Iliopoulos terms on each boundary compared to
the value of the universal one induced in the weakly coupled case
\cite{jmrfa,bifa}.

\subsection{Spontaneously broken global susy}

\label{ss:ssb}

\subsubsection{The $F$ and $D$ terms}

\label{fandd}

Global supersymmetry breaking may be either spontaneous or explicit.
Let us begin with the first case.
For spontaneous breaking, 
the lagrangian 
is supersymmetric as given in the last subsection.
But the generators $Q_\alpha$ fail to annihilate the
vacuum. Instead, they produce a spin-half field, which may be either
a chiral field $\psi_\alpha$ or a gauge field $\lambda_\alpha$.
The condition for spontaneous susy breaking is therefore to have 
a nonzero vacuum expectation value
for $\left\{ Q_\alpha,\psi_\beta \right\}$
or $\left\{ Q_\alpha,\lambda_\beta \right\}$. 

The former quantity
is defined by \eq{psitransform}, and the latter by
\eq{lamtran}.
The quantities $\partial_\mu \phi$ and $F_{\mu\nu}$ contain derivatives 
of fields, and are supposed to vanish in the vacuum.
It follows that susy is spontaneously broken if, and only if,
at least one of the auxiliary fields $F_n$ or $D$ has a non-vanishing vev.

In the true vacuum, one 
defines the
scale $M\sub S$ of global supersymmetry breaking by
\be
M\sub S^4= \sum_n |F_n|^2 + \frac12 D^2 \,,
\label{msdef98}
\ee
or equivalently
\be
M\sub S^4 = V \,.
\ee
(In the simplest case $D$ vanishes and there is just one $F_n$.)

When we go to supergravity, part of $V$ is still generated by the 
supersymmetry breaking terms, but there is also 
a contribution $-3|W|^2
/\mpl^2$. This allows $V$ to vanish in the 
true vacuum as is (practically) demanded by observation.

During inflation, $V$ is positive so the negative term is smaller
than the susy-breaking terms. In most models of inflation
it is negligible. In any case, $V$ is at least as big as the susy
breaking term, so the search for a model of inflation is also
a search for a susy-breaking mechanism in the early Universe.

Spontaneous symmetry breaking can be either tree-level
(already present in the lagrangian) or dynamical (generated
only by quantum effects like condensation).
The spontaneous breaking in general breaks the equality between
the scalar and spin-$\frac12$ masses,
in each chiral supermultiplet. But at tree level the
breaking satisfies a simple relation, which
can easily be derived from the lagrangian
(\ref{componentla}). Ignoring mass mixing for simplicity,
one finds in the case of symmetry breaking by an $F$-term,
\be
\sum_n \( m_{n1}^2 + m_{n2}^2 - 2m_{n\rm f}^2 \) =0 \,.
\label{strzero}
\ee
Here $n$ labels the chiral supermultiplets, 
$m_{n\rm f}$ is the fermion mass while 
$m_{n1}$ and $m_{n2}$ are the scalar masses.\footnote
{More generally, if the mass-squared
matrix is non-diagonal the left hand side of \eq{strzero} is
the supertrace defined in 
\eq{strzero}.}
In the case of 
symmetry breaking by a $D$ term, coming from a $U(1)$,
the right hand side of \eq{strzero} becomes 
$D {\rm Tr} {\rm \bf Q}$. But in order to cancel
gauge anomalies, it is often desirable that ${\rm Tr}
{\rm \bf Q}=0$
which recovers \eq{strzero}.

\subsubsection{Tree-level spontaneous susy breaking with an $F$ term}

Models of tree-level spontaneous susy-breaking where 
only $F$ terms have vevs are called O'Raifearteagh models.
We consider them now, postponing until Section \ref{sss:dterm}
the case of $D$-term susy breaking.

The simplest O'Raifearteagh model involves a single field $X$,
\beq
W=m^2 X + \cdots \,,
\eeq
where the dots represent terms independent of $X$.
The potential is given by $V=m^4+\cdots$,  and $F_X=m^2$;
thus supersymmetry is broken for nonvanishing $m$. 
Some models of inflation invoke such
a linear superpotential.

We shall encounter more complicated O'Raifearteagh models
for inflation later. At this point let us give the following
example, which is probably of only pedagogical interest. 
It involves three
singlet fields, $X,\phi$ and $Y$, with superpotential:
\beq
W= \lambda_1 X(\phi^2 - \mu^2) + \lambda_2 Y \phi^2.
\label{oraif}
\eeq
With this superpotential, the equations
\beq
F_X={\partial W \over \partial X}=\lambda_1 (\phi^2 - \mu^2)=0, \quad \quad
F_Y={\partial W \over \partial Y}=\lambda_2\phi^2=0
\label{susybreaking}
\eeq
are incompatible.   Note 
that at this level not all of the fields are fully determined,
since the equation
\beq
{\partial W \over \partial \phi}=0
\label{xequation}
\eeq
can be satisfied provided
\beq
\lambda_1 X + \lambda_2 Y =0.
\label{pseudoflat}
\eeq
This vacuum degeneracy is accidental and is lifted by quantum corrections.
Since either $\langle F_X\rangle$ or  $\langle F_Y\rangle$ are
nonvanishing, supersymmetry is broken at the tree-level.

\subsubsection{Dynamically generated superpotentials}

\label{sss:dsb}

It has been known for a long time that global, renormalizable supersymmetry 
may be dynamically broken in four dimensions \cite{ads,nelsonrev}.
There already exist excellent reviews of this subject
and
the reader is referred to 
\cite{reviewdsb,reviewdsb1,reviewdsb2,reviewdsb3,nelsonrev,gr} 
for more details. Several mechanisms have been proposed, but only two 
have so far been invoked for inflation model-building. These are
a dynamically generated superpotential, and a quantum moduli space,
which we look at now starting with the former.

In some cases, 
the dynamically generated superpotential 
occurs in a theory characterized by many classically flat directions.
Typically, the potentials generated along these flat
directions fall down to zero at large values of the fields.
These potentials, however, must be stabilized by some mechanism
and so far no compelling model has been proposed. 

Alternatively, models are known in which supersymmetry is broken
without flat directions \cite{ads} and no need of complicated
stabilization mechanisms.
In some directions, non-perturbative effects might raise the potential
at small field values, while tree-level terms raise it at large values.
If some $F$-term is nonzero
in the ground state, supersymmetry is spontaneously broken. 

To provide an explicit example,
let us consider the model discussed in \cite{ex} in which 
the tree-level terms are non-renormalizable.
The gauge group is $SU(6)\otimes U(1)\otimes U(1)_{\rm m}$
and the chiral superfields are $A(15,1,0)$,
$\bar{F}^{\pm}(\bar{6},-2,\pm 1)$, $S^0(1,3,0)$ and $S^{\pm}(1,3,\pm 2)$.
$U(1)_{\rm m}$ is irrelevant for supersymmetry breaking
but may play the role of messenger hypercharge.
The gauge symmetries forbid a cubic superpotential in the model.
At the level of dimension five operators,
the unique term allowed 
$W=\frac{1}{M}A\bar{F}^{+} \bar{F}^{-}S^0$, where $M$ may
 be identified with $\MPl$.
Along the $SU(6)$ and $U(1)$ $D$-flat directions
the gauge symmetry is broken down to $Sp(4)$.
Gluino condensation at the scale $\Lambda$ leads to a nonperturbative
superpotential whose  form follows uniquely {}from
symmetry considerations: $W_{{\rm np}}=\frac{\Lambda^5}{{\cal O}^{1/3}}$,
where ${\cal O}=\bar{F}_i^{+}\bar{F}_j^{-}
A^{ij}\epsilon_{klmnop}A^{kl}A^{mn}A^{op}$.
Turning on the
nonrenormalizable superpotential lifts the flat
directions and the value of the potential at the
minimum turns out to be $V_0\sim \Lambda^5/M$
and $F$-terms are of order of $\Lambda^{15/6}M^{-1/2}$ signalling
the breaking of supersymmetry. 

A generic prediction of dynamical supersymmetry breaking models
is the appearance of 
a superpotential $W\simeq \Lambda^{3+q}/\phi^q$,
leading to a potential $V(\phi)=(\Lambda^{p+4})/(|\phi^p|)$, 
where the index $p$ and the scale
$\Lambda$ depend upon the underlying gauge group. 

\subsubsection{Quantum moduli spaces}

\label{sss:qms}

Recent developments have also shown that many supersymmetric theories may have
other types of non-perturbative dynamics which lead to degenerate quantum 
moduli spaces of vacuum instead of dynamically generated
superpotentials 
\cite{reviewdsb,reviewdsb1,reviewdsb2,reviewdsb3,thomasintri,yanagida}. 
The quantum deformation of a classical moduli space 
constraint may lead to supersymmetry breaking.  This happens 
because the patterns of breakings of global and gauge 
symmetries on a quantum moduli space may differ {}from those 
on the classical moduli space and the quantum deformed constraint associated 
with the moduli space is inconsistent with a stationary superpotential.  
Indeed, moduli generally transform under global
symmetries and there is a point on the classical moduli space 
at which {\it all} the fields have zero vev and global 
symmetries are unbroken. However, at the quantum level points which 
are part of the classical moduli space may be removed. If tree-level
interactions have vanishing potential, and auxiliary fields, only 
at points on the classical moduli space which are not part of 
the quantum deformed moduli space, supersymmetry gets broken. 

We consider the following simple example.
The gauge theory considered is  an  $SU(2)$ gauge theory with
matter  consisting
of four doublet  chiral superfields $Q_I, \bar Q^J$, where $I,J = 1,2$ are
flavour indices. The theory also contains a  singlet superfield $S$ and  the 
superpotential reads 
\begin{equation}
W = g S(Q_1\bar Q_1+Q_2\bar Q_2), 
\end{equation}
where $g$ is a Yukawa coupling constant. 
At the classical level, in the absence of this superpotential  
($g=0$), the 
space of vacua ($D$-flat directions) is parameterized by a set
of complex fields consisting of
$S$ plus the following  6 $SU(2)$ invariants
 (mesons and baryons)
\begin{equation}
  M_I^J = Q_I\bar Q^J, ~~ B = \epsilon^{IJ}Q_IQ_J, ~~ \bar B =  
\epsilon_{IJ}
\bar Q^I\bar Q^J.
\end{equation}
The invariants are however subject to the constraint
\begin{equation}
 \det M - \bar BB = 0
\label{class}
\end{equation}
so that in the end the space of vacua at $g=0$ has complex dimension  
6. In the
presence of the superpotential, the classical moduli space has two  
branches:
{\it a)} $S\not = 0$, with $M_I^J=B=\bar B=0$. On this branch the quarks
 get a mass $\sim g S$ {}from the superpotential and the gauge symmetry
is unbroken;  
{\it b)} $S=0$, with  non-zero mesons and baryons satisfying two  
constraints. One 
is Eq. (\ref{class}) while the other is  $F_S={\rm Tr} M=0$. Here the  
gauge  group is broken. 

This moduli space is however reduced by quantum effects. In  
particular
a non-zero vacuum energy is generated along  the $S\not = 0$ branch.
This is  established by  
considering
the effective theory far away along $S\not = 0$.
Here the quark fields get masses of order $S$ and decouple.
The effective theory consists of the (free) singlet $S$ plus a pure 
$SU(2)$ gauge sector.
The effective scale $\Lambda_L$ of the low-energy
$SU(2)$ along this trajectory is given to all orders by the 1-loop  
matching
 of the gauge couplings  at the quarks' mass $gS$ and 
reads
\begin{equation}
\Lambda_L^6 =  g^2S^2\Lambda^4,
\end{equation}
where $\Lambda$ is the scale of the original theory with massless  
quarks.
In the pure $SU(2)$ gauge theory gauginos condense  
and an effective superpotential $\sim \Lambda_L^3$ is generated
\begin{equation}
W_{eff} = g S \Lambda^2.
\label{linear2}
\end{equation}
Thus
\begin{equation}
 F_S = g\Lambda^2 \label{Fs}
\end{equation}
and supersymmetry is broken, with a vacuum energy density $F_S^2$  
which
is independent of $S$.
As we mention later, it has been suggested \cite{ddr}
that $|S|$ is the inflaton.

\subsection{Soft susy breaking}

In the effective theory, which describes the interactions of the
Standard Model particles and their superpartners at energies
$\lsim 1\TeV$, supersymmetry is taken to be broken explicitly.
In order to preserve the theoretical motivation for 
supersymmetry (the absence of
quadratic divergences and the naturalness of the 
theory) only certain `soft' susy-breaking terms are allowed.
These are 
\begin{itemize}
\item Masses (and mass-mixing terms) for scalars, whose typical value will be 
denoted by $\widetilde{m}$.
\item Masses for gauginos, whose typical value will be
denoted by $m_{\tilde g}$.
\item Cubic terms in the scalar field potential, of the
form $(A_{ijk}\phi_i\phi_j\phi_k \,+\,{\rm c.c.})$. The typical value of 
the couplings $A_{ijk}$ will be denoted by $A$.
\end{itemize}
There are no soft chiral fermion masses, nor any soft quartic
terms. Both of these have their unbroken susy values; in particular,
the quartic term vanishes in
a flat direction of unbroken susy.
For susy to do its job one requires
that the mass scales
 $m_{\tilde g}$, $\widetilde{m}$ and $A$ are all $\lsim 1\TeV$. 

The squark and slepton masses come almost entirely from the soft susy 
breaking (except for the stop), and to have escaped detection they have
to be $\gsim 100\GeV$. So at least $\widetilde m$ should be in the 
range roughly $100\GeV $ to $1\TeV$.

The effective theory, with explicit soft susy breaking,
describes only
the `visible' sector of the 
theory that consists of the fields possessing the Standard Model gauge
interactions. In the full theory, spontaneous susy breaking is supposed 
to take place, but in a `hidden' sector,
consisting of fields which do not possess the Standard Model gauge
interactions. When the 
hidden sector is 
integrated out (footnote \ref{intout}) one obtains the effective theory
in the visible sector.

The spontaneous breaking is usually of the $F$-term type.
Models are classified as `gravity-mediated' if the interaction between 
the two sectors is only of gravitational strength, or as
`gauge-mediated' if it is stronger (usually involving 
a gauge interaction). 
In the gauge-mediated case, the entire theory including the mechanism of 
spontaneous susy breaking 
is supposed to be describable in terms of global susy. 
In the gravity-mediated case, the mechanism of spontaneous susy 
breaking is usually 
supposed to involve supergravity in an essential way, since that theory
is anyhow needed to describe the interaction between the two 
sectors. (One is however free to suppose that in this case too, 
the mechanism of spontaneous susy breaking
is describable in terms of global supersymmetry
\cite{raby}.)

\subsubsection{Soft susy breaking from a $D$ term}

\label{sss:dterm}

Before dealing with the gauge-mediated case, we look at a proposal 
\cite{dudas,pomarol,riotto1,riotto2,nelson,faraggi,berezhiani,%
ramond,murayama,ella}
that invokes a $D$ term.
The $D$ term comes from a $U(1)$ 
with a Fayet-Iliopoulos term,  which 
is usually considered to have  a stringy origin as 
described in Section \ref{globs}. As we shall see, such a term 
has also been widely used for building models of inflation, but for now 
we are concerned with the true vacuum.

The hidden sector consists of two fields $\phi_\pm$.
The part of the superpotential depending only on them is
$W=m\phi_{+}\phi_{-}$. Ignoring the rest of the superpotential for the 
moment, 
the potential is
\begin{eqnarray}
V &=& m^2(|\phi_+|^2+|\phi_-|^2)\nonumber\\
&+&\frac{g^2}{2}
\left(\sum_i q_i |\widetilde{Q}_i|^2+|\phi_+|^2-|\phi_-|^2+
\xi\right)^2.
\end{eqnarray}
The scalar fields of the visible sector are denoted by 
$\widetilde{Q}_i$, and we shall see in a moment that they have
masses of order $m$. Accordingly, we take 
$m\simeq (1-10)$ TeV, without enquiring into the origin of $m$.

Let us consider the
part of $V$ setting $\widetilde{Q}=0$. It is easy to see that its 
minimum
breaks supersymmetry as well as the anomalous $U(1)$ gauge symmetry
with \cite{pomarol}
\begin{eqnarray}
\langle \phi_-\rangle &=&\left(\xi-{{m^2}\over{g^2}}\right)^{1/2},\:\:
\langle \phi_+\rangle = 0 \label{phi98}\\
\langle F_{\phi_+} \rangle &=&m\left(\xi-{{m^2}\over{g^2}}\right)^{1/2},
\:\: \langle D\rangle = m^2.
\end{eqnarray}
If we parameterize $\xi=\epsilon M^2_{{\rm Pl}}$,  we have
$\langle \phi_-\rangle\simeq \epsilon^{1/2}M_{{\rm Pl}}$ and
$\langle F_{\phi_+}
\rangle \simeq \epsilon^{1/2}m M_{{\rm Pl}}$.
In weakly coupled string theory, $\epsilon $ is given by \eq{xi}
and is of order $10^{-1}$ to $10^{-2}$.
Integrating out\footnote
{See footnote \ref{intout}}
$\phi_{\pm}$ generates
soft susy breaking mass terms of order $m$ for the scalar fields charged
under $U(1)$\footnote
{The term $\langle F_{\phi_+}\rangle$
will give a gravity-mediated contribution
which is smaller by a factor $\epsilon$.}
\begin{equation}
\widetilde{m}^2_{\widetilde{Q}_i}=q_i\langle D\rangle =q_i m^2.
\end{equation}

The  charges $q_i$ are required to   be positive to avoid color/charge
breaking. Invariance under the anomalous $U(1)$ will require, therefore,
that  terms in the superpotential involving visible-sector fields with 
nonzero charges are multiplied by appropriate powers of $\phi_{-}/M_{{\rm Pl}}$
\cite{riotto1,riotto2}.

 If $m$ is large enough
and if the first two generations of squarks are (equally)
charged under the $U(1)$, the harmful
flavour-changing neutral currents (FCNC's) are
suppressed and trilinear soft breaking mass terms are also
suppressed by powers of $\epsilon$ so that large
supersymmetric CP-violating phases
pose no problem \cite{pomarol,riotto1,riotto2}. 

If the Fayet-Iliopoulos
has a stringy origin, it is directly proportional to 
$g^2_{{\rm str}}$, see Eq. (\ref{xi}).  As such, it  depends on the 
vacuum expectation value of (the real part of) the dilaton field $s$,
$g^2_{{\rm str}}=M_{{\rm P}}/({\rm Re}\:s)$ 
(see subsection \ref{dilgs} for more details). 

It has been recently argued \cite{aa} that, within  some particular  
mechanisms for stabilizing the dilaton in string theories, the   
supersymmetry breaking contribution to the soft masses of sfermions  
coming from the the dilaton $F$-term always dominates over the $D$-term  
supersymmetry breaking contribution from the anomalous $U(1)$. 

However, other mechanisms for stabilizing the dilaton may not have this  
effect. For instance, if the dilaton is stabilized by the contributions  
to the superpotential, the dilaton  $F$-term vanishes and the soft  
supersymmetry breaking mass terms only comes from the $D$-term.

Finally, we would like to point out that the  class of model with  
$D$-term supersymmetry breaking  may have some problems on the  
cosmological side,
as far as the dark matter abundance
is concerned \cite{grr}. 

\subsubsection{Gauge-mediated susy breaking}

\label{gmed}

Global susy models involving only the $F$ term are called 
gauge-mediated models \cite{gmsb,g1,g2,g3,g4,g5,g6,g7,g8,g9},
because communication between the hidden and visible sectors
is usually through a gauge interaction.
A review of these models is given in Reference \cite{gr}.

The minimal gauge mediated supersymmetry breaking
models are defined by three sectors: {\it (i)} a hidden
sector (often called a secluded sector in this context)
that breaks supersymmetry; {\it (ii)} a messenger sector that serves
to communicate the SUSY breaking to the standard model and {\it (iii)} the
 standard model sector.
The minimal messenger sector consists of a single
${\bf 5}+\bar{{\bf 5}}$ of $SU(5)$
(to preserve gauge coupling constant unification),
{\it i.e.} color triplets, $q$ and $\bar{q}$,
and weak doublets $\ell$ and $\bar{\ell}$
with their interactions determined by the following superpotential:
\begin{equation}
W=\lambda_1 X\bar{q}q+\lambda_2 X\bar{\ell}\ell.
\end{equation}
When the field $X$ acquires a vacuum expectation value
for both its scalar and auxiliary components,
$\langle X\rangle$ and $\langle F_X\rangle$ respectively,
the fields $q\pm q^*$ acquire masses
$\lambda_1^2\langle X\rangle^2 \pm \lambda_1 \langle F_X\rangle$,
and similarly for the fields $\ell\pm \ell^*$.
This supersymmetry breaking in the messenger sector
gives gaugino
masses at one loop and scalar masses at two loops
(with messengers and gauge bosons in the loops).
At the scale $\langle X\rangle$, the gaugino masses are given approximately by
by
\begin{equation}
M_j(\langle X\rangle)
=k_j\frac{\alpha_j(\langle X\rangle)}{4\pi}\Lambda,\:\:\:j=1,2,3,
\label{123}
\end{equation}
where $\Lambda\equiv \langle F_X\rangle/\langle X\rangle $,
$k_1=5/3$, $k_2=k_3=1$ and 
$\alpha_i$ are the three standard model gauge couplings
in \eq{70}. The scalar
masses are given approximately by
\begin{equation}
\widetilde{m}^2(\langle X\rangle)=2 \sum_{j=1}^3\:C_j
k_j\left[\frac{\alpha_j(\langle X\rangle)}{4\pi}
\right]^2\Lambda^2,
\end{equation}
where $C_3=4/3$ for color triplets, $C_2=3/4$ for weak doublets
(and equal to zero otherwise) and $C_1=Y^2$ with $Y=Q-T_3$.
To have squarks and gaugino masses of order 100 GeV, 
we need 
\be
\Lambda\equiv \langle F_X\rangle/\langle X\rangle
\sim 10^5\GeV \,.
\label{lambeq}
\ee

Because the scalar masses are functions of only the gauge quantum 
numbers,
the flavour-changing-neutral-current processes are
naturally suppressed in agreement with
experimental bounds.
The reason for this suppression is that the gauge interactions
induce flavour-symmetric supersymmetry-breaking terms
in the visible sector at the scale $\langle X\rangle$ and,
because this scale is usually much smaller than the Planck scale, only  
a slight asymmetry
is introduced by renormalization group extrapolation to low energies.
This is in contrast to the supergravity scenarios
where one generically needs to invoke additional flavor symmetries to
achieve the same goal.

Notice that there is no need to have
$\sqrt{\langle F_X\rangle}\sim \langle X\rangle$.
The only requirement is 
\eq{lambeq}, and the
hierarchy 
\be
\Lambda \ll \sqrt{\langle F_X\rangle} \ll \langle X\rangle
\label{heir}
\ee
is certainly allowed
\cite{raby,rrr}.
In fact, $\sqrt{\langle F_X\rangle}$ can take any 
value between $10^4$, and $10^{10}$ to $10^{11}\,GeV$. 
This corresponds to
$10^4\lsim \langle X\rangle \lsim 10^{15}$ to $10^{17}\GeV$.
If the upper bound is saturated, the gravity-mediated susy breaking that 
is always present (Section \ref{gravmedtext})
becomes of the same order as the gauge-mediated
susy breaking; if it were exceeded, gravity-mediated susy breaking would
make the soft susy breaking parameters too big ($\gg1\TeV$).
The upper bound is also required
from considerations about nucleosynthesis \cite{nucleobound}.

To obtain the hierarchy $\sqrt{\langle {F_X}\rangle}
\ll \langle X
\rangle$, one 
can suppose that
nonrenormalizable operators are involved, as in Section \ref{sss:tree},
or \cite{raby} that $X$ has a soft susy breaking
 mass which runs, as in
Section \ref{onel}.

In the latter case, the mass may come from supergravity corrections. 
Alternatively, $\widetilde{m}_X^2$ may receive contribution {}from 
one-loop Yukawa interactions. To illustrate this idea,  we can consider the 
following toy model
\begin{equation} W = \lambda_1  A \bar{\Psi}\Psi + B\left( \bar{\Psi}\Psi + 
\lambda_2 \Phi^+ \Phi^- + \lambda_3 B^2\right) 
 \end{equation}
where $A$ and $B$ are  singlets, $\Phi^{\pm}$ have 
 charge $\pm 1$ under a messenger $U(1)$  and $\bar{\Psi}$ and $\Psi$ are 
charged  under some gauge group $G$. We assume that some  susy breaking 
occurs in a hidden sector dynamically and is transmitted  directly to 
$\Phi^{\pm}$ via the messenger $U(1)$ resulting in a negative 
mass squared $m^2$ for these two 
states. Minimizing the potential, one can show that there is a flat direction 
represented by $X\equiv \lambda_1 A+B$ whose VEV is undetermined at 
the tree-level and that supersymmetry is broken with 
$F_X = {m^2 \over \lambda_2} {1 \over (2 - \lambda_2/3 
\lambda_3)}$. $\widetilde{m}_X^2$ gets a one-loop contribution proportional 
to $\lambda_2^2 m^2$ through  the Yukawa interaction 
$W=\lambda_2 B\Phi^+ \Phi^-$. 

Arguably, the cosmological constant problem is worse in the case of
gauge-mediated susy breaking, than in the gravity-mediated case.
To achieve the (practically) vanishing potential
that is required by observation,
the global supersymmetry
result $V=\sum|W_n|^2$ must be cancelled by a term 
$-3|W|^2/\mpl^2$ in the full supergravity theory.
But if $W$ is dominated by 
the sector of the theory responsible for gauge-mediated susy breaking,
one will typically have $|W_n|\sim |W|/|\phi_n|$ with
$|\phi_n|\ll \mpl$. The conclusion is that $|W|$ must come from some
other sector of the theory, or else be identified with the
constant $W_0$ in the expansion of $W$ (\eq{wexp}) which might perhaps
come from a string theory. In contrast,
with gravity-mediated susy breaking the sector of the theory responsible
for susy breaking usually gives $|W|$ of the right order,
because the relevant $\phi_n$ are usually of order $\mpl$.

\subsection{Loop corrections and running}

\label{loopc}

This is a good place to discuss the loop corrections in more detail.

Perhaps the most convincing reason for believing supersymmetry
is its 
solution to the hierarchy problem \cite{wilson}.
In
a theory where the largest interesting energy scale is the Planck
mass or unification scale, light fundamental scalars
(like a single Higgs doublet) get quadratically divergent
contributions to their masses via one-loop diagrams
where other heavy scalar or gauge fields are running in the loop.    
The scalar  mass is given by $
m_\phi^2
= (m_\phi^2)_0 + c \Lambda_{{\rm UV}}^2$, 
where $(m_\phi^2)_0$ is the tree-level mass term, 
 $\Lambda_{{\rm UV}}$ is the ultraviolet cutoff
scale of the theory to be identified with  some extremely
large scale and $c$ is a loop suppression factor.  The 
Higgs mass can only be small if there is a delicate
fine-tuning between classical and quantum
effects. The only known symmetry
which can suppress the quadratically divergent corrections
is supersymmetry. Indeed, the way supersymmetry works is to
cancel the leading $\Lambda_{{\rm UV}}^2$ contribution by
adding extra degrees of freedom into the game.
The cancellation works because
the number
of degrees of freedom is basically doubled in a
supersymmetric theory: each spin 0 or 1 field  is
accompanied by its fermionic partner.
This amounts to adding an extra contribution to $m_\phi^2$ which
is equal in magnitude, but opposite in sign to the original one.
The cancellation is exact in the limit of exact supersymmetry.  

\subsubsection{One-loop corrections}

\label{sss:loop}

Let us address this issue more formally and
imagine one is interested in the computation of
the one-loop effective potential $V_{1-{\rm loop}}(\phi)$
of a given scalar field of the supersymmetric theory.
In  the  dimensional reduction with modified minimal 
subtraction ($\overline{DR}$) scheme of renormalization, it reads
\beq
V_{1-{\rm loop}}(\phi)=\frac{Q^2}{32\pi^2}\:{\rm Str}{\cal M}^2+
\frac{1}{64\pi^2}\:{\rm Str}\left[{\cal M}^4(\phi)
\left({\rm ln}\frac{{\cal M}^2(\phi)}{Q^2}-\frac{3}{2}\right)\right],
\label{pot}
\eeq
where ${\cal M}^2(\phi)$ is the field dependent
mass-squared matrix for the particles contributing to the loop 
correction. These particles will in general have spins $j=0,1/2$
or $1$, and the supertrace is defined as 
\beq
{\rm Str}\:A=\sum_j(-1)^{2j}(1+2j)\:{\rm Tr}\:A_j
%=(n_B A_B- n_F A_F\right)
,
\label{strc}
\eeq
Here, $A$ denotes either ${\cal M}^2$ or the square bracket,
and $A_j$ is the ordinary trace for particles of 
spin $j$.
%$n_B(n_F)$ is the number of boson (fermion) states in the 
%theory, and $A_B(A_F)$ is the ordinary trace of the mass matrix for
%bosons (fermions).

The scale 
$Q$ is the renormalization scale, at which all the
parameters (masses, gauge and Yukawa couplings, etc.)
entering the tree-level  and the one-loop potential
(\ref{pot}) must be evaluated. 

In   Eq. (\ref{pot}) we have explicitly written
the quadratic divergent piece
proportional to   ${\rm Str}\:{\cal M}^2$.
In non-supersymmetric theories this term is field dependent
and is the source of the divergent corrections to
the squared mass $m_\phi^2$. On the contrary, in
supersymmetric (and anomaly free) theories,
this term is {\it independent} of the fields and proportional
to the soft breaking masses of the fields
contributing to the effective potential. 
It therefore contributes only to the cosmological constant, and we drop 
it giving
\beq
V_{1-{\rm loop}}(\phi)=
\frac{1}{64\pi^2}\:{\rm Str}\left[{\cal M}^4(\phi)
\left({\rm ln}\frac{{\cal M}^2(\phi)}{Q^2}-\frac{3}{2}\right)\right],
\label{cwpot}
\eeq
 
With unbroken supersymmetry, the loop correction vanishes,
and the tree-level
scalar potential of the field $\phi$ is not renormalized
at all (in particular, there is no one-loop contribution
to the squared mass $m_\phi^2$).
Notice that, in the case of global supersymmetric
theory, this property is true at any order of
perturbation theory as a result of the nonrenormalization theorem. 
If supersymmetry is 
broken,
the supertrace as well as the one-loop potential usually
no longer vanish.  

As an example, we consider a simple situation that can give
\eqs{spont}{soft}. The 
loop 
correction comes from a single complex field $\psi$ 
(with masses $m_1$ and $m_2$ for the real and imaginary parts)
and its fermionic 
partner (with mass $m_f$).
The interaction is supposed to be
$\frac12 \lambda \phi^2 |\psi^2|$.
When $\phi$ (taken to be real)
is much bigger than the masses the total loop correction is
\be
\Delta V \simeq \frac1{32\pi^2 } \left[
        \sum_{i=1,2} \left(m_i^2 + \frac12 \lambda\phi^2\right)^2
        - 2\left( m_f^2 + \frac12\lambda \phi^2\right)^2 \right]\:
        \ln \frac{\phi}{Q}
\ee
The coefficient of $\phi^4$ vanishes by virtue of the supersymmetry.
Two cases commonly arise for the other terms.

The first  case occurs when there is soft susy breaking in the relevant 
sector, with zero (or negligible) fermion masses.
Then the quadratic term dominates 
and one has
\be
\Delta V\simeq \frac1{32\pi^2} \lambda (m_1^2+m_2^2) \phi^2 
\ln(\phi/Q) \,.
\ee

The second case occurs when there is spontaneous susy breaking in the 
relevant sector, giving
$2m_f^2 = m_1^2 + m_2^2$. Then the coefficient of $\phi^2$ vanishes
leaving
\be
\Delta V \simeq \frac{(m_1^2-m_2^2)^2}{64\pi^2} \ln \frac{\phi}{Q} \,.
\label{vdval}
\ee

Including more chiral supermultiplets and/or gauge supermultiplets
gives similar results; softly broken susy makes $\Delta V\propto
\phi^2\ln(\phi/Q)$, but spontaneously broken susy makes
$\Delta V\propto \ln(\phi/Q)$
because ${\rm Str\,}{\cal M}^2$ vanishes.

\subsubsection{The Renormalization Group Equations (RGE's)}

\label{rges}

In the perturbative regime, the potential $V$
is given by the tree-level expression, plus 1-loop, 2-loop etc quantum 
corrections,
\be
V= V\sub{tree} (Q) + V\sub{1-loop} (Q) + V\sub{2-loop}(Q)
+ \cdots
\ee
It depends on the parameters appearing in the
Lagrangian (masses and couplings), but in addition each 
individual term depends on the
the renormalization scale $Q$. This amounts to a choice of energy unit,
which has to be made within any renormalization scheme.
Physical quantities like $V$ do not 
depend on $Q$, and this is ensured by a set of
linear
differential equations for the parameters, known as Renormalization  
Group Equation's (RGE's). 

The 1-loop correction, for a given particle in the loop, 
was displayed in \eq{cwpot}. If 
$\phi$ is much larger than
any relevant mass scales, the typical contribution to $\cal M$ will
be of order $\phi$ (the only relevant scale). As a result, the 
loop correction will vanish for
some choice $Q\sim \phi$.
The potential is then given just by the tree-level contribution,
\be
V(\phi)\simeq V_{{\rm tree}}(\phi,Q=c\phi),
\ee
where the coefficient $c\sim 1$ depends upon the details of the theory. 

\subsection{Supergravity}
\label{s6}

So far we have considered global supersymmetry, taken to be 
renormalizable except possibly for terms in the superpotential. 
In the usual context of collider physics, particle detectors and
astrophysics, this is adequate
for most purposes. But during inflation one needs to consider 
supergravity, which contains within it the most general 
non-renormalizable version of global susy.

A non-renormalizable field theory is an effective one, valid below
some ultra-violet cutoff $\mpltil$.
With all of the
fields and interactions in Nature included,
$\mpltil$ is generally identified with $\mpl$
(Section \ref{rvnon}), and we shall do this in the end.
But for clarity of exposition we initially leave $\mpltil$
unspecified.

\subsubsection{Specifying a supergravity theory}

\label{specsug}

In Section \ref{ss:lag} we defined the chiral and
gauge supermultiplets, and their supersymmetry transformations.
These formulas remain valid in supergravity, but the lagrangian is
different. In addition to the 
superpotential $W$ one now needs two more functions.
These are the K\"{a}hler potential $K$, 
and the gauge kinetic function $f$.\footnote
{Presumably there are also functions specifying terms
involving second and 
higher spacetime derivatives. It is reasonable to suppose that such 
terms are negligible compared with the kinetic term unless the spacetime 
derivatives are of order 1 in Planck units. But then field theory will 
break down anyway.}
Both $W$ and $f$
are holomorphic
function of the complex scalar fields, but the real function 
$K$ is not holomorphic;
it is regarded as a function of the fields and their complex 
conjugates.

Only the combination 
\be
 G\equiv \mpl^{-2} K+ \ln\frac{|W|^2}{\mpl^6}
\ee
is physically significant. So we have invariance
under the K\"{a}hler transformation
$\mpl^{-2} K\to \mpl^{-2} 
K-X-\bar X$, $W\to e^X W$ where $X$ is any holomorphic function
of the fields.

We shall adopt the following conventions \cite{wessbagger}.
The scalar components $\phi^n$ and auxiliary components 
$F^n$ of  chiral supermultiplets are 
labelled by a superscript.
A subscript $n$ denotes 
$\pa/\pa\phi^n$, and a subscript $n_*$ 
denotes $\pa/\pa\phi^{n*}$. 
(Note that $K_{nm^*}=G_{nm^*}$.)
Occasionally one lowers components, $\phi_n\equiv K_{nm^*}\phi^{m*}$
and $F_n\equiv K_{nm^*} F^{m^*}$; 
the inverse matrix of $K_{nm^*}$, which 
raises components, is denoted by 
$K^{m^* n}$. A summation over repeated indices is implied.

We first consider the expansion of $W$, $K$ and $f$ about a suitable
origin in 
field space. It may be chosen 
to be the position of the vacuum or, in the case of matter fields,
to be the 
fixed point of the symmetries.

\paragraph{The superpotential}

We already considered the superpotential, in the context of global susy.
Since it is holomorphic in the fields, it is of the form
\be
W = W_0 + \Lambda^2W_1(\phi^n) + m W_2(\phi^n)
+ W_3(\phi^n) + \mpltil^{3-d} \sum_{d=4}^\infty W_d(\phi^n) \,.
\label{wexp}
\ee
Each quantity $W_d$ is the sum of dimension $d$ terms; in other words,
it is a sum of terms, each of which is a product of $d$ fields
times a coefficient. 
For the non-renormalizable terms ($d\geq 4$),
the coefficient is expected to be of order 1,
{\em unless} it is forbidden by internal symmetries. 

As we noted on page \pageref{holow}, $W$ is strongly constrained by
internal symmetries, because it is holomorphic.
For a given
field, if one starts with an expression in which the field only occurs
at low order, one can forbid additional 
terms up to a finite order by imposing a discrete $Z_N$
symmetry, and one can forbid additional terms up to all orders
by imposing a continuous symmetry. 
However, in the case of a gauge singlet 
the continuous symmetry would have to be global, 
and as we noted on page  \pageref{glvdis}
global continuous symmetries do not seem to exist in string theory.
Therefore, in the case of a gauge singlet, it may
be unreasonable to forbid additional terms to all orders. As we shall
see, this is a problem for models of inflation where the inflaton
field has a value of order $\mpl$.

\paragraph{The K\"{a}hler potential}

The K\"{a}hler potential determines the 
 kinetic terms of the scalar fields, according to 
the formula
\be
{\cal L}_{\rm kin} = (\partial_\mu\phi^{n^*})
K_{n^* m}
(\partial^\mu\phi^m).
\ee
It is a function of the fields and their complex conjugates, and
can be chosen to have
the expansion
\be
K=K_{nm^*}\phi^n\phi^{m*}
+ \mpltil^{2-d}\sum_{d=3}^{\infty} 
K_d\(\phi^n,\phi^{n*} \) \,,
\label{Kalerexp}
\label{kinetic}
\ee
where $K_{nm^*}$ is evaluated at the origin.
For simplicity we have assumed that any constant or linear term
has been absorbed into the superpotential by a K\"ahler transformation, 
which is always possible.
One can choose the scalar fields to be canonically normalized at the 
origin, corresponding to $K_{nm^*}=\delta_{nm^*}$.

As in the previous expression, each $K_d$ is a sum with each term in the 
sum a product of $d$ fields, times a coefficient which is expected to be 
of order 1 unless it is forbidden by a symmetry.
As $K$ is not holomorphic, symmetries do not constrain it 
very strongly.
It can, for instance, be an arbitrary function of 
the $|\phi^n|^2$, and the coefficient of a monomial built out of such
terms will generically be of order 1. As we shall see, this 
is a problem for inflation model-building.

\paragraph{The gauge kinetic function}

The gauge kinetic function determines the kinetic terms of the gauge and 
gaugino fields. One can choose them to be canonically normalized
when the scalar fields are at the origin, 
which corresponds to
\be
f= 1+\mpltil^{-d} \sum_{d=1}^\infty f_d(\phi^n) \,.
\label{gkin}
\ee
As is the case with $W$, symmetries powerfully constrain the form of $f$ 
because it is homomorphic.
We need to consider $f$ because it appears in the scalar field potential.

\subsubsection{The scalar potential and spontaneously broken supergravity}

Supergravity can be broken only spontaneously, not explicitly like
global susy. The transformation equations 
\eqs{psitransform}{lamtran} hold in supergravity, so it remains
true that the condition for spontaneous breaking is 
a non-vanishing vev for
one or more of the auxiliary fields $F^n$ and $D$.

In contrast with the case of global susy, the vevs of 
$F^n$ and $D$ can receive contributions from
fermion condensates as well as from scalar fields.
A favoured possibility for susy breaking (in the vacuum)
is gaugino condensation, but as discussed later one can in that case add an
effective non-perturbative contribution to $W$ 
instead of including the 
condensate explicitly. Assuming that this has been done,
the auxiliary fields are given by\footnote
{The K\"ahler invariance of the first expression is guaranteed by
the gauge invariance. Indeed, one can replace $K_n$ by $G_n$,
because the gauge invariance requires $\sum_n q_nW_n\phi^n=0$.}
\bea
D &=& - g\( q_n K_n \phi^n
+\xi \right) \,, \\
F^n &=& -e^{K/2} K^{n m^*} 
\left( W_{m}+\mpl^{-2}WK_{m}\right)^* \,. \label{fdef}
\eea

The tree-level potential is given by
\be
V = V_D + F^2 - 3\mpl^{-2} e^{K/\mpl^2} |W|^2 \,,
\label{sugpot}
\ee
where
\be
V_D \equiv \frac 12 ({\rm Re}\, f)^{-1} g^2 
\left(q_n K_n \phi^n
+\xi \right) ^2 \,,
\label{sgdterm}
\ee
and
\bea
F^2 &\equiv& F^n K_{nm^*} F^{m^*} 
= F_n K^{m^* n} F_m \\
&=& 
 e^{K/\mpl^2} \left( W_m+\mpl^{-2}WK_m\right)^* 
K^{m^* n} 
\left( W_n+\mpl^{-2}WK_n \right)
\,.
\eea
In the second line, we defined $F_n\equiv K_{nm^*} F^{m^*}$,
and $K^{m^* n}$ is the inverse of the matrix $K_{nm*}$.

As in global supersymmetry, $V_D$ is proportional to $D^2$, 
while $F^2$ is equal to $\sum|F_n|^2$
if we choose $K_{nm^*}=\delta_{nm}$.
The last term in \eq{sugpot}
allows the true vacuum energy to vanish, as is
(practically) demanded by observation. 

It is usual to define
\bea
V_F &=& F^2 - 3e^{K/\mpl^2} \mpl^{-2} |W|^2 \\
&=& e^{K/\mpl^2} \[ \left( W_n+\mpl^{-2}WK_n \right)
K^{m^* n} 
\left( W_m+\mpl^{-2}WK_m\right)^* 
- 3\mpl^{-2} |W|^2 \]\,.
\label{vexp}
\eea
Then \be
V=V_D + V_F \,,
\ee
and one calls $V_F$ the $F$ term even though is does not come
only from the auxiliary fields $F_n$.

Taking $\mpl$ to infinity with $\mpltil$ fixed gives 
\be
V_F =  W_n K^{m^* n} (W_m)^*.
\ee
We then have non-renormalizable global supersymmetry.
Renormalizable global supersymmetry is obtained by
taking $\mpltil$ to infinity as well.

The other possible limit is $\mpltil\to\infty$ with $\mpl$ fixed.
This is minimal supergravity, characterised by canonical kinetic terms.
It has no motivation {}from string theory.

In the usually-considered case that
$\mpltil$ is identified with $\mpl$, one simply says that (renormalizable)
global supersymmetry is obtained {}from supergravity in the limit
$\mpl\to \infty$. From now on, we make this identification except where 
stated.

The scale of susy breaking in the true vacuum is denoted by
$M\sub S$ and defined  by
\be
M\sub S^4=F^2 + V_D \,. \label{ms1}
\label{MSdef1}
\ee
An equivalent definition is
\be
V = M\sub S^4 - \mpl^{-2}e^{K/\mpl^2} \vert W \vert^2 \,.
\label{MSdef2}
\ee
Since $V$ (practically) vanishes in the 
true vacuum, this
is equivalent to
\be
M\sub S^4 = 
\mpl^{-2}e^{K/\mpl^2} \vert W \vert^2 \,.
\label{mssugra}
\ee
One can show that the gravitino mass is given by
\be
M\sub S^4 = 3m_{3/2}^2\mpl^2 \,.
\label{gravmass}
\ee

\subsection{Supergravity from string theory}

\label{sugfstr}

One hopes that the lagrangian describing
field theory, will eventually be derivable from some more fundamental
theory. Candidates under consideration
at present include weakly coupled (heterotic) string theory
\cite{bookstring}
and Horava-Witten M-theory \cite{mtheory,horwit}.
In this section we look at the form of supergravity 
predicted by weakly coupled string theory.
Then we briefly mention the case of M-theory, which has not so far been 
invoked for inflation model-building.

A crucial role is played by special fields, 
namely the dilaton and the 
bulk moduli.
The dilaton, usually denoted by 
$s$, specifies the 
gauge coupling at the string scale, and the bulk moduli
specifying the radii of the compactified dimensions.
(Weakly coupled strings live in nine space dimensions,
so six of them have to be compactified.)
We consider the cases where there is just one bulk modulus 
$t$, and where there are three bulk moduli $t^I$.

For simplicity, we ignore
the Green-Schwarz term needed to cancel the modular anomaly induced
by field theory loop corrections, and initially we ignore the dilaton as 
well.
In this section, we set $\mpl=1$ unless otherwise stated.

\subsubsection{A single modulus $t$}

The simplest case corresponds to 
compactification on a six-torus \cite{w}. 
It should be regarded as a toy model, since it permits only one 
generation in the Standard Model.
In units of the string scale (slightly below $\mpl$, see
footnote \pageref{sscale})
the radius of the six-torus is 
$(2x)^{-1/2}$ where
\be
x\equiv t+ t^* - \sum_n |\phi^n|^2 \,.
\label{single1}
\ee
Here $t$ is a bulk modulus, and 
$\phi^n$ are a subset of the matter fields, 
called the untwisted sector. (The other matter fields are said to belong 
to the twisted sector.)
The K\"ahler potential derived from string theory is 
\be
K = - 3\ln x \,.
\label{single2}
\ee

If we ignore the twisted sector, and assume that 
$W$ is independent of $t$, \eq{vexp} 
takes the remarkably simple form
\be
 V_F=\frac3{x^2}\sum_n|W_n|^2 \,.
\label{onet}
\ee
It is assumed that the vacuum of the globally supersymmetric theory 
(minimum of its potential) is at $W_n=V=0$,
corresponding to unbroken global supersymmetry. Then, the vacuum of the 
supergravity theory is also at $V=0$, as is required by observation,
but supersymmetry is now in general spontaneously broken. 
At the tree level under consideration here,
the scale of supersymmetry breaking given by \eq{mssugra}
is undetermined. ($V$ in the vacuum is independent of $x$
and therefore $e^K$ is undetermined.)
This corresponds to what is called a no-scale supergravity 
theory \cite{noscale}. 

Although supersymmetry is broken, the scalar masses given by this 
tree-level expression  do not feel the 
effect of the breaking as is clear from the fact that the 
potential $V$ has the same form as in global susy. 

The no-scale model is a consequence of the 
assumptions about $W$. In general one expects that $W$ will depend on $t
$, and the twisted sector may be important. 
We shall look at these issues in the next subsection, in the context of 
the more realistic model that has three bulk moduli.

For future reference, we note that if the $D$ term 
\eq{sgdterm}  involves only
the untwisted sector it is of the form
\be
V_D = \frac 12 ({\rm Re}\, f)^{-1} g^2 
\left(x^{-1} \sum_n q_n |\phi^n|^2
+\xi \right) ^2 \,,
\label{onetd}
\ee

\subsubsection{Three moduli $t^I$}

Compactification on the six-torus is not phenomenologically viable, 
because it allows only one generation in the
Standard Model. To obtain the three generations that are observed, one 
can use \cite{iban,font2,dkl,flt,dkl2,ant,iban2,kobayashi}
orbifold compactifications with three tori.

There are now three moduli $t^I$ ($I=1$ to $3$).
This theory possesses invariance under the modular transformations.
Acting on the moduli, these transformations are generated by
$t^I\to 1/t^I$ and $t^I\to t^I\pm i$. A matter
field $\phi^\alpha$ transforms like $\prod_I\eta^{-2q^\alpha_I}(t^I)$,
where $\eta$ is the Dedekind function and $q^\alpha_I$ are the
weights of the field. Modular symmetry has a fixed point
(up to modular transformations) at which the matter fields vanish and
$t^I= e^{i\pi/6}$. At this point, the derivative of $V$ with respect to 
every field vanishes.

The matter fields are divided into fields $\phi^{AI}$, that belong to 
the untwisted sector, $\phi^{AJ}$ having
modular weight $q^{AJ}_I=
\delta^J_I$, and fields 
$\phi^A$ belonging to the twisted sector, that have 
weights $q^A_I>0$ (typically less than 1).
In units of the string scale, the radius of the $I$th torus is
$(2x_I)^{-1/2}$, where
\be x_I= t^I + t^{I*} - \sum_A|\phi^{AI}|^2 .
\label{xidef2}
\ee
We expect $|t^I|\sim 1 $ with all matter fields $\ll 1$,
both in the true vacuum and during inflation.

The superpotential
has a power series expansion in the
matter fields, of the form
\be W=\sum_m \lambda_m \prod_\alpha (\phi^\alpha)^{n_m^\alpha}
\prod_I\eta(t^I)^{2(\sum_\alpha n_m^\alpha q_I^\alpha -1)} \,,
\label{15} 
\ee
where $n_m^\alpha$ are positive integers or zero.
The $t^I$ dependence of each coefficient
is dictated by modular invariance, which requires that $W$
transforms like $\prod_I \eta^{-2}(t^I)$ (up to a
modular-invariant holomorphic function, which we do not consider
because it would have singularities). Using this expression one sees 
that
\be \frac{\pa W}{\pa t^I}\equiv W_I  = 2\xi(t^I)\left( \sum_\alpha 
q^\alpha_I\phi^\alpha W_\alpha
-W\right) . \ee

The K\"ahler potential is 
\be K= -\sum_{I=1}^{3} \ln x_I + \sum_A 
\( \prod_I x_I^{-q^A_I} \) |\phi^A |^2 + \cdots \,.
\label{threetk}
\ee
The first term comes directly from string theory, and it gives 
the part of $K$ that
is independent of the twisted fields.
The second term comes from
an expansion of the $S$-matrix as a
power series in matter fields.
The additional terms are restricted by modular invariance, but they 
could in general include 
terms like 
\be
\( \prod_I x_I^{-q^A_I} \) |\phi^A |^2  \frac{|\phi^{BI}|^2}
{t^I + t^{I*}}  \,.
\label{extraterm}
\ee
Such terms would generically have coefficients of order 1, and as we 
shall see they could 
spoil the flatness of the inflationary potential.
They can be eliminated
if we assume that $K$ depends
on the moduli and untwisted fields
only through the combinations $x_I$, as advocated in  \cite{p97berk}.

If the twisted fields and the $W_A$ are negligible, 
the
potential \eq{vexp} becomes
\be V_F=e^K\left[\sum_I \left(x_I \sum_A
|W_{AI} + \phi^{AI*} W_I|^2 +
| x_I W_I - W|^2 \right) - 3|W|^2 \right] . \label{9} \ee
In this expression, $W_I\equiv \pa W/\pa t^I$. 

If $W$ is a sum of cubic terms, each containing just one field from each 
untwisted sector, then $W$ does not depend on the moduli and we have simply
\be V_F=\frac{\sum_I x_I \sum_A |W_{AI}|^2 }{x_1x_2x_3} \,.
\label{1098}
\ee
This expression is similar to 
\eq{onet}, that we wrote down earlier. It has all the 
properties that we described then, and is also called a no-scale model.

For future reference, we note that with \eq{threetk}
the $D$ term \eq{sgdterm} becomes
\be
V_D = \frac 12 ({\rm Re}\, f)^{-1} g^2 
\left(\sum_\alpha q_\alpha \(\prod_I x_I^{-q^\alpha_I} \)|\phi^\alpha
|^2 
+\xi \right) ^2 \,.
\ee
Here, $\alpha$ runs over both twisted and untwisted fields.

\subsubsection{The dilaton}

\label{dilgs}

At the the string scale,\footnote
{\label{sscale} The string scale is the one below which, in weakly coupled
string theory,
field theory will become a valid approximation. At this scale,
the gauge
couplings in the true vacuum are supposed to have a common value
$g\sub{str}$, presumably of order 1. The scale and coupling
are related by $M\sub{str}\simeq g\sub{str}\mpl$.
The value
$g\sub{str}^2\simeq 0.5$ would correspond 
to the value $\alpha\sub{str}\equiv g\sub{str}^2/4\pi
\simeq 1/25$, which with naive running of the couplings is suggested
by observation at a scale of order $10^{16}\GeV$.}
the gauge coupling 
is related to the dilaton field $s$ by
\be
g^2\sub{str}= \mpl/({\rm Re\,}s) \,.
\label{gvac}
\ee
This expression takes the 
real part of the gauge
kinetic function to be 1. Equivalently, $g\sub{str}$ can be absorbed into 
$f$. Then at the string scale
\be
f(s) = s/\mpl \,.
\label{fstr}
\ee

Ignoring Green-Schwarz terms, the contribution of the dilaton 
to the K\"ahler potential is
\be
\Delta K = -\ln(s+ s^*) \,..
\label{delk}
\ee
This gives an extra contribution to the potential 
\bea
\Delta V&=&\frac{|F^s|^2}{(s+s^*)^2} 
\\
&=& 
e^K |(s+s^*)W_s-W|^2
\,. 
\label{vwiths}
\eea
(Of course it also contributes an overall factor $(s+s^*)^{-1}$
from the $e^K$
in front of everything in \eq{vexp}.)

In the true vacuum \eq{gvac} requires $s\sim 1$ to $10\mpl$,
and during inflation the order of magnitude of $s$ is presumably not 
very different, so as to be within the domain of attraction of the true
vacuum.

The contribution of $s$ to the superpotential is non-perturbative, and 
very model-dependent. It is often supposed to be something like
\be
W(s) = \mpl^3 e^{-s/(b\mpl)} \,.
\label{wofs}
\ee

Since $e^K\propto 1/{\rm Re\,} s$, these expressions make 
${\rm Re\,}s$ run away to infinity at least with a single term in
\eq{wofs}. There is no consensus about what stabilizes the dilaton 
either in the true vacuum or  \cite{bs93,paul,kr} during inflation.
The simplest possibility is to invoke an additional 
(non-perturbative) contribution to the K\"ahler potential.

All this assumes that the dilaton is part of a chiral supermultiplet,
like the other scalar fields.
An alternative description \cite{bgw}
puts the real part of the dilaton in a linear supermultiplet.
The situation then is qualitatively similar to the one 
that we have described, but different in important details.

\subsubsection{Horava-Witten M-theory}

In Horava-Witten M-theory \cite{mtheory,horwit}, 
$K$ receives an extra 
contribution \cite{low,li}. For the untwisted fields, this is
\be
\Delta K = \frac13 \frac1{s+s^*} \( \sum_I \alpha_I(t^I+t^{I*})
\)\( \sum_I \frac {\sum_A |\phi^{IA}|^2}{t^I + t^{I*}} \) \,.
\label{delkm}
\ee
The parameters $\alpha_I$ are expected to be roughly of order 1.
The gauge coupling in the visible sector (at the string scale)
becomes
\be
f = s + \sum_I \alpha_I t^I \,.
\label{gmth}
\ee
The `string' scale at which this expression is valid 
will be lower than in weakly coupled string theory.

\subsection{Gravity-mediated soft susy breaking}

\label{gravmedtext}

This is a good place to give a brief account of gravity-mediated 
soft susy breaking.

\subsubsection{General features}

The basics features are the same as for gauge-mediated susy breaking
(Section \ref{gmed}).
The softly broken global susy, that describes the visible sector,
is supposed to be only an effective theory. In the full theory,
supersymmetry is spontaneously broken. The spontaneous breaking 
takes place in a hidden sector, whose fields do not possess the 
Standard Model gauge interactions. 
The spontaneous breaking mechanism is supposed to involve 
an $F$-term. 

In contrast with the gauge-mediated case, the mechanism of
spontaneous susy breaking in the hidden sector 
is usually supposed to involve supergravity in an 
essential way. The defining difference, though,
is that the mechanism of {\em transmission} of susy breaking to the
visible sector comes only from interactions of gravitational
 strength.
In other words, each interaction term is multiplied by a power of
$\mpl^{-1}$. Some interaction terms of this type will 
be present as non-renormalizable terms in the expansions
\eqs{wexp}{Kalerexp}; for instance, 
no symmetry 
can prevent the appearance of a term in $K$ like
\be
K= \cdots \lambda \mpl^{-4} |\phi|^2|y|^2 \cdots \,,
\ee
where $\phi$ belongs to the hidden sector and $y$ belongs to the
visible sector, and the coupling $\lambda$ of such a term will generically
be of order 1. Additional interaction terms will 
arise in the potential because of the form of
the supergravity expression \eq{vexp}.

Given the values of the auxiliary fields that spontaneously
break susy, and those of the fields themselves,
one can calculate the soft susy masses-squared $m_n^2$ (or more generally
the soft mass-matrix) and the $A_{nm\ell}$ parameters that define the soft
 trilinear terms. One finds generically $m_n^2\sim A_{nm\ell}^2
\sim M_S^4/\mpl^2 (= 3m_{3/2}^2)$, 
where $M_S$ is the susy breaking scale defined by
\eq{MSdef1}, (\ref{MSdef2}) or (\ref{mssugra}), and $m_{3/2}$ is the 
gravitino mass defined by \eq{gravmass}.
One can see this by making rough estimates, as in the similar analysis
of Section \ref{infmass}. A classic explicit calculation, with some
specific assumptions, is given in  Section \ref{swgm}.

The gaugino masses are given
by
\be
m_{1/2} = \sum_n \frac{\pa f}{\pa \phi^n} \frac{F^n}{2{\rm Re} f}
\label{ginom}
\ee
where $f$ is the gauge kinetic function for the visible sector.

The simplest example of gravity-mediated susy breaking was given
in an unpublished paper 
by Polonyi \cite{pol}. The superpotential 
is split into the sum of two functions
\beq 
W=W(\phi)+W(y^a),
\eeq
where $y^a$ denote the visible sector fields and $\phi$ denotes 
a hidden sector field which is a gauge singlet. Its superpotential
is taken to be
\beq
W(\phi)=M\sub S^2(\phi+\beta).
\eeq
If gravitational effects are ignored, $W(\phi)$ 
leads to  a flat potential independent of $\phi$
\beq 
V=M\sub S^4,
\eeq
susy is broken, but the vev of $\phi$ is undermined. 
Once gravity is turned on, the presence of the negative terms 
produces a minimum of the potential at 
\begin{eqnarray}
\langle \phi\rangle &\sim& \mpl,\\
F_\phi&=&\frac{\partial W}{\partial \phi}\sim M\sub S.
\end{eqnarray}
 The constant $\beta=(2-\sqrt{3})\mpl$ is chosen to make the cosmological 
constant vanishing in the true  vacuum,  $V(\langle \phi\rangle)=0$. 

\subsubsection{Gravity-mediated susy breaking from string theory}

\label{gmedstr}

The nonvanishing auxiliary fields of the hidden sector 
are usually taken to be those of the 
dilaton and/or the bulk moduli. Also, the bulk moduli $t^I$
and their auxiliary 
fields $F^{t^I}$ are usually set equal to common values,
$t$ and $F^t$.
Finally, the weakly coupled string theory expression
\eq{threetk} is assumed.
Then the scalar masses are 
\cite{kl,bim}
\be
m_n^2 = m_{3/2}^2 \[(3+ q_n \cos^2\theta) C^2 -2 \] \,,
\label{mnmasses}
\ee
where $q_n=\sum_I q_I^n$ and $\tan^2\theta 
=(K_{ss^*}/K_{tt^*})|F^s/F^t|^2$. The constant $C$ is given by
$C^2-1
= V_0/(3\mpl^2m_{3/2}^2)$, and it is equal to 1 in the true vacuum 
case that we are dealing with at the moment.
As usual $m_{3/2}^2 = e^{K /\mpl^2} |W|^2/\mpl^4$. 

At a deeper level, the vevs of the auxiliary fields are usually supposed to
mimic some dynamical effect, often originating in string theory
with extra space dimensions. A favoured mechanism is gaugino 
condensation,  which is supposed to generate a superpotential $W(s)$
looking something like \eq{wofs}.
(With several hidden sectors there is a sum of such terms.)
The value of $b$ has to be such that
\be
W\sim \Lambda\sub c^3 \sim (10^{13}\GeV)^3 \,.
\ee
This gives the right soft susy breaking scale, $M\sub S ^2\sim 
\Lambda\sub c^3
/\mpl\sim (10^{10}\GeV)^2$.
With this mechanism
$F^s$ vanishes, since once $s$ is stabilized the perfect square
\eq{vwiths} is driven to zero. In weakly coupled string theory,
\eqs{fstr}{ginom} then make 
the gaugino masses vanish at the
string scale. This is probably forbidden by observation,
but it is avoided in Horava-Witten M-theory where \eq{fstr} is replaced by 
\eq{gmth}.

A particular version of gravity-mediated susy breaking is 
the no-scale theory, corresponding to
\eq{1098}. In this case, the masses of untwisted fields
vanish at tree level, though running them from the string scale 
can still give masses of order $100\GeV$ at the 
electroweak scale.\footnote
{To be more correct, the relevant coefficients in the expansion
\eq{Kalerexp} are supposed to be of order 1 at the Planck scale.
They run, which is equivalent to running the susy breaking parameters
even though the latter may really be defined only below a lower scale where
supersymmetry breaks.}

In the context of weakly-coupled string theory, no-scale gravity 
corresponds to the assumption that the superpotential in the relevant 
sector of the theory is independent of
the bulk moduli. Because of the modular invariance encapsulated in 
\eq{15}, this may be difficult to arrange in the true vacuum under 
consideration at present, 
\label{sremark} since $W$ is necessarily nonzero.
(During inflation 
the no-scale form is easier to achieve as we discuss later,
provided that $W$ is negligible.)

In Horava-Witten M-theory, no-scale gravity will presumably 
be a valid 
approximation only if some of the $\alpha_I$ in \eq{delkm} are
significantly below 1.

\subsubsection{Formalism for gravity-mediated supersymmetry breaking}

\label{swgm}

This subsection is more technical, 
and can be skipped by the general reader.
It gives a formalism for calculating the soft scalar
terms explicitly, with some assumptions, and an example of how gaugino
condensation can generate an effective contribution to $W$. 

For the formalism, we
follow the original notation
\cite{soni,gm}, in which the complex conjugate of a field
is labelled by a subscript.
The visible sector fields are $y^a$ (collectively $y$)
and the hidden sector fields are $\phi^i$ (collectively $\phi$).
It is supposed that $\phi\gg y$, and 
$\xi^i\equiv \phi^i/\mpl$ is defined (collectively $\xi$).
The soft susy breaking parameters are calculated in the limit
$\mpl\to\infty$, with $\xi$ fixed.

Requiring that the low-energy lagrangian for the visible sector
is not multiplied by powers of $\mpl$ defines
the dependence on $\mpl$ of $W$ and $K$  \cite{soni} 
\begin{eqnarray}
W(\xi,y)&=&\mpl^2 W^{(2)}(\xi)+\mpl W^{(1)}(\xi)+ W^{(0)}(\xi,y),\nonumber\\
K(\xi,\xi^\dagger,y,y^\dagger)&=& \mpl^2 K^{(2)}(\xi,\xi^\dagger)
+\mpl K^{(1)}(\xi,\xi^\dagger) + K^{(0)}(\xi,\xi^\dagger,y,y^\dagger)
\,.
\end{eqnarray}
In addition, the $y^a$ are supposed to be canonically normalized 
\cite{soni},
\begin{equation}
 K^{(0)}(\xi,\xi^\dagger,y,y^\dagger)=y^a\Lambda_a^b(\xi,\xi^\dagger)y_b^\dagger
+\left(\Gamma(\xi,\xi^\dagger,y)+{\rm h.c.}\right),
\end{equation}
with the vacuum expectation value $\langle 
\Lambda_a^b\rangle=\delta_a^b$. Finally, the $\phi^i$ 
fields are gauge singlets, so that gauge invariance requires
$\Lambda_a^b$ to be diagonal. 

If there are no mass scales in the theory other than
$\mpl$ and those induced by some spontaneous symmetry breaking
(this is what happens in string-inspired theories),
the renormalizable self couplings of the 
light fields $y^a$ is of the form \cite{gm}
\begin{eqnarray}
W^{(0)}(\xi,y)&=&\sum_n\: c_n(\xi) g_n^{(3)}(y),\nonumber\\
\Gamma(\xi,\xi^\dagger,y)&=&\sum_m\: c_m^\prime(\xi,\xi^\dagger)g_m^{(2)}(y),
\end{eqnarray}
where $g_n^{(3)}(y)$ and $g_m^{(2)}(y)$ are, respectively,
the trilinear and bilinear terms in $y^a$
allowed by the symmetries of the theory. 

After taking the 
limit $\mpl\rightarrow\infty$ \cite{gm}, we obtain for the visible 
sector  a renormalizable global susy theory, with explicit soft breaking 
terms. The scalar potential is of the form
\begin{eqnarray}
 V &=&\left|\frac{\partial \hat{g}}{\partial y^a}\right|^2 
   + m_{3/2}^2 y^a S_a^b y_b^\dagger 
   + [m_{3/2}^\dagger (y^a R_a^b \frac{\partial \hat{g}}{\partial y^b} 
   + \sum_n (A_n - 3) \hat{g}^{(3)}_n  \nonumber \\
&& + \sum_n (B_m - 2) \mu_m g_m^{(2)}) + {\rm h.c.}] + {\rm D-terms}\ .
\end{eqnarray}
The first term is the unbroken susy result, the second term is the soft
mass matrix for the complex fields, and the term in square brackets
contains soft trilinear terms, as well as bilinear terms 
that complete the  specification of the mass matrix
of the real fields.

We have imposed the constraint $V=0$ appropriate for the vacuum,
and the gravitino mass is the modulus of 
\begin{equation}
 m_{3/2} \equiv \langle e^{K^{(2)}/2} W^{(2)}\rangle  \,.
\end{equation}
The soft parameters are determined by the following formulas.
\begin{equation}
 S_a^b = \delta^b_a +  \left\langle\rho^{\dagger i}\left(\frac{\partial
         \Lambda_c^b} {\partial \xi^\dagger_j} \frac{\partial \Lambda_a^c} 
         {\partial \xi^i} - \frac{\partial^2 \Lambda_a^b} 
         {\partial \xi^\dagger_j \partial \xi^i} \right)\rho_j\right\rangle , 
 \qquad 
 R_a^b = \delta_a^b - \mpl \left\langle\rho^{\dagger i} 
  \frac{\partial \Lambda_a^b}{\partial \xi^i} \right\rangle ,
\end{equation}
where
\begin{equation}
 \rho_j \equiv \left( \frac{\partial^2 K^{(2)}} {\partial \xi^i \partial
  \xi^\dagger_j}\right)^{-1} 
 \frac{\partial}{\partial \xi^j}(\ln W^{(2)}+K^{(2)})\ .
\end{equation}
Here $\hat{g}$ is the superpotential for the light fields defined by
\begin{equation}
 \hat{g}(y) = \sum_n \hat{g}^{(3)}_n (y) + \sum_m \mu_m g_m^{(2)} (y)\ ,
\end{equation}
with 
\begin{equation}
 \hat{g}^{(3)}_n (y) = \langle e^{K^{(2)}/2} \rangle 
  c_n (\langle\xi\rangle) g_n^{(3)} (y)\ , \qquad 
 \mu_m = m_{3/2} \left\langle 
  \left(1 - \rho_i\frac{\partial}{\partial \xi^\dagger_i}\right)
  c_m^{\prime}(\xi, \xi^\dagger) \right\rangle .
\end{equation}
Also, 
\begin{eqnarray}
 A_n &=& \left\langle \rho^{\dagger i} \frac{\partial}{\partial \xi^i}  
         [K^{(2)} + \ln c_n (\xi)] \right\rangle , \nonumber\\ 
 B_m &=& \left\langle \left[2 +  \left(
           \rho^{\dagger i} \frac{\partial}{\partial \xi^i}
         - \rho_i \frac{\partial}{\partial \xi^\dagger_j}\right) 
         -  \rho^{\dagger i} \rho_j \frac{\partial^2}{\partial \xi^i 
            \partial \xi^\dagger_j}\right] 
         \frac{c'_m (\xi, \xi^\dagger)} 
          {(1 - M \rho_i \frac{\partial} 
          {\partial \xi^\dagger_i} c'_m(\xi, \xi^\dagger) 
          } \right\rangle.
\end{eqnarray}

Identifying the ultra-violet cutoff $\Lambda\sub{UV}$ in 
\eqs{wexp}{Kalerexp} with $\mpl$, one will have generically
$|S^b_a|\sim |R^b_a|\sim |A_n|
\sim |B_n| \sim 1$, making the soft susy breaking 
mass matrix-squared of order $m_{3/2}^2$ and the trilinear terms of 
order $m_{3/2}$. 

Next we see how gaugino condensation can give an effective 
superpotential. We consider an extension of susy-QCD based on the 
gauge group $SU(N_c)$ in the hidden sector 
with $N_f\leq N_c$ flavors of ``quarks'' $Q^i$ in 
the fundamental representation and ``antiquarks'' $\widetilde{Q}_{\bar{i}}$ 
in the antifundamental representation 
of $SU(N_c)$ \cite{ads}. The gauge kinetic function may be chosen to be  
$f=k s$, where $s$ is the dilaton superfield and $k$ is the Kac-Moody 
level of the hidden gauge group. 

Because of the gauge structure, the gauge group $SU(N_c)$ 
enters the strong-coupling regime at the scale
\beq
\Lambda_{{\rm c}}=\mpl \:e^{-\frac{k s}{2 b_0}},
\eeq
where $b_0=(3 N_c-N_f)/(16 \pi^2)$ is the one-loop beta function for the 
hidden sector gauge group. 

Below the scale $\Lambda\sub c$ the appropriate degrees of freedom for 
$N_f<N_c$ are the mesons $M^i_{\bar{i}}=Q^i\widetilde{Q}_{\bar{i}}$. 
The effective superpotential
is fixed uniquely by the global symmetries as follows \cite{ads}
\beq
W=(N_c-N_f) \langle \lambda\overline{\lambda} \rangle \,,
\eeq
where the gaugino condensation scale is
\beq
\langle \lambda\overline{\lambda} \rangle=
\left(\frac{\Lambda_{{\rm c}}^{3N_c-N_f}}
{{\rm Det} M}\right)^{\frac{1}{N_c-N_f}} \,.
\eeq

\section{$F$-term inflation}
\label{s18}

\subsection{Preserving the flat directions of global susy}

Let us recall the discussion of 
Section \ref{sss:tree}. We saw there that in any model of inflation,
the quartic term of the potential $V(\phi)$
should be small. One can ensure this by choosing
the inflaton to be a flat direction of global supersymmetry,
but one still has to ensure that the the mass term and 
non-renormalizable terms
are sufficiently small. At least for the mass term, this
does not happen in a generic supergravity 
theory.
The following strategies have been proposed to get around this problem.
\begin{enumerate}
\item
The potential is dominated by the $F$ term,
but the inflaton mass is suppressed because 
$K$ and $W$ have special forms.
\item
The potential is dominated by the $F$ term,
whose form is generic. However, 
the inflaton mass is suppressed because of an accidental 
cancellation between different terms.
\item
The potential is dominated by the $F$ term,
whose form at the Planck scale is generic. However, 
the inflaton mass is suppressed in the regime where inflation takes 
place, because it runs strongly
with scale.
\item
The potential is dominated by a Fayet-Iliopoulos
$D$ term.
\item
The potential is dominated by the $F$ term, whose form is generic.
However, the kinetic term of the inflaton field becomes singular
near the region where inflation takes place,
 so that after going to 
a canonically-normalized the potential becomes flat even though it
was not originally.
\end{enumerate}
We mentioned the last possibility in Section \ref{kininf}
and it will not be considered further. We consider in this section the 
three $F$-term possibilities, and then go on to the $D$-term.

\subsection{The generic $F$-term contribution to the inflaton potential}

\label{ss:flat} 

In this section we show that in a generic model
of $F$-term inflation, the flatness parameter 
$\eta\equiv \mpl^2V''/V$ of the would-be inflaton potential
is at least of order 1, in contrast with the requirement 
$|\eta|\lsim 0.1$.
We are continuing the discussion of Section \ref{sss:tree},
and supposing that the inflaton is the radial part of a matter field.

The full potential is given by \eq{vexp}, and as
it contains more than one complex field we cannot adopt the 
assumption of Section \ref{sss:tree} of exact canonical normalization;
this would correspond to the condition
$K_{nm^*}= \delta_{nm^*}$ which will be impossible to arrange
for all field values.
However, we assume this condition at the origin for the inflaton field
($n=m=i$),
in order to calculate the inflaton mass-squared.
We also assume that it provides at least a rough approximation
for all of the fields, and that in addition 
$|K_n|\lsim \mpl$ and 
$e^{K/\mpl^2}\sim 1$.
These assumptions are valid in the string theory examples that 
are usually considered.

By analogy with \eq{MSdef2}, we define the scale $\minf$ of susy
breaking during inflation by
\be
V\simeq V_0 = \minf^4 -\mpl^{-2} e^{K/\mpl^2} |W|^2 \,.
\label{minfdef}
\ee
($V_0$ is the first term of \eq{power}, which we taking to dominate
during inflation.)
In contrast with the case for the true vacuum, 
there is no need for a strong cancellation between the
first and second terms. If there is no strong
cancellation, 
\be
V_0 \simeq M\sub{inf}^4\,.
\label{vordmag}
\ee

\subsubsection{The inflaton mass}

\label{infmass}

We are mainly concerned with the contribution to $\eta$ 
of the quadratic term in \eq{power}, which is $\eta
=m^2\mpl^2/V$. Purely for
simplicity, we suppose that $V$ depends only on $|\phi_i|$
so that $m^2=V_{i i^*} $ evaluated at the origin.

To get off the ground, we first assume that all field values
are $\ll \mpl$, with $K_{nm^*}=\delta_{nm}$ at the origin.
Then we find from Eq.~(\ref{vexp}),
assuming that the inflaton is a flat direction corresponding to
$W_{ni}=0$,
\be
m^2 = \mpl^{-2} V_0 - \mpl^{-2} |W_i|^2 
+ \sum_{nm}K^{m^* n}_{i i^*} W_n^* W_m \,.
\label{mphi98}
\ee
The right hand side is evaluated with all fields at the origin.
The contribution of the first term to $\eta$ is precisely 1.
For the other terms, take first the case $V_0\sim M\sub{inf}^4$.
Then the
(negative) contribution of the second term to $\eta$ is at most of order 
1. For the third term, we use \eq{Kalerexp}, and 
set $\mpltil=\mpl$. 
Then $K^{\bar n m}_{i\bar i}$ will be of order $\mpl^{-2}$,
and the contribution of the third term to $\eta$ is also of order 1
(with either sign).
Generically, there is no reason to expect an accurate cancellation of 
the contribution $+1$ coming from the first term.

The case that one or more fields have values of order
$\mpl$ is more model dependent, but the generic contributions 
to $\eta$ are still at least of order $V_0/\mpl^2$.
In particular, one gets a contribution to $m^2$ analogous to the third one
of \eq{mphi98}, $\sum_{nm}K^{m^* n}_{i i^*} F_n^* F_m$, that is 
generically of this order.

If we abandon the assumption
$V_0\simeq M\sub{inf}^4$, the estimate becomes
bigger;
\be
m^2 \sim \frac{\minf^4}{\mpl^2} \sim
\( \frac{V_0}{\mpl^2} \) \( \frac{\minf^4}{V_0}\) 
\,.
\label{genmmag}
\ee

\subsubsection{The quartic coupling and non-renormalizable terms}

The expansion \eq{wexp} of $W$ will generically give 
coefficients $\lambda\sim \lambda_d\sim 1$ in \eq{power}.
According to Section \ref{keepflat}, $\lambda\lsim 10^{-9}$. To achieve
this, the inflaton is chosen to be  a 
flat direction, so that the relevant renormalizable terms of \eq{wexp}
vanish. Repeating the above discussion one then finds $\lambda
\sim V_0/\mpl^4$.

At least the first few $\lambda_d$ should also be suppressed, by 
eliminating the relevant non-renormalizable
terms in \eq{wexp}. These coefficients are 
then also of order $\sim V_0/\mpl^4$.

As before, these estimates assume
$V_0\simeq M\sub{inf}^4$ and more generally we have
\be
\lambda\sim\lambda_d \sim \frac{\minf^4}{\mpl^4}
\sim \(\frac{V_0}{\mpl^4} \) \(\frac{\minf^4}{V_0}
\) \,.
\ee

\subsection{Preserving flat directions in string theory}

\label{ss:sstring}

\subsubsection{A recipe for preserving flat directions}

 A strategy for keeping the $F$-term flat 
was given by
Stewart \cite{ewansgrav} (see also \cite{cllsw}). 

The basic idea is to ensure that the potential has almost the same
form as in global susy.
This is done by imposing some simple conditions on $W$
and the fields, and choosing a rather special form for $K$. 
The required form occurs in weakly coupled string theory, though 
apparently not in Horava-Witten M-theory.

The fields are divided
into three classes, which we shall label $\phi$, $\psi$ and $\chi$.
During inflation, 
it is required that the following relations are satisfied to sufficient 
accuracy
\bea
W &=& W_\phi = W_\psi = \chi = 0 \nonumber \\
W_\chi &\neq& 0 \label{genw} \,.
\eea
The inflaton is going to be one of the $\phi$ fields, which means that 
the others are constant during inflation; as a result the requirement
$\chi=0$ can always be imposed by a choice of origin, though it may not 
be a natural one. With these assumptions, the potential \eq{vexp}
becomes during inflation 
\be
V_F = e^K \sum_{nm} W_n K^{m^* n} (W_m)^* \,,
\ee
where the sum goes only over the $\chi$ fields.
The required form for $K$ is
\be
K= -\ln\[ f(\phi,\phi^*) - \sum_{nm}\chi_n^* C_{nm}(\psi,\psi^*) 
\chi_m \]
+ \tilde K(\psi,\psi^*) 
+O(\chi^2,\chi^{*2}) \,,
\label{genk}
\ee
where $f$ and $\tilde K$ are arbitrary functions, and
$C$ is a matrix which might be the unit matrix. Then the 
potential during inflation is 
\be
V(\phi) = 
e^{\tilde K} \sum _n W_n (C^{-1})_{nm} (W_m)^* \,.
\ee
We see that the dependence of $V$
on the fields $\phi$ comes only from the $W_n$.
For such fields, flat directions of global susy are preserved, provided
that they are not spoiled by 
fields that are displaced from the origin.
We can have viable inflation by choosing the inflaton to be one of these
flat directions.
Note, though, that the $\psi$ fields have to be stabilized
in the presence of the $\tilde K$ and $C_{nm}$ factors.

One can quantify \cite{ewansgrav}
the required accuracy of the assumptions by looking at 
\eq{vexp}.
A slight violation of 
the conditions on $W$ typically gives $|\eta|\sim \mpl^{-2}|W/W_\chi|^2$,
$|W_\phi/W_\chi|^2$ or $|W_\psi/W_\chi|^2$. A small contribution
$\delta K(\phi,\chi)$ to $K$ gives, assuming $\chi=0$, 
\bea
\epsilon &\sim& \mpl^{-2} \delta K'{}^2 \label{eofkp}\\
|\eta- 2\epsilon |&\sim& \delta K'' \,, \label{eofkpp}
\eea
where the prime denotes a typical partial derivative of $\delta K$.

\subsubsection{Preserving the flatness in weakly coupled string theory}

In weakly coupled string theory, ignoring Green-Schwarz terms,
$K$ given by \eqs{threetk}{delk} 
is of the required form if the $\phi$ and $\chi$ fields constitute
a single untwisted sector (with the modulus a $\phi$ field),
and the twisted fields vanish to sufficient accuracy.
Accordingly we can require the following
conditions, to sufficient accuracy during inflation
\cite{cllsw,p97berk}.
\begin{enumerate}
\item All derivatives of $W$ with respect to matter fields vanish,
except for the one corresponding to a single untwisted field, say
$W_{C3}$. (One could allow more untwisted
fields from the $I=3$ sector without changing anything,
and of course the choice $I=3$ is arbitrary.)
\item $W=W_s=W_I=\phi_{C3}=0$. (The 
easiest way of ensuring $W_I=0$ is to suppose 
that every
term in the expansion (\ref{15}) 
of $W$ vanishes.)
\item The twisted fields vanish.
\end{enumerate}
{}From \eqs{eofkp}{eofkpp} it is actually enough to have
the twisted fields fixed at values 
$\ll\mpl$. Also, condition 2 is accurate enough if
$|W|/\mpl$, $|W_s|$ and $|W_I|$ are all $\ll|W_{C3}|$.
These conditions are straightforward to achieve if one ignores the 
dilaton, which is reasonable 
for models with $V^{1/4}\gg10^{10}\GeV$, provided that the dilaton 
contribution $W(s)$ is
the same during inflation as it is in the true vacuum.
The present scheme may not work for models with $V^{1/4}\lsim
10^{10}\GeV$.

With these conditions in place, \eq{vexp} gives
\be
V = \frac{|W_{C3}|^2}{x_1 x_2} \,.
\label{tmod}
\ee
Flat directions in the untwisted $I=3$ sector are preserved,
if their flatness is not spoiled by coupling to fields with nonzero
values, and one of them can be the inflaton.
It could also be $t_3$, or a combination.
Note that the analogous procedure in the case of a single
modulus would not work, because of the factor 3 in front of \eq{single2}.

The above strategy preserves 
the flatness of the globally supersymmetric potential at all values of 
the inflaton field. This is possible because the inflaton is supposed to 
belong to an untwisted sector, and string theory gives 
the part of the K\"ahler potential depending only on the sector
for all field values. 
If the inflaton field is small it may be enough to keep the 
inflaton mass small, and this can be achieved provided that one knows
the relevant part of the K\"ahler potential up to quartic terms.
For the twisted fields, \eq{threetk} gives the required information
if we assume that $K$ depends on the untwisted fields and bulk moduli
only through the $x_I$. In general \eq{threetk} gives the usual result
$m^2 \gsim V_0/\mpl^2$, but an exception has been noted
\cite{cg,cgr}; if \eq{mnmasses} applies, with $F^s=0$ and
$m_{3/2}^2\equiv e^K|W|^2$ negligible, then 
$m^2$ vanishes provided that the inflaton field has weight $q_n=3$.

All this is in weakly coupled string theory. 
In Horava-Witten M-theory, $K$ receives an extra contribution
\eq{delkm}. If the $\alpha_I$ are of order 1 this
contribution will presumably give
us back the generic result
$m^2\gsim V_0/\mpl^2$.

\subsubsection{Case of a linear superpotential}

Returning to \eq{tmod}, we
have to ensure the stability of $t_1$ and $t_2$.
This is achieved \cite{cllsw} if
$W_{C3}$ comes from a term $\Lambda^2 \phi_{C3}$,
with $\Lambda$ independent of the matter fields.
Then, modular invariance requires 
$\Lambda^2\propto \eta^{-2}(t_1) \eta^{-2}( t_2)$, and
\be V\propto \left[|\eta (t_1) \eta( t_2)|^4 x_1 x_2
\right]^{-1} \,.
\label{vmod} \ee
To discuss the stability of the moduli, we can set the matter fields
equal to zero so that  $x_I=t_I+\bar t_I$.
As shown in \cite{cllsw}, $V$ is stabilized at
$t_1=t_2=e^{i\pi/6}$ up to modular
transformations. The masses-squared of the canonically normalized
$t_1$ and $t_2$ turn out to be precisely $V/\mpl^2$, 
which presumably hold them 
in
place during inflation.

The value $t_I= e^{i\pi/6}$ corresponds to a fixed point\footnote
{The other
fixed point in the fundamental domain, namely $t_I = 1$, is a saddle 
point of
potential (\ref{vmod}); see eg.~\cite{font}.
(To be more precise, $t_I=1$ is a fixed point if in addition to modular
invariance there is symmetry under ${\rm Im}\,t_I\to-{\rm Im}\,t_I$,
which is the case in the present model.)}
of the modular transformations. Since it must be an extremum of the
potential, it is not particularly surprising to find that it represents
the minimum during inflation.  In the model of
\cite{bgw}, it also represents a possible true vacuum value.
In that case, the moduli stabilized at this point
during inflation will remain there, and will not
be produced in the early Universe.

If our assumptions are exactly satisfied, the linear 
superpotential will make the tree-level potential absolutely flat during 
inflation. The slope might come from loop corrections or from
the assumptions not being exactly satisfied. (A contribution to $K$ 
from Green-Schwarz terms has been shown \cite{cllsw} to give a slope
corresponding to $n$ significantly less than 1.) The slope might also
come from the nonzero $D$-term that we are about to invoke, through
the $K_n$ factors in \eq{sgdterm}.

\subsubsection{Generating the $F$ term from a Fayet-Iliopoulos
$D$-term}

\label{fromd}

Instead of putting in the mass scale $\Lambda$ by hand,
one might generate it using a Fayet-Iliopoulos $D$-term
\cite{p97berk}.\footnote
{This was considered in \cite{cllsw,ewansgrav}, but
the factors $K_n$ in the $D$ term \eq{sgdterm} were not
considered whereas they are in fact crucial.}

Suppose that $W=\lambda\phi_1\phi_2\phi_3$, with each field from a different
untwisted sector. We suppose that $\phi_1$ and $\phi_2$
acquire vevs when the $D$ term is driven to (practically) zero.\footnote
{The ratio $|\phi_1|/|\phi_2|$ can be fixed,
for instance, by gauging a non-anomalous U(1) symmetry under
which $\phi_1$ and $\phi_2$ have opposite and equal charges.}
                 {}From \eq{onetd}, one sees that the
vevs $|\phi_1|^2$ and $|\phi_2|^2$ will be 
proportional to respectively $x_1$ and $x_2$, making $V$ 
given by \eq{tmod} independent of 
these quantities. 

Flat directions are now preserved in all of the untwisted sectors,
provided that they are not spoiled by the displacement of fields from 
the origin. Any of them is a candidate for the inflaton, and so are each 
of the moduli $t_I$.

The same thing actually works \cite{p97berk} in 
the toy model with a single modulus $t$; taking all three fields to 
belong to the (single) untwisted sector one eliminates the $x$
dependence appearing in \eq{onet}.
Other authors using \eq{onet} for inflation
suppose
that $x$ is fixed, either by an 
ad hoc functional form for $K(x)$ \cite{abook,olive,hitinf,steve3},
or by a loop correction
\cite{gmo}. The first option seems unsatisfactory,
and in the second option the
status of the loop correction during inflation is not clear.

We have not yet considered the stability of the dilaton, either in the
$D$-term 
model or in the one with a linear superpotential. This has been 
investigated \cite{p97berk} with 
the (real part of the) dilaton in a 
linear multiplet, using a model \cite{bgw} which
stabilizes the dilaton in the true vacuum. 
The dilaton is stabilized in the model with a linear superpotential, but
not in the $D$-term model in the simple form given above.
However, the vevs induced by the $D$ term can then induce additional
vevs through the $F$ term. It  was shown \cite{p97berk} that this
can stabilize the dilaton, while preserving the
flatness in one or more of the untwisted directions.
(By `stability' we mean existence of a minimum in the potential,
with all fields except the dilaton fixed. Starting from a wide variety of 
initial conditions, the dilaton 
will typically settle down to the minimum \cite{bcc}.)

To have a complete model, one also needs to end inflation,
and because of the form we are imposing on $W$ this will probably
require hybrid as opposed to single-field inflation.
No complete example has yet been given for the particular superstring-derived
theory that we are considering, but one can  presumably be 
constructed along the lines of the following model \cite{ewansgrav}.

The model works with a superpotential that has the general
properties listed at the beginning of the last subsection.
The K\"ahler potential is assumed to be of the form \eq{genk},
with for simplicity $C=e^{\widetilde K}=1$ 
so that the potential is the same as in 
global susy, but its detailed form is not considered.
Also, $K_n=\phi_n^*$ is used when calculating the
$D$ term.
The model contains 
one $\chi$ field and three $\psi$ fields.

Working with units $\mpl=1$, the
superpotential is
\be
W = \lambda_1\phi\psi_1 \psi_3 + \lambda_2 \psi_2^n \chi
\ee
with $n\geq 2$. The 
$D$-term is taken to be\footnote
{\label{samesym}We use the same symbol
for the square and the modulus-squared of a field since it is obvious 
from the context which is intended.}
\be
V_D = \frac12 g^2 \( \xi -\psi_1^2 - \psi_2^2  +
\psi_3^2 + n\chi^2 \)^2 \,,
\ee
and it is assumed that 
\be
\lambda_2\xi^{n-2} \ll g \,.
\label{smalllam}
\ee

It is assumed that during inflation $\psi_1^2+\psi_2^2 <\xi$. Then
$\chi$ and $\psi_3$ will be driven to zero, and so will 
the derivatives of W with respect to $\phi$, $\psi_1$ and $\psi_2$.
The potential then becomes
\be
V= \lambda_1^2 \phi^2 \psi_1^2 + \lambda_2^2 \psi_2^{2n}
+\frac12 g^2 \( \xi - \psi_1^2 - \psi_2^2 \)^2  \,.
\ee
During inflation it is assumed that
\be
\phi^2 > \frac{g^2}{\lambda_1^2}\( \xi- \psi_2^2 \) \,.
\ee
Then $\psi_1$ is driven to zero, along with the derivative of $W$ with 
respect to $\psi_3$. {}From \eq{smalllam}
$\psi_2$ has a constant value given approximately by
\be
\xi - \psi_2^2 \simeq \frac{n \lambda_2^2 \xi^{n-1}}{g^2} 
\ll \xi \,.
\ee

Restoring $\mpl$, we conclude that 
during inflation, there is an exactly flat
potential with magnitude given by
\be
V_0^{1/4} \simeq  \sqrt\lambda_2\(\frac{\xi}{\mpl^2}\)^{\frac{n-2} 4}
\sqrt \xi 
\label{v0mag}
\,,
\ee
in the regime
\be
\phi>\phi_c\equiv (\lambda_2/\lambda_1) \sqrt n  
(\xi/\mpl^2)^{\frac{n-2}{2}} \sqrt\xi \,.
\ee

This scheme is similar to the scheme of $D$-term inflation that we 
consider later, but differs from it in two crucial respects.
One is that the loop correction is much smaller, because the $D$-term
is much smaller. As a result, there is no need for the inflaton field
to have the dangerous value $\phi\sim\mpl$.
The other is that the the COBE normalization $V_0^{1/4}\lsim 10^{-2}\mpl$
can be achieved without supposing  that $\sqrt\xi$ is so small.

\subsubsection{Simple global susy models of inflation}

In these examples we took seriously the requirement of modular
invariance. We end by considering a couple of models that ignore
this requirement, while using a superpotential of the required form
\ref{genw}. It would not be difficult to generalize them
so that modular invariance is satisfied, though the stability of the 
moduli and dilaton may require care.

The mutated hybrid inflation model \eq{mutpot} is generated by 
\cite{ournew}
\be
W= \Lambda^2 \chi_1 \( 1-\frac12 \psi/M \) + \sqrt\frac\lambda 4
\phi\psi\chi_2 \,.
\ee
The $\chi$ fields are driven to zero, giving \eq{mutpot} with
$V_0=\Lambda^4$.
The COBE normalization \eq{cobemut},
with $M\sim\mpl$,
corresponds
to $\Lambda\sim 10^{13}$ to $10^{14}\GeV$. It was suggested 
\cite{mutated} that $\Lambda$ could be identified with gaugino 
condensation scale, though it is not clear how that might be achieved.

To obtain inverted hybrid inflation one can use \cite{ournew}
\be
W= \Lambda^2 \( 1- \frac{\lambda \phi^2 \psi^2}{\Lambda^4} \) \chi \,.
\ee
This drives $\chi$ to zero, 
and after adding 
suitable mass terms one 
obtains \eq{first}. The mass terms can come from the supergravity 
corrections. (A more complicated superpotential was given much earlier
\cite{burt2}, but the inflaton trajectory turned out to be unstable 
\cite{olive}.)

\subsection{Models with the superpotential linear in the inflaton}

\label{linearmod}

We next turn to models where the superpotential during inflation is
linear in the inflaton field
\cite{cllsw,qaisar,dvaliloop,qaisarlatest,linderiotto,ddr},
or linear except for small corrections
\cite{kumekawa,izawa,izawa2}. The first case gives 
hybrid inflation with a potential whose slope is dominated
by a loop correction. The second case gives single-field
inflation with an inverted quadratic, or higher-order, potential.

This paradigm has been widely regarded as a way of keeping 
supergravity corrections under control. 
Unfortunately, the analysis leading to that viewpoint is
likely to be incorrect, since it assumes that all of the fields in 
Nature have values $\ll\mpl$ during inflation.
To see what is going on, first suppose that this assumption is correct.
Then the inflaton mass-squared is given by
 \eq{mphi98}, and one can see \cite{cllsw} that 
indeed the first two terms cancel if the superpotential is linear in the 
inflaton field. So to achieve a sufficiently small mass one need
only tune down the coefficient of the relevant quartic term in $K$,
below its natural value $\sim \mpl^{-2}$. Arguably, this is preferable
to arranging an accurate cancellation.
But now suppose that there are fields $\phi_n$, with values 
$\mpl$. One sees from
\eq{vexp} that with the minimal form
$K=\sum|\phi_n|^2$, each such field contributes
$\eta =\mpl^{-2}|\phi_n|^2
\simeq 1$. There is no reason to suppose that the non-minimal
form presumably holding in reality 
will give a much smaller contribution.
So one is back with a cancellation and the 
paradigm has no special virtue.
Specific examples of fields with  values of order $\mpl$
are the dilaton and bulk moduli that emerge
from string theory.

We focus on the hybrid inflation model, where the superpotential
during inflation is exactly linear in the inflaton field.
The field whose radial part will be the inflaton is a gauge
singlet, denoted by $S$. 
The original version of the 
model \cite{cllsw} used the superpotential
\be
W=  S(\kappa\psi \psi- \mu^2) \,,
\ee
where $\kappa$ is a dimensionless coupling.
This form does not allow $\psi$ to be charged under any symmetry, 
but one can change it to \cite{qaisar}
\begin{equation}\label{1}
W=  S(\kappa\bar\psi \psi- \mu^2) \,.
\end{equation}
Here, $\psi$ and $\bar{\psi}$ are oppositely charged under all symmetries
so that their product is invariant.
The absence of additional terms involving $S$
is enforced to all orders if $S$ is charged under a global $U(1)$ 
R-symmetry, and up to a finite order if it is charged under
a discrete ($Z_N$) symmetry. As we noted before, only the latter 
seems to be allowed in the context of string theory.

Instead of putting in the scale $\mu$ by hand, one may derive it
\cite{ddr}
from dynamical supersymmetry breaking by a quantum moduli space
(Section \ref{sss:qms}).

The canonically-normalized inflaton field is $\phi\equiv\sqrt 2|S|$,
and the global susy potential is
\begin{equation}\label{2}
V =  {\kappa^2 \phi^2\over 2}\bigl(|\psi|^2 +|\bar\psi|^2\bigr)
+|\kappa\bar\psi\psi-\mu^2|^2 \,.
\end{equation}
This has the same general form as original tree-level hybrid 
inflation potential \eq{fullpot}, with zero inflaton mass.
The interaction with $\phi$ then gives $\psi$ and $\bar\psi$ identical 
$2\times 2$ mass matrices for their real and imaginary parts,
and after diagonalizing one finds masses
\be
m_\pm^2 = \frac12\kappa^2\phi^2  \pm \mu^2\kappa \,.
\ee
The critical value is therefore given by $\phi\sub c^2=2\mu^2/\kappa$.
For $\phi>\phi\sub c$, $|\psi|=|\bar\psi|=0$ and we have
slow-roll inflation.

The potential is exactly flat at tree-level, but the 
loop correction gives a significant 
slope \cite{qaisar}. Indeed, using \eq{vdval} and remembering that 
there are two chiral multiplets, one finds
the potential we wrote down in \eq{vsmallloop}, with $\Lambda=\mu$
and $Cg^2=\kappa^2$.
We already worked out the prediction of this potential for $n$,
and its COBE normalization.

Some authors \cite{panag,linderiotto}
have considered the possibility that quadratic
and quartic tree-level terms are significant, with 
the former assumed to come only from the quartic term of the K\"ahler potential
and the latter assumed to come only from the factor $e^{K/\mpl^2}$.
According to the analysis of Section \ref{ss:flat}, neither of these 
assumptions is very reasonable.

\subsection{A model with gauge-mediated susy breaking}

\label{ss:gmed}

Now we consider a 
global susy model \cite{dr,rrr}  in which $W$ does
not have the form (\ref{genw}). 
Our discussion somewhat extends the original one.

In this model, the
supergravity corrections are 
presumed
to be small because of an accident.
As we shall see, a very severe cancellation is required.
The model 
assumes that there is 
gauge-mediated susy breaking in the true 
vacuum, which also operates during inflation.
It uses a particle physics model \cite{dnns}
which replaces the $\mu$ parameter of the MSSM by
a term $\lambda_\mu \mpl^{-n} S^{n+1}$.
The field $\phi\equiv \sqrt2 {\rm Re\,} S$ becomes the inflaton.
In this model, gravitinos do not  pose a cosmological problem, while the moduli
problem is ameliorated.

The superpotential is supposed to be
\be
W= - \frac{\beta XS^{2+p}}{\mpl^p} + \frac {S^{m+3}}{\mpl^m}
+\lambda_\mu \mpl^{-n} S^{n+1} H_U H_D + \cdots \,.
\label{dinew}
\ee
This structure can be enforced
by discrete symmetries. 
The dots represent the contributions to $W$ that do not involve
$S$. They generate among other things the vev $F_X$, which we take to be 
real and positive, and close to the vacuum value discussed in 
Section \ref{gmed}.
The third term generates the $\mu$
term, but plays no role during inflation. The case 
$p=m=2$ is considered.

Adding a negative mass-squared term 
that is supposed to come from supergravity,
the potential along the real component of $S$ (denoted by the same 
symbol) is
\be
V \simeq V_0 -m^2S^2 - \frac14 \lambda S^4 + \( S^4 - 4\beta X S^3\)^2
\,,
\ee
with
\be
\lambda = 8\beta \mpl^{-2} F_X \,.
\ee
The constant term $V_0$ is given by
\be
V_0 = F_X^2 - 3\mpl^{-2} e^{K/\mpl^2} |W|^2 \,,
\ee
and as is usual in gauge-mediated models the origin of the last term is 
not unspecified.

One can determine the vacuum value of $S$ by minimizing this potential,
and using the vacuum value $X\simeq F_X/\Lambda$ with $\Lambda
\sim 10^5\GeV$. Assuming $X\ll S$, one finds
$S^4\sim \beta \mpl^2 F$ and 
$F_X^3\lsim \mpl^2\Lambda^4$ or
$\sqrt{F_X}\lsim 10^9\GeV$.
By setting $V=0$ in the vacuum, one finds that in this case
$V_0\sim \beta^2 F^2$.
In the opposite case $X\gg S$, one finds
$S^2\sim \mpl^2\Lambda^2/F_X$
and $\sqrt{F_X}\gsim 10^9\GeV$. Then,
\be
V_0\sim \frac{\Lambda \mpl^2}{\beta X^3} F_X^2 \,.
\label{vofx}
\ee

      One can check that $X$ is negligible during inflation (assuming that 
like $F_X$ it has almost the same value as in the true vacuum). The
potential then becomes the one 
that we analyzed in Section \ref{cubhigh}.
We found that the COBE normalization requires $\sqrt{\beta F_X}\sim
10^{11}\GeV$, marginally consistent with the upper limit
for gauge-mediated susy breaking if $\beta$ is close to 1.
This corresponds to $X\sim 10^{17}\GeV$.

Using \eq{vofx} one finds 
\be
V_0\sim 10^{-10} F_X^2 \,.
\ee
The generic supergravity contributions to $m^2$ are of order 
$F_X^2/\mpl^2\sim
10^{10} V_0/\mpl^2$. In contrast with the usual situation, the generic 
contributions have to be suppressed to at one part in $10^{
11}$, even if $n$ is significantly different from 1
($n-1 = m^2\mpl^2/V_0\simeq 0.1$).

\subsection{The running inflaton mass model revisited}

\label{ss:loopcorr}

Now we look in more detail at the theory behind the
running mass model of Section
\ref{hybrunning}. 

\subsubsection{The basic scenario}

The fundamental assumption of the model is that the sector of the theory
occupied by the inflaton is hidden from the sector where supersymmetry
is spontaneously broken, and communicates with it only through 
interactions of gravitational strength. We shall call the former the 
{\em inflaton sector},
and the latter the {\em inflationary SSB sector}.
The inflaton sector is supposed to be described
by a renormalizable global susy theory, with soft susy breaking terms.
In other words, there is supposed to be
gravity-mediated supersymmetry breaking,
in the inflaton sector during inflation.

It is not necessary, for the viability of the model,
to assume anything about the inflationary SSB sector.
But the simplest thing is to
identify the inflationary SSB sector with the one that generates susy 
breaking in the true vacuum, which we call the
{\em vacuum SSB sector}. Also, one might suppose that the
susy-breaking scales are the same, $\minf\sim M_S$.
In that case, we shall have $\minf\sim 10^{10}\GeV$
if there is gravity-mediated susy breaking in the visible sector,
and $10^5\lsim \minf\lsim 10^{10}\GeV$ if there is gauge-mediated
susy breaking in the visible sector.
(Presumably the inflaton sector is different from the visible sector,
though they might conceivably be identical if there is
gravity-mediated susy breaking in the visible sector.)

Even if the inflationary and vacuum SSB sectors are identical,
it is not inevitable that the susy-breaking scales are the 
same.
Take for instance the case of gaugino condensation, where that
scale is determined by $W(s)$. Even if $W(s)$ has the same functional
form in the two cases, it will not have the same value because
$s$ will be different. But $W(s)$ might be a 
different function during inflation. For
instance, gaugino condensation might 
occur only after inflation. If it does occur, the value
of $b$ might be 
different, because during inflation some of the fields which contribute to 
the running of the gauge coupling and  are light in the true vacuum,  
become heavy  and no longer contribute to the renormalization group equation  
of the gauge coupling \cite{kr}.

At the Planck scale, 
the inflaton mass-squared $m^2(\phi)$ (along with 
other soft susy breaking parameters in the inflaton sector)
is supposed to have its generic magnitude 
given by \eq{genmmag}
with $\minf^4\simeq V_0$,\footnote
{To be more correct, the relevant coefficients in the expansion
\eq{Kalerexp} are supposed to be of order 1 at the Planck scale,
in Planck units.
They run, which is equivalent to running the susy breaking parameters
even though the latter may really be defined only below a lower scale where
supersymmetry breaks.}
\be
|m^2(\mpl)| \sim V_0/\mpl^2\,.
\ee
The mass-squared is supposed to run strongly with scale, so that it 
becomes small which allows inflation to occur.

\subsubsection{Directions for model-building}

Although a complete model is far from being written down,
one can see some basic features that will be needed.

The complete potential might look roughly
like \eq{fullpot}. If the mass $m_\psi$ is also generated
by soft susy breaking, then as we
noted in Section \ref{lisamodel} $\psi$ would have a vev
of order $\mpl$; it might be 
something like
the dilaton or a bulk modulus, or a matter field with
non-renormalizable terms suppressed to high order by a discrete 
symmetry. On the other hand, $m_\psi$ might be bigger and come from some 
other mechanism, in which case $\psi$ could be a more ordinary field.

The quartic coupling of \eq{fullpot} could come from a term
$\sqrt{\lambda'}S\phi\psi$ in
the
superpotential, with $S$ some
field
that vanishes during inflation. The alternative coupling in
\eq{otherphipsi}, plus an identical term with $\phi\to\psi$
                                                           that we did 
not consider for simplicity, could come
\cite{lisa} from
a term $\phi^2\psi^2/\mpltil$ in the superpotential.

One will have to avoid the strong cancellation between the 
terms of \eq{minfdef}, that is present in the true vacuum.
In the case of gauge-mediated susy breaking in
the true vacuum, this might require an understanding of the origin
of the sector that generates the magnitude of $W$ in the true vacuum,
which is so far something of a puzzle. In the opposite case, the
situation maybe under better control, since one could use an
explicit model (such as the one of Reference \cite{bgw})
which already specifies all of the relevant quantities in the true
vacuum.

\subsubsection{Running with a gauge coupling}

Following \cite{ewanloop2,p98laura}, 
we calculate the running inflaton mass, on the 
assumption that the inflaton is charged under a gauge group
and that its Yukawa couplings have a negligible effect.

The RGE's have the same
form as the well-known
ones that describe the running of the squark masses with only QCD 
included, 
\bea
{d \alpha \over dt} &=& {b \over 2 \pi } \alpha^2 
\label{RGE-alpha}
\\
{d \over dt} \left( {\tilde m \over \alpha} \right) &=& 0 
\label{RGE-gaugino}
\\
{d m^2_{\phi} \over dt} &=& - {2 c \over \pi} \alpha \tilde m^2 
\label{RGE-inflaton}
\eea
Here 
$\alpha $ is related to the gauge coupling by
$\alpha = g^2/4\pi$, $\tilde m$ is the gaugino mass, and 
$t \equiv \ln (\phi/\mpl) < 0$. The numbers $b$ and $c$ depend on the group;
$c$ is the Casimir 
quadratic invariant of the inflaton representation under the gauge group,
for instance 
$c = (N^2-1)/2N $ for any fundamental representation of $SU(N)$,
and 
$b = - 3 N + N\sub{f} $ 
for a supersymmetric $SU(N)$ with $N\sub{f}$ pairs of fermions
in the fundamental/antifundamental representation.

The Renormalization Group Equations can easily be solved.
The result is 
\be
m^2(\phi) = m^2_0 + {2 c\over b}\tilde m^2_0
{ \left[ 1- {1 \over \left[ 1-{b\alpha_0\over 2\pi}
\ln(\phi/\mpl) \right]^2}\right] } \,.
\ee
Here $m_0$ is the inflaton mass,
$\tilde m_0$ is the gaugino mass, $\alpha_0$ is the gauge 
coupling, all evaluated at the Planck scale.

 We want the magnitude of $m^2$ to 
decrease as one goes down from the Planck scale.
This requires $m^2_0<0$, corresponding to model (i) or model (ii)
of Section \ref{hybrunning}.

We evaluate $c$, $\sigma$ and $\tau$ to leading order in $\alpha$,
which is presumably all that is justified in a one-loop calculation.
It is convenient to use the following definitions.
\bea
\mu^2 &\equiv& -m^2\mpl^2/V_0 \,,\\
A &\equiv & - {2 c\over b} {\tilde m^2\mpl^2\over V_0}
\label{param} \,,\\
\tilde \alpha &\equiv& {-b\alpha \over 2\pi } \,,\\
y&\equiv & \[1+\tilde \alpha_0 \ln(\phi/\mpl)\]^{-1} \,,\\
y_{**} &\equiv& \sqrt{1+{\mu^2_0\over A_0}} \,.
\label{ystar}
\eea
Applying the linear approximation one finds \cite{p98laura2}
\bea
c &=& 2 y_{**}^3 A_0 \tilde \alpha_0 
\label{i-c}\\
\tau &= & 2 A_0 y_{**}^2 (y_{**} -1 ) \,.
\label{i-tau}
\eea
If $m^2$ continues to run until the end of slow-roll inflation,
$\sigma$ is given by 
\be
\ln\sigma = 2 y_{**}^2 \( y_{**}^{-1} - y\sub{end}^{-1} \)
+\ln \[ \frac{4A_0 y_{**}^2 \vert y\sub{end} - y_{**} \vert }
{y\sub{end} + y_{**} } \] \,,
\ee
where $y\sub{end}=(y_{**}^2\pm A_0^{-1})$, with the plus
sign for model (i) and the minus sign for model (ii).

Using this result, one can calculate the COBE normalization, and the 
spectral index $n(N)$. There are four cases to consider, corresponding
to asymptotic freedom or not, and models (i) or (ii). Except for the 
case of model (ii) and no asymptotic freedom, there is
\cite{p98laura2}
a region of
parameter space that is allowed by the observational constraints
described in Section \ref{hybrunning}, and includes 
the theoretically favoured values
$\alpha_0\sim
10^{-1}$ to $10^{-2}$, $|\mu_0|\sim |A_0|
\sim 1$ and $10^5\GeV\lsim V^{1/4}\lsim 10^{10}\GeV$.

\subsection{A variant of the NMSSM}

\label{stevemod}

The model we just considered supposes that there is soft susy breaking
in the inflaton  sector, and that 
the relevant soft susy breaking parameters all have their 
natural values at the Planck scale. In particular, the inflaton mass
is supposed to satisfy $|m^2|\sim V_0/\mpl^2$ there.
We now consider a model \cite{steve1,steve2,steve3}
which also assumes soft susy breaking in the 
inflaton sector, with all relevant parameters {\em except the inflaton 
mass} at natural values (and actually negligible running).
But the inflaton mass is supposed to vanish at the Planck scale,
presumably because it occupies a special subsector
in which no-scale supergravity holds.
This last feature may be difficult to arrange, since
the model requires an accurate cancellation in \eq{minfdef}
and therefore a nonzero value for $W$ 
(see the remarks in
Section \ref{gmedstr}). (The specific proposal \cite{steve3} invokes
the weakly coupled string theory expressions of Section \ref{sugfstr},
but it requires a field with vanishing modular weight whereas one 
expects nonzero weights.)

In this model, the inflaton sector is actually (part of) the visible
sector, and it is assumed that gravity-mediated susy-breaking holds
with $\minf\simeq M\sub S$.

The model \cite{steve1,steve2,steve3}  
works with a variant of the 
next-to-minimal Standard Model
\cite{fayet,nsw,ds}. 
The relevant part of the
the superpotential is
\beq
W=\lambda N H\sub U H\sub D -k\phi N^2,
\label{nmssm}
\eeq
where $H\sub U$ and $H\sub D$ 
are the usual Higgs 
fields and $N$ and $\phi$ 
are two standard model  gauge singlet fields.

The actual next-to-minimal Standard Model is recovered if 
the last term of \eq{nmssm} becomes $-kN^3$, which leads to a $Z_3$
symmetry and possible
cosmological problems with domain walls.
 In the variant, the $Z_3$ becomes a global $U(1)$, which is in fact the
Peccei-Quinn symmetry commonly invoked to ensure the $CP$ invariance of 
the strong interaction. This symmetry
is spontaneously broken in the true vacuum because
$\phi$ and $N$ acquire vevs. The axion is the Pseudo-goldstone boson of this 
symmetry, and axion physics requires $\langle \phi\rangle\sim
\langle N\rangle \sim 10^{10}\GeV$ to $10^{13}\GeV$ or so.
(Higher values are allowed in some models, but not this one.)
The latter value is adopted to make the inflation model work.

The axion is practically massless, and by a choice of the axion field
one can make $\phi$ real.\footnote
{We take this real $\phi$ to be canonically normalized, which means
that the original complex $\phi$ is $\sqrt2$ times the 
canonically normalized object.}
It is going to be the inflaton, and during 
inflation $H\sub U H\sub D$ is negligible. Writing $\sqrt 2N=N_1 + i N_2$,
and including a soft susy breaking trilinear
term $2A_k k\phi N^2 + {\,\rm c.c}$
(with $A_k $ taken to be real) as well as soft susy breaking 
mass terms, the potential is
\beq
V=V_0+ k^2 |N|^4+ \frac12\sum_i m_i^2(\phi)N_i^2+
\frac12 m_\phi^2 \phi^2,
\eeq
where 
\bea
m_1^2(\phi) &=& m_1^2 -2 k A_k \phi + 4 k^2 \phi^2 \,, \label{n1m}\\
m_2^2(\phi) &=& m_2^2 + 2 k A_k \phi + 4 k^2 \phi^2 \,.
\label{n2m}\\
\eea
The parameters 
$m_i$, $A_k$ and $m_\phi$ are supposed to be generated by a 
gravity-mediated mechanism,
which is the same as in the true vacuum,
and it is supposed that the susy breaking scale is also the same.
This is supposed to give generic values $m_i\sim A_k\sim 1\TeV$
for all of these parameters except $m_\phi$.
The latter is supposed to vanish at the Planck scale, 
being generated by radiative corrections as described 
in a moment.

The constant term $V_0$ comes from some other sector of the theory,
and it is supposed to dominate 
the potential. 

The true vacuum corresponds to 
\bea
\langle \phi\rangle &=& \frac{A_k}{4k}, \\
\langle N_1\rangle &=& \frac{A_k}{2\sqrt 2 k}\sqrt
{1-4\frac{m_1^2}{A_k^2}} \\
\langle N_2\rangle &=& 0 \,.
\eea
We ignored the tiny effect of $m_\phi$ in working out the 
nonzero vevs. It is assumed that $4m_1^2$ is somewhat below
$A_k^2$, so that 
\be
A_k \sim k\langle N_1\rangle \sim k
\langle \phi\rangle \sim 1\TeV \,.
\ee
To have the vevs at the axion scale, say $10^{13}\GeV$,
we require $k\sim 10^{-10}$. Also, $\lambda$ should have a 
similar value, since $\lambda \langle N_1\rangle$ will be the 
$\mu$ parameter of the MSSM. 

The tiny couplings $k$ and $\lambda$ are supposed to be 
products of several
terms like $(\psi/\mpl)$ where $\psi$ is the vev of a field that is 
integrated out. The structure of such terms may be enforced through
discrete symmetries derived from string theory. The same terms can
ensure Peccei-Quinn symmetry to sufficient accuracy, without actually
invoking a global symmetry.
In the example given \cite{steve1}, $\phi$ is charged under 
a $Z_5$ as well as the $Z_3$ already encountered, which forbids
terms $\phi^d$ up to $d=15$ in the superpotential. 
One must in any case 
forbid them up to $d\simeq 8$, to satisfy 
the constraint \eq{f3}.

During inflation, the fields $N_i$ are trapped at the origin,
and 
\be
V=V_0 +\frac12m_\phi^2\phi^2 \,.
\ee
The field $N_1$ is destabilized if $\phi$ lies between the values
\be
\phi\sub c^{\pm} = \frac{A_k}{4k}\( 1\pm \sqrt{1-4\frac{m_N^2}{A_k^2} }
\)
\sim A_k/k
\,.
\ee

If $m_\phi^2$ is positive the model gives ordinary hybrid inflation
ending at $\phi\sub c^+$, but if it is negative it gives inverted 
hybrid inflation ending at $\phi\sub c^-$. We shall see that the 
radiative corrections actually give the latter case.
The height of the potential is 
$V_0^{1/4}\sim A_k/\sqrt k \sim 10^8\GeV$.
The COBE normalization, \eq{vofn} or \eq{cobenormhyb2}, is therefore
\be
A_k = 2.5\times 10^{-4} | n-1 | e^{\pm x} \mpl \,,
\ee
where $x\equiv \frac12|n-1|N$. This requires $n$ to be completely 
indistinguishable from 1, $|n-1|\sim 10^{-12}$. The corresponding 
inflaton mass, given by $|n-1|=2m_\phi^2 V_0/\mpl^4$ is
\be
m_\phi\sim 100\(\frac {A_k}{\mpl} \)^{2.5} \mpl
\sim \eV \,.
\ee

The loop correction generating $m_\phi$ comes from the
$N_i$, and their
fermionic partner which has mass-squared $4k^2\phi^2$. 
In the regime $\phi\gg A_k/k (\sim m_i/k)$, one finds
\cite{steve1}  using
\eqss{n1m}{n2m}{vdval} 
\be
\Delta V \simeq \frac{k^2 A_k^2}{4\pi^2}\phi^2 \ln(\phi/Q) \,,
\label{delv}
\ee
where $Q$ is the renormalization scale. The requirement
$\pa V/\pa Q=0$ at $\phi\sim Q$ gives, in the regime
$Q\gg A_k/k$, the RGE \eq{mrge},
\be
\frac{dm_\phi^2}{d\ln Q} = \frac{k^2A_k^2}{2\pi^2} \,.
\ee
The running of $kA_k$ is negligible because $k$ is small,
and setting $m_\phi=0$ at $Q=\mpl$ we obtain
\be
m^2_\phi(Q) = - \frac{k^2 A_k^2 \ln(\mpl/Q)}{2\pi^2}
\sim - k^2 A_k^2 \,.
\label{m2ofq}
\ee
    
Strictly speaking the derivation of this result holds only
for $Q\gg A_k/k$.
Inflation occurs at $\phi\sim\phi\sub c^\pm\sim A_k/k$,
and in this regime we should take $Q\sim A_k/k$ to minimize the loop 
correction.\footnote
{The argument of the log in the loop correction is actually
$2k\phi/Q$, so one might argue that the appropriate scale is
$Q\sim k\phi\sub c\sim A_k$ which is much lower.
But the effect of including $k$ here
is the same order of magnitude as the effect of 
including the two-loop correction, and is presumably negligible. This
is because making $k$ small also makes the running slower.}
Thus we are somewhat below the regime where the result is valid, but it
hopefully gives a rough approximation. If
so
we are dealing with an inverted hybrid inflation model. 
Somewhat remarkably, the
magnitude $m_\phi\sim kA_k$ agrees with the COBE normalization
$m_\phi\sim \eV$, within the uncertainties of $A_k$ and $k$.

Finally, we note that because the running of the inflaton mass is weak,
its use is optional; instead of using it, we
could set $m_\phi=0$, and generate the slope of the inflaton potential
from the loop correction \eq{delv} with $Q=\mpl$.

\section{$D$-term inflation}

\label{s19}

$D$-term inflation
can preserve the flat directions of global susy
(and in particular keep the inflaton potential flat)
provided that
one of the contributions to 
$V_D$ contains a 
Fayet-Iliopoulos term as in \eq{fivd}, and that all 
fields charged under the Fayet-Iliopoulos $U(1)$
are driven to negligible values so that $V=(g^2/2) \xi^2$.
This was first pointed out by Stewart \cite{ewansgrav},
who exhibited a hybrid inflation model which uses the $F$ term to
drive all of the charged fields to zero.\footnote
{A slightly different version of the model,
which actually was the main focus of his paper, gives
the $F$-term inflation model mentioned in Section
\ref{fromd}.
A single-field model of inflation with a Fayet-Iliopoulos 
$D$-term, and the inflaton 
charged under the relevant $U(1)$, had been considered earlier
\cite{casas,casas1}. It gives the inverted quadratic potential 
considered in Section \ref{invq}, and is viable only under the unlikely
assumption that the inflaton charge is $\ll 1$.
In any case it does not preserve the flat directions of global
susy.} 
He considered only the 
tree-level potential without any definite proposal for its
slope. Significant progress came when Bin\'etruy and Dvali \cite{bindvali} and 
Halyo \cite{halyo} 
pointed out, 
in the context of a somewhat simpler tree-level potential,
that the loop correction gives a 
well-defined slope. 
This lead to an explosion of interest in $D$-term inflation
\cite{j2,dvaliriotto,casas2,lr,mydterm,rtalk,km,err,bdr}.

\subsection{Keeping the potential flat}

\label{kflatd}

We initially
make the usual assumption that the fields charged under the $U(1)$
vanish exactly. Then
\be
V_D = \frac 12 ({\rm Re}\, f)^{-1} g^2 \xi^2 \,.
\ee
In this tree-level potential, the
only dependence of $V_D$ on the fields comes from 
the gauge kinetic function $f$.\footnote
{We are here taking $g$ to be a constant, and $f=1$ at the origin as in 
\eq{gkin}. 
Later we adopt the 
convention that $({\rm Re} f)^{-1}$ is absorbed into $g^2$, making 
the latter a function of the fields.}
It has non-renormalizable terms, and so does $W$ that appears 
in the supposedly negligible $F$-term. 
If $|\phi|\ll \mpl$, only low-dimensional terms are dangerous,
and they can be eliminated using a suitable discrete symmetry
\cite{mydterm,km}. Unfortunately, we shall find that
$|\phi|$ is of order $\mpl$, and maybe bigger. This
makes the non-renormalizable terms difficult to control
\cite{km}, as well as those of $K$. (As well as directly 
affecting the potential,
the latter can give a non-trivial kinetic 
term, which alters the potential after going to a 
canonically-normalized field).
We proceed on the assumption that non-renormalizable terms 
of $W$ and $f$ turn
out to be negligible; in particular we assume $f=1$.

With $W$ is under control, and the fields charged under the 
$U(1)$ exactly zero, the K\"ahler potential $K$ can have no effect,
and the coefficients $\lambda$ and $\lambda_d$ can be much
smaller than $V_0/\mpl^4$. This will be crucial, in view of the
fact that the inflaton field is of order $\mpl$.

In some versions of $D$ term inflation the charged fields are not
driven to zero. If their contribution to $V_D$ is a significant fraction 
of the total, the terms $K_n$ in \eq{sgdterm} (and similar ones from
$D$ terms involving other gauge groups under which they are charged)
will
generically spoil inflation \cite{bdr}. 
The conclusion seems to be that the charged fields should be driven
to sufficiently small values, even if they do not vanish. 

Finally, let us mention  that, if the Fayet-Iliopoulos term is to come 
from string theory, see Eq. (\ref{xi}), the corresponding $D$-term scales 
like $g\sub{str}^6\propto
({\rm Re}\,s)^{-3}$. The problem here  is that, assuming that the $D$-term  
dominates over any other $F$-term,
the potential during inflation appears to prefer 
${\rm Re}\,s\rightarrow \infty$ and therefore  $V_D\rightarrow 0$. 
This is the $D$-term inflation equivalent of the
dilaton runaway problem that appears in string theories in the true vacuum. 
However, it has been argued \cite{kr} that 
 the physics of gaugino condensation in ten-dimensional 
$E_8\otimes E_8$ superstring theories is likely to be modified during the 
inflationary phase in such a way as to enhance the gaugino condensation scale. 
This may allow the dilaton to be stabilized by the $F$ term
\cite{kr}, though one 
has to check that the latter does not generate dangerous supergravity
corrections to the inflaton potential.

\subsection{The basic model}

\label{simplest}

At least one of the charged fields should have
negative charge $q_n$, so that the $D$ term is driven to
zero in the vacuum (or at least to a value much smaller than
$(g^2/2)\xi^2$, as in Section \ref{sss:dterm}).
One has to give such negatively-charged fields couplings which
drive them to small values during $D$-term inflation. 
The proposal of refs. \cite{ewansgrav,bindvali,halyo}
is that every negatively-charged
field has a partner with all charges opposite. It can then couple
to the inflaton in the $F$ term, and acquire a positive mass-squared
during inflation.

Suppose for simplicity that there is just one pair, $\phi_\pm$.
The inflaton is supposed to be the radial part of an
uncharged field $S$ ($\phi=\sqrt2 |S|$).
There is a term in the superpotential 
\beq
W = \lambda S\phi_+\phi_-.
\end{equation}
Since $\phi_\pm$ are going to be driven to zero, it will be
enough to use the global susy expression for the $D$-term, giving
\begin{equation}
V = \frac12 \lambda^2 \phi^2 \left(|\phi_-|^2 + |\phi_+|^2 \right) +
\lambda^2|\phi_+\phi_-|^2 + 
{g^2 \over 2} \left(|\phi_+|^2 - |\phi_-|^2  + \xi \right)^2 \,.
\end{equation}
The global minimum is supersymmetry conserving, but
the gauge group $U(1)$ is spontaneously broken
\bea
\langle \phi  \rangle  &=& \langle \phi_+  \rangle = 0, 
\label{z} \\
\langle\phi_- \rangle &=& \xi
\eea
However, if we minimize the potential, for  fixed values of $
\phi$, 
with respect to
other fields, we find that for  $\phi$ bigger than
\be
\phi\sub{c} \equiv {g \over \lambda} \sqrt{2\xi} 
\label{phicd}
\ee
the minimum is at $\phi_+ =\phi_- = 0$. Thus, for
$\phi>\phi\sub{c}$ and $\phi_+ =\phi_- = 0$ the tree level potential
has a vanishing curvature in the $\phi$ direction and large positive
curvature in the remaining two directions 
\be m_{\pm}^2 = \frac12\lambda^2\phi^2 \pm g^2\xi .
\label{mpmsq}
\ee
For $\phi>\phi\sub{c}$, the tree level value of the potential 
has the constant value $V = {g^2 \over 2}\xi^2$.
%As stated above, the charged fields get very large masses
%due to the $D$-term supersymmetry breaking,
%whereas the gauge singlet field is massless at the tree-level. 

This is a hybrid inflation model. At tree level, the potential
$V(\phi)$ is perfectly flat, and its $\phi$ dependence comes {}from the loop 
correction. Supersymmetry is spontaneously broken,
and inserting \eq{mpmsq} into
\eq{vdval} gives 
\begin{equation}
V=
V_{{\rm 1-loop}} \equiv
 {g^2 \over 2}\xi^2 \left( 1 + {g^2 \over 16\pi^2} {\rm ln}
{\lambda^2 \phi^2 \over 2\mu^2}\right),
\end{equation}
where $\mu$ is the renormalization scale. 

We can generalize the model by including more than one pair of 
fields $\phi_{n\pm}$, with charges $q_n$ 
and superpotential couplings $\lambda_n$.
Then the one-loop potential becomes 
\be
V=
V_{{\rm 1-loop}} \equiv
 {g^2 \over 2}\xi^2 \left( 1 + C{g^2 \over 16\pi^2} {\rm ln}
{\lambda^2 \phi^2 \over 2 \mu^2}\right) .
\label{vloop}
\ee
where 
\be
C=\frac12\sum_n q_n^2. \label{cdef}
\ee
(In the log we took all $\lambda_n$ to have a common value $\lambda$
but this is not essential since $\lambda$ does not affect the 
slope of the potential.)

Since the $U(1)$ generated by string theory is anomalous,
corresponding to $\sum q_n\neq 0$, there have to be some 
unpaired charges. If they are positive, they will be driven to zero
and be irrelevant, and that is assumed in the paradigm under 
consideration. However, in weakly coupled string theory 
one actually expects unpaired negative charges  which might ruin 
this paradigm \cite{err}. That case is discussed in Section
 \ref{strmodel}.

Inflation with this potential was discussed in Section \ref{smallloop}.
As noted there, slow-roll
inflation will end when $\phi\sub c$ is reached, or when
it gives way to fast roll, whichever is sooner. 
In the latter case, fast roll begins 
when
$\eta\sim1$, at
\be
\phi\sub{fr} =\sqrt\frac{Cg^2}{8\pi^2}\mpl \,.
\ee
This is about the same as 
$\phi\sub c$ given by \eq{phicd}, so it depends on the parameters which
happens first. If fast roll begins first inflation will continue for 
an $e$-fold or so, ending when the oscillation amplitude falls below 
$\phi\sub c$. 

According to \eq{vv}, $\phi$ is
comparable with the Planck scale, and maybe bigger.
If we increase the slope of $V$, by assuming that a tree-level contribution 
dominates the loop correction \cite{lr}, this will increase 
$\phi$ (see \eq{nint}). 
The only hope of reducing it would be a 
cancellation between the loop and a tree-level contribution, which seems
unlikely over a range of $\phi$.
As mentioned earlier, the large value of $\phi$ means that 
non-renormalizable terms in the potential and the kinetic function
are not under good control, but we proceed on the assumption that they
turn out to be harmless.

The COBE normalization is
\be
\sqrt\xi = 8.5\times 10^{15}\GeV \(\frac{50C}{N} \)^{1/4} 
\ee

The scale impose by COBE  is clearly lower than the prediction 
\eq{xi} of weakly coupled string theory, which is a second worry
for the model. Indeed, \eq{xi} requires
\be
g\sub{str}^2 = \frac{192\pi^2}{\Tr {\rm\bf Q}} \(\frac{50C}{N}\)^{1/2}
\(\frac{5.9\times 10^{15}}{2.4\times 10^{18}} \)^2 
\lsim 10^{-6}\,.
\ee
Such a value is unreasonable, since
the dilaton during inflation would
presumably have to be far away from the true vacuum value,
placing it outside the domain of attraction of that value.

How can we get around this problem? The most obvious 
possibility is that weakly coupled string theory is replaced by
something else, such as Horava-Witten M-theory, which might give a lower value 
for $\xi$. 
At the time of writing, it is not clear 
whether this is an open  possibility or not \cite{jmrfa,bifa}.  

Another possibility is to make $\xi$ lower by decoupling its
origin {}from string theories. But to avoid putting it in by hand,
one should generate it
in some  low-energy effective theory after some degrees of 
freedom have been integrated out. But to do this,
one has presumably to break supersymmetry by some $F$-terms present in the 
sector which the heavy fields  belong to and to generate the $
D$-term by loop corrections. As a result,  it turns out that    
$\langle D\rangle\ll \langle F^2\rangle$, unless some fine-tuning 
is called for, and large supergravity corrections to $\eta$ appear again. 
Let us give an example. Consider 
the following superpotential 
where a $U(1)$ symmetry has been imposed \cite{rrr}
\begin{equation}
W=\lambda X \left({\bar \Phi}_1 \Phi_1 - m^2 \right)
+M_1 {\bar \Phi}_1 \Phi_2 + M_2 {\bar \Phi}_2 \Phi_1 ~.
\label{orafeq}
\end{equation}
 For $\lambda^2 m^2 \ll M_1^2,M_2^2$, the vacuum of this model is such
that $\langle \phi_i \rangle =\langle {\bar \phi}_i \rangle =0$
($i=1,2$), where  $\bar{\phi}_i$ and $\phi_i$ are the scalar 
components of the superfields $\bar{\Phi}_i$ and $\Phi_i$,  respectively. 
Supersymmetry is broken and $F_X=-\lambda^2 m^2$. This means that in the 
potential a term like $V=(F_X{\bar \phi}_1 \phi
_1+{\rm h.c.})$ will appear. It is easy to show that, integrating out the 
$\phi_{i}$ and $\bar{\phi}_i$ scalar fields, induces a a nonvanishing 
Fayet-Iliopoulos $D$-term 
\begin{equation}
\xi\simeq \frac{F_X^2}{16\pi^2(M_1^2-M_2^2)}\ln
\left(\frac{M_2^2}{M_1^2}\right),
\end{equation}
which is, however,  smaller than $F_X$ and inflation, 
if any, is presumably dominated by the $F$-term.   

Staying with the high value of $\xi$, one might consider increasing the 
COBE normalization by 
supposing that the 
slope of the potential
is bigger than the 
loop contribution. For instance, it might come from a term $\frac12m^2\phi^2$,
generated by the $F$ term \label{ourd} \cite{lr}. In general one has
\be
g\sub{str}^2 =1.9 \(\frac{100}{{\rm Tr}\,{\bf  Q}} \)^{1/3} 
\(\frac{V^{1/4}}{\mpl} \)^{4/3} \,. \label{gstrofv}
\ee
But even with the maximum allowed value $V^{1/4}\sim 10^{-2}\mpl$,
$g^2\sub{str}$ is still unreasonably low.
This is a second problem for $D$-term inflation, though unlike
the large-field problem it depends on details of the underlying
string theory.

\subsection{Constructing a workable model {}from string theory}

\label{strmodel}

The  presence of the Fayet-Iliopoulos $D$-term (\ref{xi}) in 
weakly coupled string theory leads
to the breaking of
supersymmetry at the one-loop order at very high scale, the string scale.  
This option  
is  generically not  welcome from the phenomenological point of view  because 
it  induces too large soft susy breaking masses via gravity effects, 
$\widetilde{m}\sim\xi/\mpl$. The 
standard solution to this puzzle
is to give a nonvanishing vev to some of
the scalar fields which are present in the string model and are 
negatively charged
under the anomalous $U(1)$. In such a way, the Fayet-Iliopoulos $D$-term is
cancelled and supersymmetry is preserved. In the context of string
theory, this procedure is called ``vacuum shifting'' since it amounts
to moving to a point where the string ground state is 
stable\footnote{Notice, however, that if  the  vacuum shifting 
in the true vacuum is not 
complete, because of the presence of some non-vanishing $F$ terms, it  may 
give rise to interesting phenomenological implications. This is what happens 
for the model described in Section \ref{sss:dterm}.}
While
maintaining the $D$- and $F$-flatness of the effective field theory,
such vacuum shifting may have important consequences for the
phenomenology of the string theory. Indeed, the vacuum shifting not
only breaks the $U(1)$, but may also break some other gauge symmetries
under which the fields which acquire a vev are charged. This is because
the anomalous $U(1)$ is usually accompanied by a plethora of
nonanomalous $U(1)$'s.

In the true vacuum, the 
vacuum shifting can generate effective superpotential mass terms
for vector-like\footnote
{A vector-like set of fields is one having zero total charge.} states 
that would otherwise remain massless or may even
be responsible for the soft mass terms of squarks and sleptons at the
TeV scale. 

In string theories the
protection of supersymmetry against the effects of the anomalous
$U(1)$ is extremely efficient.  If we now apply a sort of ``minimal
principle'' \cite{dr,dvaliriotto} requiring that a successful scenario
of $D$-term inflation should arise {}from ``realistic'' string models
leading to the $SU(3)_C\otimes SU(2)_L\otimes U(1)_Y$ gauge structure at
low energies, the cancellation of the  Fayet-Iliopoulos $D$-term 
by the vacuum shifting
mechanism may represent (and usually does)  a serious problem. 
Indeed, one has to make sure that  during inflation the Fayet-Iliopoulos 
$D$-term is
not cancelled by one of the many scalar fields which are negatively
charged under
the anomalous $U(1)$ and are not coupled to 
the inflaton. This usually leads to the conclusion that   a
successful $D$-term inflationary scenario in string theory require many
inflatons to render the vacuum shifting mechanism inoperative and  it is
clear that only a systematic analysis of flat directions in any
specific model may answer these and similar questions. This requires the
identification of possible inflatons and $D$- and $F$-flat
directions for a large class of perturbative string vacua.  This
classification is a prerequisite to address systematically the issue of
%$D$-term 
inflation in string theories as well as the phenomenological
issues at low energy \cite{jose,florida}. 

As an illustrative example of the possible complications one
has to face in building up a successful
 model of $D$-term inflation  in the framework of  4D string models \cite{err},
one may  consider the massless spectrum of a compactification on a Calabi-Yau
manifold with Hodge numbers $h_{1,1}$, $h_{2,1}$,  etc. The 
four-dimensional
gauge group is $SO(26)\times U(1).$ There are then $h_{1,1}$ 
left-handed
chiral supermultiplets transforming as $({\bf 26},\surd
\frac{1}{3}$)$\oplus ({\bf 1 },-2\surd \frac{1}{3})$ and $h_{2,1}$
supermultiplets transforming as $({\bf 26},-\surd \frac{1}{3}$)$\oplus
({\bf 1},2\surd \frac{1}{3}).$ In this case
the $U(1)$ is 
anomalous because $h_{1,1}$ and $h_{2,1}$ are not equal. 
Indeed, suppose that  $h_{1,1}-h_{2,1}>0$. In such  a case, out of the total 
$h_{1,1}+h_{2,1}$ chiral supermultiplets, 
there are only $2\:h_{2,1}$ left-handed chiral supermultiplets which may form 
$h_{2,1}$  vector-like pairs under the $U(1)$ and give a vanishing 
contribution to ${\rm Tr}\:{\bf Q}$. The remaining 
$h_{1,1}+h_{2,1}-2\:h_{2,1}=h_{1,1}-h_{2,1}$ fields will give
a nonvanishing contribution to the Fayet-Iliopoulos
$D$--term.

 Taking into account the multiplicity of the fields, the 
one-loop $D$-term (\ref{xi})  is therefore given by
\begin{equation}
\label{jj}
\xi =
\frac{g\sub{str}^2 \Mpl^2}{192\pi
^{2}}\cdot 2 \cdot
\frac{24}{\surd 3}(h_{1,1}-h_{2,1}).
\end{equation} 
We suppose 
the
model has a gauge singlet field $S$ which will play the role of the
inflaton. Further we assume that there is a 
discrete $R$-symmetry that 
ensures
$S$-flatness. These assumptions are quite ad hoc and in a realistic 
model we
would have to demonstrate the existence of such a field,  but we use this
simple example to illustrate another problem that must be overcome if 
one is
to obtain a realistic string model of $D$-term inflation.

With this field one may try to construct an inflationary potential. 
Gauge symmetries and the fact that $h_{1,1}-h_{2,1}>0$ impose that  
one can  generate masses only for the 
$h_{2,1}$
vectorlike combinations of the $SO(26)$ singlet and non-singlet fields 
via
the couplings in the superpotential of the form
\begin{equation}
W=\lambda S\left[({\bf 26},\surd \frac{1}{3})\cdot({\bf 26},-\surd
\frac{1}{3})+({\bf
1},-2\surd \frac{1}{3})\cdot({\bf 1},2\surd \frac{1}{3})\right].  
\label{mass}
\end{equation}
Therefore only $2\:h_{2,1}$ fields get a mass $\lambda\langle S\rangle$ and  
become very  massive during inflation. This means that they decouple from the 
theory and do not contribute anymore to the Fayet-Iliopoulos term (\ref{jj}). 
On the other hand, 
the remaining 
 $(h_{1,1}-h_{2,1})$ fields transforming as $({\bf 26}
,\surd \frac{1}{3}$)$\oplus ({\bf 1},-2\surd \frac{1}{3})$ 
remain light because the cannot couple  to the inflaton and give a 
contribution to (\ref{jj}), which remains, therefore, 
unchanged.  
The $(h_{1,1}-h_{2,1})$ $SO(26)$ singlet fields with $U(1)$ 
charge $-2\sqrt{1/3}$, let us denote them by  $\phi
_{i},$ are now available to cancel the anomalous $D$-term 
because :$\sum_{i}$ $Q_{i}|\phi 
_{i}|^{2}<0$, as is expected if supersymmetry is not to be broken by the 
Fayet-Iliopoulos $D$-term.\footnote
{In this section the charges are represented by upper case letters.}
However
this prevents one from implementing $D$-term inflation because the 
scalar
potential dependence on the $\phi _{i}$ fields arises only through the
anomalous $D$-term. 
The vacuum expectation values  of the fields $\phi _{i}$ will rapidly
flow to cancel the $D$-term preventing inflation from occurring.

This example illustrates the problem in implementing $D$-term inflation 
in a
string theory. It arises because the minimum of the potential 
does not generically 
break supersymmetry through the anomalous $D$-term and so there must be
light fields (here the $\phi _{i})$ with the appropriate $U(1)$ charge 
to
cancel it. To implement $D$-term inflation these fields must acquire a 
mass
for large values of $S$ but this was not possible in this example 
because
the $\phi _{i}$ were protected by chirality from acquiring mass by 
coupling to
the $S$ field. 

Thus we conclude that it is crucial to consider {\it all} fields with
non-trivial $U(1)$ quantum numbers when discussing the possible
inflationary potential in the framework of string theories.

We will consider now further examples to capture  other possible
aspects of  $D$-term inflation in string theories \cite{err}. For 
illustrative purposes, 
we will use the specific string models, discussed in \cite{CHL,AF1} 
whose space of
flat directions was recently analyzed in \cite{jose}. The 
emphasis will
be on exploring the different possibilities that may be realized rather 
than
proposing  a working model of inflation.  In so doing we will often
restrict the analysis to some subset of the fields present in the model
and ignore the rest. In view of what we concluded above, this is not
consistent, but the examples that follow should only be considered
as toy models attempting to capture some of the stringy characteristics
one should expect  when trying to construct  a fully realistic model of 
$D$-term inflation
in string inspired scenarios. 

The presence of several (non-anomalous) additional $U(1)$ factors is a
generic property of string models.  For the discussion of $D$-term
inflation, the relevant objects are thus no longer single elementary 
fields but
rather multiple-field directions in field space along which the $D$-term 
potential of
the non-anomalous $U(1)$'s vanishes \cite{jose2}. These directions would
 be truly 
flat if
an anomalous $U(1)_A$ (or some $F$-terms) were not present.  To study
whether a given direction remains flat in the
presence of the anomalous $U(1)_A$, the important quantity is the 
anomalous
charge $Q_A$ along the direction. If the sign of this charge is opposite to
that of the Fayet-Iliopoulos term, VEVs along the flat direction will
adjust themselves to cancel the Fayet-Iliopoulos $D$-term and 
give a zero potential.  If 
the
charge has the same sign of the Fayet-Iliopoulos $D$-term, the 
potential along that
direction rises steeply with increasing values of the field. The
interesting case
corresponds to zero anomalous charge, in which case the potential
along the given direction is flat and equals, at tree level,
$g_A^2\xi^2/2$. 
In that case, the direction can be the inflaton.

The condition $Q_A=0$, ensuring tree-level flatness of the inflaton
potential, 
is not by itself
sufficient. We must also require that the direction is stable for large
values of the inflaton, that is, all non-inflaton
masses deep in the inflaton direction
must be  positive (or zero). However the Fayet-Iliopoulos $D$-term in the
scalar potential will give a negative contribution to the masses-squared
of those fields which have
a negative anomalous charge:
\be
\label{massxi}
\delta m_\phi^2 = g_A^2Q^A_i\xi.
\ee
To ensure that masses are positive in the end one can use $F$-term
contributions (to balance the negative FI-induced masses) coming from
superpotential terms of the generic form
\be
\label{wf}
\delta W=\lambda I'\Phi_+\Phi_-,
\ee
where $I'$ stands for some product of fields that enter the inflaton
direction while $\Phi_\pm$ do not. Fields of type $\Phi_+$ and
$\Phi_-$ which couple to the inflaton
direction in the  superpotential terms get a large $F$-term
mass, $ \lambda \langle I' \rangle$.

Consider the simplest example, a toy model with two chiral fields $S_1$ 
and
$S_2$ of opposite $U(1)$ charges, so that the direction 
$|S|=|S_1|=|S_2|$ can
play the role of the inflaton. Assume that deep in this direction 
($S\gg
\sqrt{\xi}$) the masses of all 
other
fields are positive (or zero) and thus 
no other VEVs are
triggered. Then we can minimize the $D$-term scalar 
potential\footnote{In writing
this potential we are assuming for simplicity that kinetic mixing of 
different
$U(1)$'s is absent. For this to be a consistent assumption the 
vanishing of
${\rm Tr}(Q_A Q_\alpha)$ and ${\rm Tr}(Q_\alpha Q_\beta)$ is a necessary
condition.}
\bea
V_D&=\frac{1}{2}g_A^2\left[ Q_1^A\left(|S_1|^2-|S_2|^2\right)+
\sum_iQ_i^A|\phi_i|^2+\xi\right]^2\nonumber\\
&+\frac{1}{2}\sum_\alpha g_\alpha^2\left[ 
Q_1^\alpha\left(|S_1|^2-|S_2|^2\right)+
\sum_iQ_i^\alpha|\phi_i|^2\right]^2,
\eea
[where $\alpha=1,...,n$ counts the additional D-term contributions of 
the non-anomalous $U(1)$'s] for
$S_1$ and $S_2$ only.

If $\xi=0$, $|S_1|=|S_2|$ is flat and necessarily stable, as 
$V=0$. For $\xi>0$
however, the flat direction is slightly displaced and lies at
\be
\delta S^2\equiv |S_1|^2-|S_2|^2=-\frac{g_A^2}{G_{11}^2}Q_1^A\xi,
\label{s22}
\ee
where $G_{ij}^2=g_A^2Q_i^AQ_j^A+\sum_\alpha
g_\alpha^2 Q_i^\alpha Q_j^\alpha$. This displacement is the result of 
the
destabilization effect of $\xi$ referred to above and occurs  when the
fields in
the inflaton direction carry anomalous charge: as the inflaton 
direction must have
zero
anomalous charge, the fields forming it have anomalous charges of 
opposite signs
and one of them will get a negative mass of the form (\ref{massxi}).

Taking into account this displacement, the value of the potential along 
the
inflaton direction is, at tree level
\be
V_0=\frac{1}{2}\frac{g_A^2}{G_{11}^2}\xi^2\sum_\alpha 
g_\alpha^2(Q_1^\alpha)^2
\equiv \frac{1}{2}g_A^2\xi\sub{eff}^2\leq\frac{1}{2}g_A^2\xi^2.
\ee
As noted in Section \ref{kflatd}, $\xi\sub{eff}$ should be
very close to $\xi$ in order not to spoil inflation.

For a viable inflationary model we should ensure that the one-loop
potential is appropriate to give a slow roll along the inflaton 
direction.
Thus,  we must consider the one-loop corrections proportional to the
Yukawa couplings introduced in the terms of eq.~(\ref{wf}).
The field-dependent masses for the scalar components of the chiral 
fields
$\Phi_\pm$ along the inflaton direction are
\bea
m^2_\pm&=\lambda^2\langle I'\rangle^2 + g_A^2 Q^A_\pm(Q_1^A\delta S^2
+\xi)+\sum_\alpha g_\alpha^2Q^\alpha_\pm Q_1^\alpha \delta
S^2\nonumber\\
&=\lambda^2\langle I'\rangle^2 +G_{1\pm}^2\delta S^2+g_A^2Q^A_\pm\xi
\equiv \lambda^2\langle I'\rangle^2 + g_A^2a_\pm \xi,
\eea
while the fermionic partners have masses-squared equal to 
$\lambda^2\langle
I'\rangle^2$.
For large values of the field $\langle I'\rangle$, the one-loop
potential takes the form
\be
32\pi^2\delta V_1=2g_A^2(a_++a_-)\lambda^2\langle 
I'\rangle^2\xi\left(
\log\frac{\lambda^2\langle
I'\rangle^2}{Q^2}-1\right)
+g_A^4(a_+^2+a_-^2)\xi^2\log\frac{\lambda^2\langle
I'\rangle^2}{Q^2}.
\label{v1}
\ee
In this more complicated model the
scalar direction transverse to the
inflaton gains a very large mass deep in the inflaton direction. In
addition, the gauge boson corresponding to the broken $U(1)$ symmetry 
and
one  neutralino also become massive. These fields arrange
themselves in a massive vector supermultiplet, degenerate even if
$\xi\neq 0$, and their contribution to the one-loop
potential along the inflaton direction cancel exactly. The potential of 
Eq. (\ref{v1})
can be also rewritten as a RG-improved\footnote{In doing so, a careful 
treatment
of the possibility of kinetic mixing of different $U(1)$'s is required. 
The
details of our analysis are modified in the presence of such mixing but 
the
generic results are not changed.} 
tree-level potential with gauge
couplings evaluated at the scale $\lambda\langle I'\rangle$.

The term quadratic in $\lambda\langle I'\rangle$ would spoil the 
slow-roll
condition necessary for a successful inflation, but it drops out 
because
\bea
g_A^2(a_++a_-)&=&(G_{1+}^2+G_{1-}^2)\delta S^2 + g_A^2
(Q_+^A+Q_-^A)\xi\nonumber\\
&=&-G_{1I'}^2\delta S^2 - g_A^2Q^A_{I'}\xi\propto 
G_{11}^2\delta S^2 + g_A^2Q^A_1\xi=0,
\eea
where we have made use of the $U(1)$ invariance of $I'\Phi_+\Phi_-$ to 
write
the third expression which vanishes by Eq. (\ref{s22}).

The results just described for the simplest inflaton direction  
containing
more than one field are generalizable to more complicated inflatons. 
One could
have inflatons containing more than two elementary fields while still
having only a one-dimensional flat direction. Another possibility is 
that
the flat direction has  more than one free VEV (multidimensional
inflatons). It is straightforward to verify that the results obtained 
above for two mirror
fields are generic provided the inflaton does not contain some 
subdirection
capable of compensating the Fayet-Iliopoulos $D$-term.

As the next step in complexity one can  examine the case in which, 
besides
the inflaton VEVs $|S_1|$ and $|S_2|$, some other field $\varphi_i$ is
forced to take a VEV (this can be triggered by $\xi$ in the anomalous
$D$-term of the potential or by $\delta S^2$ in any $D$-term). In 
general,
the new VEV can induce further VEVs too. For simplicity, we assume that
this chain of destabilizations ends with $\langle\varphi_i\rangle$.
By minimizing the $D$-term potential, all VEVs are determined to be
\be
\delta S^2 = |S_1|^2-|S_2|^2=-{\displaystyle \frac{g_A^2}{{\rm det}
G^2}}(G_{ii}^2
Q_1^A-G_{1i}^2Q_i^A)\xi\nonumber
\ee
\be
\langle\varphi_i^2\rangle = -{\displaystyle \frac{g_A^2}{{\rm det}
G^2}}(
-G_{1i}^2Q_1^A+G_{11}^2Q_i^A)\xi ,
\ee
with ${\rm det}$ $G^2=G_{11}^2G_{ii}^2-G_{1i}^4$. The tree level
potential along this direction is
\be
V_0=\frac{1}{2}g_A^2{\displaystyle \frac{\xi^2}{{\rm det}
{}G^2}}\sum_{\alpha,\beta}
g_\alpha^2g_\beta^2Q_1^\alpha
Q_i^\beta(Q_1^\alpha
Q_i^\beta-Q_1^\beta
Q_i^\alpha)
\leq \frac{1}{2}g_A^2\xi^2.
\ee
In this background, the masses of the scalar components of $\Phi_\pm$
appearing in the superpotential (\ref{wf}) are
\be
m^2_\pm=\lambda^2\langle I'\rangle^2 + g_A^2 Q^A_\pm \langle 
D_A\rangle
+\sum_\alpha g_\alpha^2Q^\alpha_\pm \langle D_\alpha \rangle =
\lambda^2\langle I'\rangle^2 + g_A^2a_\pm \xi,
\ee
and again, one finds $a_++a_-=0$.

To illustrate the above discussion, consider the following example of a
string model \cite{AF1} that satisfies the conditions required for
$D$-term inflation, at least when we restrict the analysis to a subset 
of
the fields. The $U(1)$ charges of these fields are listed in the
table (we
follow the notation of ref.~\cite{jose} 
with charges rescaled). For every listed field 
$S_i$, a
"mirror" field $\overline{S}_i$ exists with opposite charges. At 
trilinear order the
superpotential is
\be
\label{sup}
W=\overline{S}_{11}(S_5\overline{S}_8+S_6\overline{S}_9
+S_7\overline{S}_{10}+S_{12}S_{13})
+S_{11}(\overline{S}_5S_8+\overline{S}_6S_9+\overline{S}_7S_{10}
+\overline{S}_{12}\overline{S}_{13}).
\ee

\begin{center}
\begin{tabular}{|c|rrrrrr|}
\hline\hline
Field   &$Q_A$&$Q_3$&$Q_4$&$Q_5$&$Q_6$&$Q_7$\\
\hline\hline
$S_5$&    $-$1&    1&    0&    0& $-$2&  2   \\
$S_6$&    $-$1&    1&    0&    1&    1&  2   \\
$S_7$&    $-$1&    1&    0& $-$1&    1&  2   \\
$S_8$&    $-$1& $-$1&    0&    0& $-$2&  2   \\
$S_9$&    $-$1& $-$1&    0&    1&    1&  2   \\
$S_{10}$& $-$1& $-$1&    0& $-$1&    1&  2   \\
$S_{11}$&    0&    2&    0&    0&    0&  0   \\
$S_{12}$&    0&    1& $-$3&    0&    0&  0   \\
$S_{13}$&    0&    1&    3&    0&    0&  0   \\
\hline
\hline
\end{tabular}
\end{center}
\noindent {\footnotesize  List of non-Abelian singlet fields 
with
their charges under the $U(1)$ gauge groups. The charges of these 
fields
under $U(1)_{1,2,8,9}$ are zero and not listed.}
\vspace{0.5cm}

The role of the inflaton direction can be played by $\langle S_{11}
\overline{S}_{11}\rangle$, formed by fields with zero anomalous charge. 
However for this to be viable there should be no higher order terms in the 
superpotential involving just the inflaton directions fields (or 
terms involving just a single non-inflaton direction field) for these 
will spoil the $F$-flatness of the inflaton direction. 
Given that slow-roll is expected to end at values of the inflaton 
field not much smaller than   $\Mpl$, see Eq. (\ref{vv}), 
only very high 
dimension terms will be acceptable in the superpotential. $\langle S_{11}
\overline{S}_{11}\rangle$ must be invariant under continuous gauge 
symmetries 
and so the only symmetry capable of ensuring such $F$-flatness is a 
discrete R-symmetry.  Unfortunately we do not know whether the models 
considered have such a discrete R-symmetry and thus they may allow the 
dangerous terms. Henceforth we will ignore this problem and assume the 
dangerous terms are absent.

 The
rest of the fields  acquire large positive 
masses deep in the
inflaton direction due to the Yukawa couplings in (\ref{sup}),
guaranteeing
the stability of the inflaton direction 
$S=S_{11}=\overline{S}_{11}$.
One-loop corrections to the inflaton potential proportional to $S^2$ 
are
absent and only the $\sim\xi^2\log S^2$
dependence remains, providing the slow-roll condition. However, the end 
of
inflation poses a problem for the present example: no set of VEVs for 
the
selected fields can give zero potential. As is well known, a flat
direction ($V=0$) is always associated with an holomorphic, gauge
invariant monomial built of the chiral fields. To compensate the 
FI-term
and give $V=0$, this monomial should have negative anomalous charge.
However, in the considered subset $Q_A=Q_7/2$ and all holomorphic,
gauge invariant monomials must have then $Q_A=0$.
To circumvent this problem we enlarge the field subset by adding an 
extra
field, $S_1$  with
${\overrightarrow{Q}}(S_1)=(Q_A;Q_\alpha)=(-4;0,1,0,0,-2)$.
It is easy to see that, for example, the flat direction
$\langle 1^3,5,6,10,\overline{13}\rangle$ can cancel the FI-term and
give $V=0$. Other flat directions exist, but clearly all of them 
involve
$S_1$. However, the superpotential (\ref{sup}) does not provide a large
mass for $S_1$ when we are deep in the flat direction. Unless higher 
order
terms in (\ref{sup}) provide a positive mass for $S_1$, the FI-term
induces a destabilization of the inflaton direction and $S_1$ is forced 
to
take a VEV:
\be
\langle S_1^2\rangle=\frac{-g_A^2}{G_{11}^2}Q_1^A\xi,
\ee
where we use the definition $G_{ij}^2=g_A^2Q_i^AQ_j^A+\sum_\alpha
g_\alpha^2 Q_i^\alpha Q_j^\alpha$.
This is not a problem in itself because the rest of the fields are 
forced
to have zero VEVs and so the potential cannot relax to zero. The 
presence
of additional $U(1)$ factors prevents the vacuum shift that was
problematic for the example of section~{\bf 4}. The value of the 
potential
in the presence of a VEV for $S_1$ is
\be
V=\frac{1}{2}g_A^2\xi^2_{eff},
\ee
with
\be
\xi_{eff}^2=\frac{\sum_\alpha g_\alpha^2 
(Q_1^\alpha)^2}{G_{11}^2}\xi^2.
\ee
The masses of the rest of the fields are also affected and read:
\be
m_\phi^2=\lambda_i^2\langle I_i'\rangle^2 +
\frac{g_A^2}{G_{11}^2}(Q_i^AG_{11}^2-Q_1^AG_{i1})\xi,
\ee
where $\lambda_i$ are some of the Yukawa couplings in (\ref{sup}).

In general, when all the fields in the model are included, the presence 
of
the Fayet-Iliopoulos $D$-term will induce VEVs for the fields with negative 
anomalous charge
which are not forced to have zero VEV by $F$-term contributions. These
non-zero VEVs will in turn induce, through other $D$-terms, non-zero
VEVs for other fields, even if they have positive anomalous charge.
Finding all the VEVs requires the minimization of a complicated
 multifield potential that includes both $F$ and $D$ contributions. 

In many cases the field VEVs
adjust themselves to give $V=0$ and no $D$-term inflation is 
possible.
In other cases however, especially in the presence of additional $U(1)$
factors, there is a limited number of fields that must necessarily
take a VEV to cancel the Fayet-Iliopoulos $D$-term. If the inflaton 
direction provides
a large $F$-term mass for them, cancellation of the FI-term is 
prevented.
Even if many other fields are forced to take VEVs, no configuration 
exists
giving $V=0$ and $D$-term inflation can take place in principle.
To determine if that is the case, one should minimize the effective
potential for large values of the inflaton field and determine all the
additional vevs triggered by the FI-term. These VEVs, of order $\xi$
will affect the details of the potential along the inflaton direction,
both at tree level (offering the possibility of reducing the effective
value of $\xi$) and at one-loop, via their influence on the
field-dependent masses of other fields. 

\subsection{$D$-term inflation and cosmic strings}

Let us go back now to the basic model discussed in 
Section (\ref{simplest}). The  point we would like to comment on 
is the following \cite{lr}: 
when the field $\phi_{-}$ rolls down to its present day value
 $\langle\phi_{-}\rangle=\sqrt{\xi}$  to terminate inflation, 
cosmic strings may be formed since the anomalous gauge group $U(1)
$ is broken to unity \cite{j2}. As it is known, stable cosmic strings arise 
when the manifold ${\cal M}$ of degenerate vacua has a non-trivial 
first homotopy group, $\Pi_1({\cal M})\neq {\bf 1}$. 
The fact that at the end of hybrid inflationary models the formation of
 cosmic strings may occur was already noticed in Ref. \cite{j1} in the 
context of global supersymmetric theories and in Ref. \cite{linderiotto} 
in the context of supergravity theories.

It has been recently shown \cite{daff}  that (at least some of) the
strings formed at the breaking of the anomalous $U(1)$ are  local, in
the sense that their energy per unit length can be localized in a
finite region surrounding the string's core, even though this energy is
formally logarithmically infinite. This happens because the axion field
configuration may be made to wind around the strings so that any
divergence must come from the region near the core instead of
asymptotically.
Moreover, as we have seen in the previous Section (\ref{strmodel}) 
in realistic four-dimensional 
string models, there are extra local $U(1)$ 
symmetries that can be also spontaneously broken by the $D$-term. 
This happens necessarily if there are no singlet fields charged under 
the anomalous $U(1)$ only. In such a case, there may arise 
local cosmic strings associated 
with extra $U(1)$ factors.

 In $D$-term inflation the string per-unit-length is given by $\mu=2\pi\xi$. 
Cosmic strings forming at the end of $D$-term inflation are very heavy
 and temperature anisotropies may arise both {}from the inflationary 
dynamics and {}from the presence of cosmic
strings. {}From recent numerical simulations on the cosmic microwave 
background  anisotropies induced by cosmic strings \cite{a1,a2,ax}
it is possible to infer than 
this mixed-perturbation scenario \cite{linderiotto}
leads to the COBE normalized value $\sqrt{\xi}=4.7\times 10^{15}$ GeV 
\cite{j2}, which is of course smaller  than the value obtained in the 
absence of cosmic strings.  Moreover, cosmic strings contribute to 
the angular spectrum an amount of order of 75\% 
in the simplest version of  $D$-term inflation \cite{j2}, 
which might render the angular spectrum, 
when both cosmic strings and inflation contributions are summed up, 
too smooth to be in agreement with present day observations  \cite{a1,a2}. 

Thus, even though  cosmic strings produced at the end of $D$-term inflation may play a fundamental role 
in the production of the baryon asymmetry \cite{br}, 
all the previous considerations and, above all, the fact that the value of   
$\sqrt{\xi}$ is further reduced with respect to the case in which cosmic 
strings are not present,   would appear   to   exacerbate the problem of 
reconciling the value of  $\sqrt{\xi}$ suggested 
by COBE with the value inspired by weakly coupled
string theory when cosmic
 strings are present. One has to remember, however,  the condition to produce cosmic strings 
is $\Pi_1({\cal M
})\neq {\bf 1}$ and therefore  consider the structure of the {\it whole} 
potential, {\it i.e.} all the $F$-terms and all the $D$-terms. 
When this is done, it turns out that, depending on the specific models, 
some or all of the (global and local) cosmic strings may disappear. 
In general there can be models with anomalous $U(1)$ 
that have just global cosmic 
strings, just local cosmic strings, both global and local 
strings or, more important,   no cosmic
strings at all \cite{casas,casas1}. 
The latter case is certainly  the most
 preferable case since the presence of cosmic strings renders the  problem 
of reconciling the COBE normalized low value of $\xi$ with the one 
suggested by string theory even worse.   

In the case in which the Fayet-Iliopoulos $D$-term is present in 
the theory {}from the very beginning  because of  anomaly-free $U(1)$ 
symmetry and not due to  some underlying string theory, the value 
$\sqrt{\xi}\sim 10^{15}$ GeV is very natural and is not
 in conflict with the presence of cosmic strings. The only shortcoming seems 
to be a  too smooth angular spectrum because cosmic strings may provide most 
contribution to the angular spectrum. If this problem is taken seriously and 
one wants to avoid the presence of cosmic strings, 
a natural solution to it is to assume that the  $U(1)$ 
gauge group is broken before the onset of inflation  so that no cosmic 
strings will be produced when $\phi_{-}$ rolls down to its ground state.
 This may be easily achieved by introducing 
a pair of vector-like
(under $U(1)$) fields $\Psi$ and $\bar{\Psi}$ and two gauge singlets $X$ 
and $\sigma$  with a superpotential of the form 
\begin{equation}
W=X\left(\kappa\bar{\Psi}\Psi-M^2\right)+\beta\sigma\bar{\Psi}\Phi_{+}+
\lambda S\Phi_{+}\Phi_{-},
\end{equation}
 where $M$ is some high energy scale, presumably the grand unified  scale. 
It is easy to show that the scalar components of the two-vector superfields
    acquire vacuum expectation values 
$\langle\psi\rangle=\langle\bar{\psi}\rangle=M$, and 
$\langle X\rangle=\langle \sigma\rangle=0)$ which leave supersymmetry 
unbroken and $D$-term inflation unaffected. In this example, cosmic strings 
are  produced prior to the onset  of inflation and subsequently diluted. 

\subsection{A GUT model of $D$-term inflation}

\label{gutm}

A $D$-term inflationary scenario
may be constructed within the
framework of concrete supersymmetric Grand Unified Theories (GUT's)
where  realistic fermion masses are predicted and the
doublet-triplet splitting problem is naturally solved
by the pseudo-Goldstone boson mechanism in $SU(6)$ \cite{dvaliriotto}.
The presence of the  $D$-term is essential
in order to generate vacuum expectation values
and therefore simplify the structure of the superpotential.
As a by-product, the model has a  built-in inflationary trajectory
in the field space along which all $F$-terms are vanishing and
only the associated $U(1)$ $D$-term is nonzero.
In this case, the COBE-normalized scale
$\sqrt{\xi}\sim 10^{16}$ GeV appears more natural to
accept since the the GUT scale is of the same order 
of magnitude, even though it must be put in by hand
along with two similar mass scales $M$ and $M'$.

This model gives a four-component inflaton (Section \ref{s10}), instead
of the usual one-component inflaton. Its predictions depend on the 
initial conditions as well as on the potential, but for a significant 
range of initial conditions they will be the same as for the other 
$D$-term inflation models. A problem is that the field values while 
cosmological scales leave the horizon are of order $\mpl$, making it 
questionable
if the field theory is really under control.

The model is based on the $SU(6)$ supersymmetric GUT with one adjoint Higgs
$\Sigma$ and a number of fundamental Higgses 
$H_A, \bar H^A, H_A', \bar H^{A'}$. Each of these
fundamentals transforms as a doublet of a certain custodial $SU(2)_c$
symmetry that is required to solve the hierarchy problem. The index 
$A =  1,2$ is the   $SU(2)_c$-index.
The  $H_A, \bar H^A$ carry unit charges
opposite to $\xi$ and are the ones that compensate $U(1)$ $D$-term
in the present Universe. 
The superpotential reads
\begin{equation}
 W = c{\rm Tr}\Sigma^3 + (\alpha\Sigma + aX + M)H_A\bar H^{'A}
+ (\alpha'\Sigma + a'X + M')H_A'\bar H^A. 
\end{equation}
Minimizing both the $D$- and the  $F$-terms we get the following 
supersymmetric
vacuum which leaves  the Standard Model
$SU(3)\sub C \otimes SU(2)_L\otimes U(1)$
as unbroken gauge symmetry
\begin{eqnarray}
H_{Ai} &=&\bar H^{Ai} = \delta_{A1} \delta_{i1}\sqrt{{\xi \over 2}},~~~~~
H_A' = \bar H^{A'} = 0,~~~\nonumber\\
\Sigma &=& {aM' - a'M \over a'\alpha - \alpha' a}
{\rm diag}(1,1,1,-1,-1,-1),~~~~~
X = -{\alpha M' - \alpha 'M \over a'\alpha - \alpha' a}. \label{vac}
\end{eqnarray}
Here $i,k = 1,2,..6$ are $SU(6)$ indexes.
 The role of the $\Sigma$ vacuum expectation value  is crucial since it leaves
the unbroken $SU(3)\sub C\otimes SU(3)_L\otimes U(1)_Y$ symmetry,
consequently it can cancel masses of all upper three or lower three
components of the fundamentals.
The fundamental vevs are $SU(5)$ symmetric, so that the
intersection gives the  unbroken standard model symmetry group.

In this vacuum the electroweak Higgs doublets {}from 
$H_2,\bar H^2, H_2^{'}, \bar H^{'2}$
are massless. This is an effect of custodial $SU(2)_c$
symmetry. Indeed, since $H_1$ and $\bar H^1$ break  one of the
$SU(3)$ subgroups to $SU(2)_L$, their electroweak doublet components
become eaten up Goldstone multiplets and cannot get masses {}from the
superpotential due to the Goldstone theorem. This forces the vevs of
$\Sigma$ and $X$ to exactly cancel their mass terms and
those of $H_2,\bar H^2, H_2^{'}, \bar H^{'2}$ due to the  custodial
symmetry. This solves the doublet-triplet splitting problem
in a natural way \cite{solution}.

 Quarks and leptons of each generation are placed in a minimal anomaly
free set of $SU(6)$ group: $15$-plet plus two $\bar 6_A$-plets per family.
We assume that $\bar 6_A$ form a  doublet under $SU(2)_c$ so that
$A = 1, 2$ is identified as $SU(2)_c$ index.
The fermion masses are  then  generated through the
couplings ($SU(6)$ and family indices are suppressed)
$ \bar H^A\cdot 15\cdot\bar 6_A  +
\epsilon^{AB}{H_A\cdot H_B \over M_{\xi}} 15\cdot 15$, 
where $M_{\xi}$ has to be  understood as the
mass of order $\sqrt{\xi}$ of integrated-out
heavy states.  
When the large vevs of $H_1$ and $\bar H^1$ are inserted, the additional,
vectorlike under $SU(5)$-subgroup, states: $5$-s {}from $15$-s and $\bar 5$-s
{}from $\bar 6_1$, become heavy and decouple.
Low energy couplings are just the usual $SU(5)$-invariant
Yukawa interactions of the light doublets {}from $H_2$
and $\bar H^2$ with the usual quarks and leptons.

The relevant branch for inflation in the field space 
is represented by the $SU(6)$ $D$- 
and $F$-flat trajectory parameterized by the invariant ${\rm Tr}\Sigma^2$. This
corresponds to an arbitrary expectation value along the component
\begin{equation}
\Sigma = {\rm diag} (1,1,1,-1,-1,-1) \frac{S}{\sqrt{6}}.
\end{equation}
The key point here is that above component has
no self-interaction ({\it i.e.} ${\rm Tr}\Sigma^3 = 0$)
and appears in the superpotential linearly. At the generic point of this
moduli space the gauge $SU(6)$ symmetry is broken to
$SU(3)\otimes SU(3)\otimes U(1)$. All gauge-non
singlet Higgs fields are getting masses ${\cal O}(S)$  and
therefore, for large values of $S$,  $S \gg \sqrt{\xi}$,
they decouple. Part of them gets eaten up
by the massive gauge superfields. These are the  components of $\Sigma$
transforming as $(3,\bar 3)$ and $(\bar 3, 3)$ under the unbroken subgroup.
All other Higgs fields get large masses {}from the
superpotential.
The massless degrees of freedom
along the branch are therefore: 
two singlets $S$ and $X$, the massless
$SU(3)\otimes SU(3)\otimes U(1)$ super- Yang-Mills
multiplet and the massless matter superfields.

By integrating out the heavy superfields,  we can write down an effective
low energy superpotential by simply 
using holomorphy and symmetry arguments.  This superpotential,
as well as all gauge $SU(6)$ $D$-terms,
is vanishing.  Were not for the $U(1)$-gauge symmetry, the 
branch parameterized by $S$, would simply correspond to a SUSY preserving flat
vacuum direction remaining flat to all orders in perturbation theory.
The $D$-term, however, lifts this flat direction,
taking an asymptotically constant value
for arbitrarily large $S$ at the tree-level. This is because all Higgs
fields with   charges opposite to  $\xi$ gain large masses and decouple,
and $\xi$ can not be compensated any more
(notice that  heavy fields decouple in pairs with
opposite charges and therefore 
${\rm Tr} {\bf  Q}$ over the remaining low energy fields is not changed).
 As a result, the branch of interest is represented by two massless
degrees of freedom $X$ and $S$ whose vevs  set the mass scale for the heavy
particles, and a constant tree level vacuum energy density $
V_{{\rm tree}} = {g^2 \over 2}\langle D ^2\rangle = \frac{g^2}{2}\xi^2$ which
is responsible for inflation.

This result can be easily
rederived by explicit solution of the equations of motion along the inflationary
branch.
For doing this,  we can explicitly minimize all $D$- and $F$- terms subject to
large values of $S$ and $X$.  The relevant part of the potential is
\begin{equation}
V = |F_{H_A^{'}}|^2 +  |F_{\bar H_A^{'}}|^2  +  
{g^2 \over 2}D^2, \label{fandd2} 
\end{equation}
since the remaining $F$- and $D$- terms are automatically vanishing as long as
all other gauge-non singlet Higgses are zero.
We would need to include them only if the  minima of the
potential  (\ref{fandd2}) (subject to $S,X \gg
\xi$) were
incompatible with such an assumption. However for the branch of our interest
this turns out to be not  the case.

As with the simpler models that we considered earlier, the negatively
charged fields that might drive $V_D$ can acquire positive
masses-squared from the $F$ term. These fields come
purely {}from the $H,\bar H', H',\bar H$ 
superfields.\footnote
{All other states either have
vanishing charge (these are $X, \Sigma$ and
the gauge fields), 
or have no inflaton dependent mass but positive charge 
(these are matter fields).}
These are the  fragments $(1,3),(1,\bar 3)$ and $(3,1),(\bar 3, 1)$
of the  $H, \bar H'$ with masses-squared
\be
\phi_+^2 \pm g^2\xi 
\ee 
and 
\be
\phi_-^2 \pm g^2\xi \,,
\ee 
where
\be
\phi_\pm\equiv \pm\alpha S/\sqrt{6}
+ aX + M \,,
\ee
and the analogous fragments of the  $H', \bar H$ with 
masses-squared
\be
{\phi_+^2}' \pm g^2\xi 
\ee 
and
\be
{\phi_-^2}' \pm g^2\xi \,,
\ee 
where
\be
\phi_\pm'\equiv \pm\alpha' S/\sqrt{6}
+ aX' + M' \,,
\ee
For each of these four cases there 
are eight pairs of charged fields.

When $\phi_\pm^2$ and $\phi_\pm'{}^2$ are both bigger
than $g^2\xi$, there is inflation. Including the loop correction the
potential is
\begin{equation}
\label{pot1}
 V_{{\rm inf}} = {g^2 \over 2}\xi^2\left[1 + 
\frac{3g^2}{16\pi^2}
\ln \(|\phi_+|^2|\phi_-|^2|\phi_+'|^2|\phi_-'|^2 \) 
\right]\,.
\end{equation}
(To obtain this expression, we added the four contributions given
by \eq{vloop}, with $C=8$ for each of them.)
This potential is a function of four real fields, namely the real and 
imaginary parts of $S$ and $X$. As discussed in Section \ref{s10}, 
there will in general be a family of non-equivalent inflationary 
trajectories. We are dealing with a four-component inflaton,
and the predictions depend in general on the initial conditions.
However, for a significant range of initial conditions, the inflaton
trajectory after the observable Universe leaves
the horizon will be 
roughly a straight line pointing towards the origin, in the space of the 
fields. If $\phi$ is the canonically-normalized field along the 
trajectory, the inflaton potential is then given by \eq{vloop}
with $C=96$ (except for an insignificant change in $V_0$ coming
from the constant ratio of $\phi^8$ and the argument of the log).

{}From the estimate \eq{vv}, one sees that in this case, when the 
observable Universe leaves the horizon, 
$\phi$ is at least of order $\mpl$ and maybe of order $10\mpl$.
One needs the former case to have any chance of keeping the field theory 
under control.

Notice that in the usual hybrid inflationary scenarios
inflation is terminated by the rolling down of a Higgs field
coupled to the inflaton and consequent  phase transition with symmetry
breaking. Whenever the vacuum
manifold has a non-trivial homotopy, the topological defects will form
much in the same way as in the conventional thermal phase transition.
Thus,  the straightforward generalization of the hybrid
scenario in the GUT context would result in the post-inflationary
formation of the unwanted magnetic monopoles.
In the model proposed in \cite{dvaliriotto} this
disaster never happens, since the inflaton field is the GUT Higgs itself.
The GUT symmetry is broken both during and after inflation and the monopoles
(even if present at the early stages) get inevitably inflated away.
The unbroken symmetry group along the inflationary branch is
$G_{{\rm inf}} = SU(3)\otimes SU(3)\otimes U(1)\otimes
SU(2)\otimes U(1)$\footnote{If the gauge $U(1)$ is a stringy anomalous $U(1)$, 
it will be broken by the dilaton even if all other charged fields vanish.
In this case the unbroken symmetry has to be understood as a global one.}
which gets broken to
$G_{{\rm postinf}} = SU(3)\otimes SU(2)\otimes U(1)\otimes U(1)$
modulo the electroweak phase transition (extra $U(1)$ -factor is global).
Since $\pi_2(G_{{\rm inf}}/G_{{\rm postinf}}) = 0$ no monopoles are
formed\footnote{Other ways of solving the monopole problem
exist in previous papers \cite{mutated,lazpan,thermal,linderiotto}.}. 

The model described above demonstrates that  $D$-term
inflation may satisfy a  a sort of  "minimal principle" \cite{dr} 
which requires that  any successful inflationary scenario should 
naturally arise {}from models which are  entirely
motivated by particle physics considerations and  should not
involve (usually complicated and {\it ad hoc})
sectors on top of the existing structures.

\section{Conclusion}

\label{s9}

In the face of increasingly accurate observations of 
the cosmic microwave background anisotropy and of the galaxy
distribution, slow-roll inflation
seems to provide the only known origin for structure in the Universe.
In this review we have seen how to build models of inflation, and test
them against observation.

What is the point of such an exercise? To address this question, one
needs to understand what is meant by a model of inflation. 
One can think of a model as something analogous to a building.
It has an outer shell, which is visible to the casual observer, 
but hopefully also something inside.

The shell is a specification of the form inflationary potential.
In a single-field model the potential 
depends only on the inflaton field,
while in a hybrid model it depends on one or more additional fields.
Observation, notably through the spectral index of the density 
perturbation, can discriminate sharply between different shells.
Most, and perhaps all, of the present zoo of shells will be
rejected by observations in the next ten to fifteen years, 
culminating with the Planck satellite that will give an essentially 
complete measurement of the cmb anisotropy. 
One can imagine that 
eventually just one basic form for the shell is singled out by the community,
which 
by virtue of its intrinsic beauty and
its accurate description of the observations is likely to be the
one chosen by Nature.
Then, in a sense, there will be a consensus about 
the origin of all structure in the Universe.
One will have arrived at the rather boring conclusion, that
it probably comes from a certain scalar field potential!

Things are very different when 
we come to consider the interior of the 
shell. Here, one recognizes that the inflationary potential is part 
of an 
the extension of the 
Standard Model, that is supposed to describe the 
fundamental interactions at the level of field theory. The field theory
description is, hopefully, an approximation to some more fundamental
theory like weakly coupled string theory or Horava-Witten M-theory. 
Although different interiors generally have different shells, that is 
not inevitable as we have seen in more than one example.

At this point, inflation model-building becomes part of 
the enterprise that has occupied the particle physics community
for more than two decades. That is, to find the extension of the 
Standard Model that has been chosen by Nature.\footnote
{This is the usual viewpoint but one can vary it.
Maybe there is only one mathematically consistent theory that gives
anything resembling physical reality, in which case we have in principle 
little need of observation. Maybe the usual assumption that there
are many possible theories is correct, but many or all of them 
have been realized by Nature 
in different parts of the universe, that may or may not be 
connected with the homogeneous Universe around us. These variations make 
no difference for the present purpose.}
Because there is so 
little guidance from observation, this enterprise has been driven by
theoretical considerations to an extent that is unprecedented in the 
history of science. In particular, the rich structure of supersymmetry
is almost always assumed because it seems to be the only way of avoiding
a certain type of extreme fine-tuning. In the forseeable future we shall
find out whether supersymmetry and other theoretical structures have been chosen
by Nature, and therefore whether pure thought has successfully
pulled so far ahead of observation. Whether positive or negative,
this resolution will surely be a permanent landmark in the history
of the human intellect.

Assuming that current ideas are basically correct, one still has to ask
to what extent it will ever be possible to discriminate between 
different fundamental theories. Observation by itself provides, so
far, only a few 
numbers relevant to this purpose, together with
some upper and lower limits. Among them are the 
parameters of the Standard Model and, if one accepts the increasingly
strong evidence, one or two numbers relating to the neutrino masses.
There is also strong evidence for non-baryonic dark matter, which 
probably has to be in the form of one or more as-yet undiscovered 
particle species. And finally, coming to the concern of this review,
there is the magnitude of the
spectrum of the primeval density perturbation, measured on the
scales explored by COBE.

Among the quantities with crucial upper or lower limits one might mention
on the particle physics side
the Higgs masses, neutrino masses and mixing angles, the proton lifetime
and the electric dipole moment of the neutron. As we have seen,
one should add to these the limit on the departure from scale invariance
represented by the result $|1-n|<0.2$, and the upper limit
of order 50\% on the relative contribution
of gravitational waves to the spectrum of the cmb anisotropy.

These lists are incomplete but they serve to explain the 
role of inflation. It will add to the
precious collection of numbers and limits, that guide us
in a search for what lies beyond the Standard Model.
Possibly there will even be a non-trivial function, $n(k)$, that requires
explanation.

Analogously with the situation concerning the outer shell of a model of 
inflation, the hope is that the community will eventually be able to agree
that some model of the fundamental interactions is likely to be the one 
that Nature has chosen, 
by virtue of its intrinsic beauty and
accurate agreement with the few numbers provided by observation.
Because the numbers are few, this 
would hardly be possible at the level of a field theory, but it might be 
possible at the level of something like string theory where there 
are essentially no free parameters and everything is dictated by
group theoretic and topological considerations.

With this perspective, let us look at some of the models of inflation 
that are presently under consideration. 

As we have discussed at length, supersymmetry is both a blessing and a 
curse for inflation model-building. It 
is a blessing, primarily because it allows one to 
understand the existence of scalar fields. As a bonus, it 
can practically eliminate the quartic term in the inflaton potential,
which would normally spoil inflation. 
It is a curse, because in a generic supergravity theory all
scalar fields have masses that are too big to support inflation.
Let us recall ways of handling this problem. 

According to supergravity, the potential is the sum of an
$F$-term and a $D$-term. In most models the $F$-term dominates
and we consider them first. With an $F$-term of generic form, the
inflaton mass is too big. One can suppose that it is suppressed by
an accidental cancellation, but one can instead invoke a 
non-generic form, which guarantees the suppression. Such
a form can emerge from 
weakly coupled heterotic string
theory, though probably not from
Horava-Witten M-theory. Alternatively, one can suppose that while
the inflaton mass is indeed unsuppressed at the Planck scale,
quantum corrections drive it to a small value at lower scales 
so as to permit inflation after all. At the present time this
`running mass' model looks quite attractive.

A different strategy is to suppose that 
a Fayet-Illiopoulos
$D$-term dominates, with the charged fields driven to zero.
These models have received a lot of attention 
because at least in the simplest versions they have two remarkable features.
One is that supergravity corrections to the inflaton mass
are absent. The other is that there is an accurate prediction for the
spectral index, $n=0.96$ to $0.98$ which will eventually be testable.
Further investigation, though, has revealed a serious problem.
In 
contrast with the $F$-term models,
the inflaton field value has to be at least of order $\mpl$. As a result,
one has gained control of the inflaton mass, only to be in danger of 
losing it for
the quartic and higher terms of the 
potential. In string theory there are
two additional problems. One is the existence of 
fields which
are liable to drive the $D$-term to zero.
The other is that the predicted 
magnitude of the cmb anisotropy is far higher than the COBE measurement.
It is fair to say that $D$-term inflation
is under considerable pressure at the moment.

The predictions of different models for the spectral index $n$,
and for its scale-dependence,  are summarised
in the table on page \pageref{t:1}.
Remarkably, the eventual accuracy $\Delta n\sim 0.01$ offered by the Planck
satellite is just what one might have specified in order
to distinguish between 
various models, or at least between their various shells.
At the most extravagant, one might have asked for $\Delta n\sim
10^{-3}$.

In summary, observation will discriminate strongly between models of 
inflation during the next ten or fifteen years. By the end of that 
period, there may be a consensus about the form of the inflationary
potential, and at a deeper level we may have learned something valuable
about the nature of the fundamental interactions beyond the Standard 
Model. We shall also have confirmed, or practically rejected,
the remarkable hypothesis that inflation is responsible for 
structure in the Universe.

\section*{Postscript}

At the final proof-reading, observation is beginning to pin down the
cosmological parameters, and therefore the spectral index. A preliminary
estimate [R. Bond, Pritzker Symposium,
http://www-astro-theory.fnal.gov/Personal/psw/talks/bond/bond.03.gif.]
is $|n-1| < 0.05$. Looking at Table 1, this would rule out a potential
of the form $V=V_0(1-c\phi^3)$, and almost rule out one of the form
$V=V_0(1-c\phi^4)$. The latter case is practically equivalent to the
form chosen for the first viable model of inflation, Eq.~186, so that
form is almost ruled out
as well. For a number of other forms of the potential (Table 2)
the preliminary result for $n$ places a
non-trivial lower limit on $N$, the number of $e$-folds of inflation
occurring after cosmological scales leave the horizon.
It seems that we are already entering the promised land, the
golden age of cosmology!
                    
\newpage

\underline{Acknowledgements}:

DHL is  grateful to CfPA and LBL, Berkeley,
for the provision of financial 
support and a stimulating working environment when
this work was started.
He is  indebted to Ewan Stewart and Andrew Liddle for long-standing 
collaborations, and to David Wands for many useful conversations.
He has
also received valuable input {}from 
Mar Bastero-Gil, Laura Covi,
Mary Gaillard, Andrew Liddle,
Andrei Linde, Hitoshi Murayama, Hans-Peter Nilles,
Burt Ovrut, Graham Ross and Subir Sarkar. 
AR is grateful to the  Theoretical Astrophysics group at Fermilab, where
 this work was initiated,  for the 
incomparable  stimulating  atmosphere.
In particular, he is indebted to Scott Dodelson, Will
 Kinney and Rocky  Kolb for many stimulating conversations and for 
continuously spurring his efforts. 
He is also grateful to  Michael Dine,
Gia Dvali, Jose Ramon Espinosa, Steve King, Andrei Linde and Graham  Ross for   
enjoyable collaborations. 
DHL acknowledges support {}from PPARC and NATO grants, and 
{}from the European Commission
under the Human Capital and Mobility programme, contract
No.~CHRX-CT94-0423. 

\def\NPB#1#2#3{Nucl. Phys. {\bf B#1}, #3 (19#2)}
\def\PLB#1#2#3{Phys. Lett. {\bf B#1}, #3 (19#2) }
\def\PLBold#1#2#3{Phys. Lett. {\bf#1B} (19#2) #3}
\def\PRD#1#2#3{Phys. Rev. {\bf D#1}, #3 (19#2) }
\def\PRL#1#2#3{Phys. Rev. Lett. {\bf#1} (19#2) #3}
\def\PRT#1#2#3{Phys. Rep. {\bf#1} (19#2) #3}
\def\ARAA#1#2#3{Ann. Rev. Astron. Astrophys. {\bf#1} (19#2) #3}
\def\ARNP#1#2#3{Ann. Rev. Nucl. Part. Sci. {\bf#1} (19#2) #3}
\def\mpl#1#2#3{Mod. Phys. Lett. {\bf #1} (19#2) #3}
\def\ZPC#1#2#3{Zeit. f\"ur Physik {\bf C#1} (19#2) #3}
\def\APJ#1#2#3{Ap. J. {\bf #1} (19#2) #3}
\def\AP#1#2#3{{Ann. Phys. } {\bf #1} (19#2) #3}
\def\RMP#1#2#3{{Rev. Mod. Phys. } {\bf #1} (19#2) #3}
\def\CMP#1#2#3{{Comm. Math. Phys. } {\bf #1} (19#2) #3}


\begin{thebibliography}{999}

%AAAAAAAAAAAAA

\bibitem{AdamsFreese} F. C. Adams and K. Freese, Phys. Rev.  {\bf D43}, 
353 (1991);

\bibitem{natural2} F. C. Adams et al.
1993, Phys. Rev. {\bf D47}, 426 (1993).

\bibitem{grahamnew} J. A. Adams, G. G. Ross and S. Sarkar,
Phys.Lett. {\bf B391}, 271 (1997).

\bibitem{graham3} 
J. A. Adams, G. G. Ross and S. Sarkar,
Nucl. Phys. {\bf B503}, 405 (1997).

\bibitem{ads} I. Affleck, M. Dine and N. Seiberg, Nucl. Phys. 
{\bf B256}, 557 (1985) . 

\bibitem{new2}
A. Albrecht  and P. J. Steinhardt, Phys. Rev. Lett. {\bf 48}, 1220
(1982) . 

\bibitem{a1} B. Allen, R.R. Caldwell, E.P.S. Shellard, A. Stebbins and 
S. Veeraraghavan, FERMILAB-PUB-97-334-A preprint.

\bibitem{a2} B. Allen,  R.R. Caldwell, S. Dodelson, 
 L. Knox, E.P.S. Shellard 
and  A. Stebbins, Phys. Rev. Lett. {\bf 79}, 2624 (1997).

\bibitem{g5}L.~Alvarez-Gaum\'e, M.~Claudson, and M.~Wise, \NPB{207}{82}{96}.

\bibitem{couplingunity}
U. Amaldi, W. de Boer and H. Furstenau,
Phys. Lett. {\bf B260}, 447 (1991).

\bibitem{r3} G.W. Anderson, A. Linde and A. Riotto, Phys. Rev. Lett. 
{\bf 77}, 3716 (1996).

\bibitem{aadd} I.  Antoniadis, 
N. Arkani-Hamed, S. Dimopoulos and G. Dvali, hep-ph/9804398; 

\bibitem{ant} I. Antoniadis, K. S. Narain and T. R. Taylor,
\PLB{267}{91}{37}.

\bibitem{add} N. Arkani-Hamed, S.  Dimopoulos (Stanford U., Phys. Dept.), 
G. Dvali, Phys. Lett. {\bf B429}, 263 (1998),  hep-ph/9803315.  

\bibitem{murayama} N. Arkani-Hamed and H. Murayama, Phys. Rev. 
{\bf D56}, 673 (1997).

\bibitem{aa} N. Arkani-Hamed, M.  Dine, S.P. Martin,  hep-ph/9803432. 

\bibitem{fi1} J. Atick, L. Dixon and 
A. Sen, Nucl. Phys. {\bf B292},  109 (1987) .

%BBBBBBBBBBBB

\bibitem{g9} J. Bagger et al., Phys. Rev. {\bf D55}, 3188 (1997) 
for a review of the phenomenological signals of gauge mediation.

\bibitem{bailinlove}
        D. Bailin and A. Love, {\em Supersymmetric Gauge Field Theory
        and String Theory}, IOP, Bristol (1994) .

\bibitem{banks88} T. Banks, L. Dixon, D. Friedan and E. Martinec,
Nucl. Phys. {\bf B299}, 613 (1988) .

\bibitem{paul}
T. Banks et al.,
Phys. Rev. {\bf D52}, 3548 (1995).

\bibitem{bardeen80} J. M. Bardeen, 
Phys Rev D {\bf 22}, 1882 (1980).

\bibitem{bst}
J. M. Bardeen, P. S. Steinhardt and M. S. Turner,
{Phys. Rev. } {\bf D28}  679 (1983) .

\bibitem{c1}  
J.M. Bardeen, J.R. Bond and D. Salopek, in {\em Proceedings of the Second
Canadian Conference on General Relativity and Relativistic Astrophysics,
Toronto, Canada} edited by A. Coley, C. Dyer and B. Tupper
(World Scientific, Singapore, 1988).

\bibitem{vplanck} J. M. Bardeen, J. R. Bond and G. Efstathiou,
Astrophys. J. {\bf 321}, 28 (1990).

\bibitem{t1}T. Barreiro et al., hep-ph/9602263.

\bibitem{bcc} T. Barreiro, B, de Carlos  and E. J. Copeland, 
hep-th/9805005.

\bibitem{blp} J. D. Barrow, A. R. Liddle and P. Parsons, 
Phys. Rev. {\bf D50}, 7222 (1994).

\bibitem{steve1} M. Bastero-Gil and S. F. King,
hep-ph/9709502.

\bibitem{steve2} M. Bastero-Gil and S. F. King,
hep-ph/9801451

\bibitem{steve3} M. Bastero-Gil and S. F. King,
hep-ph/9806477.

\bibitem{bento} M. C. Bento and O. Bertolami, Phys. Lett. 
{\bf B384}, :98 (1996).

\bibitem{berezhiani} Z. Berezhiani and Z. Tavartkiladze, 
Phys. Lett. {\bf B409}, 220 (1997).  

\bibitem{bingall} P. Bin\'etruy and M. K. Gaillard, Phys.
Rev. D {\bf 34}, 3069 (1986).

\bibitem{bindvali}  P. Bin\'etruy and G. Dvali,  
Phys. Lett. {\bf B388}, 241 (1996) .

\bibitem{dudas} P. Bin\'etruy and E. Dudas, Phys. Lett. {\bf B389},  503 (1996).

\bibitem{bgw} P. Bin\'etruy, M. K. Gaillard and Y. Wu, Nucl. Phys. {\bf 
B493}
(1997) 27 and Phys. Lett. {\bf 412} (1997) 228 . 

\bibitem{ramond} P. Bin\'etruy et al. Phys. Lett. {\bf B403}, 38 (1997).  

\bibitem{daff} P. Bin\'etruy, C. Deffayet and P. Peter, hep-ph/9807223.

\bibitem{bdr} P. Bin\'etruy, G. Dvali and A. Riotto, in preparation.

\bibitem{bifa} P. Bin\'etruy, C. Deffayet, E. Dudas, P. Ramond, hep-th/9807079. 

\bibitem{bl}  R. Bousso and A. Linde, 
gr-qc/9803068.

\bibitem{br} R. Brandenberger and  A. Riotto,  hep-ph/9801448.

\bibitem{bim} A. Brignole, L. E. Ibanez and C. Munoz,
Nucl. Phys. {\bf B422}, 125 (1994).

\bibitem{bs93} R. Brustein and P. J. Steinhardt, Phys. Lett.
{\bf B302}, 196 (1993) . 

\bibitem{o1}M. Bucher, A. S. Goldhaber and N. Turok, Phys. Rev.  
{\bf D52}, 3314 (1995) .

\bibitem{bunn96}  E. F Bunn, A. R. Liddle and M. White, Phys. 
Rev. D {\bf 54}, 5917R (1996).

%CCCCCCCCCCC

\bibitem{burt} G. L. Cardoso and B. A. Ovrut,
Phys. Lett. {\bf B298}, 292 (1993).

\bibitem{jim} B. J. Carr, J. H. Gilbert and J. E. Lidsey,
Phys. Rev.  {\bf D50}, 4853 (1994).

\bibitem{casas} J.A. Casas and C. Munoz, Phys. Lett. {\bf B216}, 37 (1989).

\bibitem{casas1} J.A. Casas, J.M. Moreno, C. Munoz and M. 
Quiros, Nucl. Phys. {\bf B328}, 272 (1989).

\bibitem{casas2} J. A. Casas and G. B. Gelmini,  Phys. Lett. 
{\bf B410}, 36 (1997).

\bibitem{cg}
J. A. Casas and G. B. Gelmini,  Phys. Lett. {\bf B410}, 36 (1997).

\bibitem{cgr}
J. A. Casas, G. B. Gelmini and A. Riotto, to appear.
   
\bibitem{CHL} S. Chaudhuri, G. Hockney and J. Lykken, Nucl. Phys. {\bf
B469} (1996) 357.

\bibitem{choikim} K. Choi and J. E. Kim, \PLB{154}{85}{393}.

\bibitem{jose} G.  Cleaver, M.  Cveti\v c, J.R. Espinosa, L. Everett and P. 
Langacker, hep-th/9711178. 

\bibitem{jose2} G.  Cleaver, M.  Cveti\v c, J.R. Espinosa, L. Everett and  
P. Langacker, to appear. 

\bibitem{ckn} A. G.  Cohen, D. Kaplan and A. E. Nelson, hep-th/9803132.

\bibitem{cllsw} E. J. Copeland, A. R. Liddle, D. H. Lyth, E. D. Stewart
        and D. Wands, Phys. Rev.  {\bf D49}, 6410  (1994) . 

\bibitem{coughlan} G. D. Coughlan, R. Holman, P. Ramond and G. G. Ross,
        Phys. Lett. {\bf 140B}, 44 (1984). 

\bibitem{p98laura} L. Covi, D. H. Lyth and L. Roszkowski, 
hep-ph/9809310.

\bibitem{p98laura2} L. Covi and D. H. Lyth, hep-ph/9809562.

%DDDDDDDDDDDD

\bibitem{davis} R. L. Davis  et al.,
Phys. Rev. Lett. {\bf 69}, 1856 (1992); ibid. {\bf 70}, 1733
(1993) . 

\bibitem{ds} J.-P. Derendinger and C. A Savoy, Nucl Phys.
{\bf B237}, 307 (1984).

\bibitem{dkl} L. Dixon, V. Kaplunovsky and J. Louis,
\NPB{329}{90}{27}.

\bibitem{dkl2} L. Dixon, V. Kaplunovsky and J. Louis,
\NPB{355}{91}{649}.

\bibitem{noncanon}
N. Deruelle, K. Gundlach and D. Langlois, Phys. Rev D {\bf 45},
3301 (1992).

\bibitem{non1}
N. Deruelle, K. Gundlach and D. Langlois, Phys. Rev D {\bf 46},
5337 (1992).

\bibitem{dienesrev}   K.R. Dienes, Phys. Rep. {\bf 287}, 447 (1997).

\bibitem{g1}S.~Dimopoulos and S.~Raby, \NPB{192}{81}{353}.
\bibitem{g7}S.~Dimopoulos and S.~Raby, \NPB{219}{83}{479}.

\bibitem{ddr} S. Dimopoulos, G. Dvali and 
R. Rattazzi,  Phys. Lett. {\bf B410}, (1997) . 

\bibitem{g8}S. Dimopoulos et al., Nucl. Phys. Proc. Suppl. {\bf A52}, 38 
(1997) for a review of the phenomenological signals of gauge mediation.

\bibitem{dine} A useful review of these questions is given by M. Dine,
hep-th/9207045.

\bibitem{reviewdsb} M. Dine, 
{\it Supersymmetry phenomenology}, hep-ph/9612389 and refs. therein.

\bibitem{gmsb} M.~Dine, W.~Fischler, and M.~Srednicki,  
\NPB{189}{81}{575}. 

\bibitem{g2}M.~Dine and W.~Fischler, \PLB{110}{82}{227}.
\bibitem{g3}M.~Dine and M.~Srednicki, \NPB{202}{82}{238}.
\bibitem{g4}M.~Dine and W.~Fischler, \NPB{204}{82}{346}.

\bibitem{dinefisch} M. Dine, W. Fischler and D. Nemeschansky,
        Phys. Lett. {\bf 136B}, 169 (1984) .

\bibitem{dnns} M. Dine, A. E. Nelson, Y. Nir and Y. Shirman,  
Phys. Rev. {\bf D53}, 2658 (1996).

\bibitem{fi} M. Dine, N. Seiberg and E. Witten, Nucl. 
Phys. {\bf B289}, 585 (1987) .

\bibitem{fi2} 
M. Dine, I. Ichinose and N. Seiberg, Nucl. 
Phys. {\bf B293},  253 (1987). 

\bibitem{ex} M. Dine, A.E. Nelson, Y. Nir  and Y. Shirman, Phys. Rev. 
{\bf D53},  2658(1996) . 

\bibitem{dr}  M. Dine and A. Riotto, Phys. Rev. Lett. {\bf 79}, 2632
(1997) . 

\bibitem{qaisar} G. Dvali, Q. Shafi  and R. Schaefer,
Phys. Rev. Lett. {\bf 73}, 1886 (1994) . 

\bibitem{dvaliloop} G. Dvali, hep-ph/9605445.

\bibitem{solution} G. Dvali and S. Pokorski, 
Phys. Rev. Lett. {\bf 78} (1997) 807. 

\bibitem{dvaliriotto}  G. Dvali and A. Riotto,
Phys. Lett. {\bf B417}, 20  (1998) .

%EEEEEEEEEEEEEE

\bibitem{primordial}
J. Ellis, D. V. Nanopoulos, K. A. Olive and K. Tamvakis,
Phys. Lett. {\bf 127B}, 331 (1983).

\bibitem{n2} J. Ellis, S. Kelley
and D.V. Nanopoulos, Phys. Lett. {\bf B249}, 441 (1990);
{\bf B260}, 131 (1991).

\bibitem{enqvist1}
K. Enqvist, and D. V. Nanopoulos, 
Nucl. Phys. {\bf B252}, 508 (1985)
\bibitem{enqvist2} K. Enqvist, D. V. Nanopoulos, M. Quiros, and C. Kounnas, 
Nucl. Phys. {\bf B262}, 538 (1985).

\bibitem{err} J.R. Espinosa, A. Riotto and G.G. Ross, 
hep-ph/9804214.

%FFFFFFFFFFFF

\bibitem{AF1} A. Faraggi, Phys. Lett. {\bf B278}, 131 (1992).

\bibitem{faraggi} A.E. Faraggi and J.C. Pati, hep-ph/9712516.

\bibitem{fayet} P. Fayet, Nucl. Phys. {\bf B90}, 104 (1975).

\bibitem{fayil}
P. Fayet and J. Iliopoulos, Phys. Lett. {\bf 51B}, 461
(1974).

\bibitem{flt} S. Ferrara, D. L\"ust and S. Thiesen, \PLB{233}{89}{147}.

\bibitem{font}
 A. Font, L. Iba\~nez, D. Lust and F.
Quevedo, Phys. Lett. {\bf B245}, 401 (1990).

\bibitem{font2}
 A. Font, L. Iba\~nez, F. Quevedo and A. Sierra,
Nucl. Phys. {\bf B331}, 421 (1990).

\bibitem{natural} K. Freese, J. Frieman and A. V.  Olinto
Phys. Rev. Lett. {\bf 65}, 3233 (1990).

\bibitem{d8}
J. N. Fry and Y. Wang, Phys. Rev.  {\bf D46}, 3318 (1992).

%GGGGGGGGGGGG

\bibitem{gmo}  M. K. Gaillard, H. Murayama and K. A.
Olive, Phys. Lett. {\bf B355}, 71 (1995) . 

\bibitem{p97berk} M. K. Gaillard, D. H. Lyth and H. Murayama, hep-th/9806157. 

\bibitem{juan97} J. Garcia-Bellido, hep-th/9707059.

\bibitem{non3}
J. Garcia-Bellido and D. Wands, Phys. Rev.  {\bf D52}, 6739 (1995).

\bibitem{davidjuan}
J. Garcia-Bellido and D. Wands, Phys. Rev.  {\bf D53}, 5437 (1996).

\bibitem{gbgz} J. Garcia-Bellido, A. R. Liddle, D. H. Lyth, and D. Wands, 
Phys. Rev. D {\bf 52}, 6750 (1995).

\bibitem{glw}
J. Garcia-Bellido, A. Linde and D. Wands, Phys. Rev. {\bf D54}, 6040 (1996).

\bibitem{gv} J. Garriga and A. Vilenkin, 
Phys. Rev. {\bf D57}, 2230 (1998).

\bibitem{grr}  T. Gherghetta, A. Riotto and L. Roszkowski, hep-ph/9804365. 

\bibitem{nucleobound} T. Gherghetta, G.F. Giudice and A. Riotto, to appear.

\bibitem{gm} G.F. Giudice and A. Masiero, Phys. Lett. {\bf 206},  480 (1988).

\bibitem{gr}  G.F. Giudice and R. Rattazzi, hep-ph/9801271 .

\bibitem{n1}C. Giunti, C.W. Kim and U.W. Lee, Mod. Phys.
Lett. {\bf 16}, 1745 (1991).

\bibitem{open}
J. R. Gott, Nature {\bf 295} , 304 (1982) .

\bibitem{d4}
S. Gottlober, V. Muller
and A. A. Starobinsky, Phys. Rev D {\bf 43}, 2510 (1991).

\bibitem{d11}
S. Gottlober, J. P. Mucket and A. A. Starobinskii, Astrophys. J. 
{\bf 434}, 417 (1994).

\bibitem{green} A. M. Green and A. R. Liddle, 
Phys. Rev. {\bf D54}, 2557 (1996).

\bibitem{toniandrew}  A. M. Green, A. R. Liddle and A. Riotto,
Phys. Rev. {\bf D56}, 7559 (1997).

\bibitem{bookstring} See, for instance, the two volumes of the book 
{\it Superstring Theory}, 
M.B. Green, J.H. Schwarz and  E. Witten,   
Cambridge Univ. Press,  1987 . 

\bibitem{u(1)A} M.B. Green and J. Schwarz, Phys. Lett.
 {\bf B149}, 117 (1984) . 

\bibitem{grisaru79} M. Grisaru, W. Siegel and M. Rocek, 
Nucl. Phys. {\bf B159}, 429 
(1979).

\bibitem{gz} L. P. Grishchuk and Ya. B. Zel'dovich, 
Sov. Astron. {\bf 22}, 125 (1978) . 

\bibitem{guth} Guth, A. H., Phys. Rev.  {\bf D23}, 347 (1981).

\bibitem{guthpi}
A. H. Guth and S.-Y. Pi, {Phys. Rev. Lett.} {\bf 49} 1110 (1982)
.

%HHHHHHHHHHHHHH

\bibitem{haberkane} H.E. Haber and 
G.L. Kane, Phys. Rept. {\bf 117}, 75 (1985). 

\bibitem{halyo}  E. Halyo, Phys. Lett. {\bf B387}, 43 (1996) .

\bibitem{hawkingellis} S. W. Hawking and G. F. R. Ellis,
{\em The Large-Scale Structure of Space-Time},
Cambridge University Press (1973) .

\bibitem{hawking}
S. W. Hawking, {Phys. Lett.} {\bf B115} 295 (1982) . 

\bibitem{ht98} S. W. Hawking and N. Turok, 
hep-th/9802030, gr-qc/9802062 and hep-th/9803156.

\bibitem{hodges}
H. M. Hodges, G. R. Blumenthal, L. A. Kofman and J. R. Primack,
Nuc. Phys. {\bf B335}, 197 (1990).

\bibitem{blumhodges} 
H.M.  Hodges and G. R. Blumenthal,
Phys. Rev.  {\bf D42}, 3329 (1990).

\bibitem{d3}
H. M. Hodges, Phys. Rev. Lett. {\bf 64}, 1080 (1990).

\bibitem{hodpri} H. M. Hodges and J. R. Primack, Phys. Rev. {\bf D43}, 
3155 (1991).

\bibitem{d6}
R. Holman et al., Phys. Rev.  {\bf D43}, 3833 (1991).

\bibitem{horwit} 
P. Horava and E. Witten, Nucl. Phys. {\bf B475}, 94 (1996);
{\em ibid} {B460}, 506 (1996) .

%IIIIIIIIIIIIIII

\bibitem{iban} L.E. Iba\~nez, H.-P. Nilles and F. Quevedo, Phys. Lett. {
\bf
B187}, 25 (1987).

\bibitem{iban2} L.E. Iba\~nez and D. Lust, \NPB{382}{92}{305}.

\bibitem{reviewdsb2}K. Intriligator and N. Seiberg, hep-th/9509066.

\bibitem{thomasintri} K. Intriligator and S. 
Thomas, Nucl. Phys. {\bf B473}, 121 (1996).

\bibitem{florida} N. Irges and S. Lavignac,  hep-ph/9712239.  

\bibitem{yanagida} K. Izawa and T. Yanagida, 
Prog. Theor. Phys. {\bf 95}, 829 (1996). 

\bibitem{izawa} K. I. Izawa and T. Yanagida, 
Phys. Lett. {\bf B393}, 331 (1997).

\bibitem{izawa2} K. I. Izawa, M. Kawasaki and T. Yanagida, Phys
Lett. {\bf B411}, 249 (1997).

%JJJJJJJJJJJJJJJJJ

\bibitem{j2} R. Jeannerot, Phys. Rev. {\bf D56}, 6205 (1997). 

\bibitem{j1} R. Jeannerot, Phys. Rev. {\bf D53}, 5426 (1996). 

%KKKKKKKKKKKKKKKKKKKK

\bibitem{ella} D.E Kaplan, F, Lepeintre, A. Masiero,  
A.E. Nelson, A. Riotto, hep-ph/9806430.

\bibitem{kl} V. Kaplunovsky and J. Louis, Phys. Lett. {\bf B306}, 
269 (1993).

\bibitem{kawasaki} M. Kawasaki et al. hep-ph/9710259.

\bibitem{r2}S.Yu. Khlebnikov and I.I. Tkachev,
Phys. Rev. Lett. {\bf 77}, 219 (1996).
\bibitem{r5} 
S. Khlebnikov and I. Tkachev, Phys. Lett. {\bf B390}, 80
(1997).
\bibitem{r6} S. Khlebnikov and I. Tkachev, Phys. Rev. Lett. {\bf 79}, 1607
(1997).
\bibitem{r7} S. Khlebnikov and I. Tkachev, Phys. Rev. {\bf D56}, 653
(1997).

\bibitem{tom} T. W. B. Kibble, J. Phys. {\bf A9}, 1387 (1976) . 

\bibitem{kr} S.F. King and  A. Riotto, hep-ph/9806281.

\bibitem{natural4} W. H. Kinney and K. T. Mahanthappa,
Phys. Rev.  {\bf D52}, 5529 (1995).

\bibitem{kinney} W. H. Kinney and K. T. Mahanthappa,
 Phys. Rev.  {\bf D53}, 5455 (1996).

\bibitem{riottokinney}   W. H.  Kinney and A. Riotto, hep-ph/9704388. 

\bibitem{riottokinney2}  W.H. Kinney and A. Riotto, 
hep-ph/9802443, to appear in Phys. Lett. {\bf B}.

\bibitem{natural3} L. Knox and A. Olinto, Phys. Rev. {\bf D48}, 946
 (1993).

\bibitem{kobayashi} T. Kobayashi and H. Nakano,
Nucl. Phys. {\bf B496}, 103 (1997).

\bibitem{kodamasasaki}
H. Kodama and M. Sasaki Prog Theor Phys Supp {\bf 78}, 1
(1984) . 

\bibitem{Kofman} L. A. Kofman and A. D. Linde, Nucl. Phys. {\bf B282}, 
555 (1987).

\bibitem{kofpog} L. A. Kofman and D. Yu. Pogosyan, Phys. Lett. 
{\bf B214}, 508 (1988).

\bibitem{d1} 
L. A. Kofman, A. D. Linde and A. A. Starobinsky, Phys. Lett.
{\bf B157}, 361 (1985).

\bibitem{c2}
L. Kofman and D. Pogosyan, Phys. Lett. 
{\bf B214}, 508 (1988).

\bibitem{kbhp} L. Kofman, G. R. Blumenthal, H. Hodges and J. R. Primack,
        Proceedings of the Workshop on LS  Structure, Rio (1989),
        eds D W Latham and L N da Costa.

\bibitem{kls}
L. Kofman, A. D. Linde and A. A. Starobinsky,
        Phys. Rev. Lett. {\bf 73}, 3195 (1994).
\bibitem{r1}
L. Kofman, A. D. Linde and A. A. Starobinsky,
Phys. Rev. Lett. {\bf 76}, 101 (1996).

\bibitem{r8} L. Kofman, {\it The origin of matter in the Universe: 
reheating after inflation}, astro-ph/9605155,  UH-IFA-96-28 
preprint, 16pp.,  to appear in
Relativistic Astrophysics: A Conference in Honor of Igor 
Novikov's 60th Birthday,
eds. B. Jones and D. Markovic for a more 
recent review and a collection of refs.

\bibitem{r9} L. Kofman, A. D. Linde and A. A. Starobinsky, 
Phys. Rev. {\bf D56}, 3258 (1997). 

\bibitem{extended1} E. W. Kolb, Phys. Scr.
{\bf T36}, 199 (1991).

\bibitem{kt} E. W. Kolb and M. S. Turner, {\em The Early Universe},
Addison-Wesley (1990) .

\bibitem{r4}E.W. Kolb, A.D. Linde and A. Riotto, Phys. Rev. Lett. 
{\bf 77}, 4290 (1996).

\bibitem{km}  C. Kolda and J. March-Russell, 
hep-ph/9802358 .

\bibitem{kv} E. W. Kolb and S. L. Vadas, 
Phys. Rev. {\bf D50}, 2479 (1994).

\bibitem{running} A. Kosowsky and M. S. Turner,
Phys. Rev. {\bf D52}, 1739 (1995).

\bibitem{kumekawa} K. Kumekawa, T. Moroi, and T. Yanagida,
Prog. Theor. Phys. {\bf 92}, 437 (1994).

%LLLLLLLLLLLLLLLLL

\bibitem{extended} D. La and P. J. Steinhardt, Phys. Rev. 
{\bf D62}, 376 (1989) . 

\bibitem{noscale} A.B. Lahanas and D.V. Nanopoulos, 
Phys. Rept. {\bf 145}, 1 (1987). 

\bibitem{n3} P. Langacker and M.-X. Luo,
Phys. Rev. {\bf D44},  817 (1991).

\bibitem{langbein}  R. F. Langbein, K. Langfeld, H. Reinhardt
and L. v Smekal, Mod. Phys. Lett. {\bf A11}, 631 (1996).

\bibitem{shafi1} G. Lazarides and Q. Shafi, 
Phys. Lett. {\bf B308}, 17 (1993);

\bibitem{thermal2} G. Lazarides and 
Q. Shafi, Nuc. Phys. {\bf B392}, 61 (1993).

\bibitem{lazpan}
G. Lazarides and C. Panagiotakopoulos, 
Phys. Rev.  {\bf D52}, 559 (1995); 

\bibitem{l1}G. Lazarides and Q. Shafi, Phys. Lett. {\bf B372}, 20 (1996).
\bibitem{l2}G. Lazarides and Q. Shafi, Phys. Lett. {\bf B372},
20 (1996).

\bibitem{qaisarlatest} G. Lazarides, R. K. Schaefer and Q. Shafi,
Phys. Rev. {\bf D56}, 1324. (1997).

\bibitem{li} T. Li, hep-th/9801123.

\bibitem{LL1} A. R. Liddle and D. H. Lyth, Phys. Lett. {\bf B291},
391 (1992) . 

\bibitem{LL2} A. R. Liddle and D. H. Lyth, Phys. Rep. {\bf 231}, 
1 (1993) .

\bibitem{LL3} A. R. Liddle and D. H. Lyth, 
{\em Cosmological Inflation and Large-Scale Structure},
to be published by Cambridge University Press.

\bibitem{lpb} A. R. Liddle, P. Parsons and J. D. Barrow,
 Phys. Rev. D        {\bf 50}, 7222 (1994).

\bibitem{recon} 
J. E. Lidsey et al.,  Rev. Mod. Phys. {\bf 69}, 373 (1997).

\bibitem{new1}  A. D. Linde, Phys. Lett. {\bf 108B} (1982) . 

\bibitem{chaotic} A. D. Linde, Phys. Lett. {\bf B129}, 177 (1983) . 

\bibitem{nontherm} A. D. Linde, Phys. Lett. {\bf B132}, 317
(1983).

\bibitem{eternal}  A. D. Linde, Phys.
Lett. {\bf B175}, 395  (1986).

\bibitem{abook} A. D. Linde, {\em Particle Physics and Inflationary Cosmology},
        Harwood Academic, Switzerland (1990).

\bibitem{l90} A. Linde, Phys. Lett. {\bf B249}, 18 (1990).

\bibitem{LIN2SC} A. D. Linde, Phys. Lett. {\bf B259}, 
38 (1991) . 

\bibitem{LIN2SC2} 
A. D. Linde, Phys. Rev. {\bf D49}, 748 (1994).

\bibitem{o2}A. Linde, Phys.Lett. {\bf B351}, 99 (1995).

\bibitem{l98} A. D. Linde, gr-qc/9802038.

\bibitem{eternal1}
A. Linde, D. Linde, and  A. Mezhlumian, Phys. Rev.  {\bf D49},  1783  (1994).

\bibitem{o3}
A. Linde and A. Mezhlumian, 
Phys. Rev. {\bf D52}, 6789 (1995).

\bibitem{linderiotto}  
A. D. Linde and A. Riotto,  
Phys. Rev. {\bf D56}, 1841 (1997) . 

\bibitem{low} A.  Lukas, B. Ovrut and D. Waldram, 
hep-th/9710208.

\bibitem{lyth85} D. H. Lyth, Phys. Lett. {\bf 147B}, 403 (1984),
erratum Phys. Lett. {\bf 150B},  465 (1985).

\bibitem{lyth85b} D. H. Lyth, {Phys. Rev. } {\bf D31} 1792 (1985) .

\bibitem{lyth90} D. H. Lyth, Phys. Lett. {\bf B246}, 359 (1990).

\bibitem{myaxion} D. H. Lyth, Phys Rev  {\bf D45}, 3394 (1992).

\bibitem{mygwave} D. H. Lyth, Phys. Rev. 
Lett. {\bf 78}, 1861 (1997).                   

\bibitem{mydterm} D.H. Lyth, hep-ph/9710347 , 
to appear in Phys. Lett. {\bf B}. 

\bibitem{lythmuk} D. H. Lyth and M. Mukherjee, Phys Rev  {\bf D38}, 485
(1988).

\bibitem{lr} D. H. Lyth and A. Riotto,
Phys. Lett. {\bf B412}, 28 (1997).

\bibitem{lyst}
D. H. Lyth and E. D. Stewart, Astrophys J {\bf 361}, 
343 (1990).

\bibitem{ewanopen}  D. H. Lyth and E. D. Stewart, Phys. Lett.
{\bf B252}, 336 (1990) .

\bibitem{t3}D. H. Lyth and E. D. Stewart, Phys. Rev. Lett.
{\bf 75}, 201 (1995).

\bibitem{thermal} D. H. Lyth and E. D. Stewart,
Phys. Rev.  {\bf D53}, 1784 (1996) . 

\bibitem{ournew} D. H. Lyth and E. D. Stewart, 
Phys. Rev. {\bf D54}, 7186 (1996).

%MMMMMMMMMMMMMMM

\bibitem{macorra} A. De la Macorra and S. Lola,
Phys. Lett. {\bf B373}, 299 (1996).

\bibitem{map} home page at {\tt http://map.gsfc.nasa.gov/}

\bibitem{jmrfa} J. March-Russell, hep-ph/9806426.

\bibitem{riotto1} R. N. Mohapatra and A. Riotto, Phys. Rev. 
{\bf D55}, 1138 (1997).

\bibitem{riotto2} R.N. Mohapatra and A. Riotto,  Phys. Rev. 
{\bf D55},  4262 (1997).

\bibitem{n4} R.N. Mohapatra, hep-ph/9801235 and refs. therein. 

\bibitem{silvia} S. Mollerach, S. Matarrese and F. Lucchin, 
Phys. Rev. {\bf D50}, 4835 (1994).

\bibitem{mukhanov} V. F. Mukhanov, JETP Lett. {\bf 41}, 493 (1985) .

\bibitem{mc81} V. F. Mukhanov  and G. V. Chibisov, JETP Letters {
\bf 33},         532 (1981); Sov. Phys. JETP {\bf 56
},        258 (1981) .

\bibitem{d2}
L. V. Mukhanov, L. A. Kofman and D. Yu. Pogosyan, Phys. Lett.
{\bf 193}, 427 (1987).

\bibitem{d51} V. F. Mukhanov and M. I. Zelnikov, 
Phys. Lett. {\bf B263}, 169 (1991).

\bibitem{mfb} V. F. Mukhanov, H. A. Feldman and R. H. Brandenberger, 
        Phys. Rep. {\bf 215}, 203 (1992).

\bibitem{hitinf}  H. Murayama et al., Phys. Rev. D {\bf 50}, 2356 (1994).

%NNNNNNNNNNNNN

\bibitem{nagasawa} M. Nagasawa and J. Yokoyama, Nucl. Phys. {\bf B370}, 
472        (1992).

\bibitem{nak} T. T. Nakamura and E. D. Stewart, 
Phys. Lett. {\bf B381}, 413 (1996).

\bibitem{g6}C.R.~Nappi and B.A.~Ovrut, \PLB{113}{82}{175}.

\bibitem{nelsonrev}
A. E. Nelson, Talk given at 5th International Conference on
           Supersymmetries in Physics (SUSY 97), Philadelphia, PA, 27-31 
May 1997; hep-ph/9707442.

\bibitem{nelson} A.E. Nelson and D. Wright, Phys. Rev. {\bf D56}, 1598 (1997). 

\bibitem{nilles}
        H. P. Nilles, Phys. Rep. {\bf 110}, 1 (1984).

\bibitem{noy} H. P.  Nilles, M. Olechowski, and M. Yamaguchi, 
\PLB{415}{1997}{24}.

\bibitem{nsw} H. P. Nilles, M. Srednicki and D. Wyler,
Phys. Lett. {\bf B120}, 346 (1983).

\bibitem{nus} H. P. Nilles and N. Polonsky, 
Phys. Lett. {\bf B412}, 69 (1997).

%OOOOOOOOOOOOOO

\bibitem{olive} K. Olive, Phys. Rep. {\bf 190}, 307 (1990).

\bibitem{burt1} B. A. Ovrut and P. J. Steinhardt, \PLBold{133}{83}{161}

\bibitem{burt2} B. A. Ovrut and P. J. Steinhardt, \PRL{53}{84}{732}.

\bibitem{ovrut} B. A, Ovrut and S. Thomas, Phys. Lett. {\bf B267},
227 (1991); ibid {\bf B277}, 53 (1992).

%PPPPPPPPPPPPP

\bibitem{panag} C. Panagiotakopoulos, Phys. Lett. {\bf B402},
257 (1997).

\bibitem{ax} Ue-Li Pen,  U. Seljak and 
N.  Turok,   Phys. Rev. Lett. {\bf 79}, 1611 (1997). 

\bibitem{planck} home page at {\tt http://astro.estec.esa.nl/Planck}

\bibitem{d10}
P. Peter, D. Polarski and A. A. Starobinsky, Phys. Rev.
D {\bf 50}, 4827 (1994).

\bibitem{d7} D. Polarski and A. A. Starobinsky, 
Nucl. Phys. {\bf B385}, 623 (1992).

\bibitem{d9}  D. Polarski, Phys. Rev.  {\bf D49}, 6319 (1994).

\bibitem{isocurv}
D. Polarski and A. A. Starobinsky, Phys. Rev.  {\bf D50}, 6123 (1994).

\bibitem{d13}
D. Polarski and A. A. Starobinsky, Phys. Lett. {\bf B356}, 196 (1995).

\bibitem{pol} J. Polonyi, preprint no. KFKI-77-93, 1997.

\bibitem{pomarol} G. Dvali and A. Pomarol, Phys. Rev. Lett. 
{\bf 77}, 3728 (1996).

%QQQQQQQQQQQQQQQQQ

%RRRRRRRRRRRRRRRRRRR

\bibitem{raby} S. Raby, Phys. Rev. {\bf D56}, 2852(1997). 

\bibitem{lisa} L. Randall, M. Soljacic and A. H. Guth,
Nucl. Phys. {\bf B472}, 408 (1996) .

\bibitem{rrr} A. Riotto, hep-ph/9707330,  Nucl. Phys. {\bf B515}, 413 (1998).

\bibitem{rtalk} A. Riotto, hep-ph/9710329.

\bibitem{dave} D. Roberts, A. R. Liddle and D. H. Lyth,
Phys Rev D {\bf 51}, 4122 (1995).

\bibitem{graham} G. G. Ross and S. Sarkar,  Nuc. Phys. {\bf B461}, 597 (1996).

\bibitem{rubakov}
V. A. Rubakov, M. V. Sazhin and A. V. Veryaskin,
Phys. Lett {\bf B115}, 189 (1982).

%SSSSSSSSSSSSSSSSSSS

\bibitem{c3} D. S. Salopek, J. R. Bond and J. M. Bardeen, Phys. Rev.
        D{\bf 40}, 1753 (1989).

\bibitem{salopek} D. S. Salopek, Phys. Rev. Lett. {\bf 69}, 3602 
(1992).

\bibitem{salopek95} D.S. Salopek, Phys. Rev.  {\bf D52}, 5563 (1995).

\bibitem{subir} S. Sarkar, Rep. on Progress in Phys. {\bf 59}, 
1493 (1996).

\bibitem{sasaki} M. Sasaki, Prog. Theor. Phys. {\bf 76}, 1036 (1986).

\bibitem{ewanmisao} 
M. Sasaki and E. D. Stewart, Prog. Theor. Phys. 
{\bf 95}, 71 (1996).
A more accurate calculation, analogous to the one in Ref.~\cite{stly}
for a single-component inflaton, is provided by
Ref.~\cite{nak}.

\bibitem{singlet} Q. Shafi and A. Vilenkin, Phys. Rev. Lett. {\bf 52},
691 (1984).

\bibitem{reviewdsb1}M. Shifman; 
{\it  Nonperturbative dynamics in supersymmetric gauge theories}, 
hep-th/9704114 and refs. therein.

\bibitem{cobe1} G. F. Smoot et al., Astroph J. {\bf 396}, L1 (1992).

\bibitem{soni} S. Soni and A. Weldon, Phys. Lett. {\bf B126}, 
215 (1983).

\bibitem{starob82}
A. A. Starobinsky, {Phys. Lett.} {\bf B117} 175 (1982) .

\bibitem{starob82gw}
A. A. Starobinsky, in; {\sl Quantum Gravity (Proc. 2nd Seminar Quantum
Theory of Gravitation)} [in Russian], Inst. Nucl. Res. USSR Acad. Sci.,
Moscow (1982), p. 58.

\bibitem{starob} A. A. Starobinsky, Sov. Astron. Lett. {\bf 11}, 133
(1985). 

\bibitem{starob85} A. A. Starobinsky, JETP Lett. 
{\bf 42}, 152 (1985).

\bibitem{stochastic} A. A. Starobinsky, in: 
{\em  Lecture Notes in  Physics}, Vol. 242, eds H. J. de Vega
and N. Sanchez, Springer, Berlin, (1986) .

\bibitem{rsq}
A. A. Starobinsky, Phys. Lett. {\bf B91}, 99 (1990).

\bibitem{non2}A. A. Starobinsky and J. Yokoyama, gr-qc/9502002 (1995).

\bibitem{ewansgrav} E. D. Stewart, Phys. Rev. {\bf D51}, 6847 (1995).

\bibitem{mutated} E. D. Stewart, Phys. Lett. {\bf B345},
414 (1995).

\bibitem{stly} E. D. Stewart and D. H. Lyth, {Phys. Lett.} {\bf B302
}, 171 (1993).

\bibitem{t2}E. D. Stewart, M. Kawasaki, and T. Yanagida, 
Phys. Rev. {\bf D54}, 6032 (1996).

\bibitem{ewanloop1} E. D. Stewart, Phys. Lett. {\bf B391}, 34 (1997) .
\bibitem{ewanloop2}  E. D. Stewart,   Phys. Rev. {\bf D56}, 2019 (1997) .

\bibitem{iso}
R. Stompor et al.,  astro-ph/9511087.

\bibitem{wilson}
L. Susskind, Phys. Rev.
{\bf D20}, 2619 (1979).  

%TTTTTTTTTTTTTTTTTTTTT

\bibitem{reviewdsb3} S. Thomas, hep-ph/9801007. 

\bibitem{tilted} M. S. Turner, Phys. Rev. D {\bf 44}, 3737 (1991).

%UUUUUUUUUUUUUUUUUUUUUUU

%VVVVVVVVVVVVVVVVVVVVVV

\bibitem{v98} A. Vilenkin, hep-th/9803084.

\bibitem{vishniac} E. T. Vishniac, K. A. Olive and D. Seckel, Nucl.
        Phys. {\bf B289}, 717 (1987).

%WWWWWWWWWWWWWWWWWWWWWW

\bibitem{wang} Y. Wang, Phys. Rev.  {\bf D50}, 6135 (1994).

\bibitem{weinberg} S. Weinberg, Phys. Rev.  {\bf D9}, 3357 (1974).

\bibitem{wessbagger} J. Wess and
J. Bagger, {\it Supersymmetry and Supergravity}, (Princeton University
Press, Princeton, 1983) .

\bibitem{constraint2}
M. White et al., Mon. Not. Roy. Astron. Soc. {\bf 283}, 107 (1996).

\bibitem{w} E. Witten, Phys. lett. {\bf B155}, 151 (1985). 

\bibitem{witten85} E. Witten, Nucl. Phys. {\bf B258}, 75 (1985).

\bibitem{mtheory} E. Witten, Nucl. Phys. {\bf B471}, 135 (1996). 

%XXXXXXXXXXXXXXXXXXXXX

%YYYYYYYYYYYYYYYYYYY

\bibitem{yokoyama} J. Yokoyama, Phys. Lett. {\bf B212}, 273 (1988);
        Phys. Rev. Lett. {\bf 63}, 712 (1989).

%ZZZZZZZZZZZZZZZZZZZZZZ
\bibitem{d5}
M. I. Zelnikov, and V. F. Mukhanov, JETP Lett. {\bf 54}, 197 (1991).

\end{thebibliography}
\end{document}